\documentclass[letterpaper,twocolumn,10pt]{article}
\usepackage{usenix-2020-09}

\usepackage{tikz}
\usepackage{amsmath}

\usepackage{epsfig,endnotes}

\usepackage{xspace}
\usepackage{xcolor}
\usepackage{algorithm}
\usepackage{algpseudocode}
\usepackage{amsmath}
\usepackage{subcaption} 
\usepackage{graphicx}
\usepackage{multirow} 
\usepackage{array}
\usepackage{adjustbox}
\usepackage{booktabs}
\usepackage{tabularx}
\usepackage{makecell}
\usepackage{threeparttable}
\usepackage{balance}

\usepackage[scaled]{beramono}    %

\algnewcommand\Input{\item[\textbf{Input:}]} %
\algnewcommand\Output{\item[\textbf{Output:}]} %

\algnewcommand\LeftComment[1]{\Statex \(\triangleright\) #1}

\newcommand{\ours}{\emph{Sluice}\xspace}

\newcommand{\suppref}[1]{Appendix~\ref{#1}}

\newcommand{\stitle}[1]{\vspace{1ex}\noindent\textbf{#1}}

\usepackage{enumitem}
\setlist{noitemsep,topsep=0pt,parsep=0pt,partopsep=0pt, leftmargin=*}

\usepackage{circledsteps}

\usepackage[dvipsnames]{xcolor}
\makeatletter
\@ifpackageloaded{newtxmath}{}{\usepackage{amssymb}}
\makeatother
\usepackage{pifont}
\newcommand{\cpmark}{ \textcolor{YellowOrange}{\ding{51}}  }
\newcommand{\cmark}{ \textcolor{green}{\ding{51}}  }
\newcommand{\xmark}{ \textcolor{red}{\ding{55}} }

\newcommand{\ccc}[1]{#1}

\newcounter{lx}
\stepcounter{lx}

\newcounter{VasiaNOC}
\stepcounter{VasiaNOC}

\newcounter{LeiNOC}
\stepcounter{LeiNOC}

\newcounter{Prateek}
\stepcounter{Prateek}

\newcounter{yuanliready}
\stepcounter{yuanliready}

\begin{document}

\date{}

\title{\Large \bf Live Adaptation of Dataflow Stream Processing for Hybrid Edge-Cloud Analytics}
\author{
{\rm Yuanli Wang$^{1}$, Prateek Jain$^{1}$, Lei Huang$^{1}$, Le Xu$^{2}$, Vasiliki Kalavri$^{1}$}\\
$^{1}$Boston University \quad $^{2}$University of Edinburgh
}

\maketitle

\begin{abstract}

Dataflow streaming systems, like Apache Flink and Spark Streaming, are integral to modern business analytics, providing real‐time insights across industries. However, for emerging hybrid edge-cloud analytics pipelines, neither processing everything in the cloud nor static edge offloading can reliably sustain the target rate under dynamic WAN conditions and limited edge resources. In this paper, we explore how these datacenter‐centric platforms can be extended to support such hybrid analytics pipelines. Our goal is to maintain full transparency for user applications, while retaining built‐in fault tolerance, out‐of‐order processing, and existing ecosystem integrations. Achieving this is challenging because existing dataflow engines assume homogeneous resources, rely on static task placement, and do not support adaptation under dynamic network and edge conditions.

We introduce \ours, an adaptive data stream processing system for hybrid clouds, compatible with dataflow streaming engines. \ccc{\ours's key contributions are: (i) an automatic \emph{query rewriter} that decomposes applications into independent, fault-tolerant segments, and (ii) a live \emph{routing mechanism} that shifts computations between edge and cloud without pausing execution while preserving event-time semantics.} We implement \ours on Apache Flink and evaluate it on a physical testbed with Raspberry Pi devices, as well as on a large simulated testbed of 100 VMs. Across five real-world applications, our results show that \ours seamlessly shifts computations between edge and cloud, while  consistently achieving higher throughput than baseline systems.

\end{abstract}

\section{Introduction}

The wide deployment of connected smart devices has enabled a new class of continuous analytics scenarios, such as traffic management systems, air quality monitoring, optimizing energy grid consumption, healthcare interventions, and video analytics~\cite{9397280, falcon24,shepherd22, NebulaStream25,wang2018bandwidth,li2020reducto, ghosh2023react,padmanabhan2023gemel}. These emerging applications ingest data from multiple edge locations (e.g., wearable devices, cameras, drones, and mobile sensors) and aggregate the results in the cloud to power web-based visualization dashboards or drive decision-making pipelines.

However, for these types of applications, neither streaming all raw data to the cloud nor statically offloading operators to the edge can reliably sustain the target rate under dynamic WAN conditions and limited edge resources.
Recent research has explored the use of compute-enabled edge resources for analytics. Systems like DART~\cite{liu2021dart, AgileDART}, Frontier~\cite{frontiervldb18}, and NebulaStream~\cite{zeuch2020nebulastream,chatziliadis2024nemo,NebulaStream25} proposed new architectures for pure-edge processing, offering robustness under churn and scalability to hundreds of edge nodes and concurrent queries. Others, like Falcon~\cite{falcon24} and Shepherd~\cite{shepherd22}, focus on hybrid, hierarchical wide-area deployments, with optimizations for regional data aggregation and strategies to handle limited resources and dynamic conditions. 

\ccc{As summarized in Table~\ref{tab:related}, although previous work proposes innovative mechanisms for WAN- and edge-aware live adaptation, they often lack support for fault tolerance, event-time windows, monitoring and data connector integrations provided by mature engines such as Apache Flink~\cite{carbone2015apache, hueske2019stream, carbone2017state}, Spark Structured Streaming~\cite{armbrust2018structured}, and Kafka Streams~\cite{wang2021consistency}. Reimplementing these features and migrating existing applications require substantial engineering.}
\ccc{In this paper, we investigate how to extend these datacenter-centric engines to hybrid edge--cloud deployments while preserving existing applications, familiar APIs, built-in features of fault tolerance, out-of-order processing, and  ecosystem integration.}

\begin{table*}[t]
\centering
\caption{Comparison with representative prior systems.}
\label{tab:related}
\scriptsize
\setlength{\tabcolsep}{3pt}
\renewcommand{\arraystretch}{1.1}
\begin{threeparttable}

\begin{tabularx}{\textwidth}{
>{\raggedright\arraybackslash}p{2.5cm}
>{\raggedright\arraybackslash}p{2.5cm}
>{\raggedright\arraybackslash}p{2.0cm}
>{\raggedright\arraybackslash}p{2.0cm}
>{\raggedright\arraybackslash}p{1.8cm}
>{\raggedright\arraybackslash}p{1.8cm}
>{\raggedright\arraybackslash}p{2cm}
>{\raggedright\arraybackslash}X
}
\toprule
\textbf{System} &
\textbf{Target setting} &
\textbf{Compatible with existing SPE} &  %
\textbf{Live adaptation \tnote{b}} &  %
\textbf{WAN-aware policy  } &
\textbf{No data loss \tnote{c}} &
\textbf{Expressivity} &  %
\textbf{Open source}  \\
\midrule

DART \cite{liu2021dart} &
Edge  &
\xmark  &
N.S.  &
\xmark  &
N.S. &
\cpmark (Partial)  &
\cmark \\

\addlinespace[1pt]

Frontier \cite{frontiervldb18} &
Edge  &
\xmark &
\cmark &
\cmark &
\cmark &
\cpmark (Partial)  &
\cmark \\

NebulaStream~\cite{NebulaStream25, zeuch2020nebulastream, chatziliadis2024nemo} &
Edge / Fog  &
\xmark &
N.S. &
\xmark  &
N.S.   &
\cpmark (Partial)  &
\cmark \\

\addlinespace[1pt]
\hline
\addlinespace[1pt]

Shepherd \cite{shepherd22} &
Hierarchical wide-area &
\xmark &
\cpmark (Pause) &
\xmark &
\xmark &
\cpmark (Partial)  &
\xmark \\

\addlinespace[1pt]

Falcon \cite{falcon24} &
 Hierarchical wide-area &
\xmark &
\cpmark (Pause) &
\xmark &
\cmark &
\cpmark (Partial)  &
\cmark \\

\addlinespace[1pt]

WASP \cite{jonathan2020wasp} &
Geo-distributed &
\cmark: Apache Flink &
\cpmark (Pause) &
\cmark &
\cmark &
\cmark &
\xmark \\

\addlinespace[1pt]

CAPSys \cite{capsys} &
Data center &
\cmark: Apache Flink &
\xmark &
\cmark &
\cmark &
\cmark &
\cmark \\

\hline
\addlinespace[1pt]

\textbf{\ours} &
Hybrid edge-cloud &
\cmark: Apache Flink &
\cmark &
\cmark  &
\cmark  &
\cmark  &
\cmark \\

\bottomrule
\end{tabularx}

\begin{tablenotes}[flushleft]
\footnotesize
\item[a] N.S. : Not explicitly stated: the paper does not make a clear claim about this aspect.
\item[b] Whether adaptation is performed without fully or partially pausing the job.
\item[c] Event delivery guarantee.
\end{tablenotes}

\end{threeparttable}
\end{table*}

We identify three major challenges in adapting dataflow streaming systems for hybrid edge–cloud deployments: (i) \textbf{Resource homogeneity assumption}: As descendants of MapReduce~\cite{dean2004mapreduce}, dataflow engines assume uniform resources and do not distinguish between high-bandwidth intra-data\-center links and slow, variable edge-to-cloud connections. Consequently, out-of-the-box deployments may generate task placements that cause excessive WAN traffic, overload edge devices, and create performance bottlenecks. 
(ii) \textbf{Static task scheduling}: Most dataflow systems assign tasks to workers statically, assuming stable operating conditions~\cite{capsys,isard2007dryad}. Hybrid edge–cloud applications, however, require continuous adaptation: WAN links exhibit variable bandwidth, edge devices vary in capacity, and event rates shift over time~\cite{jonathan2020wasp,shepherd22}. Under such dynamics, any static placement eventually becomes obsolete. Recent efforts at dynamic placement~\cite{capsys,falcon24,shepherd22} either rely on stop-and-restart reconfiguration or lack automatic policies to detect when to adapt. 
(iii) \textbf{Disruptive recovery}: Cloud-centric dataflow systems rely on job-wide checkpoints~\cite{carbone2017state,mei2025disaggregated,siachamis2024checkmate,streamingsurvey_vldb24} for fault tolerance, so a failure in a single task triggers recovery of the entire job. This approach is ill-suited to edge environments, where hardware failures are more frequent. A single slow or failed edge task can stall the whole pipeline, causing downtime, input backlog, or source-side data loss.

We address these challenges with \ours, an adaptive data stream processing system for hybrid clouds, compatible with modern dataflow engines. \ccc{\ours dynamically reconfigures stateless preprocessing operators between edge and cloud, while stateful operators remain in the cloud. This design avoids state migration and supports common hybrid analytics pipelines, where lightweight preprocessing (e.g., filters and projections) reduces data volume before stateful computation~\cite{olston2003adaptive, cugola2012processing,babu2004adaptive}.}

\ccc{\ours makes three contributions:}%
\begin{itemize}
    \item An automatic \emph{query re-writer} that transforms dataflow programs written for datacenter settings into independent, fault-tolerant query segments that can be scheduled, deployed, monitored, and reconfigured in isolation. A major benefit of this design is that a failure or slowdown in one segment does not affect the others.
    \item An \emph{adaptive routing mechanism} that enables reconfiguration without downtime. \ccc{The re-writer generates both edge and cloud task replicas for the stateless preprocessing operators and inserts routing controllers.} \ccc{A reference \emph{routing policy} guides these controllers: it continuously monitors metrics to detect performance bottlenecks and dynamically adjusts routing decisions to sustain the required event rate under changing conditions. }
    \item \ccc{\emph{Correctness and no data loss across segments.} \ours guarantees event-time correctness and no data loss even though its query segments run as independent jobs. It propagates watermarks along both edge and cloud paths and coordinates them across segments, so that live rerouting preserves event-time semantics. It also combines per-segment checkpoints with a persistent message broker to provide at-least-once processing under transient edge and cloud failures (\S\ref{sec:guarantees}).}
\end{itemize}\vspace{0.5mm}

\noindent We implement \ours on Apache Flink and evaluate it in two environments: a physical hybrid testbed with Raspberry Pi devices and a large-scale cloud testbed using 100 VMs to emulate edge nodes. Across five real-world applications, our results show that \ours seamlessly shifts computations between edge and cloud, while its policy quickly converges to a configuration that achieves the target throughput under variable bandwidth, rate, and scarce compute conditions.

\section{Background and motivation}

We revisit fundamentals of dataflow stream processing and describe examples of analytics pipelines that ingest data at the edge to drive applications hosted in the cloud.

\subsection{Streaming dataflow concepts}\label{sec:background}
Streaming dataflow systems~\cite{akidau2015dataflow, streamingsurvey_vldb24} represent applications as directed acyclic graphs (DAGs), where vertices are operators performing transformations on streams and edges are data channels. 
Operators can be either \textbf{stateless}, if processing an event is independent of other events, or \textbf{stateful}, if processing an event depends on results of prior events. Filtering, data type transformations, and attribute projection are stateless, while aggregations, windows, and continuous joins are stateful. Every dataflow pipeline has at least one \textbf{source} operator that handles data ingestion and a \textbf{sink} that produces results to downstream applications. We say that the pipeline can sustain the \textbf{target throughput}, if it can process events faster than the source produces them. If an operator cannot sustain its input rate, it will stall its upstream stages and the pipeline will experience \textbf{backpressure}. If backpressure persists,  the source will eventually need to buffer input, leading to backlog and data loss, if buffer size is exceeded.

\stitle{Event time.}
Streaming dataflow systems implement event-time semantics~\cite{akidau2015dataflow,akidau2021watermarks} to support correct processing of out-of-order events generated at multiple distributed sources. For example, windowed aggregations group events into temporal intervals based on the timestamps that events carry. To track the progress of event time, source operators periodically emit \textbf{watermarks}. A watermark with event time $t_w$ is a guarantee that no future events with $t \leq t_w$ will arrive. Watermarks flow through the dataflow graph alongside data events, enabling distributed operators to coordinate their event-time progress.

\stitle{Fault tolerance.}
The de facto approach to fault tolerance in cloud-based streaming systems is via periodic consistent \textbf{checkpoints}~\cite{streamingsurvey_vldb24,carbone2015apache,carbone2020beyond,siachamis2024checkmate,wang2019lineage}.
The checkpoint frequency is determined by the \textbf{checkpoint interval} and can be tuned  to trade off steady-state overhead over recovery duration.
To achieve no data loss, sources must be replayable and include their input \textbf{offset} in their local checkpoint. Recovery reverts all operator states to the last successful checkpoint and replays the input streams from the corresponding offset. 

\subsection{Application scenarios}
\label{sec:usecase}

To understand the requirements of hybrid edge-cloud applications, we now examine five continuous analytics scenarios from prior work~\cite{debs2015Jerzak, tahir2021debs,faisal2023tva, tuballa2016review, liu2021dart,shepherd22,bordin2020dspbench}.

\begin{itemize}
    \item \textbf{Q1: Air quality monitoring} continuously monitors PM2.5 sensor data~\cite{air_quality_dataset} and generates alerts when air pollution levels are deemed to be potentially unhealthy. %
    \item \textbf{Q2: Taxi route monitoring} ingests streams of taxi trip reports~\cite{debs2015_dataset}
    to continuously identify the most frequent routes. The results feed downstream applications, such as real-time traffic maps and pricing control of ridesharing platforms. %
    \item \textbf{Q3: Smart grid anomaly detection} monitors smart meter data~\cite{smart_meter_datasets} to detect household energy spikes. 
    The results are transmitted to downstream energy grid optimization pipelines to trigger alerts when normal usage is exceeded.
    \item \textbf{Q4: Sentiment analysis} processes geo-tagged social media posts~\cite{jonathan2020wasp,shepherd22,bordin2020dspbench} and feeds results into downstream advertising applications. 
    \item \textbf{Q5: Video-assisted traffic monitoring} analyzes continuous video feeds from smart cameras and applies object recognition models to detect traffic violations~\cite{bordin2020dspbench,shepherd22}. 
\end{itemize}

\noindent \ccc{We make two observations about these scenarios.}

\stitle{Stateless-to-stateful processing pattern}. The dataflow graphs implementing the applications above follow a consistent structure, where initial pipeline stages consist of lightweight data transformations (filtering, projection, data cleaning, format conversion, compression) that reduce data volume, while later stages involve resource-intensive stateful operations (joins, windowed aggregations) or complex analytics (ML inference) requiring substantial compute and memory. The observation that stateless transformations often occur early also underpins classic stream processing optimization techniques, like filter pushdown and operator reordering~\cite{olston2003adaptive, cugola2012processing,babu2004adaptive}. Prior work in edge-cloud analytics has argued that offloading such early-stage operations to edge devices can  significantly lower WAN data transfer~\cite{9397280,wang2018bandwidth,9084281,li2020reducto, ghosh2023react}. 

\begin{figure}[t!]
\centering
\includegraphics[width=\linewidth]{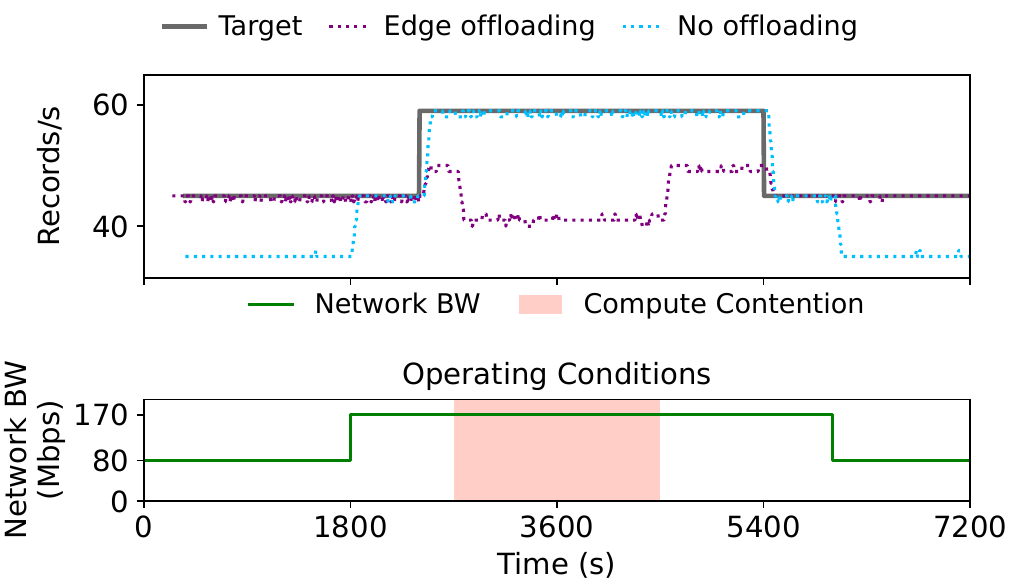}
\vspace{-2mm}
\caption{\ccc{Throughput of the traffic monitoring application (Q5) under two static deployments (top) and the operating conditions applied over time (bottom). \emph{Edge offloading} runs frame filtering and image compression at the edge. \emph{No offloading} runs all operators except ingestion in the cloud.}}
\label{fig:intro_q5}
\end{figure}

\stitle{Workload and environment dynamics.} Our second observation is that static operator offloading cannot cope with the fluctuating conditions of hybrid edge–cloud deployments. As applications ingest data from distributed edge locations and forward results to the cloud, they must adapt to variable WAN bandwidth availability~\cite{jonathan2020wasp,brito2025ai,zhang2021adaptive}. At the same time, event rates may change over time (for example, more taxis during rush hour), and edge devices can suffer interference from co-running processes or intermittent faults. To illustrate these challenges, we measure the throughput of \textbf{Q5} under varying \ccc{edge--cloud bandwidth, input rates, and edge compute contention} on a hybrid testbed with a Raspberry Pi and a cloud server (see \S\ref{sec:exp-setting} for setting). Figure~\ref{fig:intro_q5} compares two static configurations: (i) offloading the \ccc{filtering and image compression stages} to the edge and (ii) executing all operations, other than data ingestion, on the server. The results show that neither deployment sustains the target throughput throughout the experiment. \ccc{Cloud-only processing cannot sustain the target during low-bandwidth periods. Edge offloading reduces network traffic, but cannot sustain the higher input rate when edge compute capacity is limited.} These results underscore the need for dynamic operator placement that can shift computation between edge and cloud to meet performance objectives and prevent data loss.

\section{\ours system design}\label{sec:design_overview}

\begin{figure}[t]
  \centering
      \includegraphics[width=0.999\linewidth]{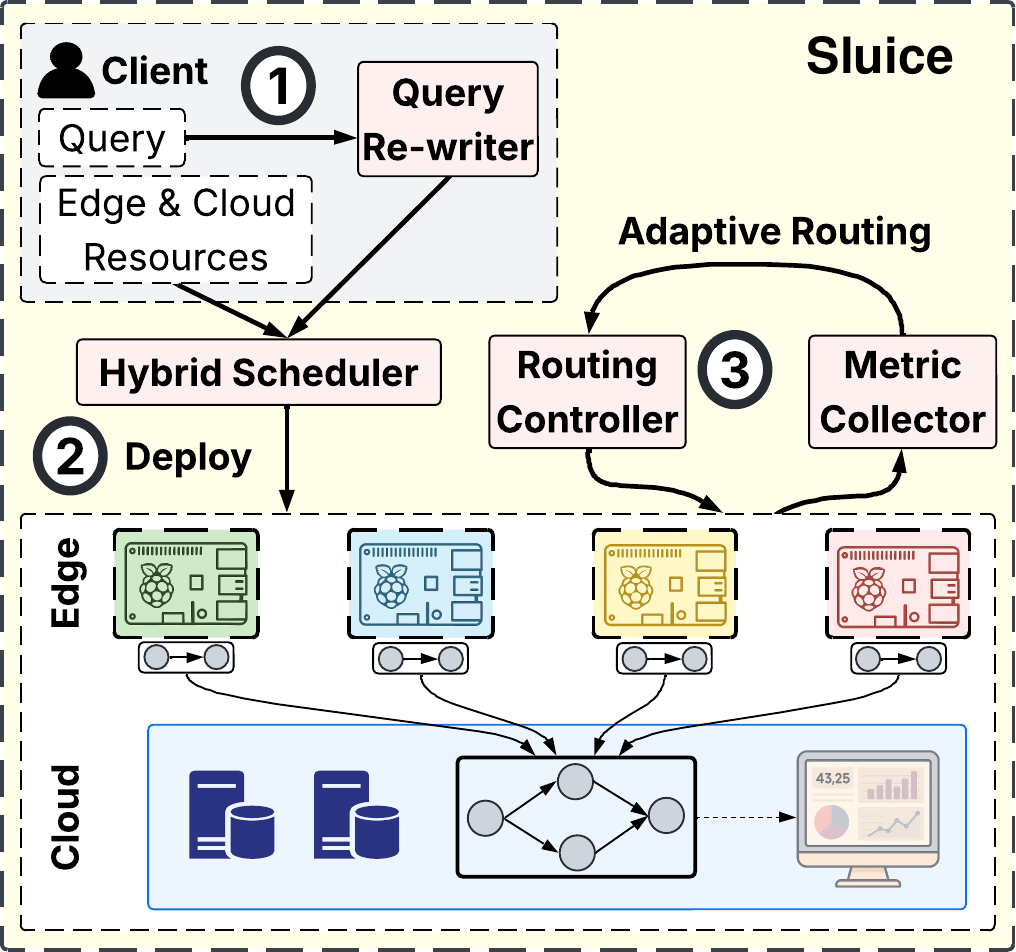}\vspace{-1mm}
  \caption{\ours system overview}
  \label{fig:sec3_sys_overview}
\end{figure}

In this section, we introduce the \ours hybrid edge-cloud stream processing system and describe how it transforms monolithic dataflow applications into independent query segments to enable zero-downtime reconfiguration under dynamic network and workload conditions. We discuss \ours's correctness and fault tolerance guarantees in \S\ref{sec:guarantees}.

\stitle{System overview.} As shown in Figure~\ref{fig:sec3_sys_overview}, user queries are parsed and transformed into segments by a \emph{Query Re-writer} module (\Circled{1}). Next, \ours's \emph{Hybrid Scheduler} deploys those segments as separate jobs on the available edge and cloud resources (\Circled{2}). The \emph{Metric Collector} continuously monitors performance metrics of the deployed jobs and feeds them to the \emph{Routing Controller}, which triggers reconfiguration as necessary, according to \ours's adaptive routing policy (\Circled{3}). Table~\ref{tab:notation} summarizes paper notation.

\begin{table}[t!]
\centering
\caption{Notation used in this paper}
\label{tab:notation}
\small
\begin{tabular}{lp{5.3cm}}
\hline
\textbf{Notation} & \textbf{Description} \\
\hline
$G=(V,E)$ & Logical DAG of user's original query \\
$ Q=|V|$ \\
\hline
$G_1=(V_1, E_1)$ & Logical DAG of the stateless query segment \\
$N = |V_1|$ & The number of operators in $G_1$ \\
$N_k = N-2$ & The number of movable operators in $G_1$ \\
$v_i \in V_1$ & Operators of $G_1$ in topological order\\ & $i \in [0,N-1]$ \\
$v_0$ & Data source operator \\ 
$v_{N-1}$ & Message broker sink operator  \\
\hline
$G_{1R}=(V_{1R}, E_{1R})$ & Logical DAG of the stateless query segment with dynamic routing mechanism \\
$v_i^E \in V_{1R}$ & Edge replica of operator $v_i \in V_1$ \\ %
$v_i^C \in V_{1R}$ & Cloud replica of operator $v_i \in V_1$ \\ %
$CP_i \in V_{1R}$ & Routing choice operator between $v_i^E$ and $v_i^C$ \\%
\hline
$G_2=(V_2, E_2)$ & Logical DAG of the stateful  query segment \\
\hline
$T$ & Target throughput  (source output rate) \\
$BW$ & Outbound bandwidth from the edge device to the cloud \\
\hline
\end{tabular}
\end{table}

\begin{figure}[t]
\centering
\begin{minipage}{0.5\textwidth}
    \centering
    \begin{subfigure}{\textwidth}
    \centering
    \includegraphics[width=0.49\textwidth]{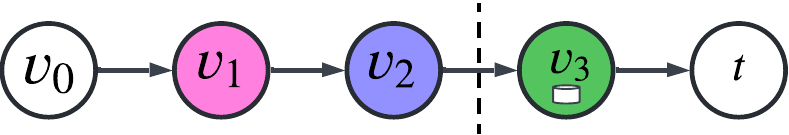}\vspace{-1mm}
    \caption{An example query $G$} 
    \label{fig:sec3_query_route0}\vspace{-0.5mm}
    \end{subfigure}    
    \begin{subfigure}{\textwidth}
    \includegraphics[width=0.97\textwidth]{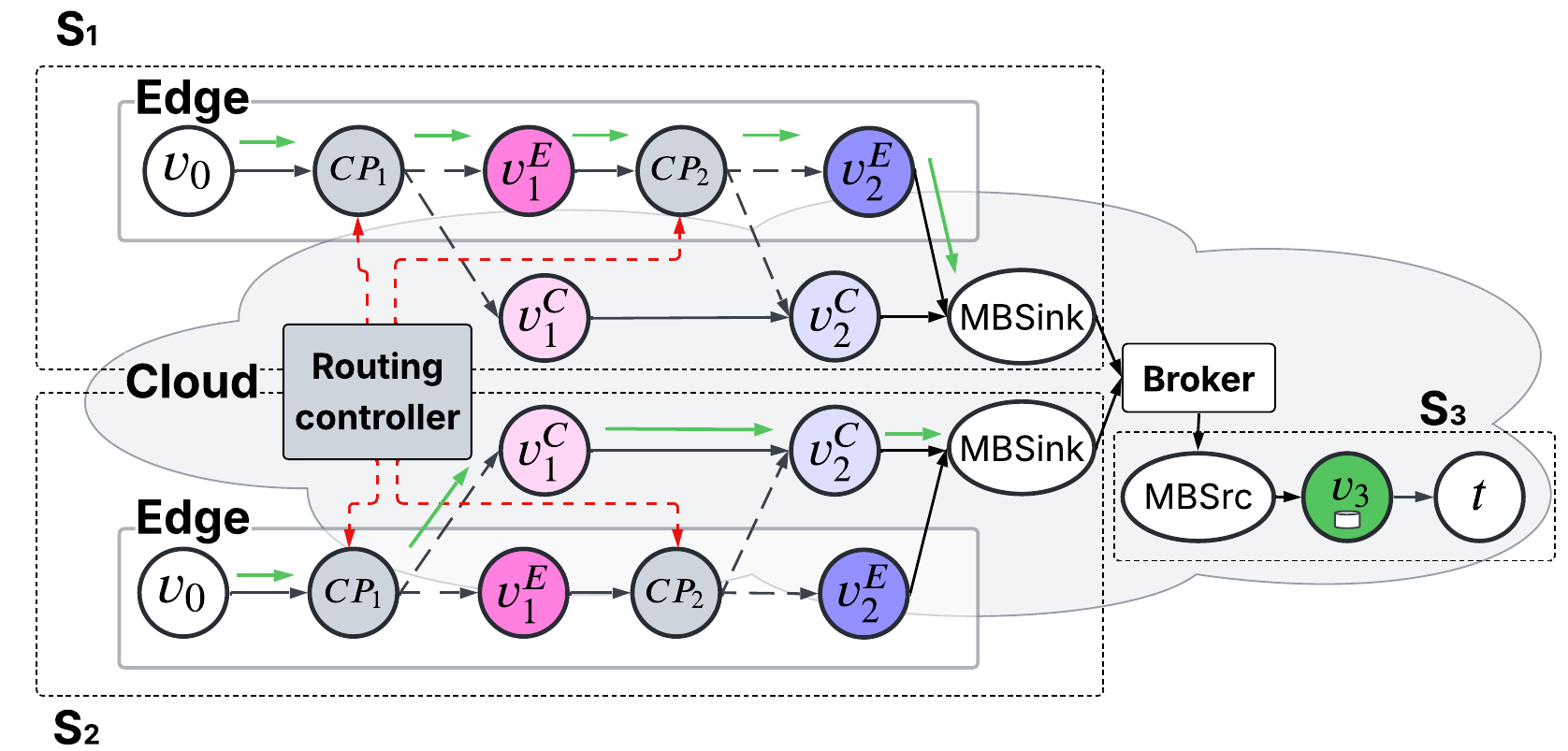}\vspace{-1mm}
    \caption{The deployed query on hybrid edge-cloud resources} 
    \label{fig:sec3_query_route2}\vspace{-1mm}
    \end{subfigure}
\end{minipage}\vspace{-1mm}
\caption{\ours transforms query $G$ into independent query segments $S1$, $S2$, and $S3$.}
\label{fig:sec3_query_route}
\end{figure}

\subsection{Adaptive routing mechanism}\label{sec:mechanism}

To enable live adaptation to dynamic operating conditions and workloads, \ours deploys operator replicas on both edge and cloud resources, confining reconfiguration to \emph{dynamic routing decisions}. Its routing mechanism seamlessly activates and deactivates replicas, effectively shifting computations between the edge and the cloud. 

Specifically, \ours identifies all stateless operators occurring before the first stateful one in the original dataflow graph. It creates one replica task for each such operator and deploys these replicas in the cloud. This design decision is informed by the consistent stateless-to-stateful pattern in the motivating use cases of Section~\ref{sec:usecase}.
Early-stage stateless operators are lightweight and reduce record size, 
making them good candidates for edge offloading in bandwidth-constrained settings. 
\ours places stateful operators exclusively in the cloud, so their state can be checkpointed to persistent storage that edge devices may lack. By restricting adaptation to stateless operators, \ours leverages edge resources and eliminates reconfiguration downtime by switching dataflow channels rather than moving tasks and state.

Consider the example query graph of Figure~\ref{fig:sec3_query_route0} with stateless operators, $v_1, v_2$. Figure~\ref{fig:sec3_query_route2} shows the resulting deployment with adaptive routing, across two edge devices and cloud resources. \ours decomposes the query into three segments, $S_1, S_2, S_3$. 
$S_1$ and $S_2$ span edge and cloud resources and  include two replicas per stateless operator. We denote $v_i^E$ as the edge-deployed task and $v_i^C$ as its corresponding cloud replica. \ccc{Both replicas implement the same logic as $v_i$.}
Upstream of each  $(v_i^E,v_i^C)$ pair, a routing operator $CP_i$  dynamically directs incoming events to either the edge or cloud replica, based on policy decisions from a centralized routing controller. The controller periodically issues routing‐switch commands to $CP_i$ tasks via a separate control channel (shown with dashed red arrows). 

\ccc{Operator replicas are organized in a mutual exclusion group: each data event is processed by only one replica of each stateless operator, so deploying replicas at both locations does not duplicate data-processing work.} Each query segment is controlled independently, allowing different execution paths across segments. Figure~\ref{fig:sec3_query_route2} shows an example of alternative paths with green solid arrows, where segment $S_1$ has offloaded $v_1$ and $v_2$ to the edge, while segment $S_2$ executes both in the cloud. %

\ccc{Inactive replicas are deployed but idle. Since these are stateless operators, they hold no state and only occupy a task thread and their channels' network buffers~\cite{flink_architecture_docs,flink_network_mem_docs}, while forwarding the watermarks propagated along both paths (\S\ref{sec:correctness}). The memory-intensive stateful operators are never replicated.}

\subsection{Query re-writer}\label{sec:rewriter}

\begin{figure}[t!]
\centering
    \begin{minipage}{\linewidth}
    \centering
        \begin{subfigure}{\textwidth}
        \centering
        \includegraphics[width=\textwidth]{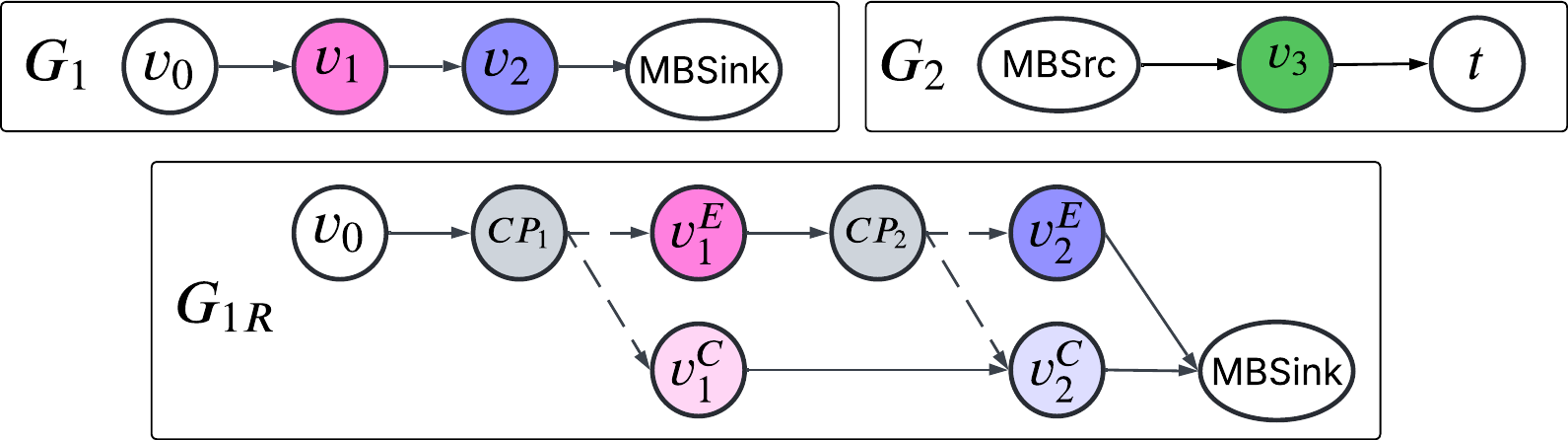}
        \end{subfigure}
    \end{minipage}\vspace{-1mm}
    \caption{Query re-writing steps.}
\label{fig:sec3_rewriting}
\end{figure}

To make \ours compatible with existing user applications, we develop a query re-writer that automatically transforms the original query into individual segments, generates the adaptive routing mechanism components, and connects everything into an end-to-end deployment, like the one shown in Figure~\ref{fig:sec3_query_route2}.

Let $G=(V, E)$ denote the logical DAG of the original user query, where nodes $v \in V$ are logical operators, and edges $e \in E$ represent directed data stream channels between operators. The re-writer first partitions the user query $G$ into two sub-queries at the boundary between stateless and stateful operators, as shown in Figure~\ref{fig:sec3_rewriting}. Sub-query $G_1$ contains the source and all stateless operators up to the first stateful operator. Sub-query $G_2$ contains that first stateful operator and all its downstream operators. Next, the re-writer transforms $G_1$ into $G_{1R}$ by creating replicas for stateless operators and inserting dynamic routing operators to enable adaptive selection of execution location. The two logical sub-queries are then connected via a message broker and sent to \ours's hybrid scheduler which deploys them on edge and cloud resources. 

Algorithm~\ref{alg:add-routing} details the transformation of $G_1$ to $G_{1R}$. After initializing the routing controller, the re-writer processes each non-source, non-sink operator $v_i$ in $G_1$ in topological order (lines 5–16), marking it as \emph{mobile} (i.e., capable of running on either edge or cloud). It then adds three components to $V_{1R}$ for each mobile operator (lines 6–8): (i) a routing operator $CP_i$, (ii) an edge replica $v_i^E$, and (iii) a cloud replica $v_i^C$. Next, the algorithm establishes control channels from the routing controller to each $CP_i$ (line 10) and data channels between $CP_i$ and its upstream and downstream replicas (lines 12–13). 
The re-writing algorithm ensures that once an event is routed to the cloud at any operator $v_i$, all subsequent operators will also be executed in the cloud. This is structurally enforced by connecting edge replica outputs to the next $CP$ operator (line 14), preserving routing flexibility, while cloud replica outputs bypass the next $CP$ and connect directly to the following cloud replica (line 15). As a result, any return path from the cloud to the edge is prevented, avoiding unnecessary data movement. Finally, the algorithm completes the transformation by adding the original source and sink operators with appropriate connections (line 17). 

\begin{algorithm}[t]
\caption{Transform $G_1$ to $G_{1R}$}
\label{alg:add-routing}
\begin{algorithmic}[1]
\Require  $G_1 = (V_1, E_1)$
\Ensure $G_{1R}=(V_{1R}, E_{1R})$

\State $CS \gets $ Routing Controller
\State $V_{1R} \gets \{CS\}$, $E_{1R} \gets \emptyset$

\State Sort operators $v_0, v_1, \ldots, v_{N-1} \in V_1$  in topological order
\State \Comment{$v_0$ is data source, $v_{N-1}$ is message broker sink}

\For{each mobile operator $v_i \in V_1$ , $i \in \{1, \ldots, N-2\}$}    %
    \State Create  operator $CP_i$    %
    \State Create replica $v_i^{E}$ and $v_i^{C}$ from $v_i$    %
    \State $V_{1R} \gets V_{1R} \cup \{CP_i, v_i^{E}, v_i^{C}\}$
    
    \State \textbf{Add control connections:}
    \State $E_{1R} \gets E_{1R} \cup \{(CS, CP_i)\}$    %
    
    \State \textbf{Add data flow connections:}
    \State Connect upstream operator outputs to $CP_i$ \Comment{From $v_{i-1}^{E}$ and $v_{i-1}^{C}$ if $i > 1$}
    \State $E_{1R} \gets E_{1R} \cup \{(CP_i, v_i^{E}), (CP_i, v_i^{C})\}$ 
   
    \State $E_{1R} \gets E_{1R} \cup \{(v_i^{E}, CP_{i+1})\}$ \Comment{Edge output can route to either replica}
    \State $E_{1R} \gets E_{1R} \cup \{(v_i^{C}, v_{i+1}^{C})\}$ \Comment{Cloud output bypasses $CP_{i+1}$, goes directly to operator on cloud}
\EndFor

\State Add original source $v_0$ and sink $v_{N-1}$ to $G_{1R}$ with appropriate edges
\State \Return $\tilde{G}_1 = (V_{1R}, E_{1R})$
\end{algorithmic}
\end{algorithm}

\subsection{Adaptive routing policy}
\label{sec:policy}

After query rewriting, \ours deploys the query segments on the available edge and cloud resources. Initially, \ours places only the source operators at the edge, which are necessary to perform event ingestion. All other operators are placed in the cloud. During execution, the routing controller collects metrics at user-specified intervals to identify performance bottlenecks and dynamically adjusts routing to ensure the deployment can sustain the rate of events generated by all edge devices without causing backpressure.

\ours automatically adjusts routing decisions by periodically evaluating a reconfiguration policy that determines whether each mobile operator in the stateless query segment should run on the edge or in the cloud. This policy is evaluated on a per‐device basis and accounts for variations in input event rates, network conditions, and compute contention. We first formulate the adaptive routing policy as a constrained optimization problem and then describe the heuristic algorithm that \ours employs to solve it.

\stitle{Problem definition.}
We define a \emph{routing plan} as a vector $X = \{x_1, x_2, \ldots, x_{N_k}\}$, with each indicator $x_i \in \{0,1\}$ corresponding to a mobile operator $v_i$ in $G_1$: $x_i = 0$ means $v_i$ executes at the edge, and $x_i = 1$ means it executes in the cloud. The objective is to dynamically identify an event routing plan $X$ that satisfies the target throughput under current conditions.

We assume that at any time there exists at least one routing plan capable of achieving the target throughput, although the optimization process may require several iterations to find it. We further assume that cloud resources are sufficient and can be scaled as needed, so that the stateful query segment will not become a bottleneck. Therefore, any observed bottleneck must arise from limited network bandwidth or constrained compute resources at the edge.

\stitle{Optimization objective.}
We formulate the routing plan decision as an optimization problem. 
We model the performance of the streaming dataflow using the concept of true and observed rates, as proposed in prior work~\cite{ds2}.
Let $\lambda_{cloud}^{i}$ and $\lambda_{edge}^{i}$ be the true processing rate of operator $i$, at the cloud and the edge respectively. 
$\ell_{o}^{i}$ is the observed output rate of operator $i$. 
Then, the true processing rate of operator $i$ under the current routing plan $X$ is given by the following equation:\vspace{-2mm}

\begin{equation}
    \lambda_{X}^{i} = x_i \cdot \lambda_{cloud}^{i} + (1-x_i) \cdot \lambda_{edge}^{i}
\end{equation}

To achieve the target throughput, every mobile operator in $G_1$ must sustain the observed output rate of its upstream operator. That is, the following inequality must hold:\vspace{-2mm}

\begin{equation}
\lambda_{X}^{i} - \ell_{o}^{i-1} \ge 0 , \forall  i \in [1, N_k]
\end{equation}

\noindent Accounting for all mobile operators, our optimization objective can be formulated as follows:\vspace{-2mm}

\begin{equation}
    maximize \sum_{i=1}^{N_k}{\frac{\lambda_{X}^{i}-\ell_{o}^{i-1}}{|\lambda_{X}^{i}-\ell_{o}^{i-1}|}}
\end{equation}\vspace{-2mm}

\stitle{Constraints.}
Let $b_{o}^{i}$ be the size per output record of operator $i$.
The first constraint states that the traffic from edge to cloud does not exceed the currently available bandwidth $BW$:\vspace{-2mm}

\begin{equation}
\sum_{i=1}^{N_k-1} (1 - x_i) \cdot x_{i+1} \cdot \ell_{o}^{i} \cdot b_o^i 
+ (1 - x_{N_k} ) \cdot \ell_{o}^{N_k} \cdot b_o^{N_k} \leq BW
\label{eq:bw}
\end{equation}\vspace{-2mm}

\noindent The second constraint ensures that once the event is routed to the cloud, it will not be routed back to the edge:\vspace{-2mm}

\begin{equation}
x_i \le x_{i+1}, \forall i, i \in [1, k]
\end{equation}

\stitle{Heuristic routing algorithm.} 
The standard way to solve an optimization problem like the one we formulate above is to use a solver. However, that would require accurate bandwidth measurements across wide-area links. Instead, \ours employs a fast heuristic policy that relies solely on system metrics. First, it detects whether the current routing plan is suboptimal by monitoring backpressure at the source and checking if the target throughput is sustained. If not, \ours iteratively adjusts the routing of one or more operators until it converges on a plan that sustains the target rate. The specific adjustment action depends on the type of bottleneck.

According to our problem definition, backpressure can originate either from limited network bandwidth between edge and cloud or from constrained compute capacity at the edge. To decide on a routing adjustment, \ours inspects the metrics of the boundary operators $v_i$ (deployed at the edge) and $v_{i+1}$ (deployed in the cloud). 
We distinguish two cases:

\begin{itemize}
\item \textbf{Case 1: Network bottleneck}. %
High backpressure at the edge-to-cloud boundary indicates that the WAN bandwidth is saturated. In this case, the effective bandwidth between edge and cloud equals the observed data transfer rate between $v_i$ and $v_{i+1}$. We therefore apply Equation~\ref{eq:bw} to all candidate operator pairs $(v_i, v_{i+1})$ and select a routing plan that satisfies the bandwidth constraint. If multiple such plans exist, \ours chooses the one with the fewest operators at the edge, to reduce the compute load of edge devices.

\item \textbf{Case 2: Compute bottleneck}. %
Low backpressure at the edge-to-cloud boundary indicates sufficient bandwidth capacity. Hence, if the deployment cannot sustain the target throughput, the bottleneck must be the edge device’s compute capacity. To resolve this, \ours initially shifts all mobile operators to the cloud. In the next decision interval, if the target throughput is still not met, the deployment has become network‐constrained again. As a result, \ours will move one or more operators back to the edge, as in the previous case. Since \ours’s mechanism incurs no downtime, taking multiple steps to converge is not a practical concern.
\end{itemize}

\noindent Note that when all mobile operators are deployed in the cloud (i.e., \(x_i = 1\) for all \(i \in [1, N_k]\)), the edge-to-cloud boundary lies immediately after the source operator $v_0$. If the target throughput is not met under this configuration, only Case 1 applies, since the source operator solely ingests data and cannot become a compute bottleneck.

\section{\ours guarantees}\label{sec:guarantees}

\ours aims for application transparency, enabling users to rely on familiar \emph{event-time} and \emph{fault-tolerance} semantics of dataflow systems. Preserving these guarantees is challenging because \ours query segments are deployed as independent jobs and any guarantees the underlying engine provides within a single job do not extend \emph{across} jobs. As a result, event time and processing semantics are not well-defined anymore, even within the same logical application. In this section, we describe the measures \ours takes to mitigate these issues and ensure correctness (\S\ref{sec:correctness}) and no data loss (\S\ref{sec:fault-tolerance}).

\subsection{Assumptions}\label{sec:assumptions}
\ours's correctness guarantees rely on standard properties provided by modern stream processing systems and message brokers. Specifically, we assume the message broker ensures (i) per-partition event order, (ii) reliable delivery of events with no loss, (iii) high-availability via broker replication, and (iv) persistence-before-acknowledgment for produced messages. Additionally, we assume the dataflow engine implements checkpoint-based fault tolerance. These features are widely supported in systems like Apache Kafka~\cite{narkhede2017kafka,wang2021consistency}, Apache Pulsar~\cite{pulsar}, Amazon Kinesis Data Streams~\cite{kinesis}, Apache Flink~\cite{flink,hueske2019stream}, and Spark Structured Streaming~\cite{spark-streaming,armbrust2018structured}.

\subsection{Semantic equivalence}
\label{sec:correctness}

\ours's rewriting module decomposes the original application graph into a system of query segments, where some operators are replicated to enable adaptive routing. Since each event can be processed by either replica, and potentially in a different order, this transformation raises a question of semantic equivalence: will the transformed query exhibit the same event-time ordering guarantees as the original? Because \ours only replicates stateless, order-agnostic operators, preserving correctness under event time reduces to two key properties: (1) ensuring that events will be processed by the same ordered sequence of operators and (2) correctly propagating watermarks across query segments.

\stitle{Execution path equivalence.}
Let $S = (e_1, e_2, ..., e_n)$ be an input stream consisting of a sequence of events, produced by a deterministic, replayable source. Let also $\pi_G(e_i)$ denote the execution path of event $e_i$ through a logical query graph $G$, defined as the ordered sequence of logical operators traversed by $e_i$ from source to sink. We define the \textbf{Execution Path Equivalence} property as follows: \emph{for every event $e_i$ at position $i$ in the input stream $S$, the execution path $\pi_G(e_i)$ in the original graph $G$ and the execution path $\pi_{G^T}(e_i)$ in the transformed graph $G^T$ are identical}:\vspace{-2mm}

\begin{equation}
    \forall i \in \{1, 2, ..., n\},\ \pi_G(e_i) = \pi_{G^T}(e_i)\label{eq:equivalence}
\end{equation}

In other words, no operator is omitted, duplicated, reordered, or introduced by the transformation.
It is straightforward to see why \ours preserves execution path equivalence between $G$ and $G^T$. First, consider any stateless operator $v_j$ (excluding the data source). In $G^T$, each such operator has two replicas, $v_j^E$ and $v_j^C$, both implementing identical logic, so for any event $x$, $v_j^E(x) = v_j^C(x) = v_j(x)$. Since exactly one routing operator ($CP_j$) is associated with each $v_j$, every event $x$ is exclusively processed by either $v_j^E$ or $v_j^C$. Second, the routing policy enforces a \emph{monotonic transition}: once event $x$ is routed to the cloud at any operator, all subsequent operators also process $x$ in the cloud. This guarantees that execution never switches back to the edge for the same event, preserving the sequence of logical operators traversed. As a result, each event follows the same execution path as in $G$, with only the physical placement of stateless operators potentially differing.

\begin{figure}[t!]
\centering
\begin{subfigure}{\linewidth}
\centering
\includegraphics[width=0.56\linewidth]{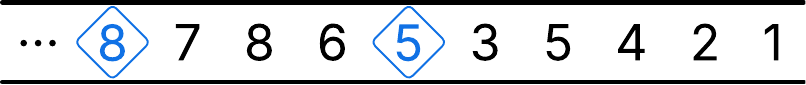}
\caption{An example stream with watermarks shown as blue diamonds} 
\label{fig:sec4a1}
\end{subfigure} 
\hfill
\begin{subfigure}{\linewidth}
\centering
\includegraphics[width=0.56\linewidth]{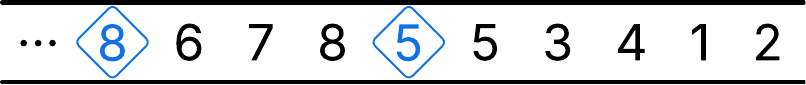}
\caption{An equivalent stream order under event-time semantics} 
\label{fig:sec4a2}
\end{subfigure}\vspace{-2mm}
\caption{A watermark $t_w$ guarantees all events with $t \leq t_w$ have already been received. Events between watermarks may be reordered, as long as they respect watermark boundaries.}
\label{fig:sec4a}
\end{figure}

\stitle{Watermark correctness.}
For pipelines composed only of stateless, time-agnostic operators, path equivalence would be sufficient for correctness. However, time-aware operators, such as windows, must also produce deterministic results based on event time. To be precise, events between two watermarks may be reordered arbitrarily but they must respect watermark boundaries. Figure~\ref{fig:sec4a} shows two stream orders that are equivalent under event time: a window firing at any timestamp would include the same set of events in both cases\footnote{This holds even for windows firing on non-watermark timestamps, as they will only be evaluated upon receiving the subsequent watermark.}. 

Given that watermark $w_k$ is a guarantee that all events with $t(e) \leq w_k$ have arrived, let $\mathcal{E}_v^{(w_{k-1}, w_k]}$ be the multiset of events with timestamps $t(e) \in (w_{k-1}, w_k]$ that are processed by operator $v$ in $G$ and $\mathcal{E}_{v^T}^{(w_{k-1}, w_k]}$ be the corresponding multiset of operator $v^T$ in $G^T$. We define the \textbf{Watermark Correctness} property as follows: \emph{for any $v$, $v^T$, the set of events processed and emitted by $v^T$ in the interval defined by any consecutive watermarks $(w_{k-1}, w_k]$ is identical to that of $v$}:\vspace{-2mm}

\begin{equation}
    \forall k,\quad \mathcal{E}_v^{(w_{k-1}, w_k]} = \mathcal{E}_{v^T}^{(w_{k-1}, w_k]}\label{eq:watermarks}
\end{equation}

\begin{figure}[t!]
\centering
\includegraphics[width=0.9\linewidth]{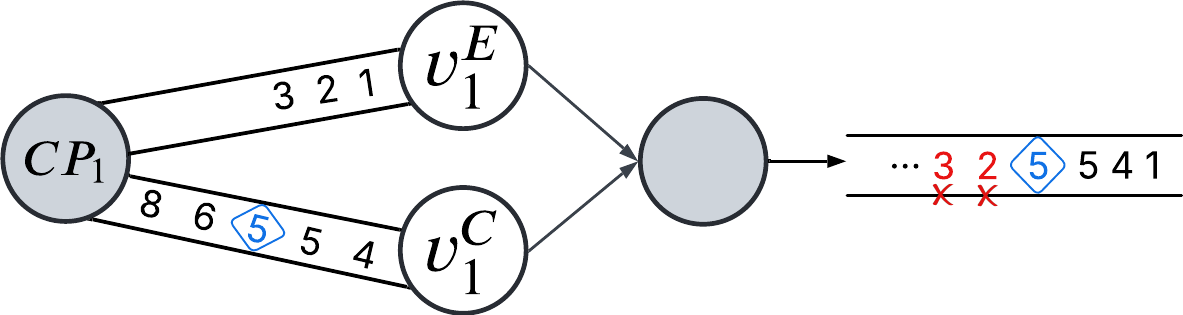}\vspace{-2mm}
\caption[Failure of naive live rerouting without watermark coordination]{\ccc{Naive live rerouting without watermark coordination. The cloud path is activated after event \texttt{3}; events \texttt{3} and \texttt{2} arriving after watermark \texttt{5} are incorrectly treated as late and dropped. \ours  prevents this correctness failure.}}
\label{fig:sec4b}
\end{figure}

\ccc{Figure~\ref{fig:sec4b} illustrates a correctness failure of naive live rerouting that \ours prevents. Without this coordination,} two problematic scenarios may occur: (i) when traffic is switched from the edge to the cloud, a watermark sent via the cloud path may reach downstream operators before prior events still in transit on the edge path, and (ii) as the stateful query source belongs to a separate job, it generates watermarks without considering the event-time progress of its upstream jobs.

To address the first problem, \ours propagates watermarks along both edge and cloud paths in the stateless query segment, regardless of which path is active, and introduces a special \emph{watermark union} operator during rewriting. This operator is added before the message broker sink and ensures that event time advances only after the corresponding watermarks have been received from all upstream channels. 

To mitigate the second scenario and ensure end-to-end event-time correctness, \ours records watermarks from edge query segments as special \emph{control} events in the message broker sink partitions, alongside regular data events. The stateful segment, $G_2$, processes these control events, tracking the highest watermark $w_p$ per partition $p$ to ensure monotonicity. The global watermark is computed as the minimum watermark across all upstream $G_{1R}$ jobs: $w_{global} = \min_{p \in [1, P]} w_p(t)$, where $P$ is the number of edge partitions. To prevent stalling of $w_{global}$, each $G_{1R}$ job keeps emitting control events during inactivity. Only true partition silence (such as failure or connectivity loss) can delay event-time progress. We handle such cases separately as described in \S\ref{sec:WatermarkToleranceMechanism}.

Execution path equivalence and watermark correctness guarantee processing semantics during failure-free execution. Next, we discuss how \ours prevents data loss under failures.

\subsection{Correctness under failures}\label{sec:fault-tolerance}

Most modern dataflow systems achieve no data loss during failures using persistent, replayable sources and consistent snapshot protocols~\cite{chandy1985distributed,streamingsurvey_vldb24,carbone2015apache,carbone2020beyond,siachamis2024checkmate,wang2019lineage}, where a special \emph{checkpoint barrier} traverses the dataflow graph, coordinating state snapshots across tasks. However, when event sources run on low-resource devices, assuming persistent and replayable streams is unrealistic. Consequently, \ours does not aim to tolerate permanent hardware failures. Instead, \ours provides at-least-once processing (no data loss) under transient failures (e.g., process crashes or bugs), as long as the source device is able to buffer events during recovery.

A key challenge is that barrier-based distributed snapshot protocols ensure consistency only within a single, connected dataflow graph. Extending the protocol across jobs would require waiting for barriers from all edge segments before taking a snapshot. In \ours, we explicitly decide not to align checkpoint barriers across query segments to avoid a single slow or disconnected edge device holding up global checkpoint progress or potentially causing checkpoints to never complete. Additionally, maintaining separate checkpoints per segment gives us the flexibility to independently tune the checkpointing interval for different parts of the dataflow. For example, we can configure less frequent checkpoints at the edge to minimize resource consumption, and more frequent checkpoints in the cloud for faster recovery. 

Below, we informally show that \ours guarantees at-least-once semantics under failures, which is sufficient for supporting exactly-once when combined with deduplication at the cloud source. Deduplication can be easily achieved with unique record IDs, as in Millwheel~\cite{akidau2013millwheel} and Kafka Streams~\cite{wang2021consistency}, though this is not yet implemented in \ours.

\stitle{Case 1: Edge job failure.} If one or more edge jobs become stragglers or fail, they can be restarted independently without disrupting query segments running on other edge devices or in the cloud. Event loss is prevented by the combination of the message broker’s persistence guarantees and periodic checkpoints. Specifically, results are durably written to the message broker sink before acknowledgments are sent to the edge segment (the producers). Because a snapshot is only acknowledged after its barrier traverses the entire dataflow graph from the source to the sink, we know that for a snapshot taken at source offset $x$, all events processed before $x$ have been successfully acknowledged and persisted in the message broker. When a failed edge segment recovers, it restarts from the last successful checkpoint offset $x$, which must be no later than the last acknowledged message offset in the broker sink. The edge source then replays all events from offset $x$ onward, ensuring that each event is processed at least once.

\stitle{Case 2: Cloud job failure.}  In this scenario, the cloud segment fails while the edge segments continue to operate normally. Importantly, the message broker remains highly available and retains all events, regardless of consumer failures (cf. \S\ref{sec:assumptions}). When the cloud job restarts, it reprocesses all events from its last successful checkpoint offset, ensuring no data loss.

\stitle{Case 3: Simultaneous edge and cloud job failures.} In this scenario, both edge and the cloud segments fail simultaneously, while the message broker continues to operate. Let $x_c$ denote the checkpoint offset of the cloud job and $x_e$ denote the checkpoint offset of the edge job. Since edge and cloud jobs checkpoint independently, their offsets may differ, which affects how recovery is coordinated. Without loss of generality, we focus on a single edge segment.
\begin{enumerate}
    \item $x_e \geq x_c$. The cloud job's last checkpoint may be behind the edge's. Suppose that before failing, the edge job has produced events up to offset $x_e+k$ to the broker. Upon recovery, events between $x_e$ and $x_e+k$ will be replayed, causing the cloud to reprocess them ($x_c < x_e+k$).
    \item $x_e < x_c$. The cloud job has already processed and checkpointed events that the edge job has not yet checkpointed. This can happen due to independent checkpoint intervals. If the edge job has produced events up to offset $x_e + k$ before failing, events between $x_e$ and $x_e + k$ will appear as duplicates in the broker. After recovery, the cloud job will reprocess all duplicates in this range, as well as any duplicates between its own checkpoint offset $x_c$ and $x_e + k$.
\end{enumerate}\vspace{-2mm}

\subsection{Availability under failures}
\label{sec:WatermarkToleranceMechanism}

Despite watermark correctness (cf. \S\ref{sec:correctness}), computation progress can stall if an edge segment fails or becomes a straggler. To address this, \ours introduces additional logic at the cloud source that advances the global watermark when a user-specified period of inactivity is detected in a broker partition. If a partition does not emit a watermark within this period, it is marked as inactive. The global watermark is then calculated as the minimum of the maximum watermarks from active partitions only. This approach prevents a single inactive partition from stalling the entire pipeline.  %

\section{Evaluation}
\label{sec:eval}

We implement \ours on Apache Flink 1.20 and Apache Kafka 3.9.0. 
However, our adaptive techniques are agnostic to the underlying system and can be integrated with any dataflow stream engine that satisfies the requirements in Section~\ref{sec:assumptions}. We discuss the implementation details in \suppref{sec:sup-implementation}.

Our evaluation aims to answer four key questions:
\ccc{(1) Can \ours reconfigure without pausing or restarting execution (\S\ref{sec:exp-mechanism})?
(2) Can \ours adapt to variable workloads and operating conditions to consistently sustain the target throughput (\S\ref{sec:exp-vary2222})?
(3) How is \ours's performance impacted by individual parameters, including fluctuating input rates, network bandwidth, and compute contention (\S\ref{sec:exp-vary1})?
(4) Can \ours scale to large hybrid deployments (\S\ref{sec:exp-100vm})?}
We further evaluate \ours's  availability under edge failures in \suppref{sec:exp-failures}.

\subsection{Experimental setting}
\label{sec:exp-setting}

We evaluate \ours on two testbeds:
\textbf{(i) Testbed-RPi} is a hybrid cluster consisting of five Raspberry Pi 4 Model B devices (4GB RAM, quad-core ARM Cortex-A72 1.8 GHz CPU) acting as edge nodes and one bare-metal cloud node equipped with an AMD EPYC 7713 64-Core Processor, 256GB memory, and 3.84TB SSD. 
\textbf{(ii) Testbed-100VMs} is a large-scale CloudLab~\cite{CloudLab} cluster with 100 VMs acting as edge devices and 45 physical nodes. Each VM emulates a resource-constrained edge device with 4 CPU cores and 4GB memory, matching the compute-to-memory ratio of Raspberry Pi devices to ensure comparable resource constraints. Each cloud node is equipped with a 16-core AMD EPYC 7302P CPU, 128GB memory, and 480GB SSD.
We deploy one Flink TaskManager per edge device, while the Flink JobManager and  Kafka run in the cloud nodes. 

\stitle{Baselines.}
We compare \ours with three baselines:  
\textbf{(i) All-Edge} deploys the application as a single job that executes both stateless and stateful operators on edge nodes, with only the sink running in the cloud. This baseline maximizes edge processing to minimize WAN data transfer. 
\textbf{(ii) All-Cloud} deploys the application as a single job that executes all operators, except the source, in the cloud, maximizing cloud usage to avoid edge resource constraints. 
\textbf{(iii) CAPSys}~\cite{capsys} is a state-of-the-art task placement controller for Flink that uses a contention-aware cost model to balance resource usage across workers. Because CAPSys assumes homogeneous workers and stable operating conditions, its initial placement plan does not change during the duration of our experiments. To ensure a fair comparison, we execute CAPSys plans directly on the \ours runtime and not on vanilla Flink. For \ours and CAPSys, all operators except the source are initially deployed in the cloud. 

\stitle{Benchmarks.}
We evaluate all systems on the five hybrid edge-cloud streaming queries described in \S\ref{sec:usecase}. For each query, we ingest a publicly available dataset \cite{air_quality_dataset, debs2015_dataset, smart_meter_datasets, twitter_datasets, video_datasets}, at the source and replay it with configurable target input rates. 

\stitle{Metrics collection.}
Unless otherwise stated, we run each experiment for 2 hours with 5min warm-up , using 5s metrics interval and 10min routing policy interval.
To assess whether a pipeline can sustain its target throughput, we measure the output rate (in records/s) of sources at each edge node. A rate below the target indicates backpressure (cf. \S\ref{sec:background}).

\subsection[Zero-downtime reconfiguration]{\ccc{Zero-downtime reconfiguration}}
\label{sec:exp-mechanism}

\begin{figure}[t!]
\centering
\includegraphics[width=0.96\linewidth]{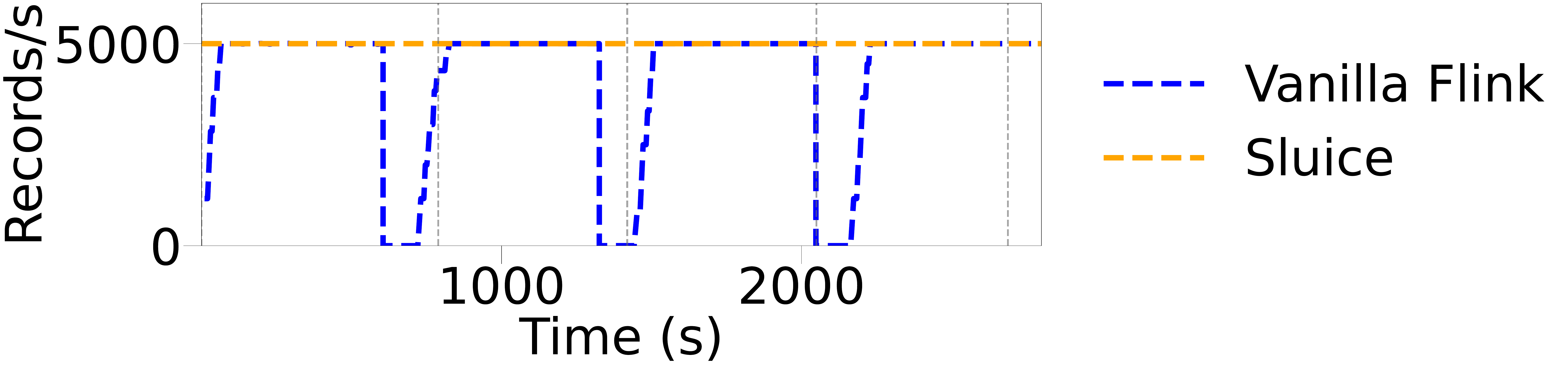}
\vspace{-0.3cm}
\caption[center]{Comparison of \ours' and Flink's reconfiguration mechanisms. \ours shifts operators without interruption.}
\label{fig:downtime}
\end{figure}

\ccc{We compare reconfiguration overhead of \ours and vanilla Flink using \textbf{Q2} on \textbf{Testbed-RPi}, triggering operator shifts every 10 minutes. In vanilla Flink, a placement change means stopping the job and restarting it with the new placement, which is also the mechanism CAPSys relies on. Figure~\ref{fig:downtime} shows the throughput achieved over time for both systems. Every shift incurs approximately 3 minutes of downtime in vanilla Flink, while \ours shows no visible throughput dips. The result shows that \ours effectively avoids service interruptions.}

\subsection{\ours under variable operating conditions}
\label{sec:exp-vary2222}

\begin{figure}[t!]
\centering

\begin{subfigure}[b]{0.8\linewidth}
\centering
\includegraphics[width=\linewidth]{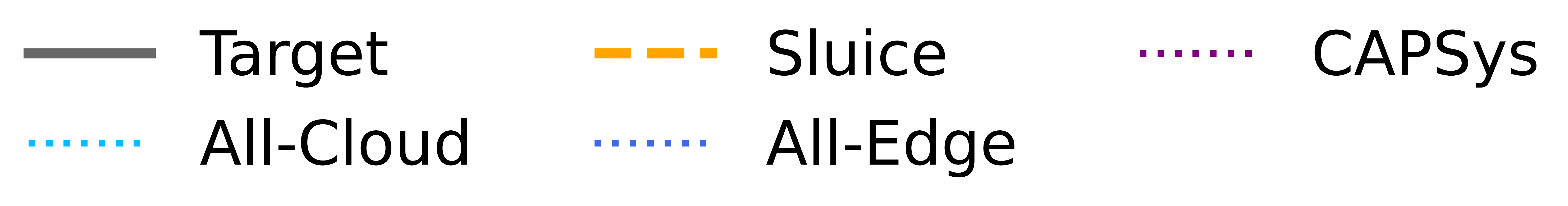}
    \vspace{-0.4cm}
\end{subfigure}\vspace{-2mm}

\setlength{\tabcolsep}{1pt}
\renewcommand{\arraystretch}{1.5}

\begin{tabular}{ccc}
    \adjustbox{valign=m}{\rotatebox[origin=c]{90}{\textbf{Q1}}} &
    \adjustbox{valign=m}{\includegraphics[width=8.0cm,trim=0 0 0 90,clip]{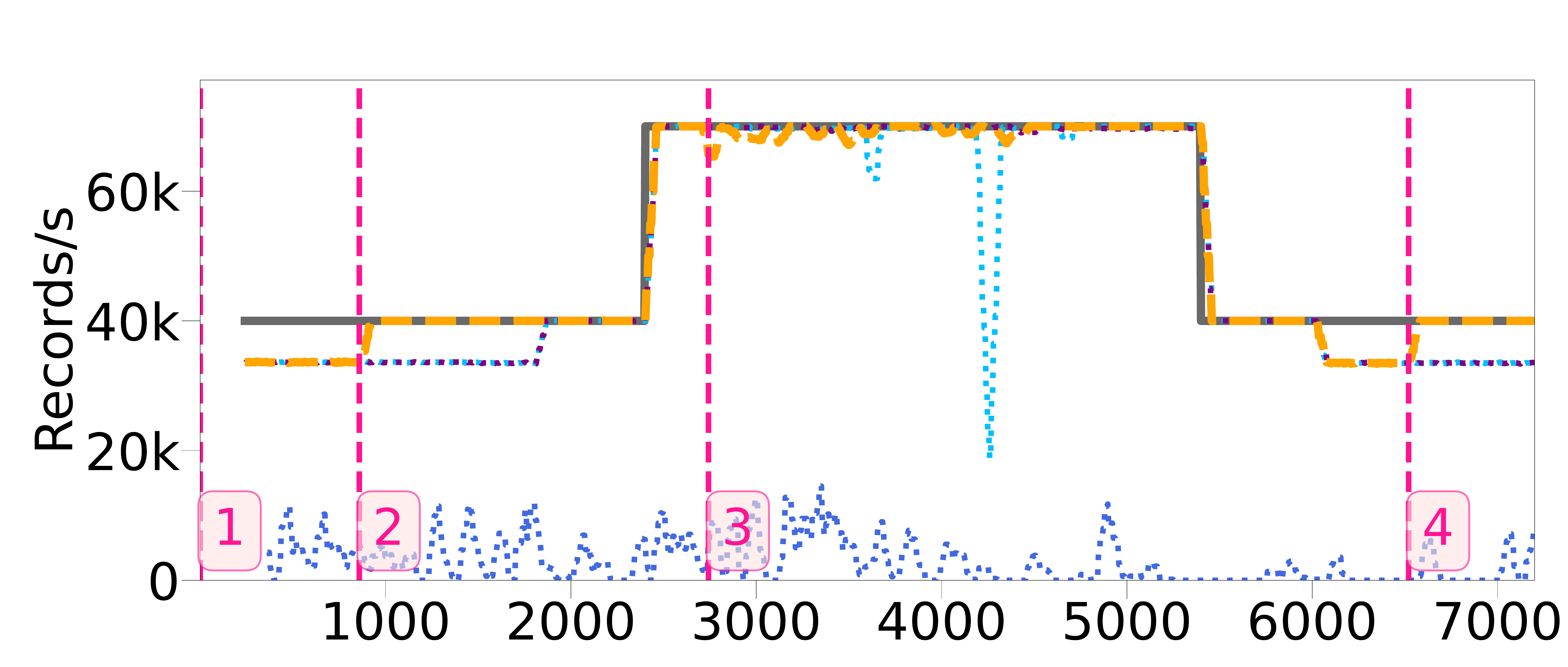}} \\
    \adjustbox{valign=m}{\rotatebox[origin=c]{90}{\textbf{Q5}}} &
    \adjustbox{valign=m}{\includegraphics[width=8.0cm,trim=0 0 0 90,clip]{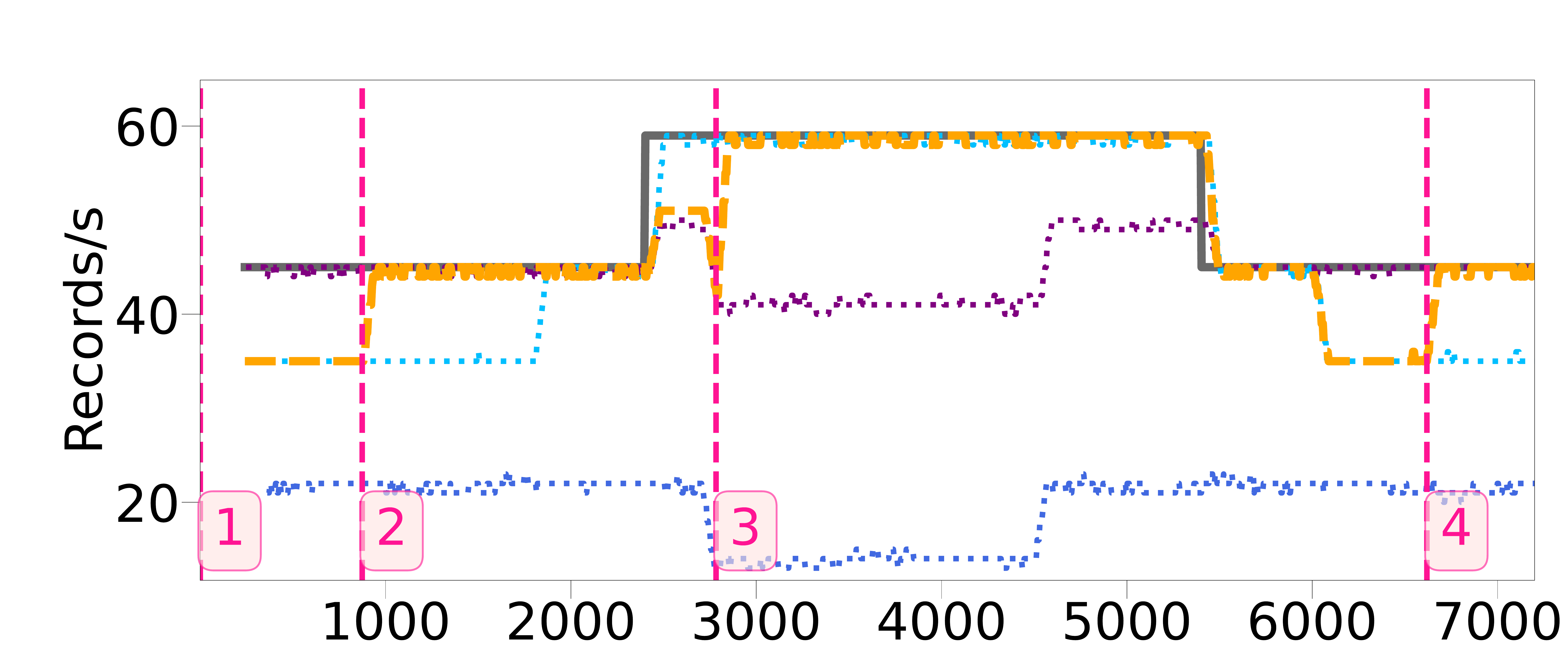}} \\
\end{tabular}

\begin{subfigure}[b]{0.8\linewidth}
\includegraphics[width=\linewidth]{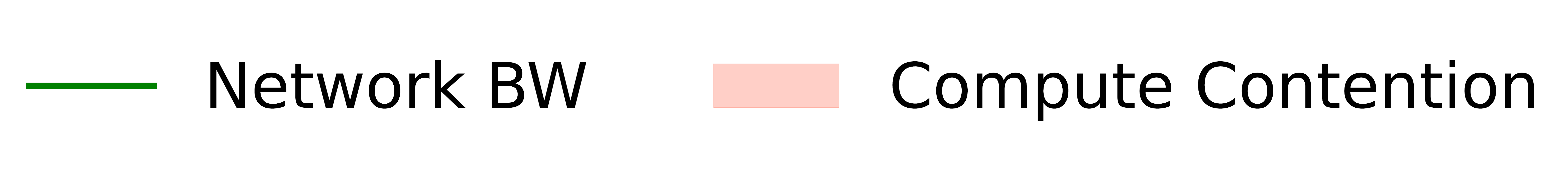}
    \vspace{-0.4cm}
\end{subfigure}\vspace{-4mm}

\begin{tabular}{ccc}
    \adjustbox{valign=m}{\rotatebox[origin=c]{90}{\parbox{2.4cm}{\centering Network BW }}} &
    \adjustbox{valign=m}{\includegraphics[width=8.06cm]{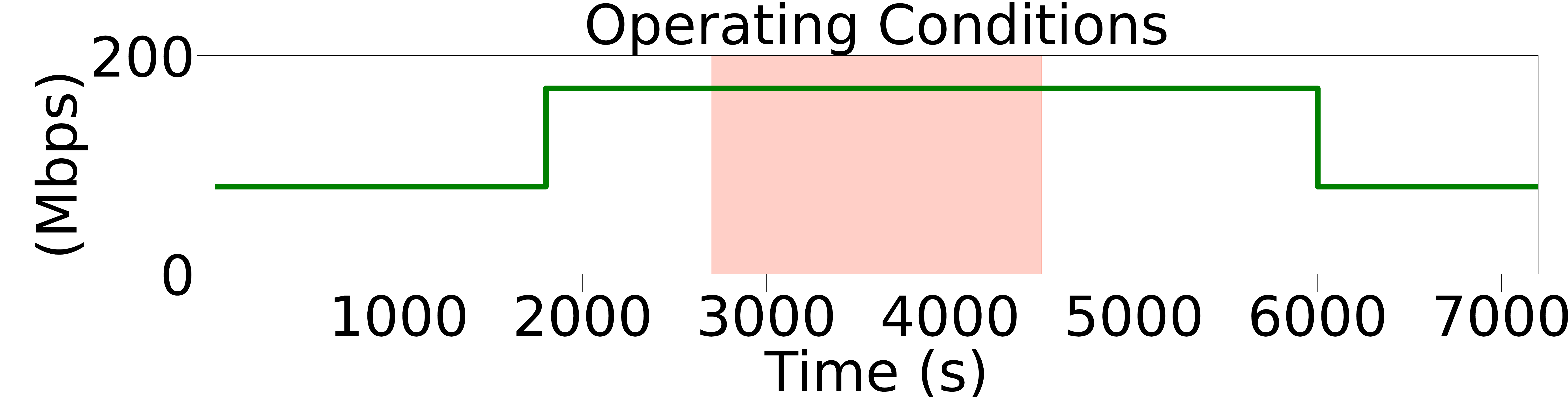}} \\
\end{tabular}
\vspace{-3mm}
\caption{Comparison of \ours and baselines under simultaneous changes in their operating environment, using  \textbf{Q1} and \textbf{Q5} on \textbf{Testbed-RPi}. The plot shows the target rate and achieved throughput over time (top) under variable operating conditions (bottom).}
\label{fig:eval_vary2222}
\end{figure}

\ccc{We next evaluate} \ours's ability to adapt to simultaneous changes in its environment: workload (target rate), network capacity (edge-to-cloud bandwidth), and edge compute capacity (compute contention). We run this experiment on \textbf{Testbed-RPi}. Figure~\ref{fig:eval_vary2222} (bottom) shows how operating conditions change over a period of 3 hours.

The target rate changes according to a step function every hour, as indicated by the gray line. We use Linux \texttt{tc} to shape the outbound bandwidth of each RPi to 80–170~Mbps, matching the range reported in WASP ~\cite{jonathan2020wasp}. To make the setting even more challenging, we execute a background application that performs arithmetic computations to consume CPU cycles during the second hour of the experiment, imposing compute contention at the edge (red shaded region). %

Figure~\ref{fig:eval_vary2222} shows the results for \textbf{Q1} and \textbf{Q5}. We plot throughput over time for one of the RPis and include the remaining results in \suppref{sec:sup-allfig}. \ours rapidly reacts to dynamics and it is the only system that can sustain the target during the entire experiment. Four distinct phases are marked on the plot. (1): Initially, \ours's deployment is equivalent to the All-Cloud strategy, which cannot sustain the target throughput due to insufficient bandwidth. (2): \ours triggers a policy decision (shown as a vertical dashed line). \ours shifts one operator to the edge in \textbf{Q1} and two operators in \textbf{Q5}. (3): When compute contention takes effect, operator processing at the edge becomes a bottleneck. \ours's policy detects the issue and shifts all operators back to the cloud. The new deployment meets the target, hence, \ours does not trigger any further shifts in the next hour. After the compute contention is lifted and bandwidth drops, the edge-to-cloud data transfer becomes the bottleneck. (4): \ours' policy reacts quickly and shifts operators to the edge, once again achieving the target.

The All-Edge deployment is bottlenecked by the limited edge device capacity, particularly during the compute contention period. On the other hand, the All-Cloud strategy is clearly restricted by the available bandwidth. CAPSys is able to avoid the bandwidth bottleneck by placing first two operators at the edge, however, it cannot meet the target rate when compute resources at the edge are limited.%

\subsection{Impact of individual parameters }
\label{sec:exp-vary1}

Next, we evaluate the behavior of \ours in a more controlled setting, where we vary a single parameter at a time: event rate, bandwidth, and compute contention. In this set of experiments, we run all five applications of \S\ref{sec:usecase} on \textbf{Testbed-RPi}. We report results for one of the RPis per query and include remaining in \suppref{sec:sup-allfig}. 

\begin{figure}[t!]
\centering

\begin{subfigure}[b]{0.8\linewidth}
\includegraphics[width=\linewidth]{figures/eval/shift/res_video6nw_0.pdf_legend.pdf}
    \vspace{-0.5cm}
\end{subfigure}
\setlength{\tabcolsep}{1pt}
\renewcommand{\arraystretch}{1.5}

\begin{tabular}{ccc}
    \adjustbox{valign=m}{\rotatebox[origin=c]{90}{\textbf{Q1}}} &
    \adjustbox{valign=m}{\includegraphics[width=4.0cm,trim={0 0 0 180bp},clip]{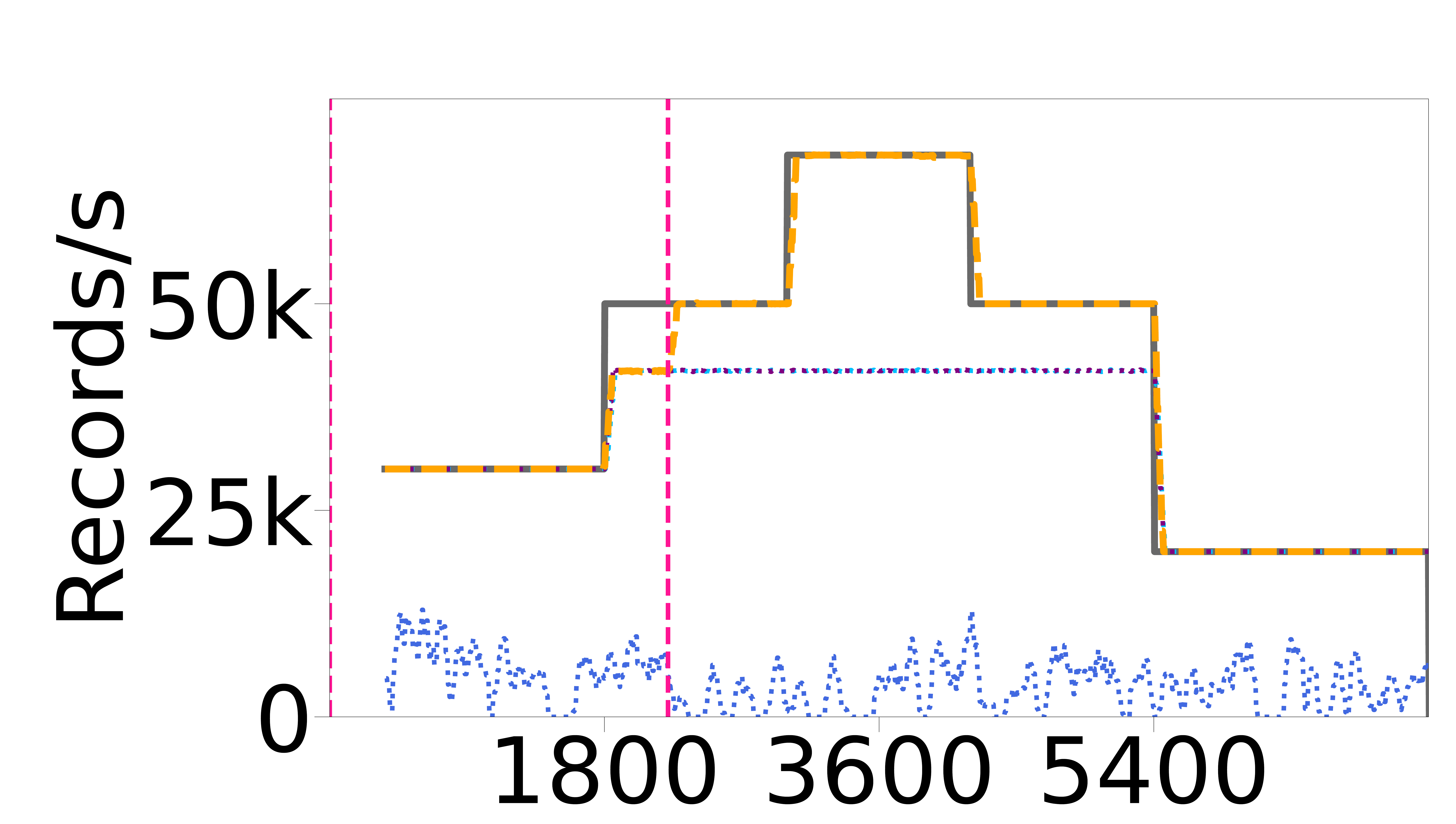}} &
    \adjustbox{valign=m}{\includegraphics[width=4.0cm,trim={0 0 0 180bp},clip]{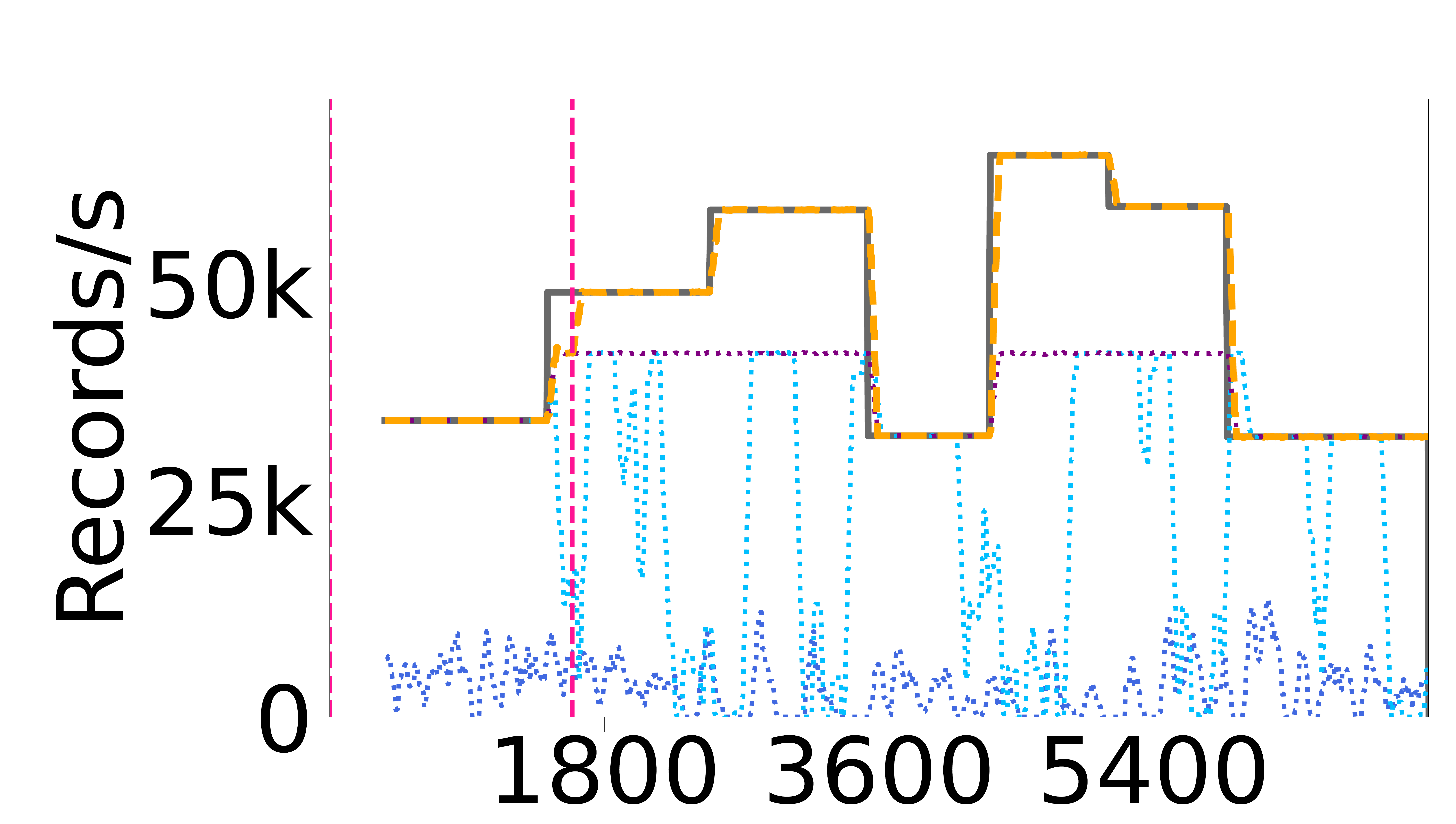}} \\

    \adjustbox{valign=m}{\rotatebox[origin=c]{90}{\textbf{Q2}}} &
    \adjustbox{valign=m}{\includegraphics[width=4.0cm,trim={0 0 0 180bp},clip]{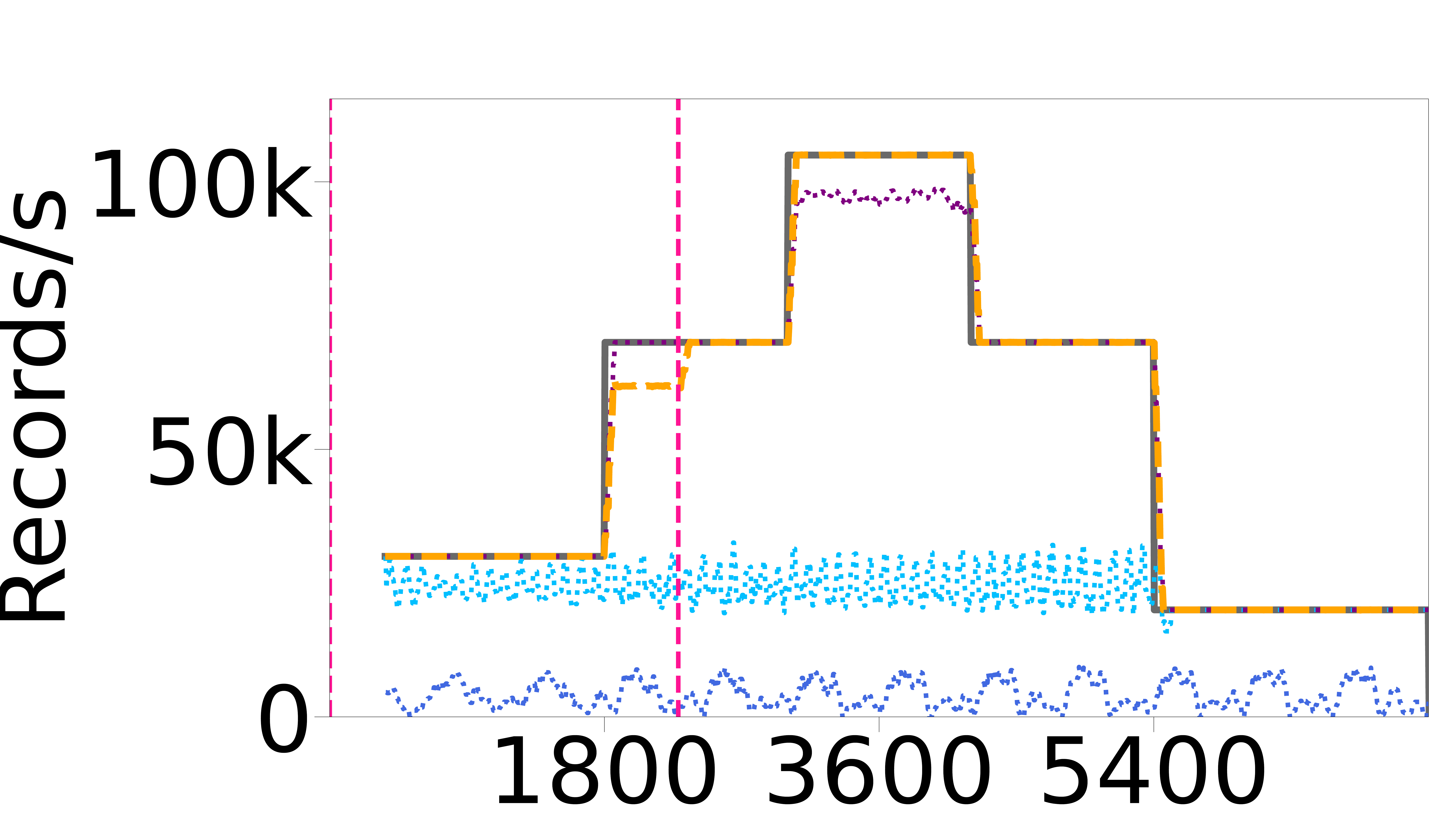}} &
    \adjustbox{valign=m}{\includegraphics[width=4.0cm,trim={0 0 0 180bp},clip]{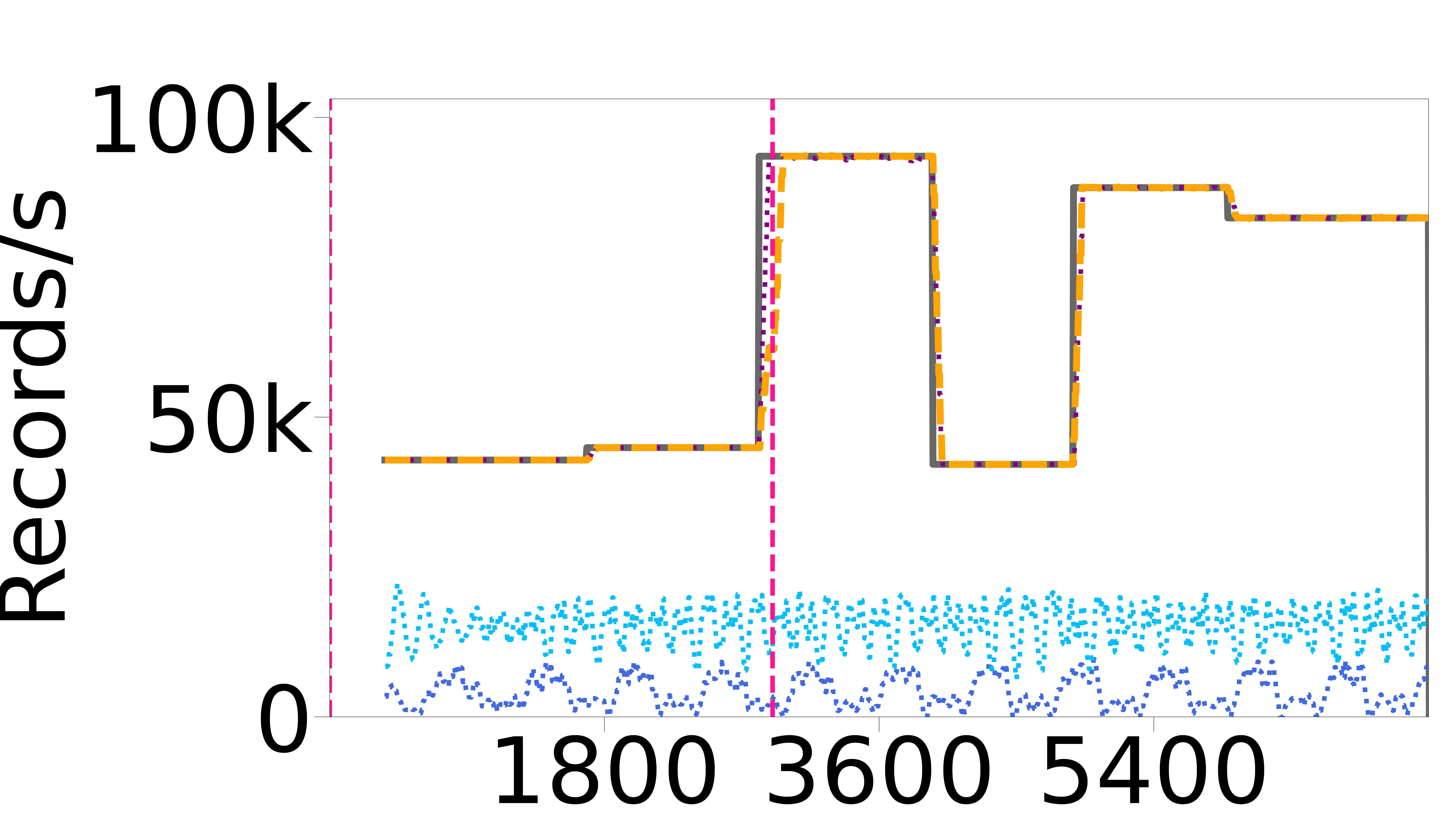}} \\

    \adjustbox{valign=m}{\rotatebox[origin=c]{90}{\textbf{Q3}}} &
    \adjustbox{valign=m}{\includegraphics[width=4.0cm,trim={0 0 0 180bp},clip]{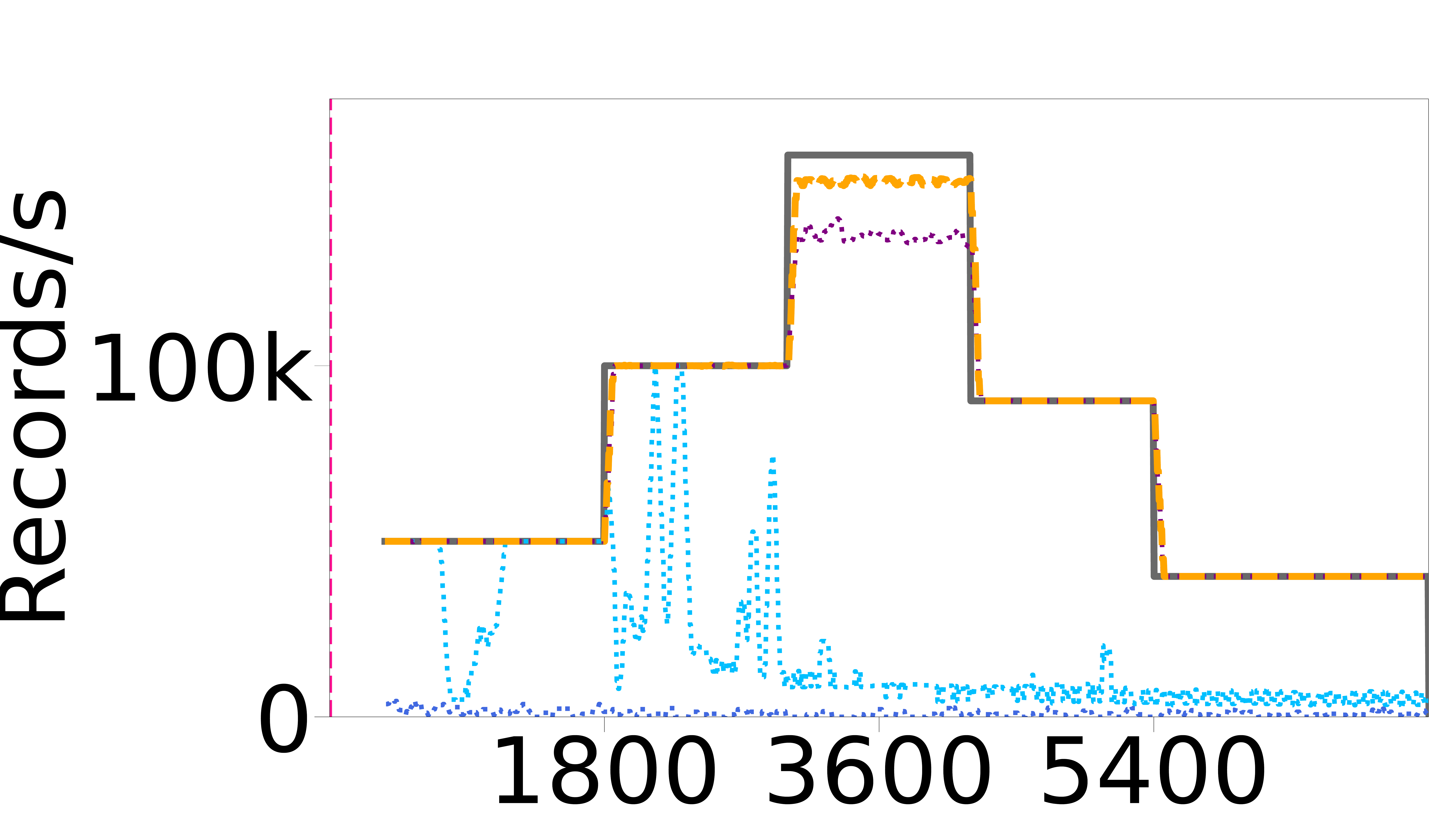}} &
    \adjustbox{valign=m}{\includegraphics[width=4.0cm,trim={0 0 0 180bp},clip]{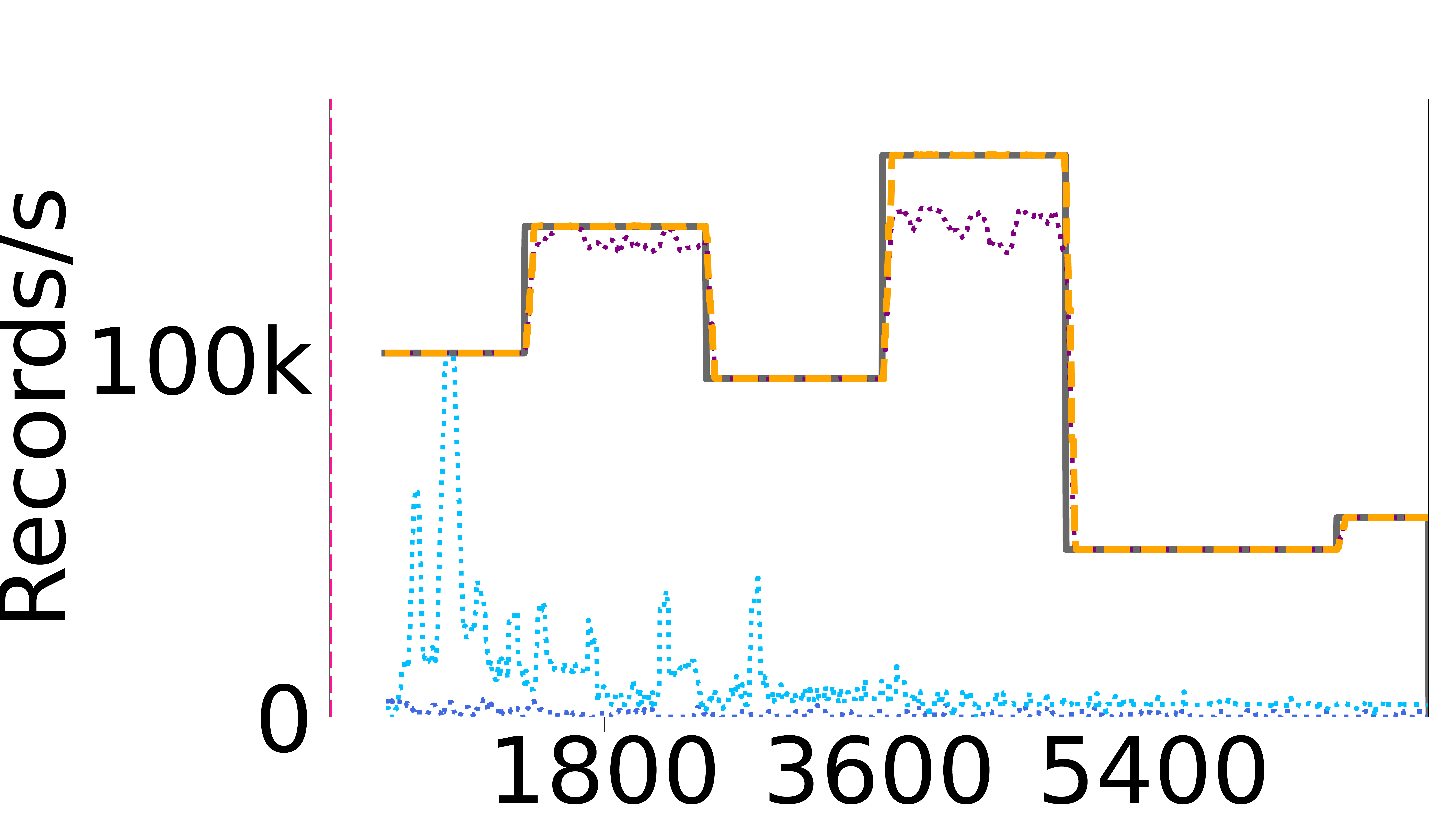}} \\

    \adjustbox{valign=m}{\rotatebox[origin=c]{90}{\textbf{Q4}}} &
    \adjustbox{valign=m}{\includegraphics[width=4.0cm,trim={0 0 0 180bp},clip]{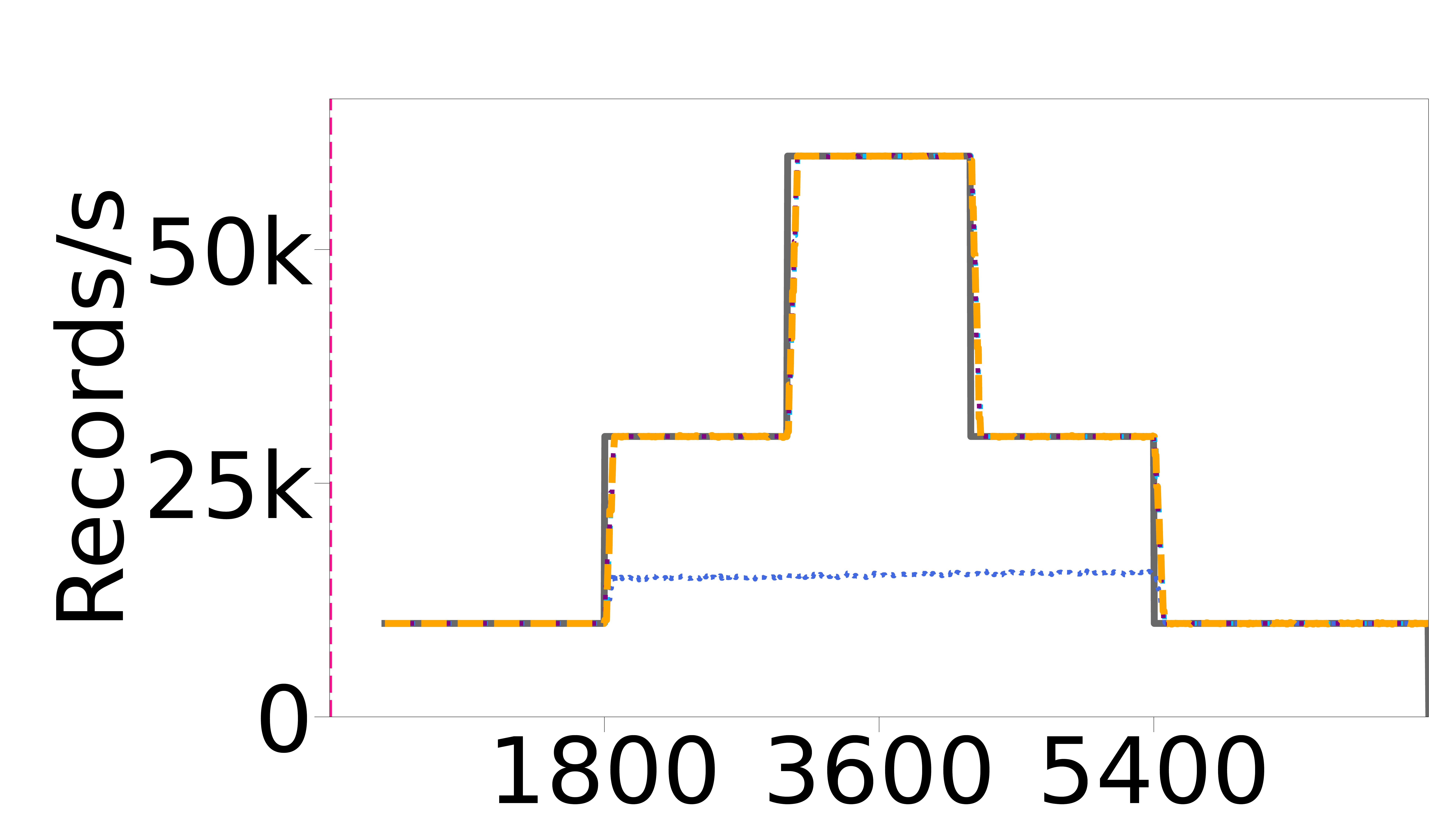}} &
    \adjustbox{valign=m}{\includegraphics[width=4.0cm,trim={0 0 0 180bp},clip]{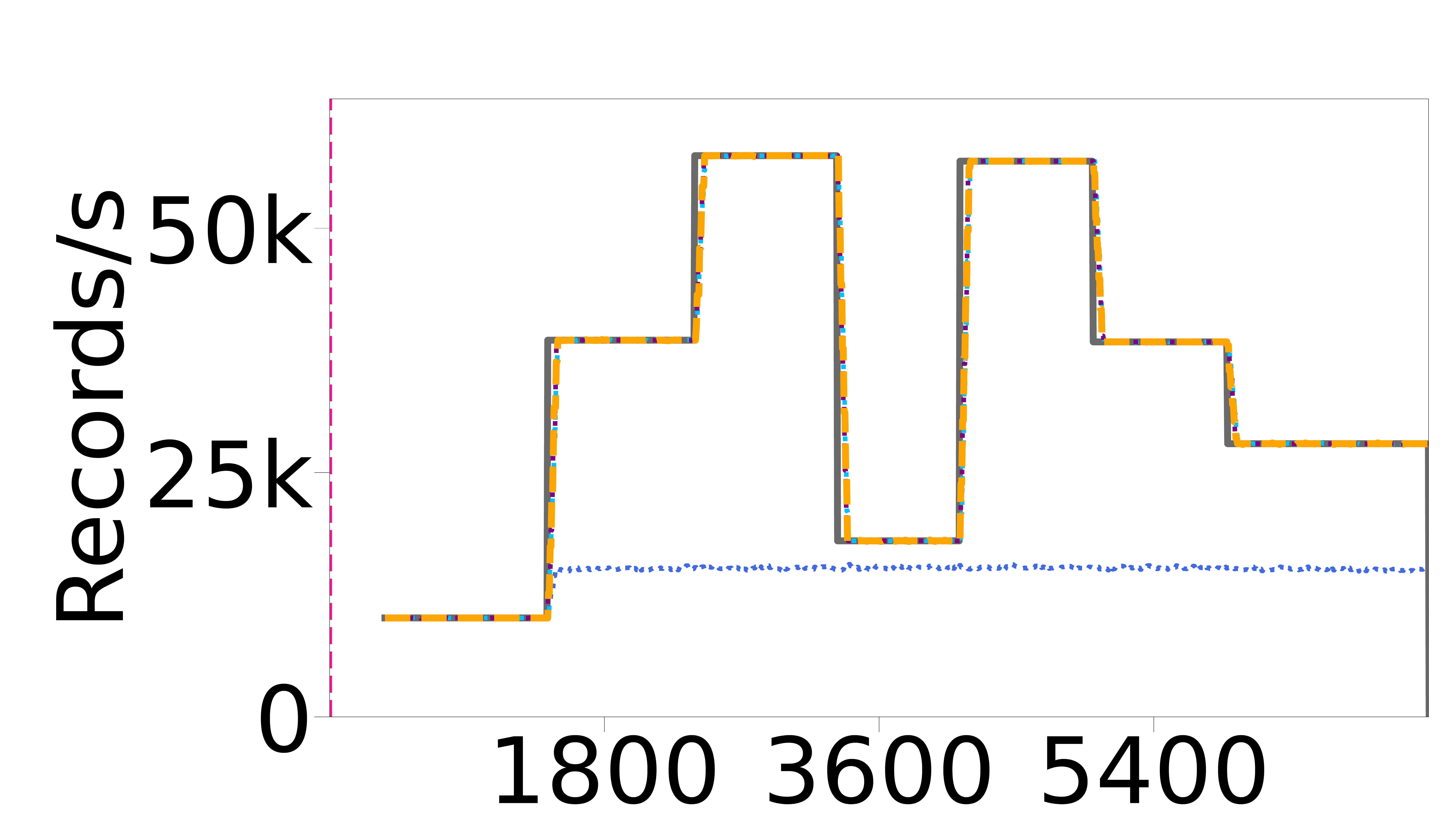}} \\

    \adjustbox{valign=m}{\rotatebox[origin=c]{90}{\textbf{Q5}}} &
    \adjustbox{valign=m}{\includegraphics[width=4.0cm,trim={0 0 0 180bp},clip]{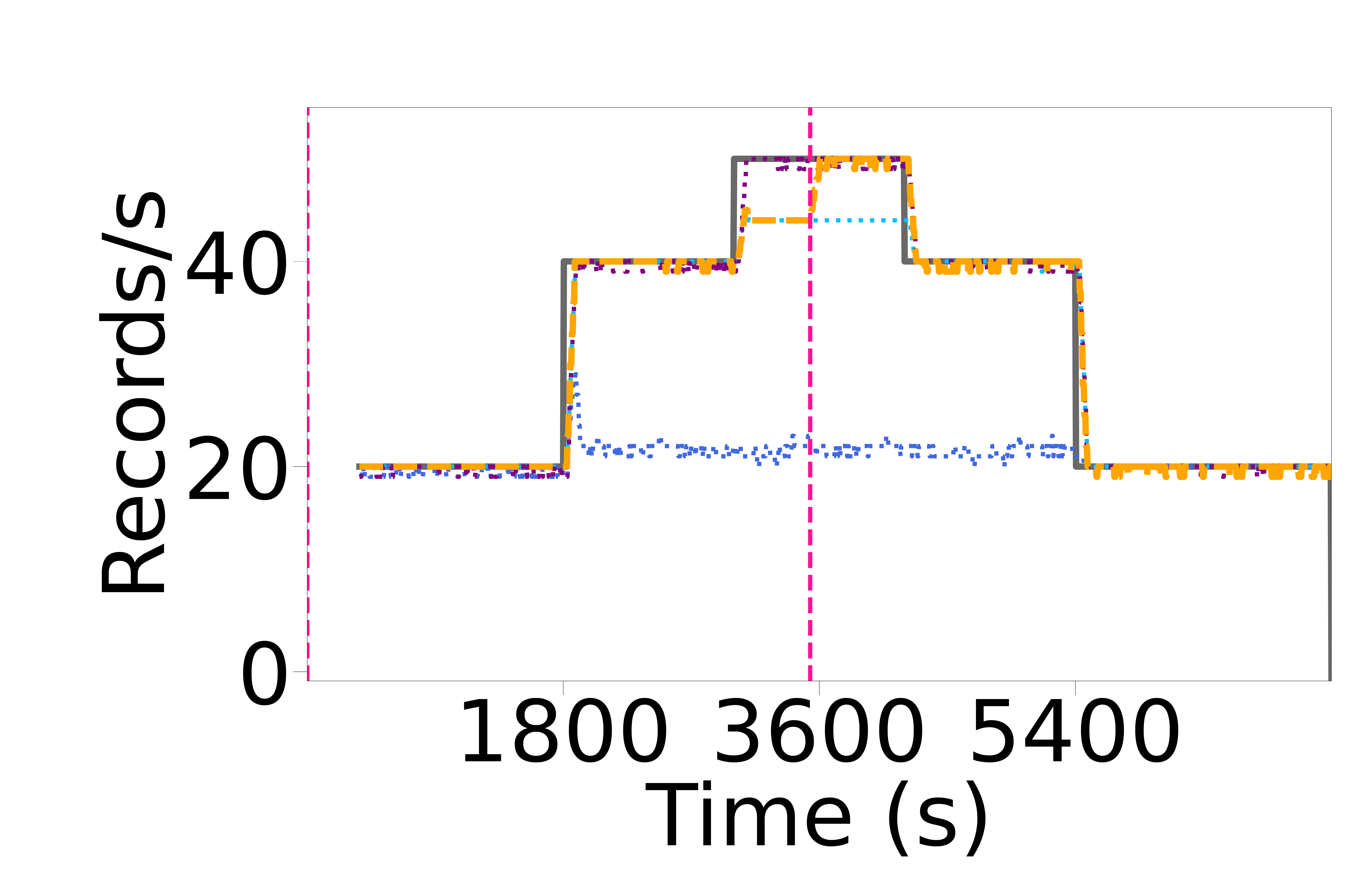}} &
    \adjustbox{valign=m}{\includegraphics[width=4.0cm,trim={0 0 0 180bp},clip]{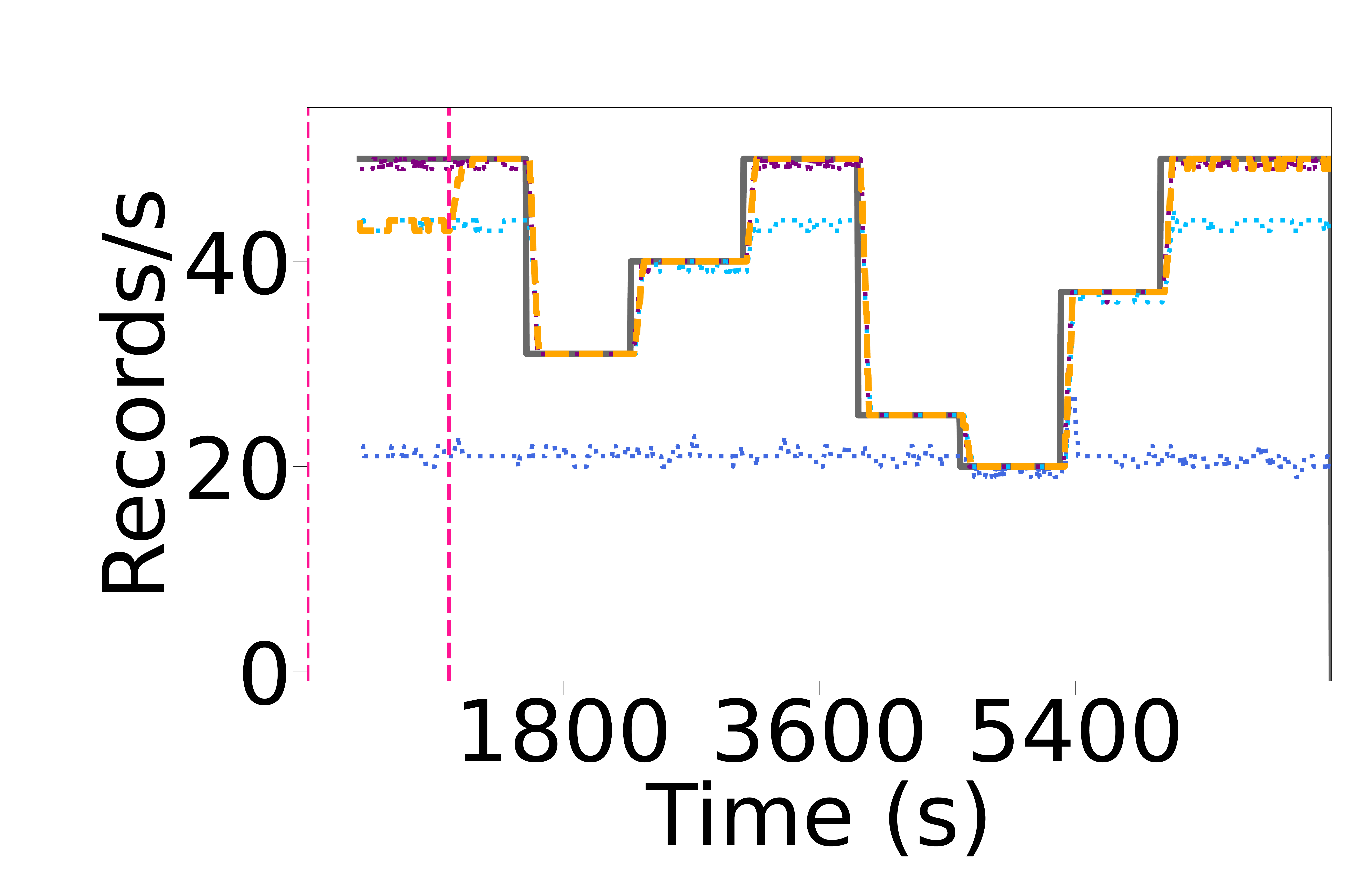}} \\
    
    & \adjustbox{valign=m}{\refstepcounter{subfigure}\label{fig:eval_vary_rate_a}  (a) gradual rate pattern} & \adjustbox{valign=m}{\refstepcounter{subfigure}\label{fig:eval_vary_rate_b}  (b) abrupt rate pattern} \\
\end{tabular}
\vspace{-3mm}
\caption{Comparison of \ours and baselines under variable input rate patterns on \textbf{Testbed-RPi}.} 
\label{fig:eval_vary_rate}
\end{figure}

\stitle{Impact of input rate.} To understand the impact of variable event rate, we evaluate all systems under two patterns: (i) \textbf{gradual pattern}, where we start from a low rate, then gradually increase it every 20 minutes, and finally, decrease to its initial value, and (ii) \textbf{abrupt pattern}, where both the change interval and the rate value are chosen at random. To ensure realistic conditions, rate targets and interval periods are set independently on different RPis; that is, variations across edge devices are not synchronized. All other operating conditions are fixed to their default values during the experiment.
Figure~\ref{fig:eval_vary_rate} shows the throughput over time.

\ours is the only system to meet the target rate across all queries and rate patterns.  For \textbf{Q1}, \textbf{Q2}, and \textbf{Q5}, which generate large event records, \ours shifts operators to the edge to reduce the network traffic. 
All-Edge is limited by the computational capacity of edge devices in all cases, while All-Cloud performs poorly for \textbf{Q2}, \textbf{Q3}, and \textbf{Q5}. %
Since All-Edge and All-Cloud execute the application as a single job, the performance of stateful operators affect the upstream stateless ones, incurring backpressure at the edge source. An example of this behavior is visible in \textbf{Q3}, where the window aggregation operator exhibits data skew, affecting the source rate of the All-Cloud strategy. In this instance, even though \ours has also placed all operators in the cloud, its decomposed query deployment shields the edge sources from backpressure, achieving the target rate when All-Cloud cannot.
CAPSys is also unaffected by data skew because we run it on \ours's runtime. However, its static policy fails to keep up for \textbf{Q1}, \textbf{Q2}, and \textbf{Q3}, during high rate periods.%

\begin{figure}[t!]
\centering
\begin{subfigure}[b]{0.8\linewidth}
\includegraphics[width=\linewidth]{figures/eval/shift/res_video6nw_0.pdf_legend.pdf}
    \vspace{-0.5cm}
\end{subfigure}

\setlength{\tabcolsep}{1pt}
\renewcommand{\arraystretch}{1.5}

\begin{tabular}{ccc}
    \adjustbox{valign=m}{\rotatebox[origin=c]{90}{\textbf{Q1}}} &
    \adjustbox{valign=m}{\includegraphics[width=3.8cm,trim={0 0 0 180bp},clip]{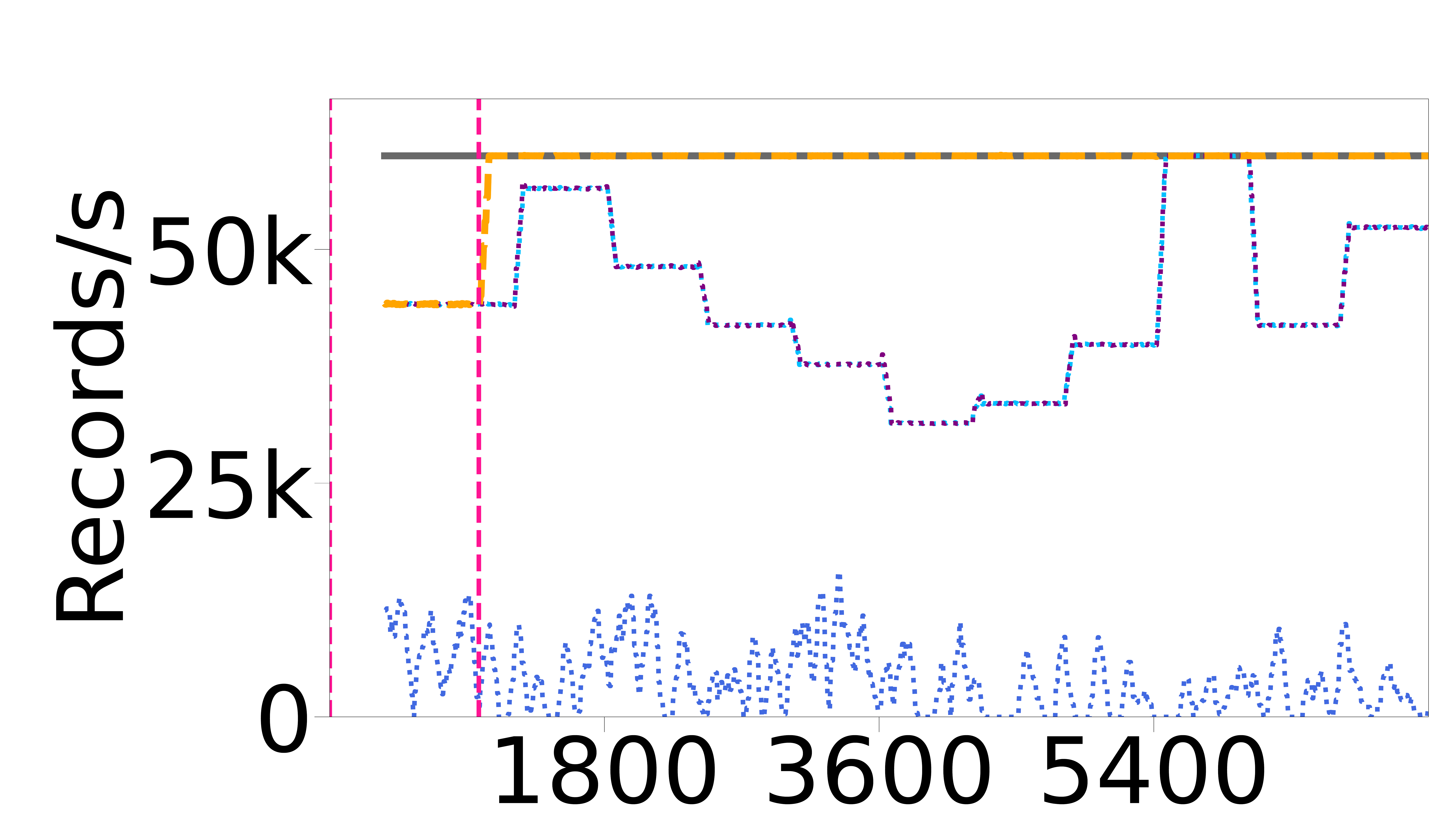}} &
    \adjustbox{valign=m}{\includegraphics[width=3.8cm,trim={0 0 0 180bp},clip]{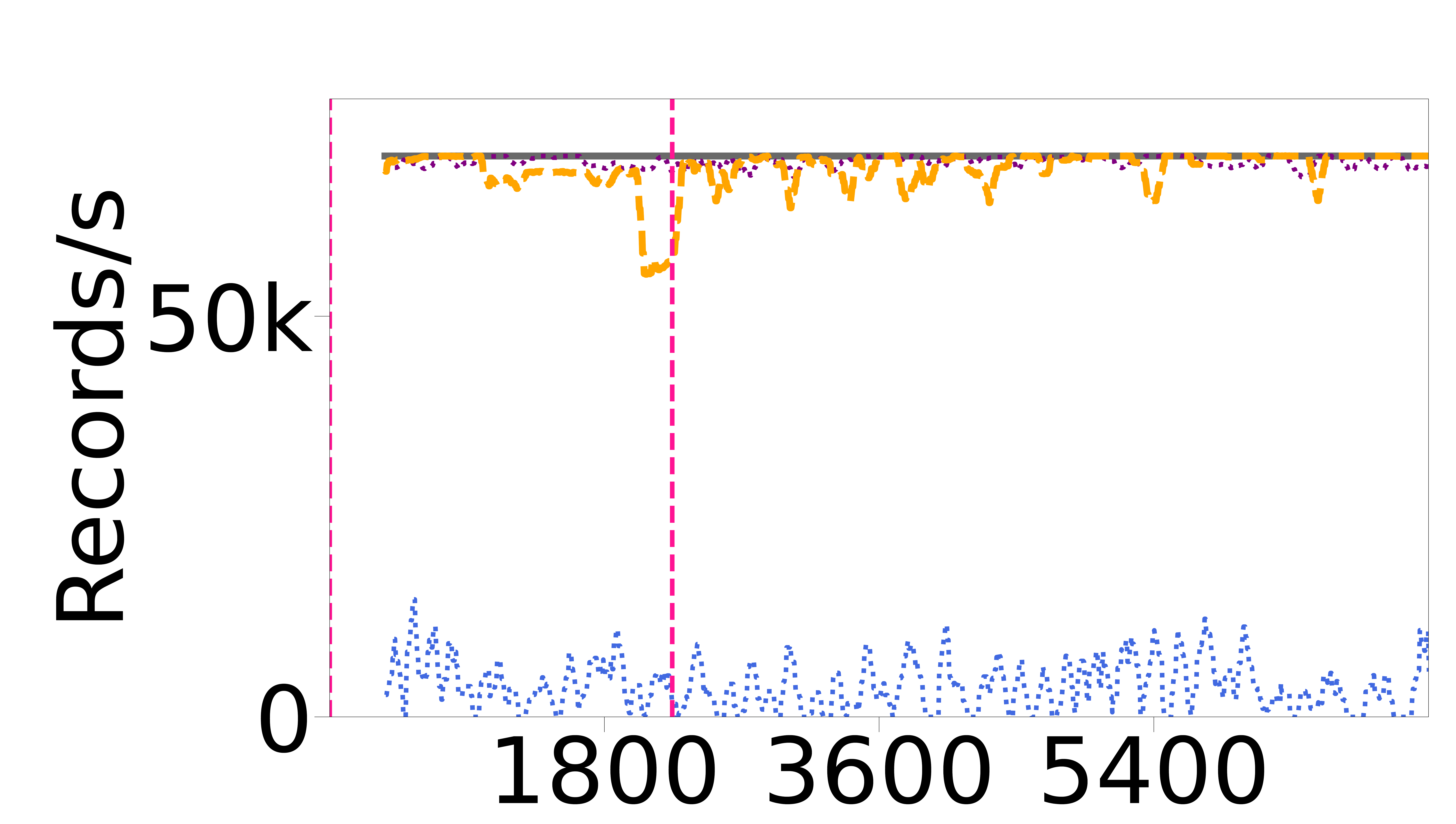}} \\

    \adjustbox{valign=m}{\rotatebox[origin=c]{90}{\textbf{Q2}}} &
    \adjustbox{valign=m}{\includegraphics[width=3.8cm,trim={0 0 0 180bp},clip]{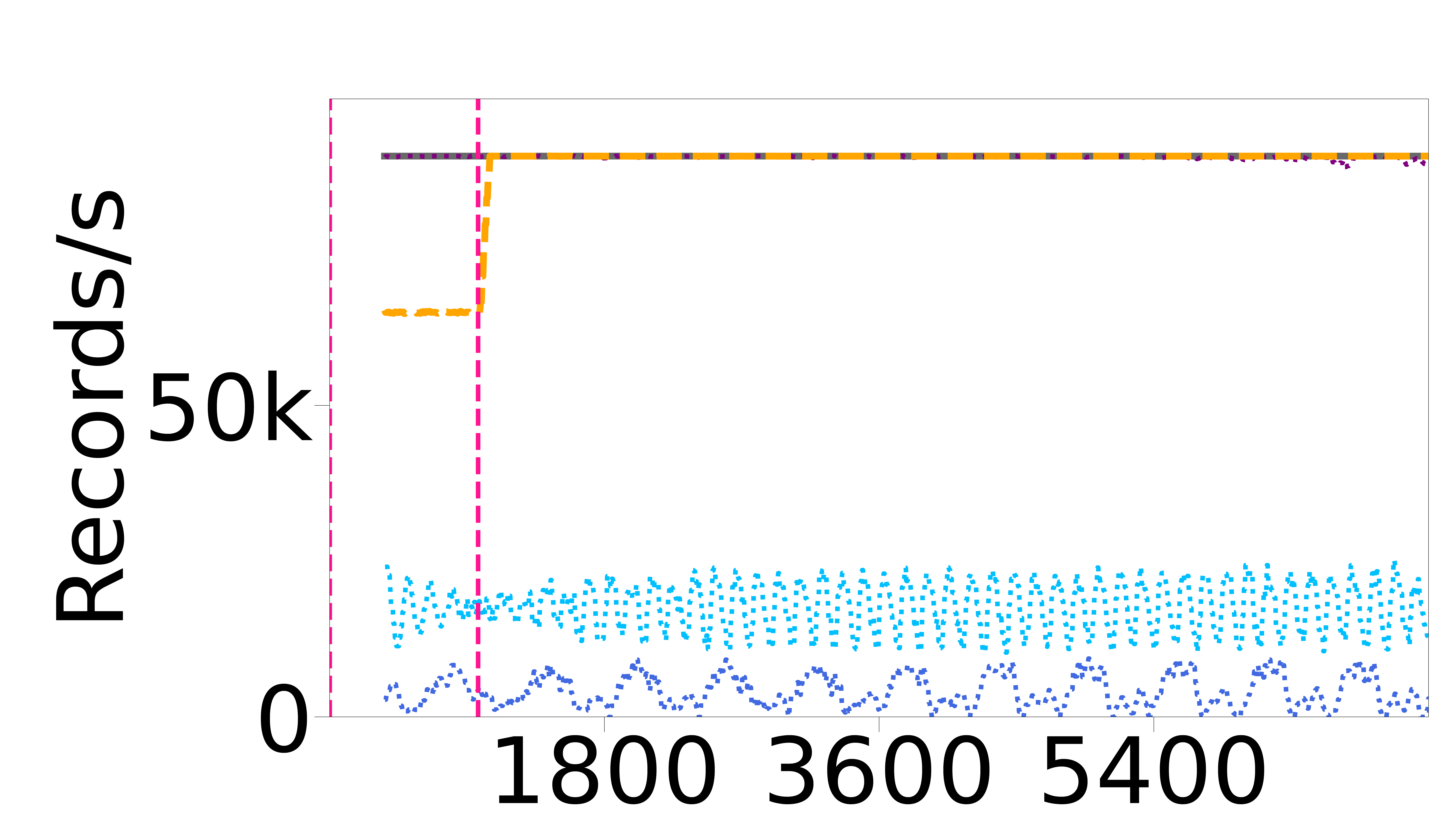}} &
    \adjustbox{valign=m}{\includegraphics[width=3.8cm,trim={0 0 0 180bp},clip]{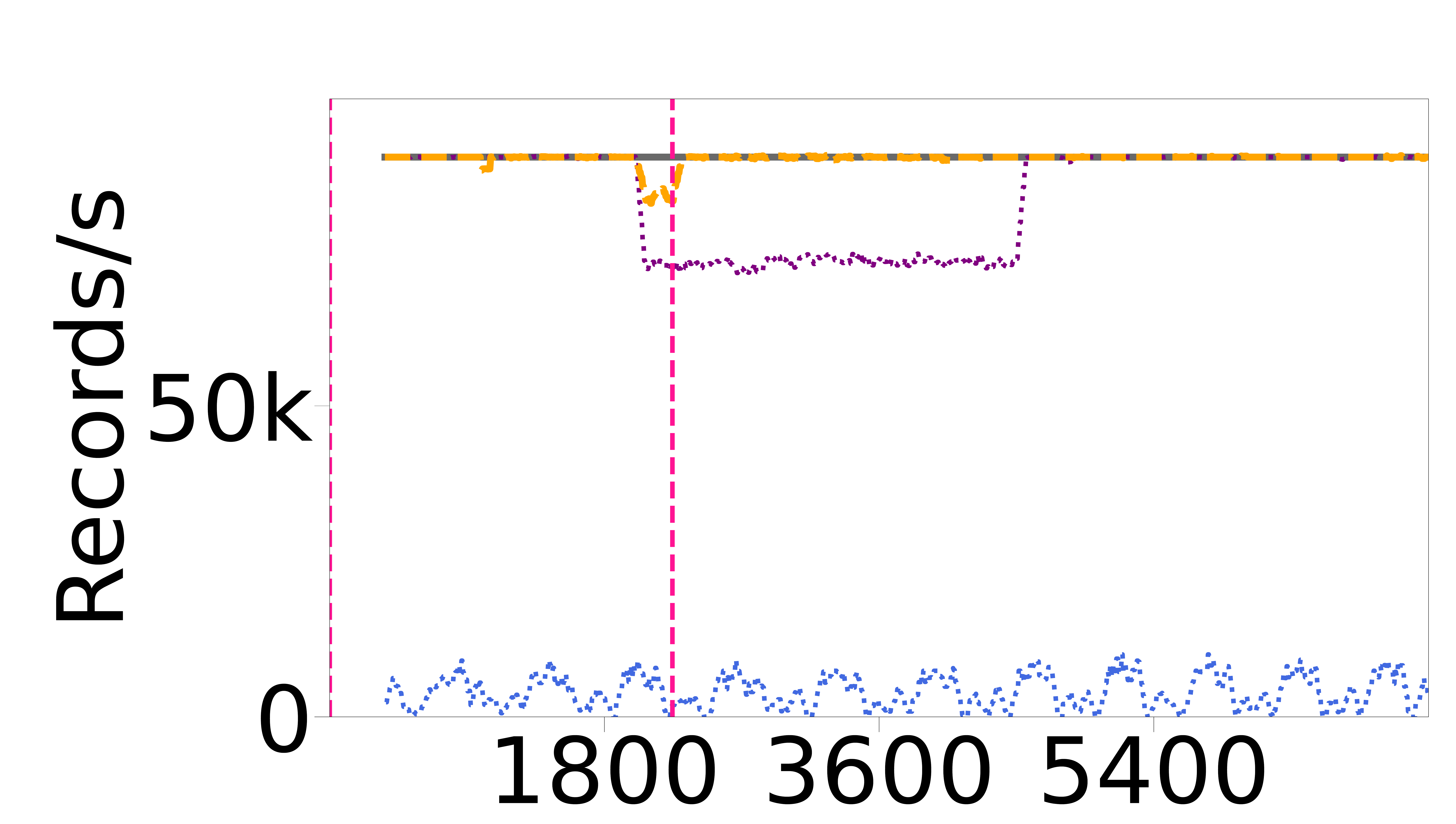}} \\

    \adjustbox{valign=m}{\rotatebox[origin=c]{90}{\textbf{Q3}}} &
    \adjustbox{valign=m}{\includegraphics[width=3.8cm,trim={0 0 0 180bp},clip]{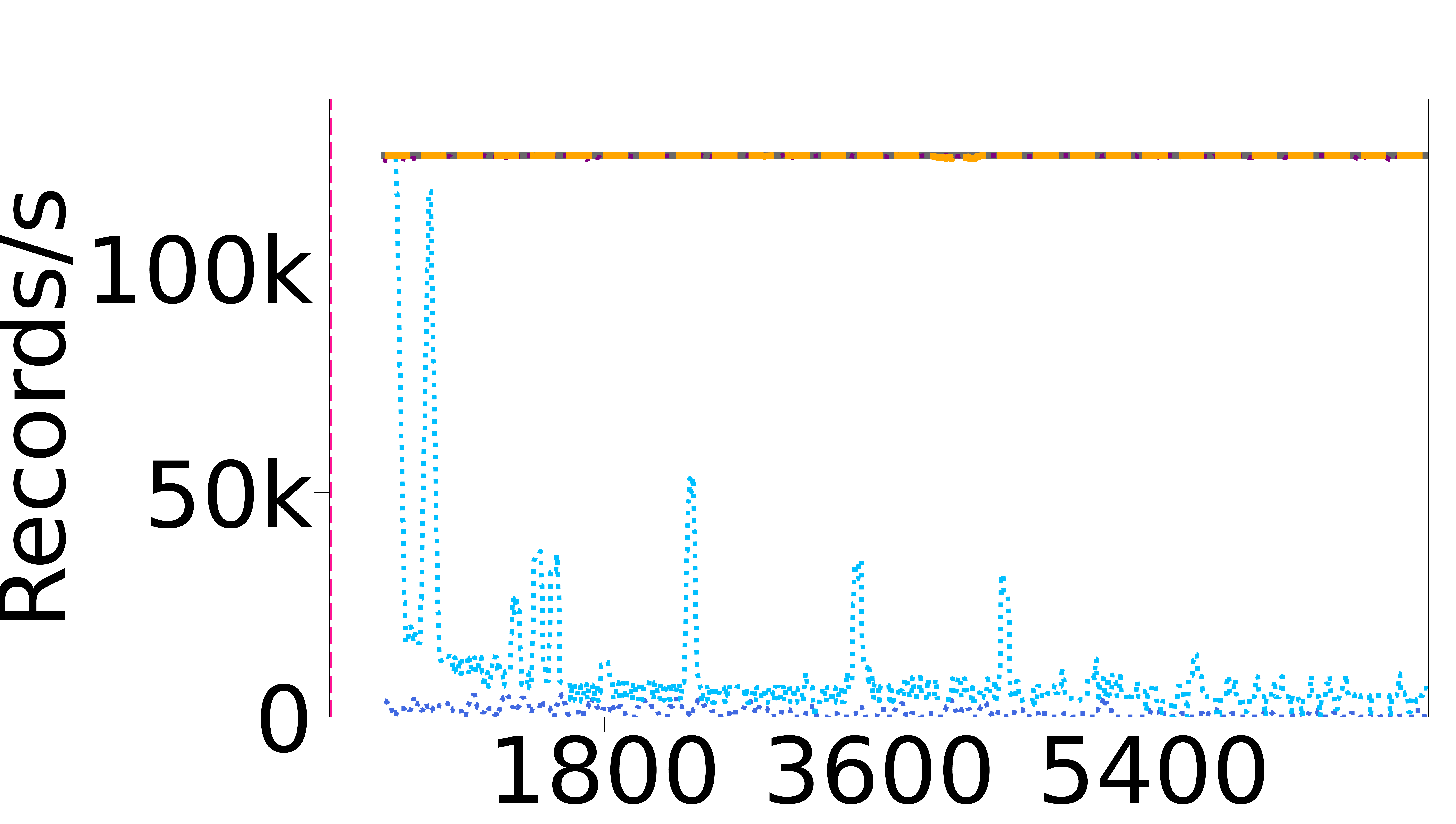}} &
    \adjustbox{valign=m}{\includegraphics[width=3.8cm,trim={0 0 0 180bp},clip]{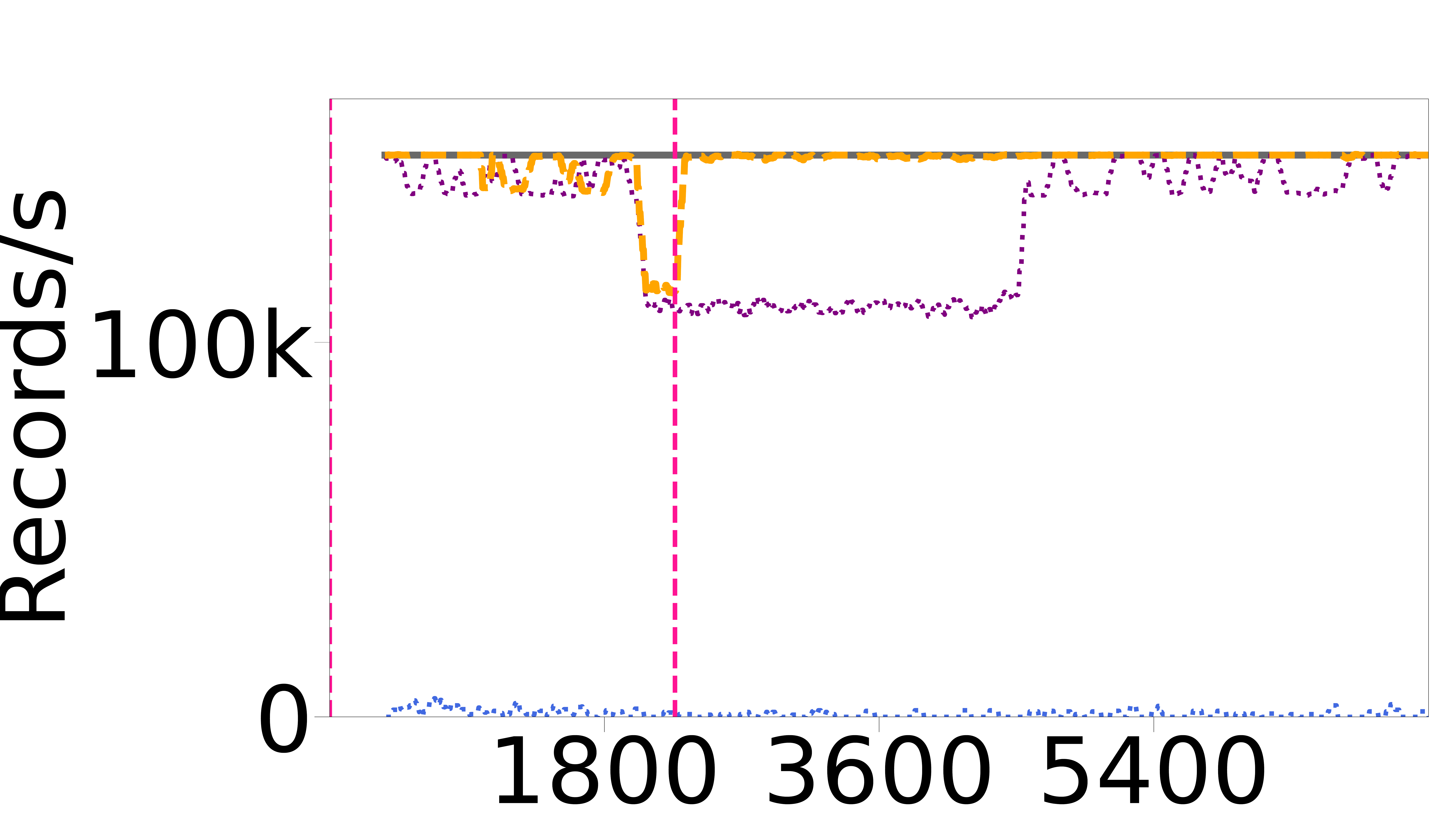}} \\

    \adjustbox{valign=m}{\rotatebox[origin=c]{90}{\textbf{Q4}}} &
    \adjustbox{valign=m}{\includegraphics[width=3.8cm,trim={0 0 0 180bp},clip]{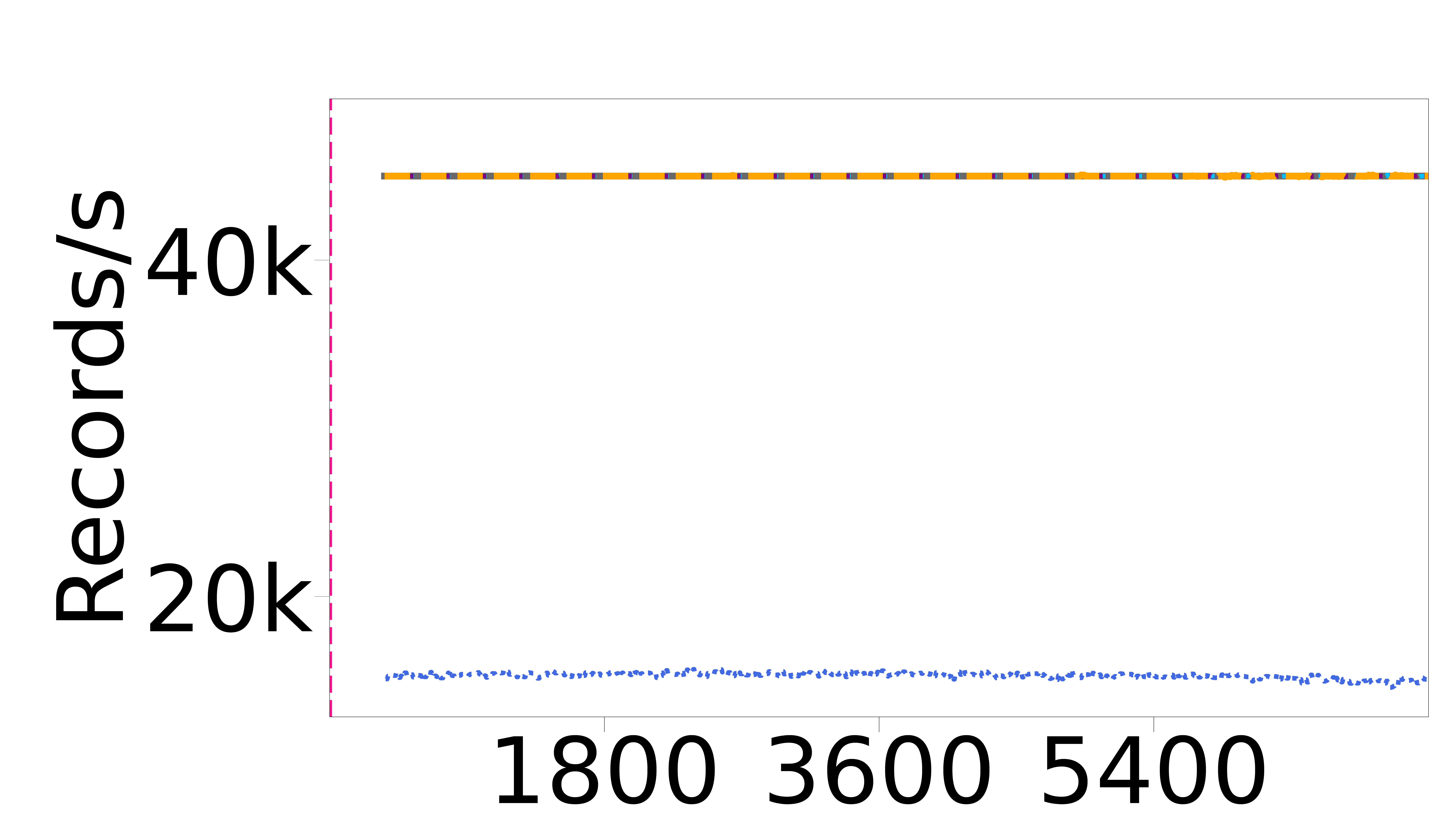}} &
    \adjustbox{valign=m}{\includegraphics[width=3.8cm,trim={0 0 0 180bp},clip]{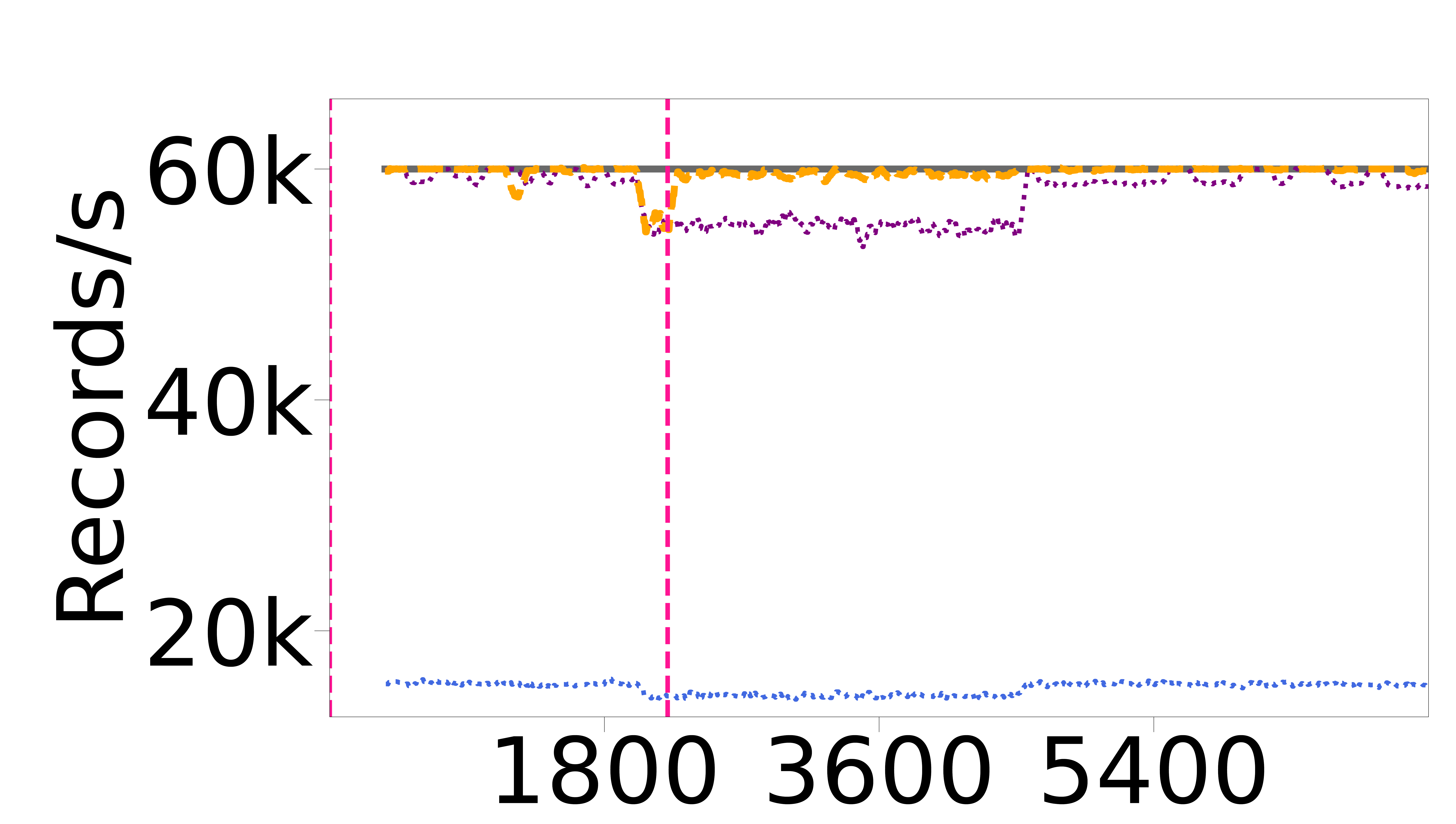}} \\

    \adjustbox{valign=m}{\rotatebox[origin=c]{90}{\textbf{Q5}}} &
    \adjustbox{valign=m}{\includegraphics[width=3.8cm,trim={0 0 0 180bp},clip]{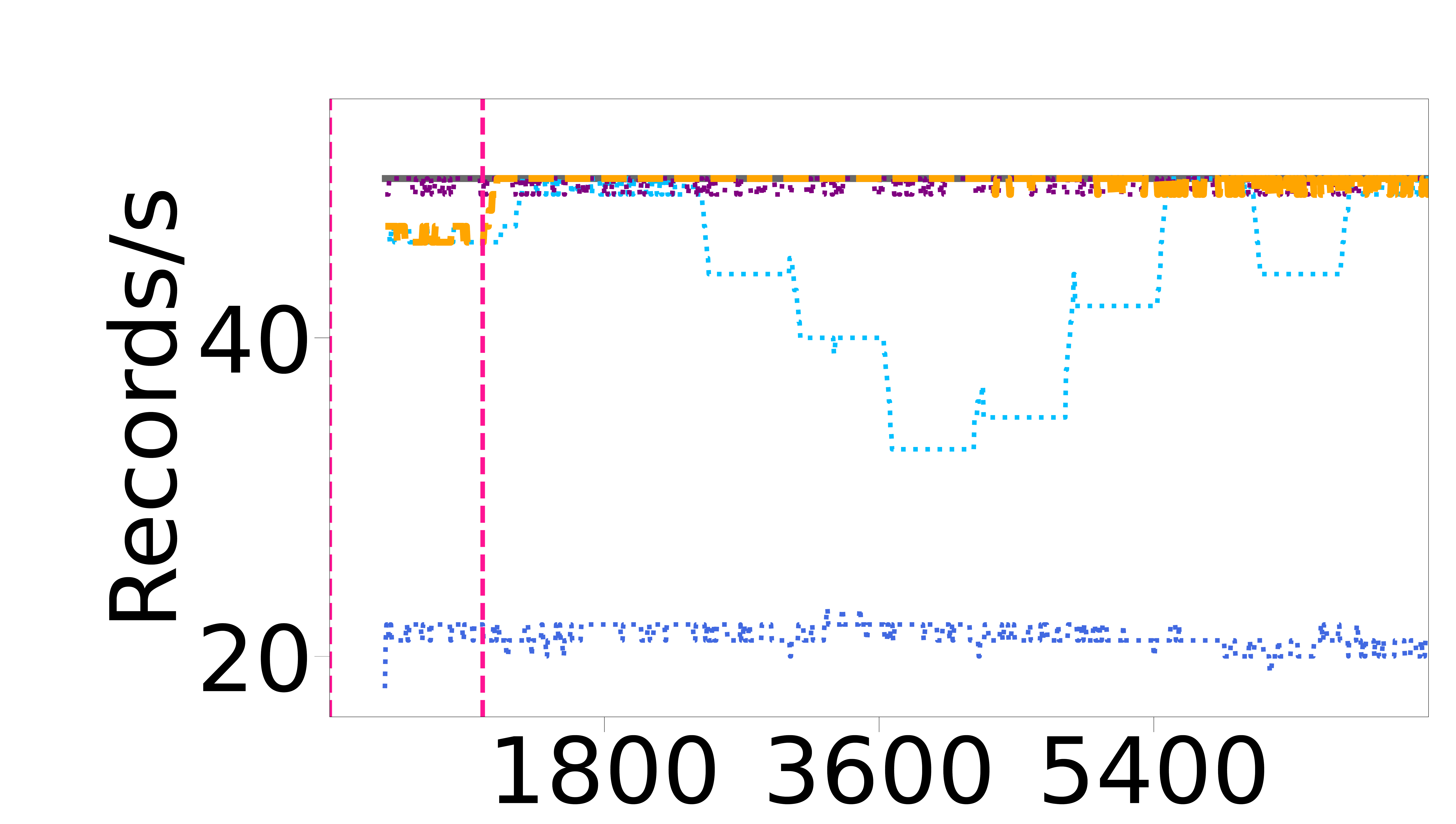}} &
    \adjustbox{valign=m}{\includegraphics[width=3.8cm,trim={0 0 0 180bp},clip]{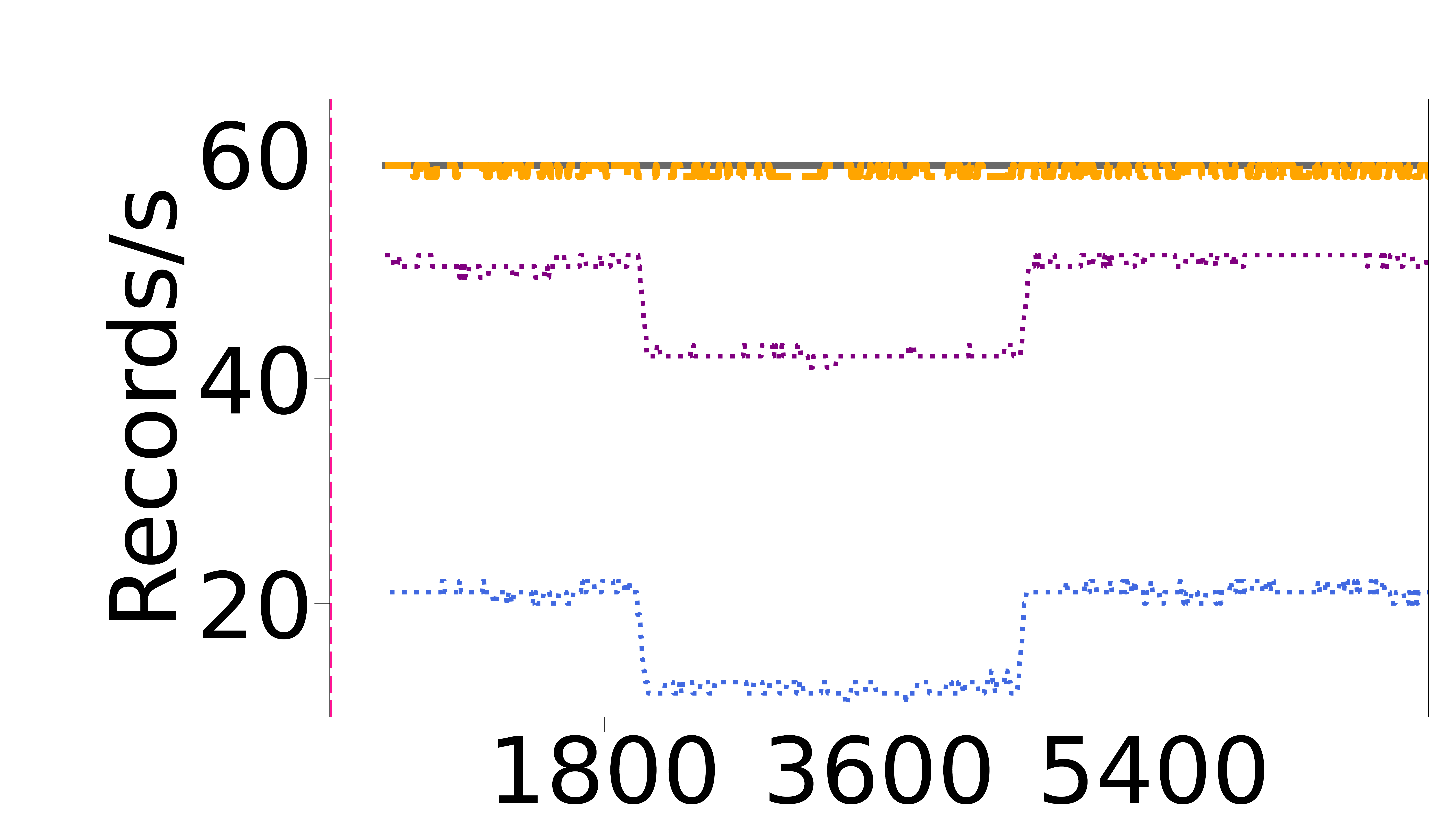}} \\
    
    \adjustbox{valign=m}{\rotatebox[origin=c]{90}{\parbox{2.5cm}{\centering Operating Conditions}}} &
    \adjustbox{valign=m}{\includegraphics[width=3.8cm]{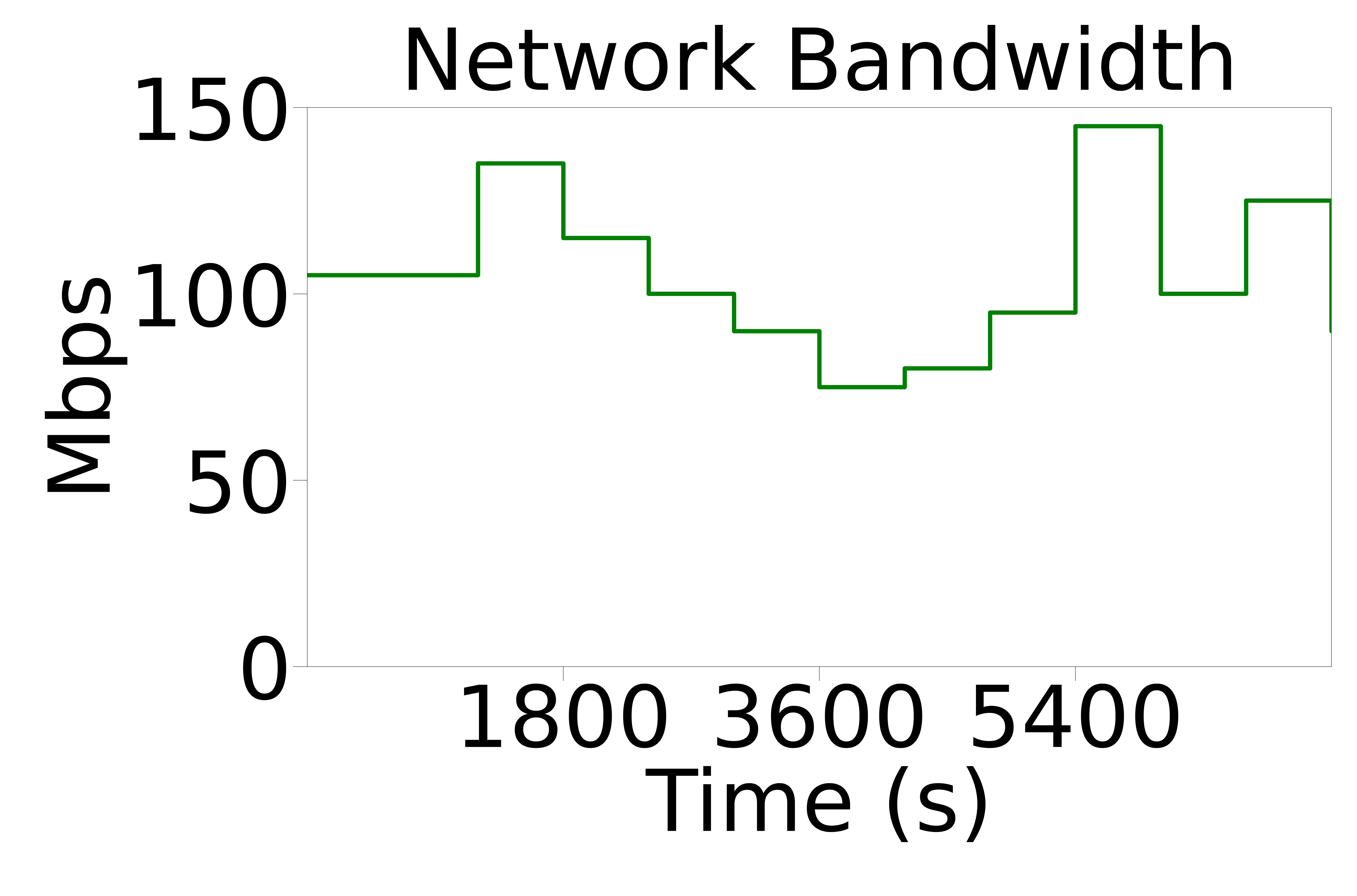}} &
    \adjustbox{valign=m}{\includegraphics[width=3.8cm]{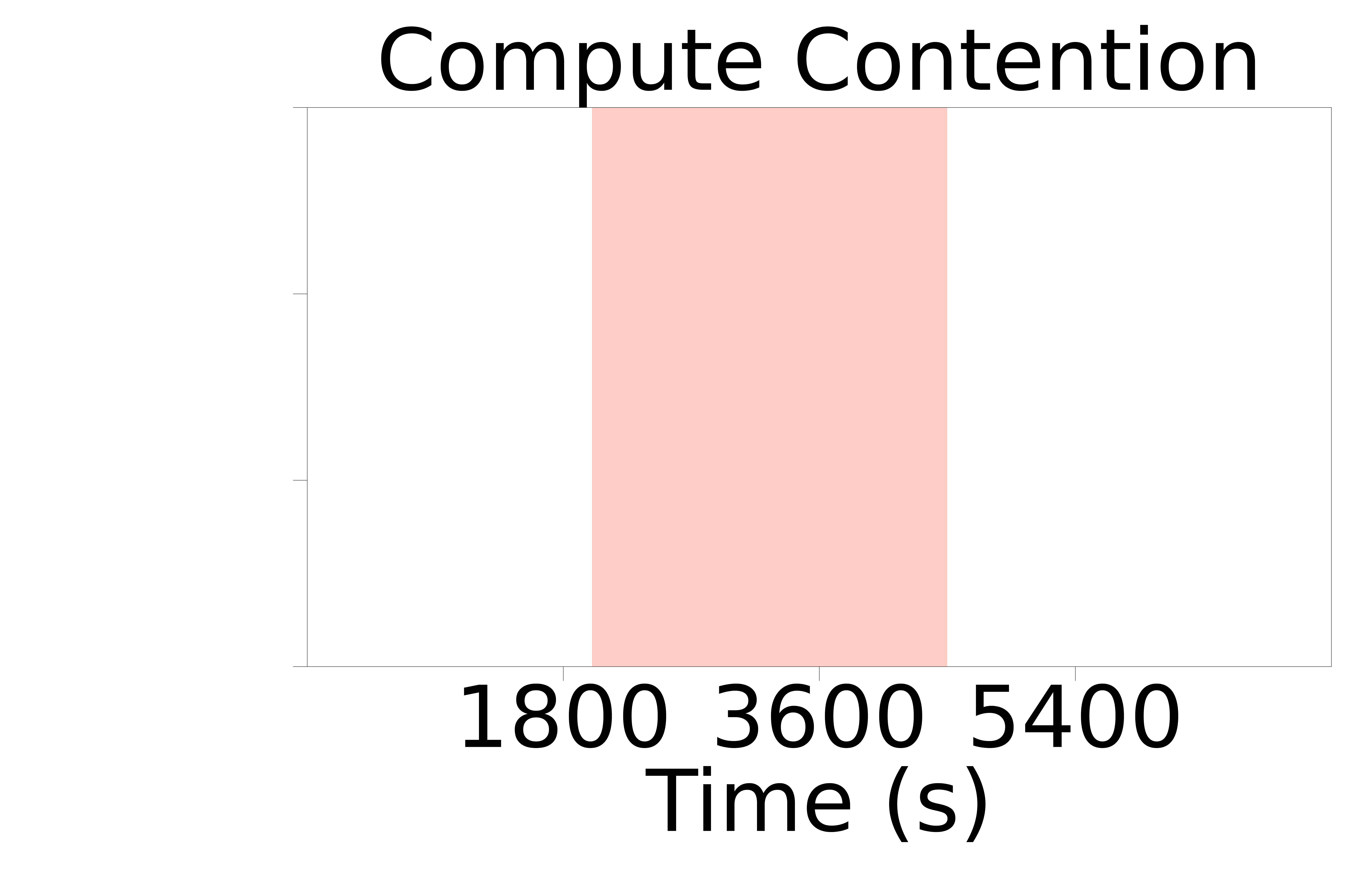}} \\

    & \adjustbox{valign=m}{\parbox{3.5cm}{\centering \refstepcounter{subfigure}\label{fig:eval_vary_env_a} (a) Network variations }} & 
    \adjustbox{valign=m}{\parbox{3.5cm}{\centering \refstepcounter{subfigure}\label{fig:eval_vary_env_b} (b) Compute contention }}  \\
\end{tabular}
\vspace{-3mm}
\caption{Comparison of \ours and baselines under network and compute contention variations on \textbf{Testbed-RPi}. The bottom plots show operating conditions over time.}
\label{fig:eval_vary_env}
\end{figure}

\stitle{Impact of network bandwidth.}
We now evaluate all systems under fixed target rate and variable network conditions. We again use Linux traffic control (\texttt{tc}) to shape the outbound bandwidth of RPis.
Figure~\ref{fig:eval_vary_env_a} shows the throughput for all queries and the operating conditions (bottom).

As bandwidth changes, only \ours can consistently sustain the target rate. In \textbf{Q1}, \textbf{Q2} and \textbf{Q5}, \ours adapts by shifting operators to the edge. On the other hand, \textbf{Q3} and \textbf{Q4} are not network-intensive, requiring no shifts during the experiment. All-Edge cannot sustain the rate for any of the queries, while All-Cloud's performance is clearly restricted by the bandwidth availability for \textbf{Q1} and \textbf{Q5}.
In the case of \textbf{Q2} and \textbf{Q3}, backpressure from stateful operators propagates upstream and affects the source operator at the edge. 
CAPSys fails to sustain the throughput for \textbf{Q1}, since it places all operators on the cloud. In contrast, \ours shifts the first operator to the edge, reducing the bandwidth usage by filtering out data early.

\stitle{Impact of compute contention.}
To evaluate the impact of imposing compute contention at the edge, we run en experiment with the target rate fixed and set the edge-to-cloud bandwidth to 200 Mbps, to ensure the network will not become a bottleneck. To observe if \ours can effectively react to compute contention, the initial deployment places the first operator at the edge.
We do not compare with the All-Cloud baseline, as it is unaffected by compute availability at the edge. Figure~\ref{fig:eval_vary_env_b} shows the results for \ours, CAPSys, and All-Edge. 

Across all applications, only \ours can always sustain the target rate. During periods of contention, \ours quickly detects the problem and shifts operators to the cloud as necessary. In the case of \textbf{Q5}, \ours's initial policy is sufficient even under contention, hence, no shift occurs. 
All-Edge performs poorly across all queries, while CAPSys performs well only for \textbf{Q1}, when its strategy runs all operators on the cloud. 
In all other applications, CAPSys has placed several operators at the edge and its performance is affected by contention. 

\subsection{\ours scalability}\label{sec:exp-100vm}

\begin{figure}[t!]
\centering
\vspace{2mm}
\includegraphics[width=\linewidth]{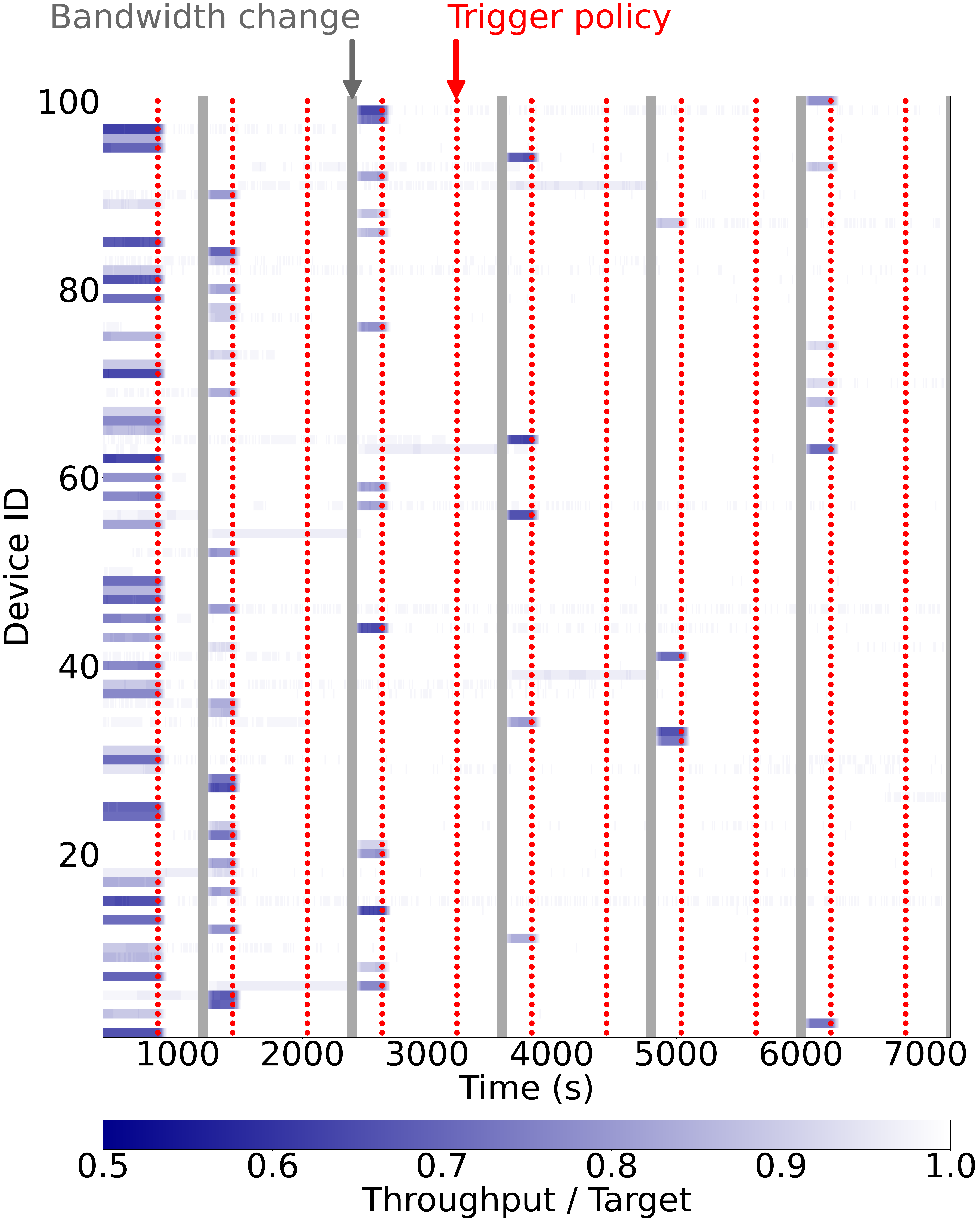} 
\vspace{-6mm}
\caption{\ours's scalability running \textbf{Q5} on \textbf{Testbed-100VMs}. Each point on the y-axis corresponds to an edge device. White color indicates the device can sustain the rate, while dark color indicates backpressure.}
\label{fig:eval_100vm}
\end{figure}

We evaluate \ours's scalability running \textbf{Q5} on \textbf{Testbed-100VMs}. The outbound bandwidth of each edge device follows an individually-set variable pattern, with values between 80 and 200 Mbps, chosen at random every 20min.
Figure~\ref{fig:eval_100vm} summarizes the performance across 100 devices, plotting the ratio of throughput over the target rate in a heatmap. 
Each point on the y-axis corresponds to the throughput-to-target ratio of an edge device, while time is shown on the x-axis. 
Bandwidth changes and policy triggers are shown with grey and red vertical lines, respectively. 

The initial deployment runs all operators on the cloud, and as a result, 39\% of the devices cannot achieve the target.
When the policy triggers at 900s, \ours shifts operators to the edge. 
Afterwards, all devices can sustain the target with an average throughput-to-target ratio of 99.8\%
At 1220s, the bandwidth changes, causing backpressure on 23\% of the devices. The policy reconfigures routing again at 1500s, ensuring all devices meet the target. This pattern repeats a few times during the experiment. In all cases, \ours converges to a backpressure-free configuration across all devices.

\section{Related Work}

\stitle{Edge stream processing engines.} 
Many recent works~\cite{liu2021dart,AgileDART,zeuch2020nebulastream,NebulaStream25,fu2019edgewise} propose novel system designs for the edge setting and aim to address challenges related to restricted resources and dynamic deployment conditions, such as frequent failures and churn. 
DART~\cite{liu2021dart} introduces a peer-to-peer system architecture that enables dynamic operators placement and scaling in volatile edge environments. EdgeWise~\cite{fu2019edgewise} proposes a congestion-aware scheduler to optimize resource usage of a single device.  
Frontier~\cite{frontiervldb18} replicates the dataflow graph across devices for fault tolerance.
In contrast to these edge-centric systems, \ours explicitly targets the hybrid setting, where data is produced at the edge but requires analysis results for downstream applications  in the cloud. \ccc{Rather than building a clean-slate system, \ours reuses mature dataflow engines and adapts them to the hybrid setting.}

\stitle{Adaptive filtering and routing.}
\ccc{Early works have explored  adaptive query execution in single-node stream processing engines ~\cite{avnur2000eddies,chandrasekaran2003continuous,madden2002continuously,babu2004adaptive,olston2003adaptive}. Recent research has revisited the concept of adaptive query processing in the context of edge computing. }
For example, Shepherd~\cite{shepherd22}'s late binding routing mechanism reroutes data tuples between operators and provides seamless adaptation with minimal downtime. Falcon~\cite{falcon24} introduces a live key migration protocol to support stateful operator reconfiguration in such environments, and Frontier~\cite{frontiervldb18}'s per-batch routing mechanism dynamically forwards events to downstream operator replicas.  
\ours takes these ideas a step further by making adaptive routing transparent to the application.  

\stitle{Adaptive placement and scheduling.} 
Various prior works study dynamic scheduling and reconfiguration for stream processing~\cite{Cameo,chi_vldb,Drizzle,capsys,trisk}.
The systems closest to \ours, CAPSys~\cite{capsys}, Trisk~\cite{trisk}, and WASP~\cite{jonathan2020wasp}, propose adaptive task placement policies to handle network and compute bottlenecks. However, they still require at least partial job stopping, resulting in downtime, backlog accumulation, and potential source-side data loss during reconfiguration. In addition, homogeneous resource assumptions, like those in CAPSys, render existing placement policies ineffective in the hybrid cloud setting. 
\ccc{APEROL~\cite{banelas2025aperol} efficiently searches large placement plan spaces across heterogeneous edge devices and execution platforms, but assumes that workflows must stop ingesting and processing data during operator migration.}
\ccc{In contrast, \ours automatically rewrites queries into independent jobs and creates edge and cloud replicas for stateless preprocessing. It switches routing between these replicas without pausing execution while preserving watermark coordination and event-time semantics.}

\section{Conclusion }\label{sec:conlcusion}
We presented \ours, an adaptive stream processing system that extends datacenter-centric dataflow engines to support hybrid edge-cloud deployments without requiring application changes. \ours enables dynamic offloading of computations to the edge, automatically adjusting to operational and workload dynamics. \ours allows users to rely on familiar event-time and fault-tolerance semantics of the underlying dataflow engine, ensuring correctness and no data loss.

\bibliographystyle{plain}
\bibliography{reference}

\clearpage
\appendix
\section*{Supplementary material}
\section{Implementation details}\label{sec:sup-implementation}

\stitle{Query re-writing}.
 We implement the re-writer using \textit{JavaParser}\cite{javaparser} and the \textit{Spoon} library\cite{spoon_library}.
Users submit Flink applications directly to \ours without performing any manual query splitting, resource management, or task placement. The re-writer parses Flink \texttt{DataStream} programs and automatically generates the query segments and adaptive routing paths. To ensure compatibility with \ours, user programs must conform to a minimal set of structural conventions: (i) the program must include a valid source operator, (ii) mobile operators must be defined as \texttt{DataStream} variables without chaining methods (e.g., \texttt{name()}, \texttt{setParallelism()}), and (iii) all user-defined objects must expose JSON-serializable fields via public getters and setters. 

\stitle{Watermark propagation.}
To enable correct watermark propagation across segments and availability under failures, \ours injects heartbeat records into Kafka, each carrying a watermark timestamp and a partition tag. On the stateful segment side, we implement a custom watermark strategy that extracts these heartbeats to compute the global watermark. Because heartbeats are ordinary Kafka records, they inherit at-least-once delivery, however, duplicates are harmless since the we only advance watermarks through a monotonic max-min reduction. To prevent any single slow partition from stalling the pipeline, we rely on Flink’s idleness mechanism~\cite{flink_idleness}. Partitions that stop producing heartbeats are marked as idle and excluded from the watermark calculation until they recover.

\stitle{Routing controller.}
\ours’s controller periodically gathers metrics from Flink Job Managers and evaluates the routing policy in real time, without relying on any offline profiling.
Policy decisions are independently evaluated for each segment, hence, the routing controller can be trivially scaled to support large deployments, as we show in \S\ref{sec:exp-100vm}. 
Routing decisions are sent via HTTP requests and each $CP$ operator listens for them on a dedicated port. 
Operators maintain a routing flag that dynamically directs incoming events to either its main output (the cloud path) or its side output (the edge path).

 \stitle{Hybrid scheduler.}
To enforce a deterministic task placement plan on edge and cloud resource at runtime, we implemented a custom Flink scheduler that deterministically assigns tasks to edge/cloud workers according to a placement plan. 
The plan is encoded as a static mapping where each task is pinned to a specific location specified  by the Task Manager IP address. 
To this end, we extended Flink’s \texttt{ResourceProfile} class to specify the target scheduling location per task. We also implemented a custom \texttt{Slot\allowbreak Matching\allowbreak Strategy} that overrides the default slot allocation logic to match the task’s specified location with the available slots on the corresponding Task Manager, ensuring that edge and cloud replicas are deployed exactly where the placement plan dictates.

\stitle{Metrics collector.}
The metrics collector periodically pulls Flink metrics from the JobManagers during runtime. It records the useful processing time, observed and true input and output rates per task \cite{ds2}.
The routing controller pulls these metrics on demand whenever it needs to recompute a routing plan.

\stitle{Kafka integration}.
In \ours, each query uses a dedicated Kafka topic and the results of each edge device are directed to a distinct partition within this topic.
The cloud-based segment then consumes data from this same topic using a \texttt{KafkaSource} operator, reading from all partitions. 

\section{Additional experiments}

\subsection{\ours under failures }
\label{sec:exp-failures}

\begin{figure}[t]
    \centering
    \includegraphics[width=\linewidth]{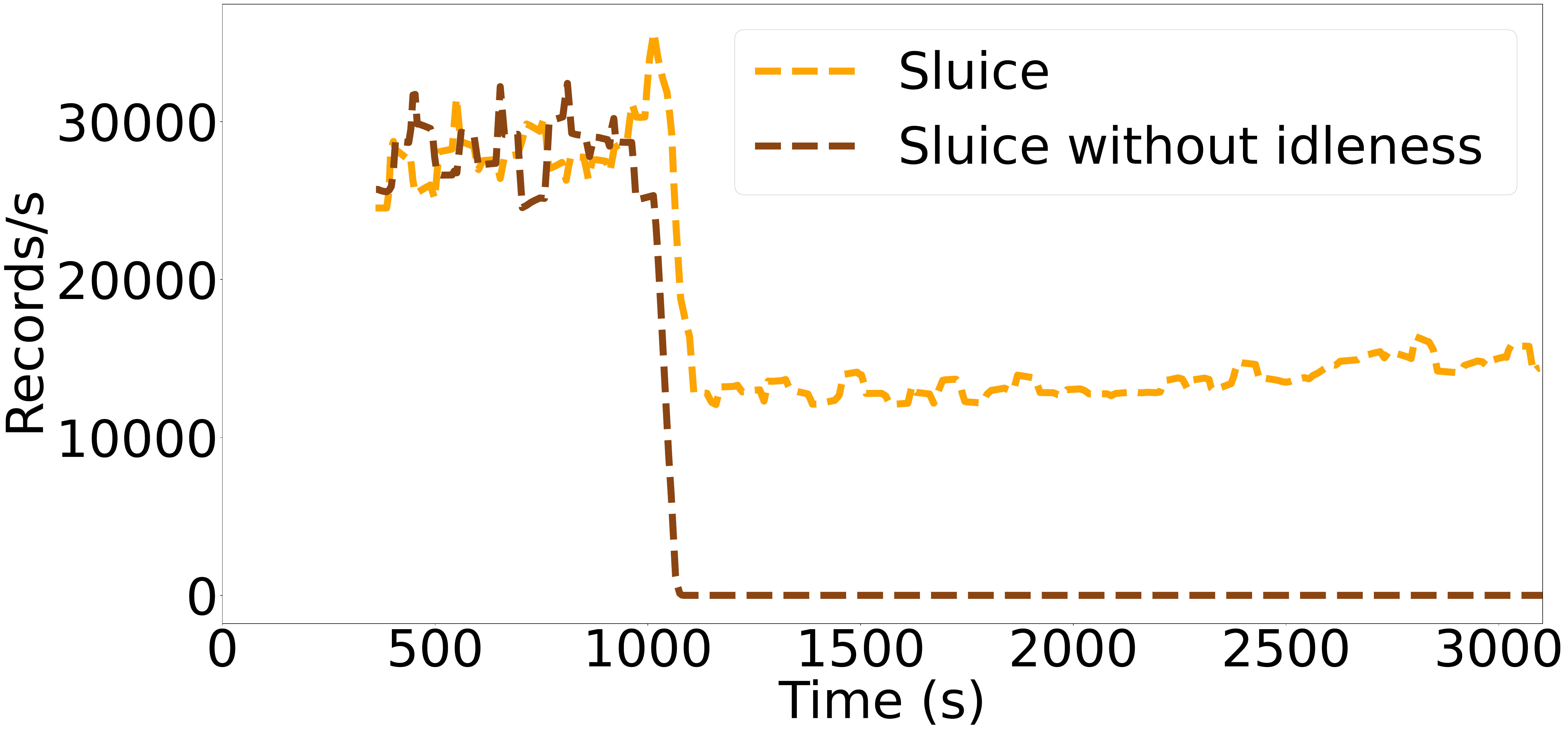}
    \vspace{-0.35cm}
    \caption[center]{Comparison of \ours with and without watermark inactivity tolerance mechanism}
    \label{fig:idleness}
\end{figure}

This experiment measures how the watermark inactivity tolerance mechanism described in Section~\ref{sec:WatermarkToleranceMechanism} affects the availability and event time progress. 
We run \textbf{Q1} on \textbf{Testbed-RPi} with a fixed input rate for 1 hour with a 5min warm-up.
We compare the \ours mechanism with and without the watermark inactivity tolerance mechanism. In \ours without idleness, we keep the \ours mechanism unchanged except to disable the watermark inactivity tolerance mechanism in the cloud source.
At time 900s, we cancel the stateless query segment on one of the 5 edge devices to simulate failure happening, and leave the remaining 4 edge devices running continuously. 

Figure~\ref{fig:idleness} shows the sliding average output throughput of the window operator on the stateful query segment, where each data point is calculated as the average throughput over the last 10 samples. We use this sliding averaging to smooth throughput spikes caused by the query logic itself, allowing us to isolate and focus on the effect of the watermark inactivity tolerance mechanism. 

The throughput is stable during the initial phase. 
When one of the edge devices fails at 900s, for \ours, the throughput briefly spikes and then becomes stable again once the tolerance period elapses, because now the global watermark advances using the minimum over the remaining 4 live broker partitions. The throughput is lower than before, since we have fewer event producers.
However, for \ours without idleness, the global watermark is blocked by the silent partition, whose corresponding stateless query segment has already been canceled. And the throughput remains 0 until the end. 
From this experiment, we show that the watermark inactivity tolerance mechanism prevents system-wide downtime and preserves availability for aggregative analytics under partial edge devices failures.

\section{Additional results}\label{sec:sup-allfig}

We report per-device performance for the remaining 4 RPi devices for all queries in Figures~\ref{fig:sup_vary2222_rpi1} - \ref{fig:sup_vary_env_rpi4}, complementing the main text where we show results for a single device per query in Section \ref{sec:exp-vary2222} and \ref{sec:exp-vary1}.
The additional devices show similar results as the one device shown in the main sections. 
Note that for \textbf{Q3}, the input dataset exhibits data skew across different RPis. In RPi 2, 3, and 4, the input events are entirely filtered out by the mobile operators in the stateless query segment, which substantially reduces workload. 
Consequently, for \textbf{Q3}, the All-Cloud and All-Edge baselines achieve better performance on RPi 2,3,4 than on the other RPis.

\noindent\begin{minipage}{\columnwidth}
\centering
\begin{minipage}[b]{0.8\linewidth}
\centering
\includegraphics[width=\linewidth]{figures/eval/shift2222/res_video6nw2_0.pdf_legend.pdf}
    \vspace{-0.4cm}
\end{minipage}\vspace{-2mm}
\setlength{\tabcolsep}{1pt}
\renewcommand{\arraystretch}{1.5}
\begin{tabular}{ccc}
    \adjustbox{valign=m}{\rotatebox[origin=c]{90}{\textbf{Q1}}} &
    \adjustbox{valign=m}{\includegraphics[width=8.2cm]{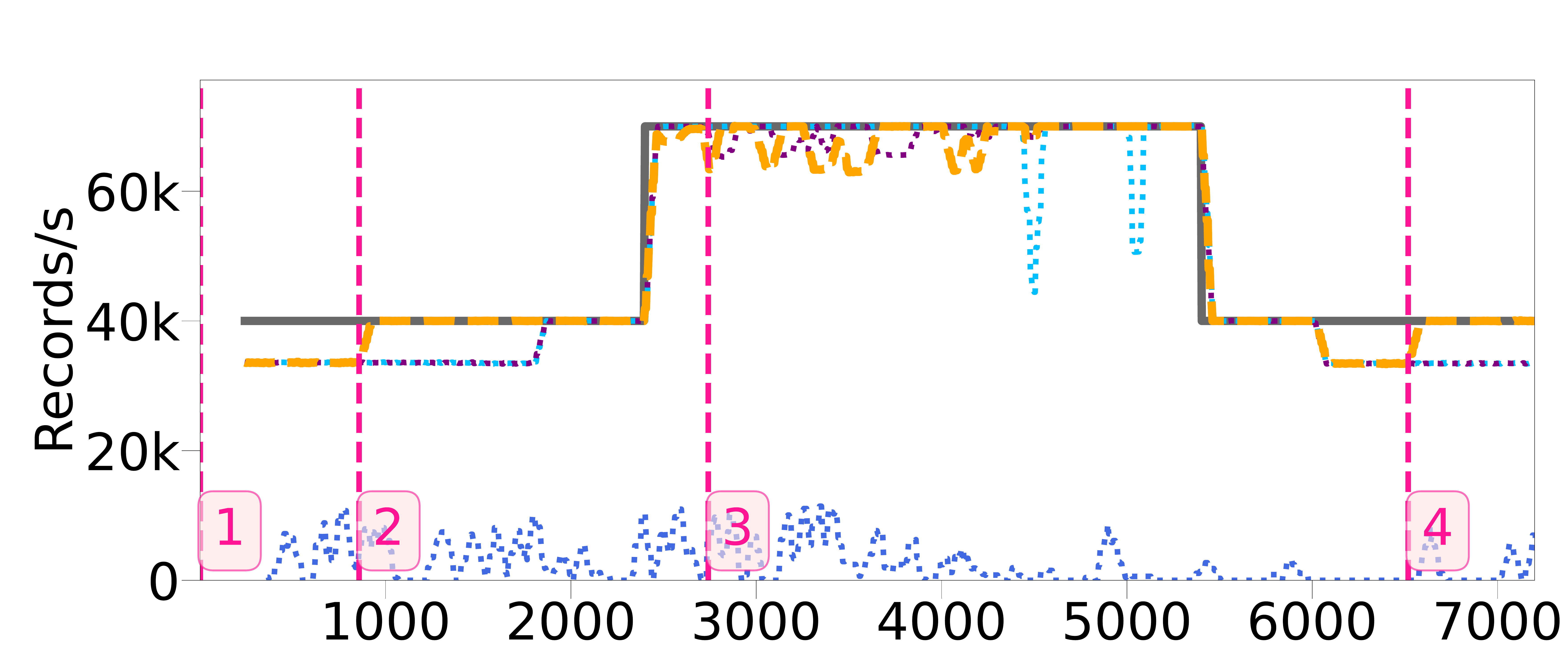}} \\
    \adjustbox{valign=m}{\rotatebox[origin=c]{90}{\textbf{Q5}}} &
    \adjustbox{valign=m}{\includegraphics[width=8.2cm]{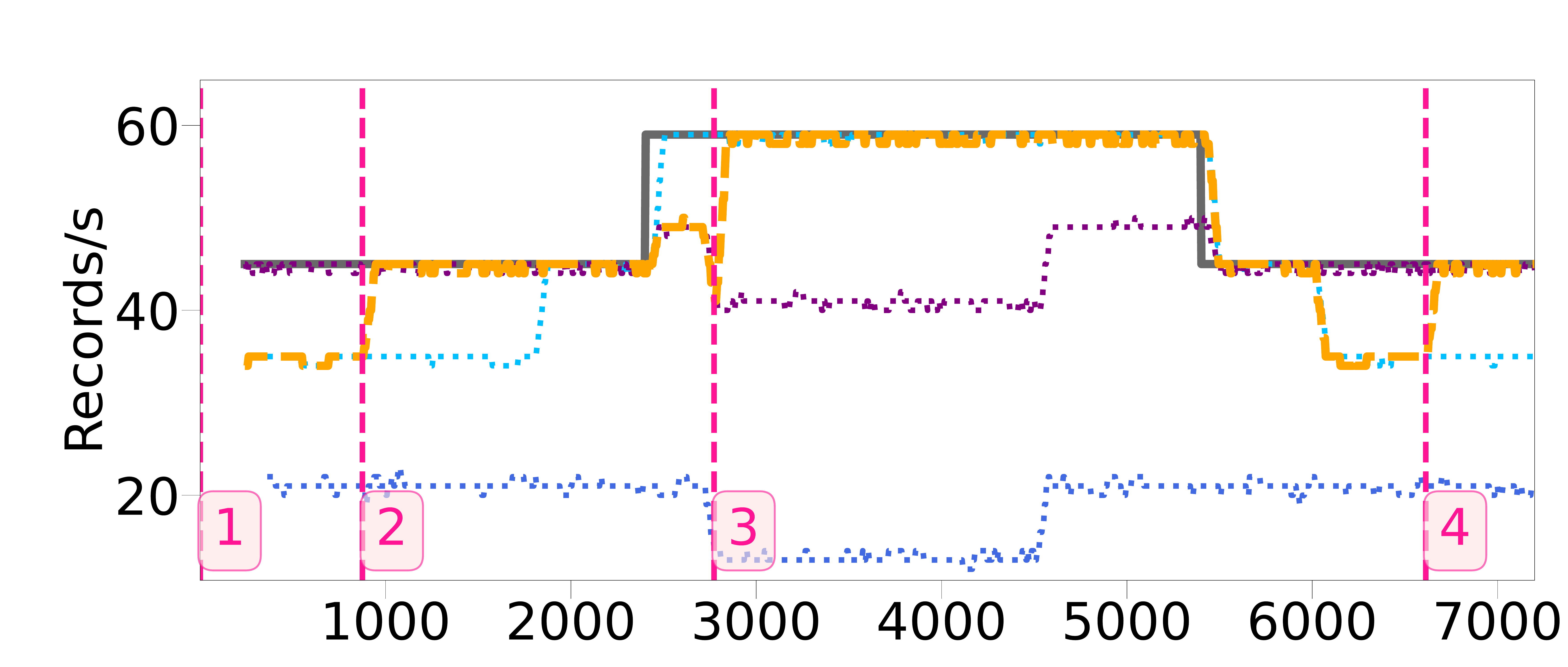}} \\
\end{tabular}
\begin{minipage}[b]{0.8\linewidth}
\includegraphics[width=\linewidth]{figures/eval/shift2222/res_video6nw2_0_lower.pdf_legend.pdf}
    \vspace{-0.4cm}
\end{minipage}\vspace{-2mm}
\begin{tabular}{ccc}
    \adjustbox{valign=m}{\rotatebox[origin=c]{90}{\parbox{2.4cm}{\centering Network BW }}} &
    \adjustbox{valign=m}{\includegraphics[width=8.26cm]{figures/eval/shift2222/res_video6nw2_0_lower.pdf.pdf}} \\
\end{tabular}
\captionof{figure}{\ours under variable operating conditions - RPi1}
\label{fig:sup_vary2222_rpi1}
\end{minipage}\par\vspace{\textfloatsep}

\noindent\begin{minipage}{\columnwidth}
\centering
\begin{minipage}[b]{0.8\linewidth}
\centering
\includegraphics[width=\linewidth]{figures/eval/shift2222/res_video6nw2_0.pdf_legend.pdf}
    \vspace{-0.4cm}
\end{minipage}\vspace{-2mm}
\setlength{\tabcolsep}{1pt}
\renewcommand{\arraystretch}{1.5}
\begin{tabular}{ccc}
    \adjustbox{valign=m}{\rotatebox[origin=c]{90}{\textbf{Q1}}} &
    \adjustbox{valign=m}{\includegraphics[width=8.2cm]{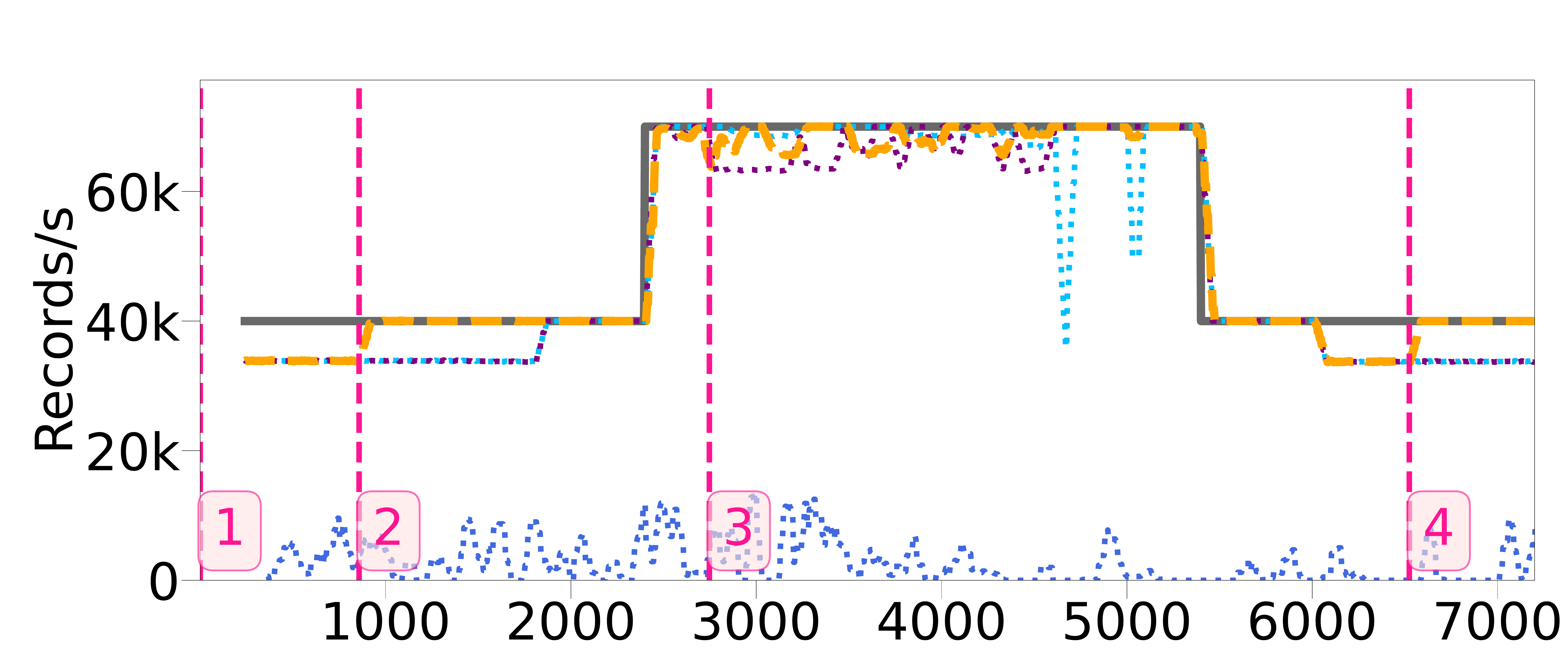}} \\
    \adjustbox{valign=m}{\rotatebox[origin=c]{90}{\textbf{Q5}}} &
    \adjustbox{valign=m}{\includegraphics[width=8.2cm]{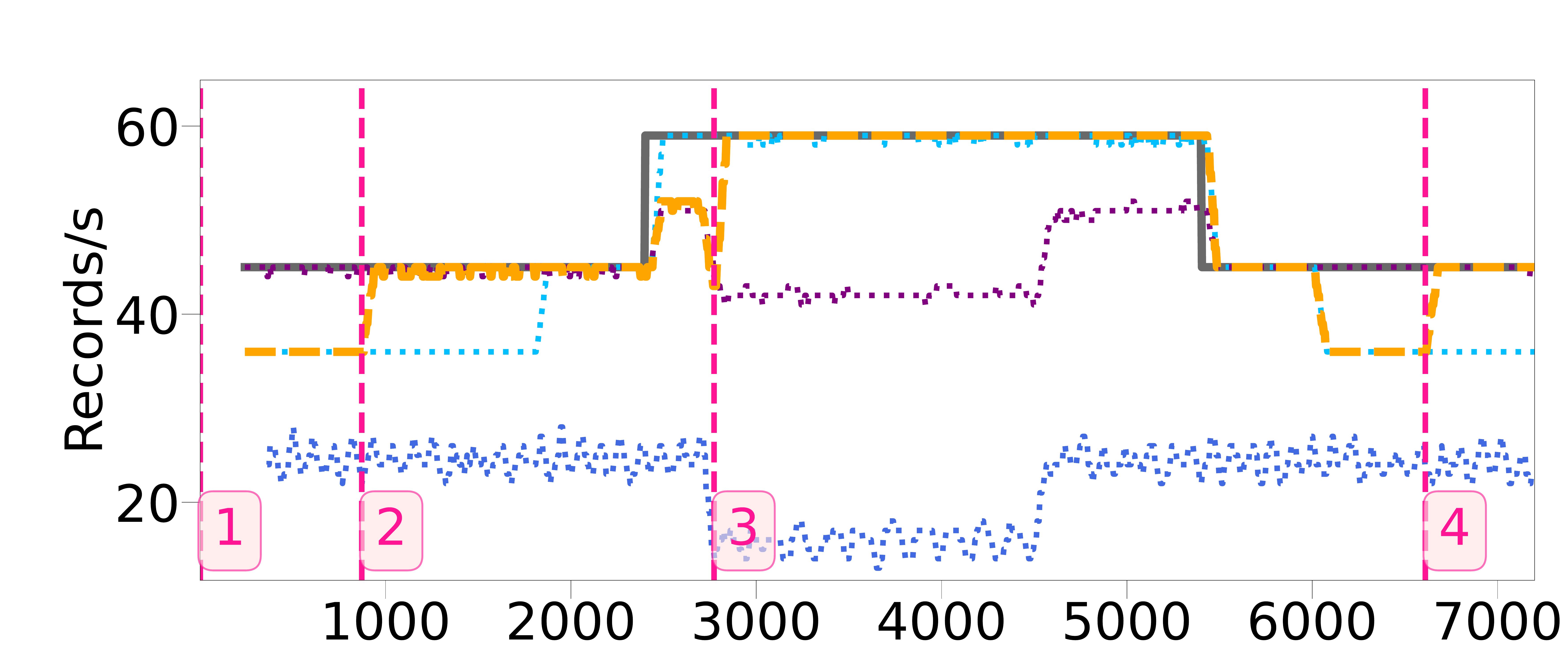}} \\
\end{tabular}
\begin{minipage}[b]{0.8\linewidth}
\includegraphics[width=\linewidth]{figures/eval/shift2222/res_video6nw2_0_lower.pdf_legend.pdf}
    \vspace{-0.4cm}
\end{minipage}\vspace{-2mm}
\begin{tabular}{ccc}
    \adjustbox{valign=m}{\rotatebox[origin=c]{90}{\parbox{2.4cm}{\centering Network BW }}} &
    \adjustbox{valign=m}{\includegraphics[width=8.26cm]{figures/eval/shift2222/res_video6nw2_0_lower.pdf.pdf}} \\
\end{tabular}
\captionof{figure}{\ours under variable operating conditions - RPi2}
\label{fig:sup_vary2222_rpi2}
\end{minipage}\par\vspace{\textfloatsep}

\noindent\begin{minipage}{\columnwidth}
\centering
\begin{minipage}[b]{0.8\linewidth}
\centering
\includegraphics[width=\linewidth]{figures/eval/shift2222/res_video6nw2_0.pdf_legend.pdf}
    \vspace{-0.4cm}
\end{minipage}\vspace{-2mm}
\setlength{\tabcolsep}{1pt}
\renewcommand{\arraystretch}{1.5}
\begin{tabular}{ccc}
    \adjustbox{valign=m}{\rotatebox[origin=c]{90}{\textbf{Q1}}} &
    \adjustbox{valign=m}{\includegraphics[width=8.2cm]{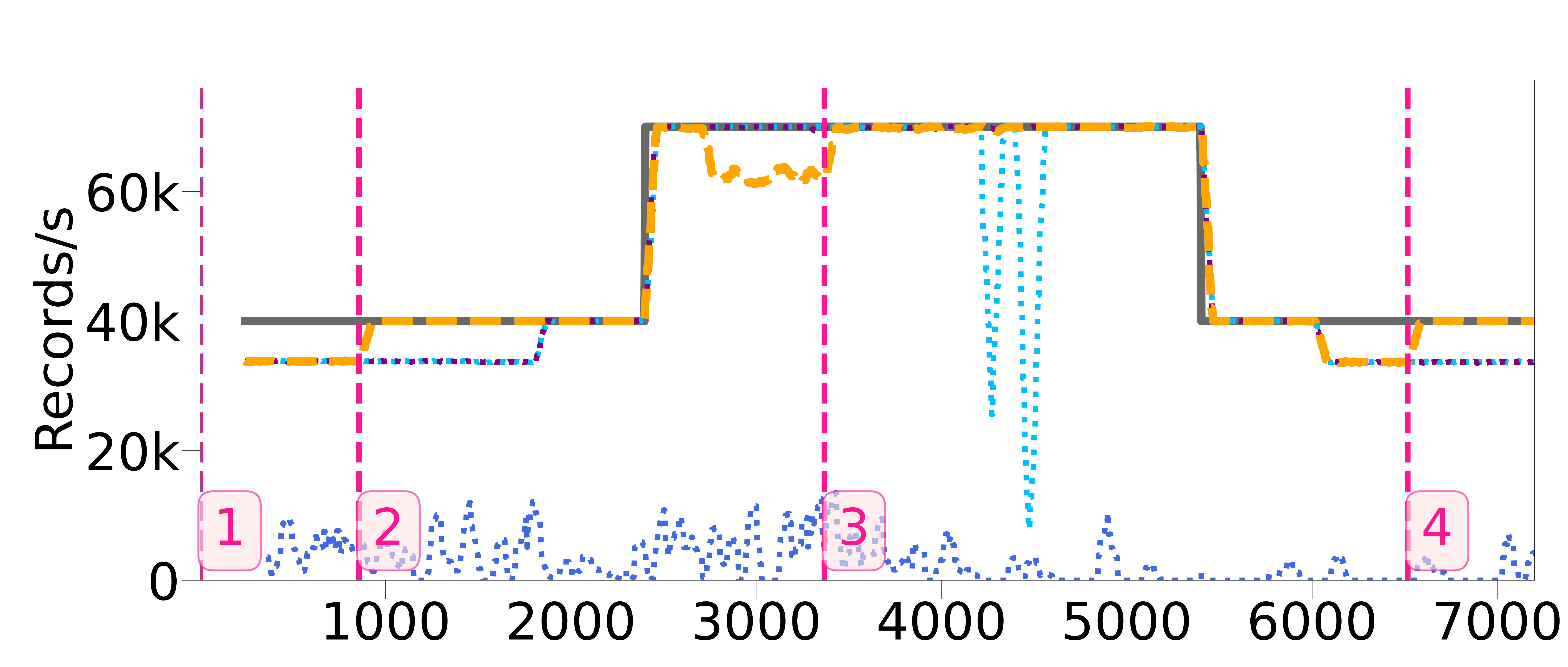}} \\
    \adjustbox{valign=m}{\rotatebox[origin=c]{90}{\textbf{Q5}}} &
    \adjustbox{valign=m}{\includegraphics[width=8.2cm]{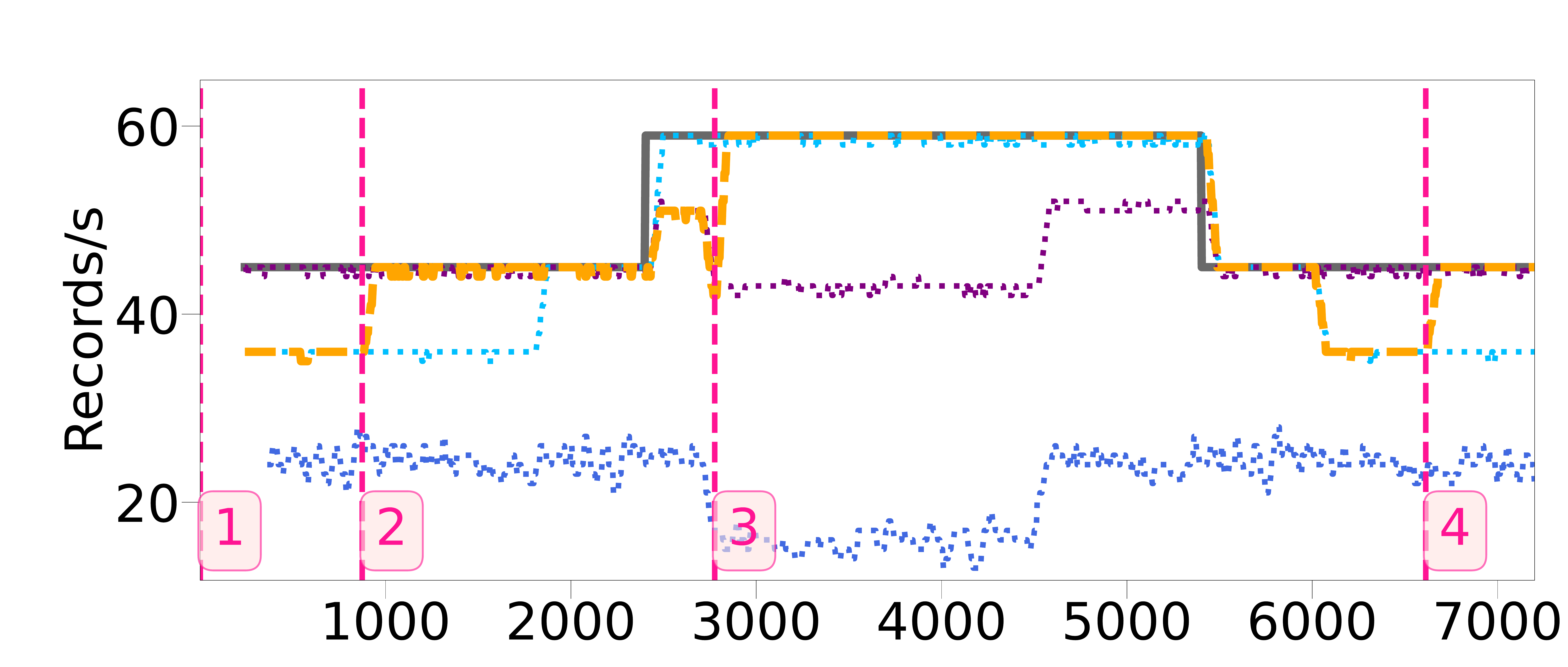}} \\
\end{tabular}
\begin{minipage}[b]{0.8\linewidth}
\includegraphics[width=\linewidth]{figures/eval/shift2222/res_video6nw2_0_lower.pdf_legend.pdf}
    \vspace{-0.4cm}
\end{minipage}\vspace{-2mm}
\begin{tabular}{ccc}
    \adjustbox{valign=m}{\rotatebox[origin=c]{90}{\parbox{2.4cm}{\centering Network BW }}} &
    \adjustbox{valign=m}{\includegraphics[width=8.26cm]{figures/eval/shift2222/res_video6nw2_0_lower.pdf.pdf}} \\
\end{tabular}
\captionof{figure}{\ours under variable operating conditions - RPi3}
\label{fig:sup_vary2222_rpi3}
\end{minipage}\par\vspace{\textfloatsep}

\noindent\begin{minipage}{\columnwidth}
\centering
\begin{minipage}[b]{0.8\linewidth}
\centering
\includegraphics[width=\linewidth]{figures/eval/shift2222/res_video6nw2_0.pdf_legend.pdf}
    \vspace{-0.4cm}
\end{minipage}\vspace{-2mm}
\setlength{\tabcolsep}{1pt}
\renewcommand{\arraystretch}{1.5}
\begin{tabular}{ccc}
    \adjustbox{valign=m}{\rotatebox[origin=c]{90}{\textbf{Q1}}} &
    \adjustbox{valign=m}{\includegraphics[width=8.2cm]{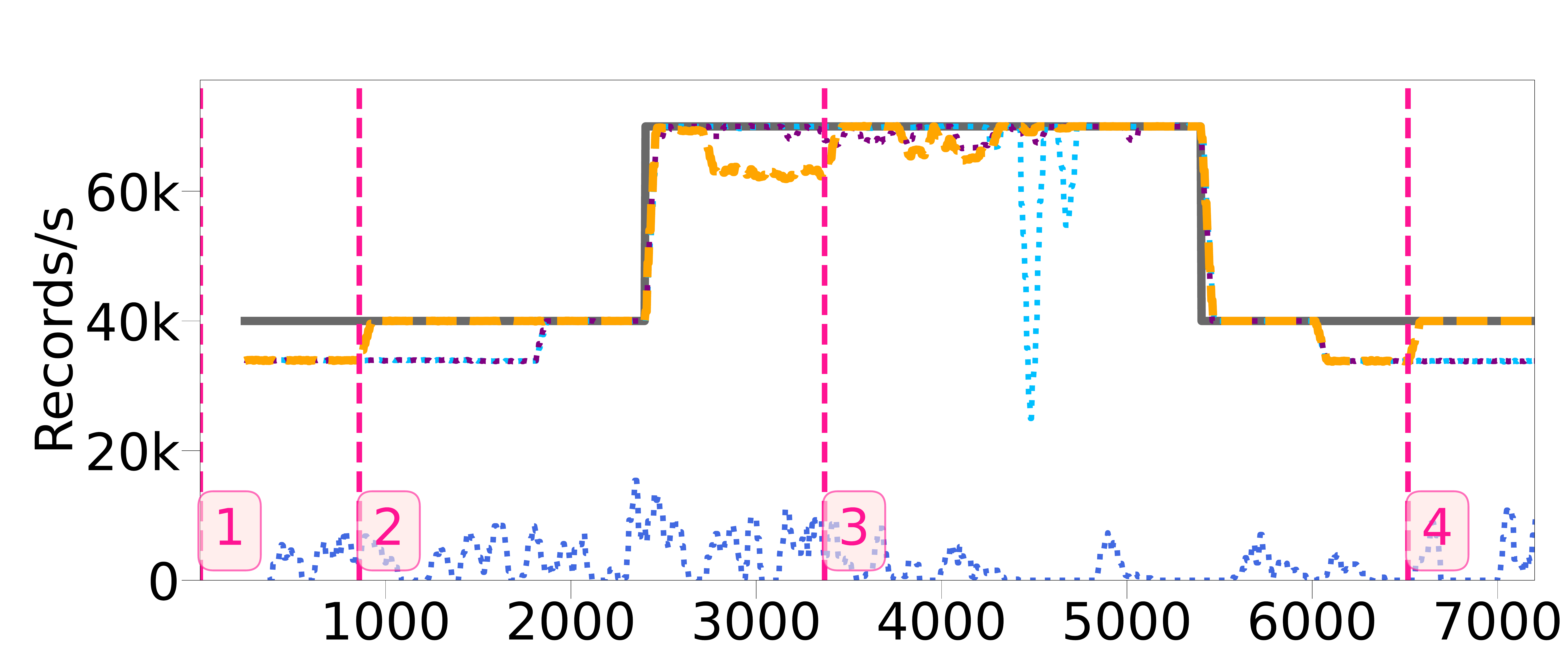}} \\
    \adjustbox{valign=m}{\rotatebox[origin=c]{90}{\textbf{Q5}}} &
    \adjustbox{valign=m}{\includegraphics[width=8.2cm]{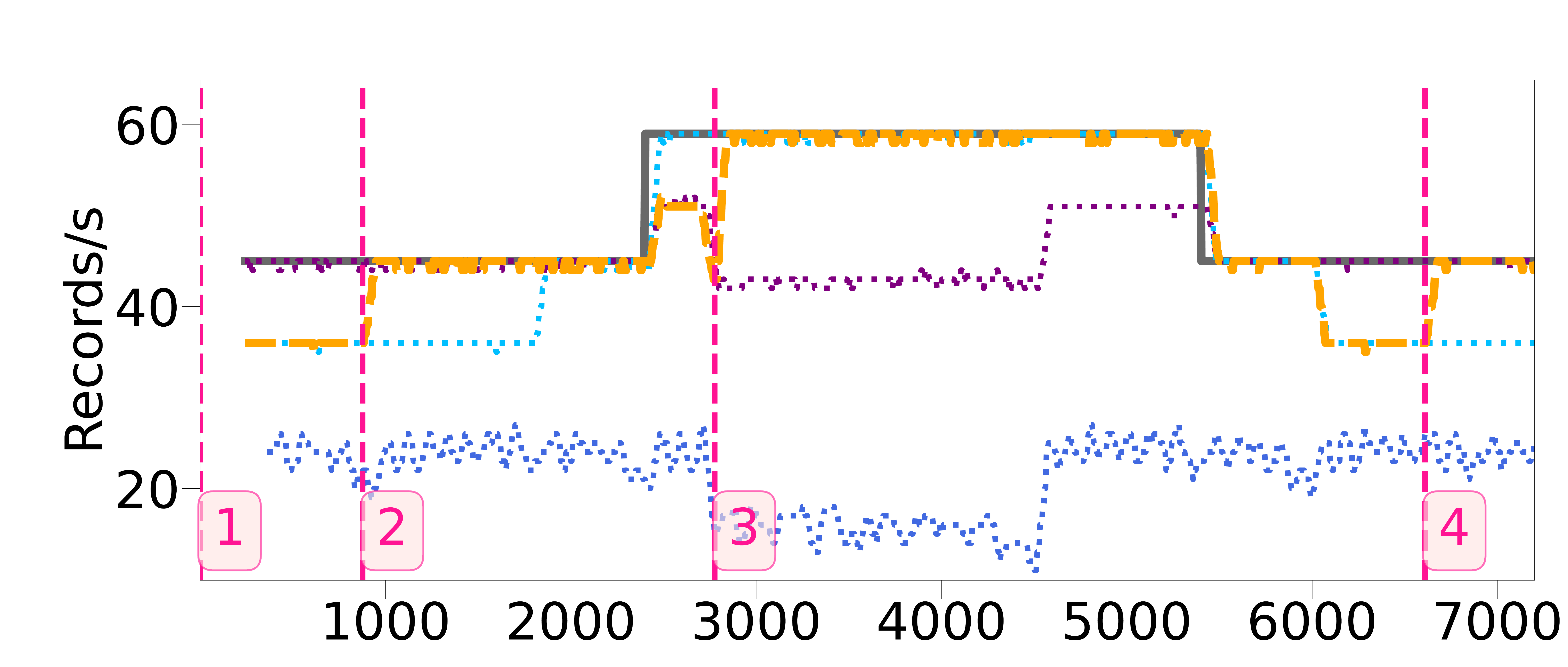}} \\
\end{tabular}
\begin{minipage}[b]{0.8\linewidth}
\includegraphics[width=\linewidth]{figures/eval/shift2222/res_video6nw2_0_lower.pdf_legend.pdf}
    \vspace{-0.4cm}
\end{minipage}\vspace{-2mm}
\begin{tabular}{ccc}
    \adjustbox{valign=m}{\rotatebox[origin=c]{90}{\parbox{2.4cm}{\centering Network BW }}} &
    \adjustbox{valign=m}{\includegraphics[width=8.26cm]{figures/eval/shift2222/res_video6nw2_0_lower.pdf.pdf}} \\
\end{tabular}
\captionof{figure}{\ours under variable operating conditions - RPi4}
\label{fig:sup_vary2222_rpi4}
\end{minipage}\par\vspace{\textfloatsep}

\noindent\begin{minipage}{\columnwidth}
\centering
\begin{minipage}[b]{0.8\linewidth}
\includegraphics[width=\linewidth]{figures/eval/shift/res_video6nw_0.pdf_legend.pdf}
    \vspace{-0.4cm}
\end{minipage}
\setlength{\tabcolsep}{1pt}
\renewcommand{\arraystretch}{1.5}
\begin{tabular}{ccc}
    \adjustbox{valign=m}{\rotatebox[origin=c]{90}{\textbf{Q1}}} &
    \adjustbox{valign=m}{\includegraphics[width=4.0cm]{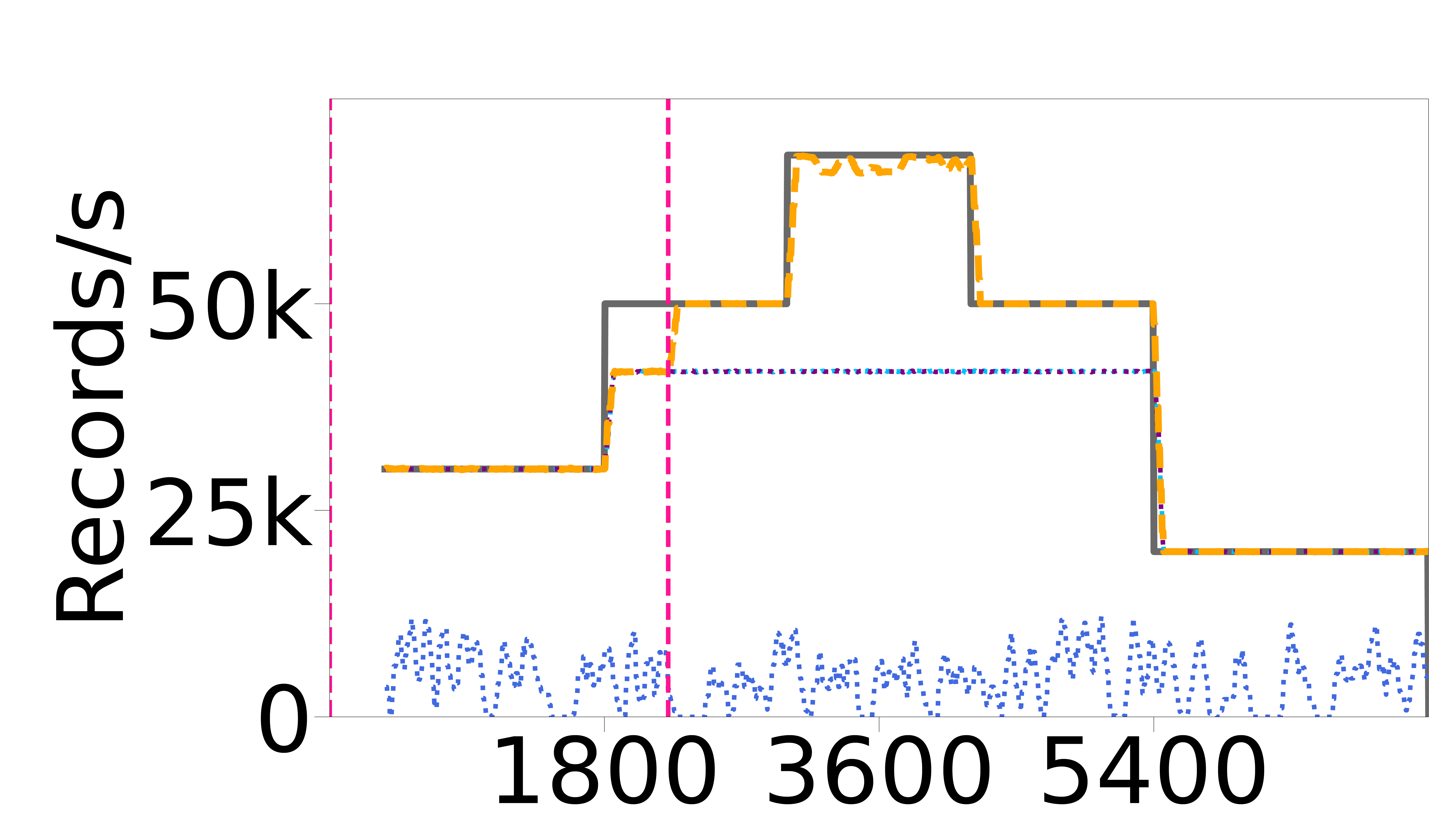}} &
    \adjustbox{valign=m}{\includegraphics[width=4.0cm]{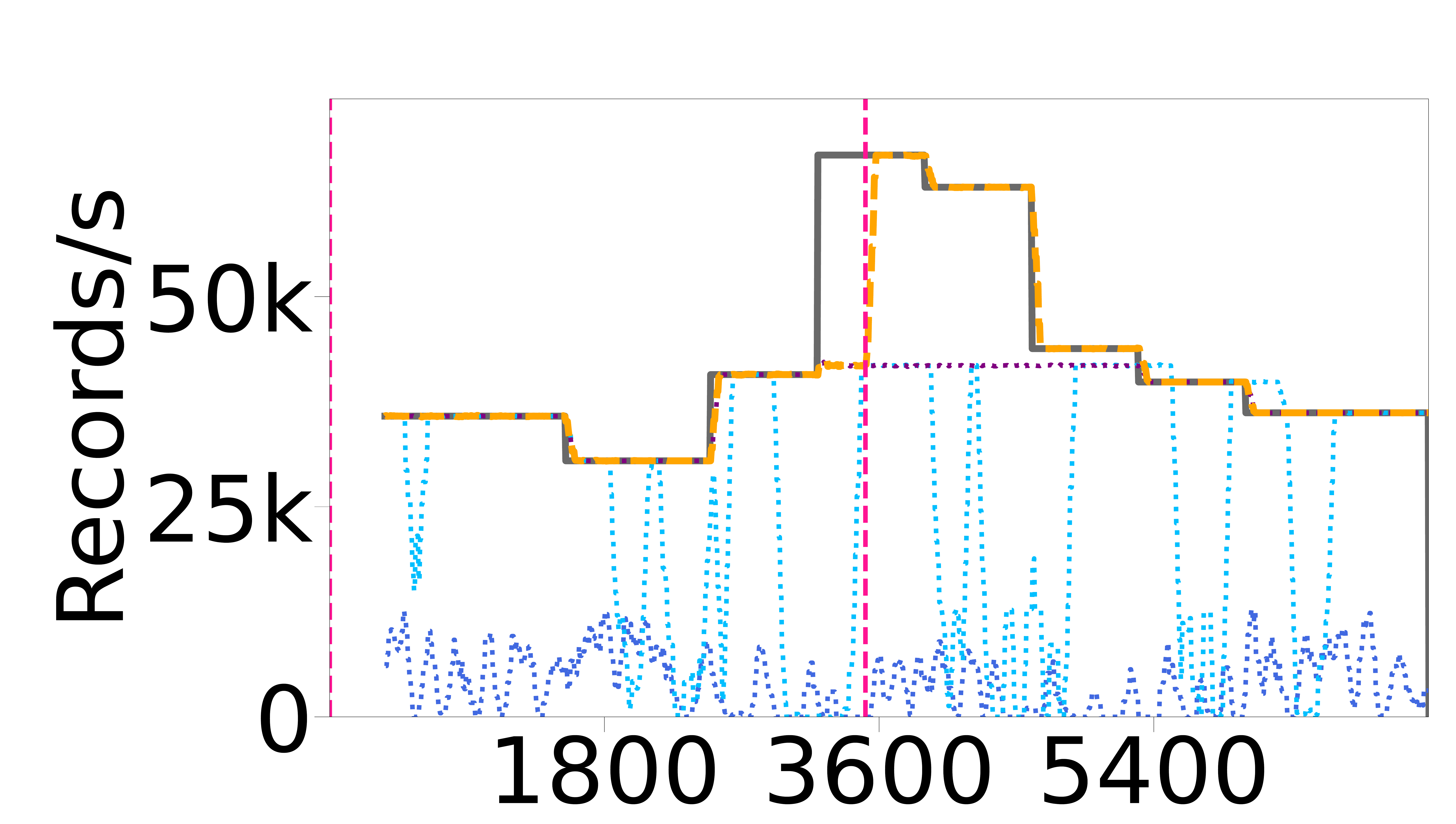}} \\
    \adjustbox{valign=m}{\rotatebox[origin=c]{90}{\textbf{Q2}}} &
    \adjustbox{valign=m}{\includegraphics[width=4.0cm]{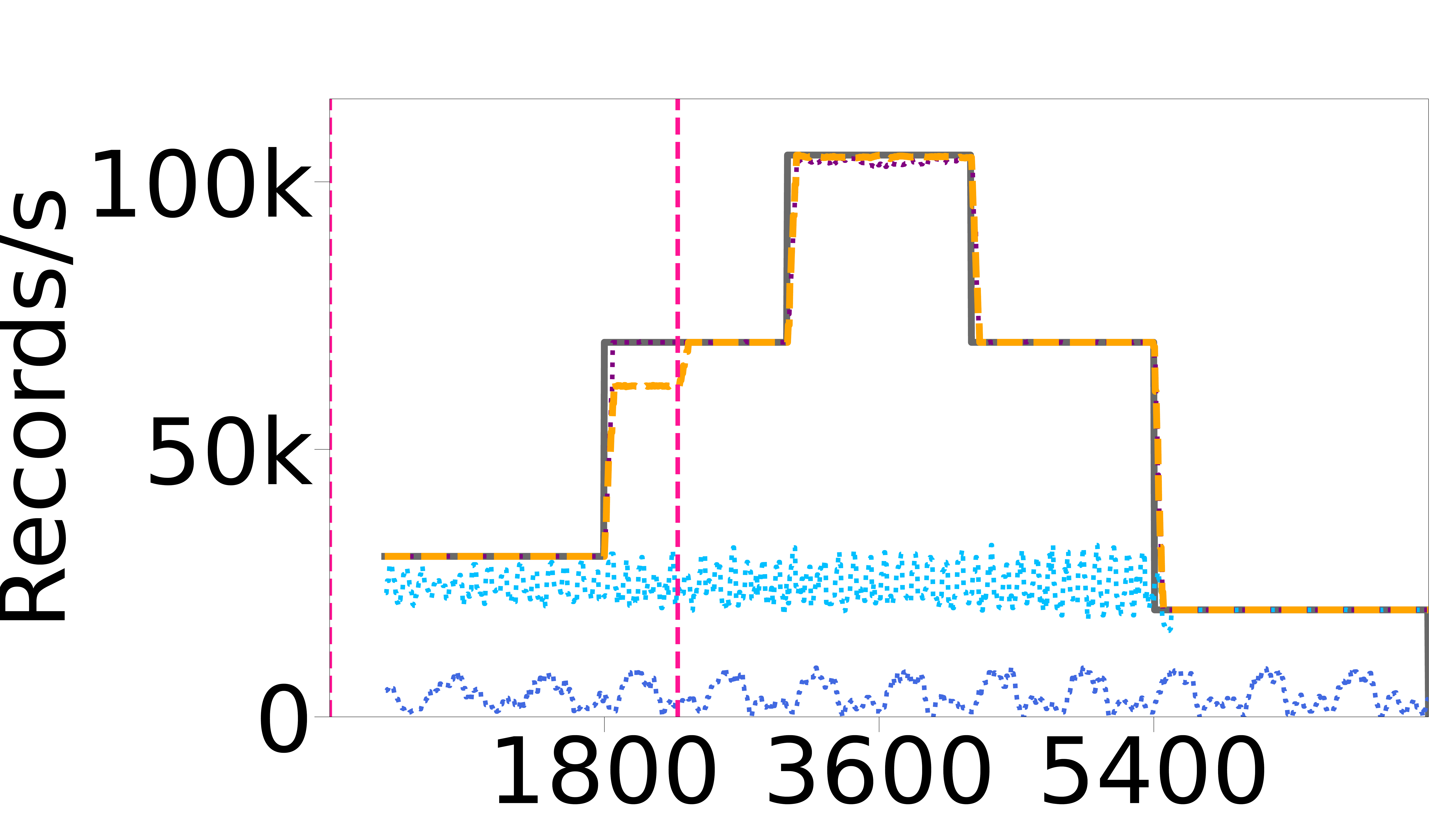}} &
    \adjustbox{valign=m}{\includegraphics[width=4.0cm]{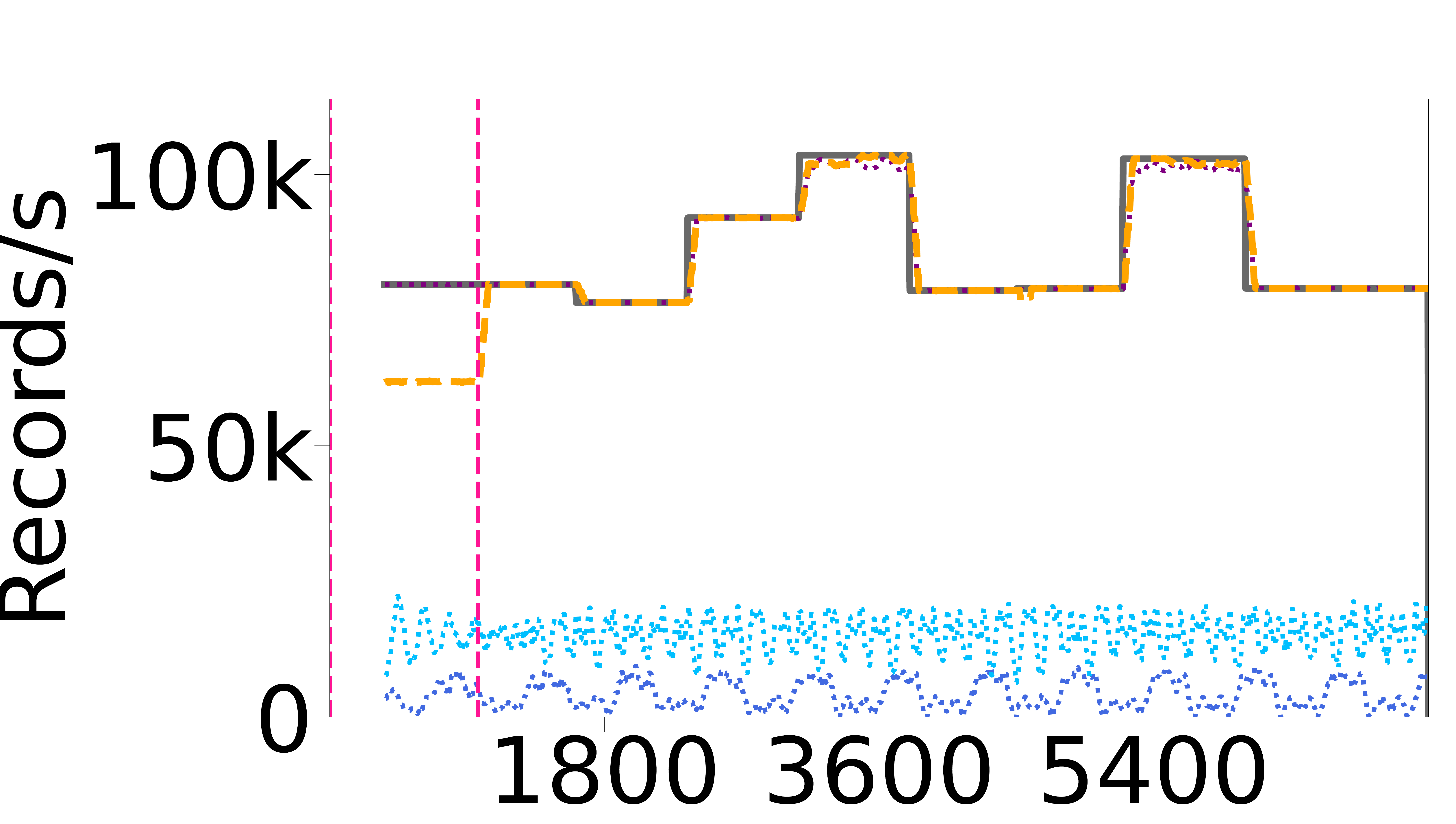}} \\
    \adjustbox{valign=m}{\rotatebox[origin=c]{90}{\textbf{Q3}}} &
    \adjustbox{valign=m}{\includegraphics[width=4.0cm]{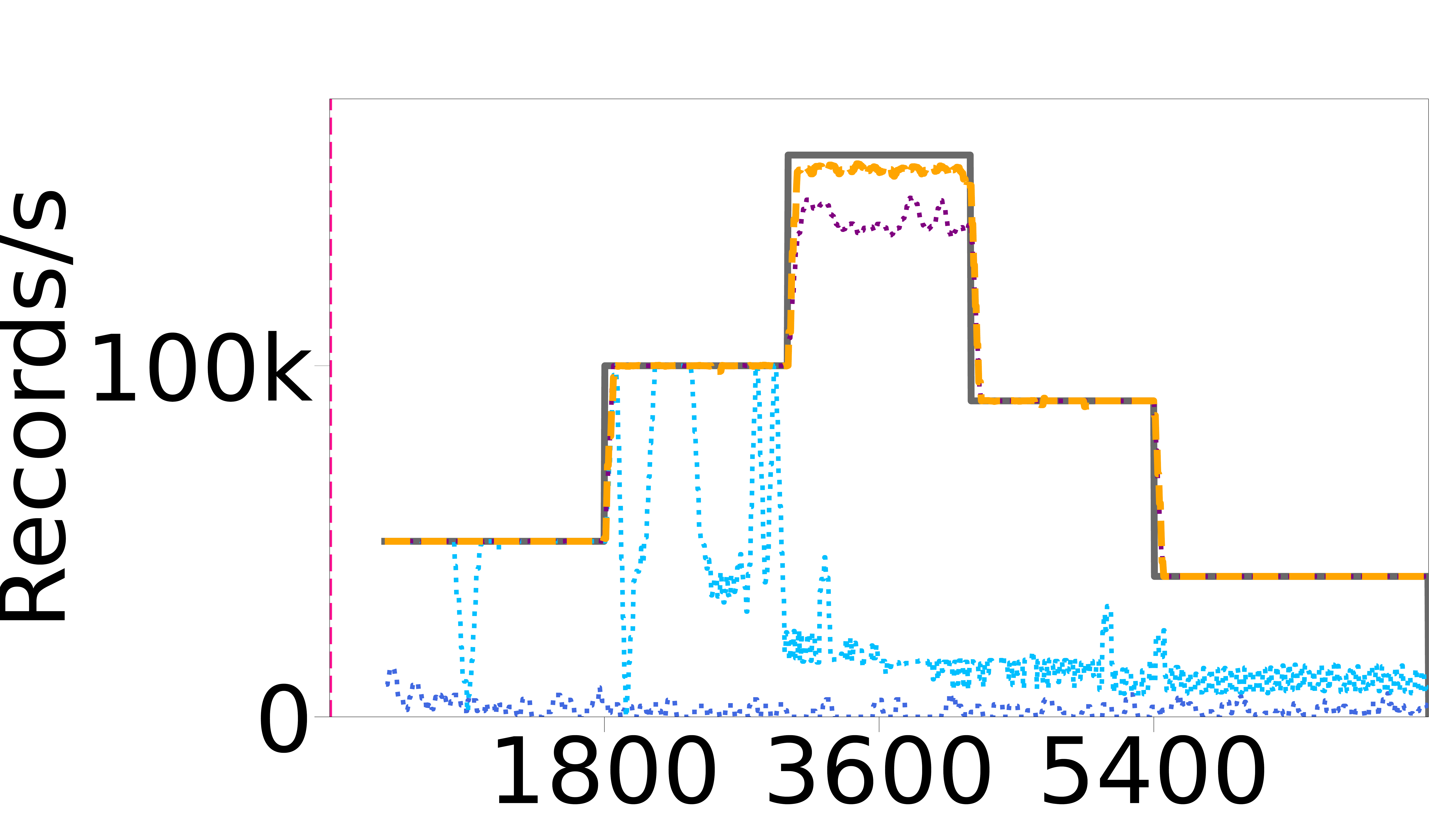}} &
    \adjustbox{valign=m}{\includegraphics[width=4.0cm]{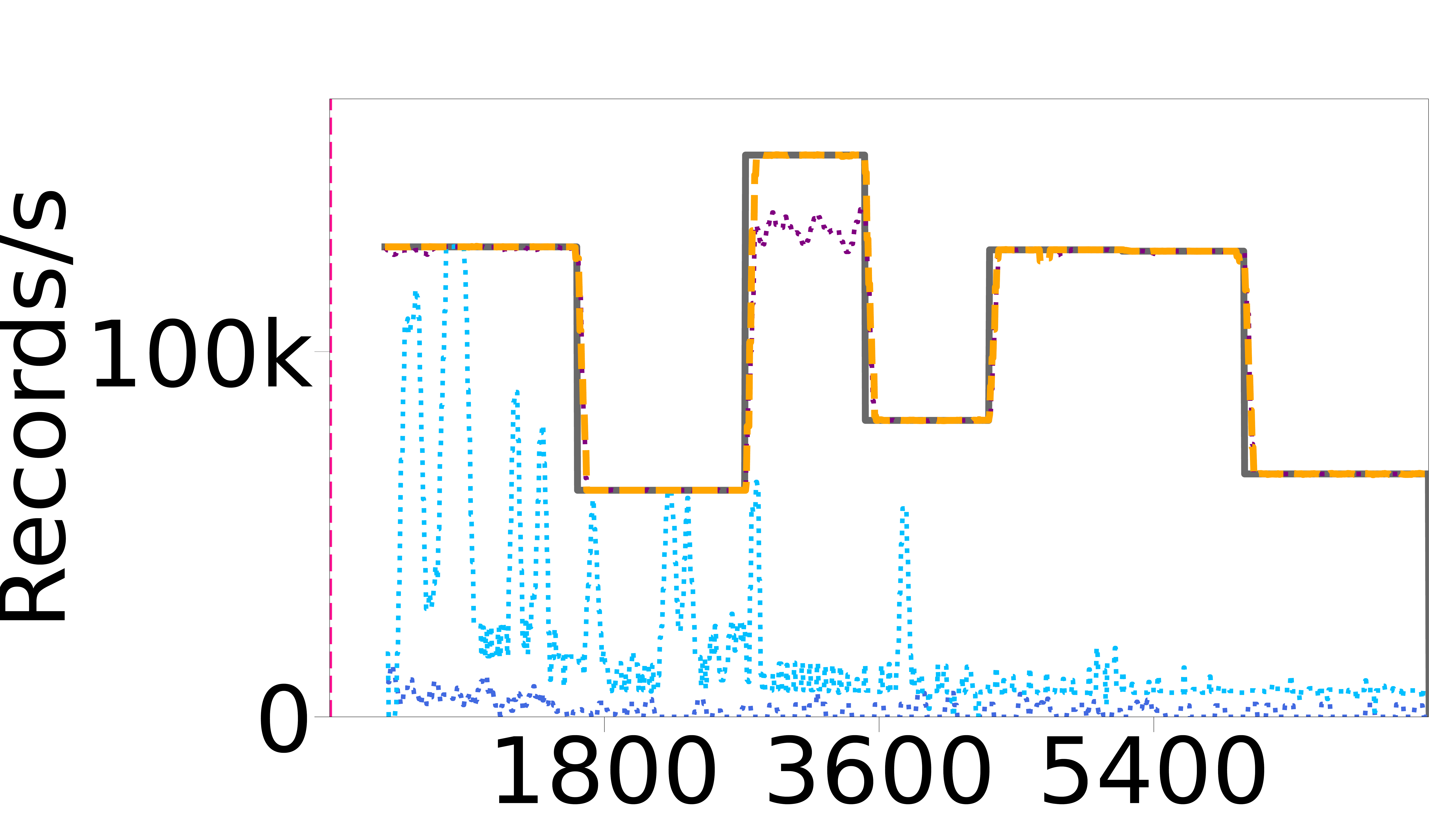}} \\
    \adjustbox{valign=m}{\rotatebox[origin=c]{90}{\textbf{Q4}}} &
    \adjustbox{valign=m}{\includegraphics[width=4.0cm]{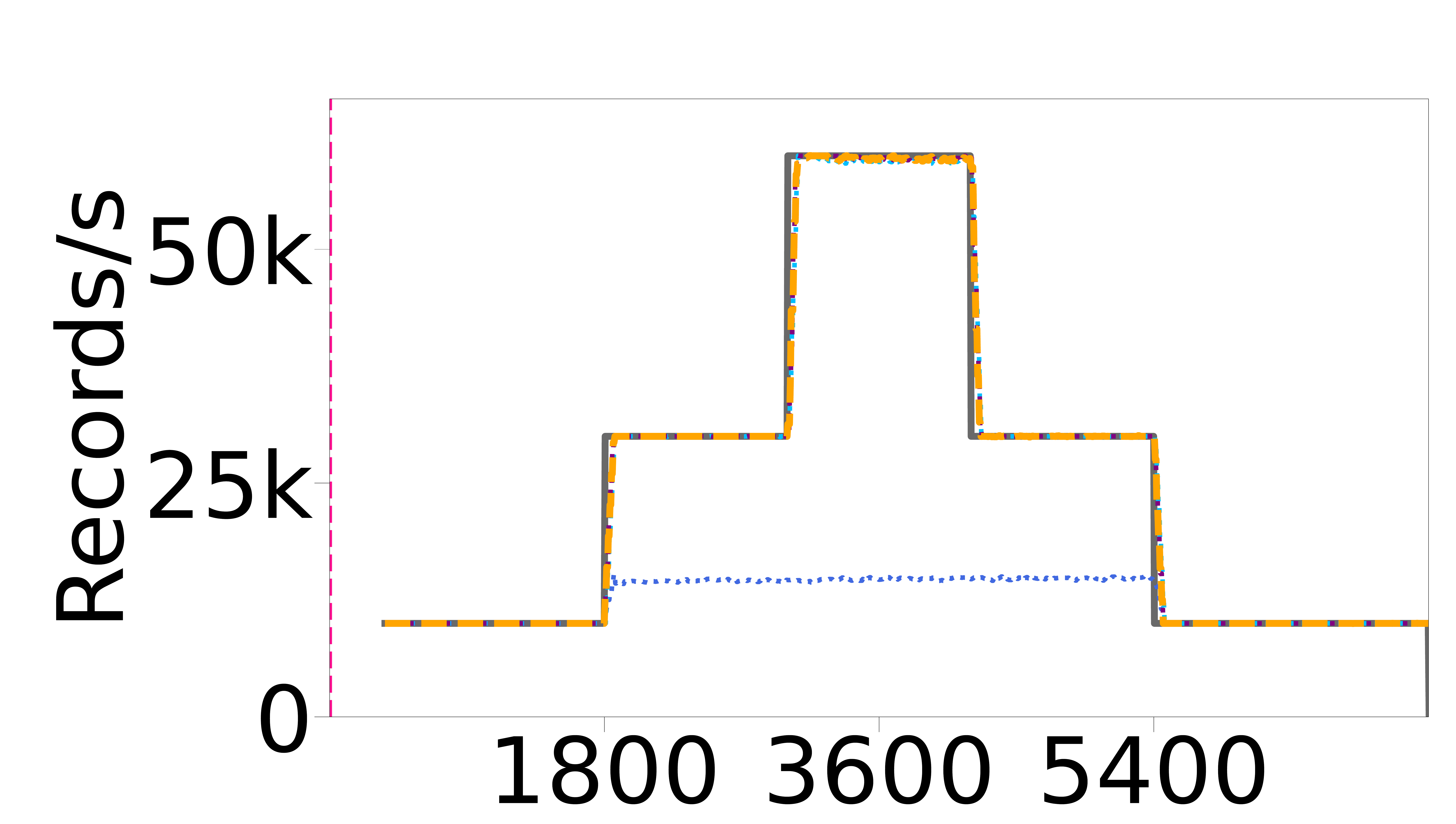}} &
    \adjustbox{valign=m}{\includegraphics[width=4.0cm]{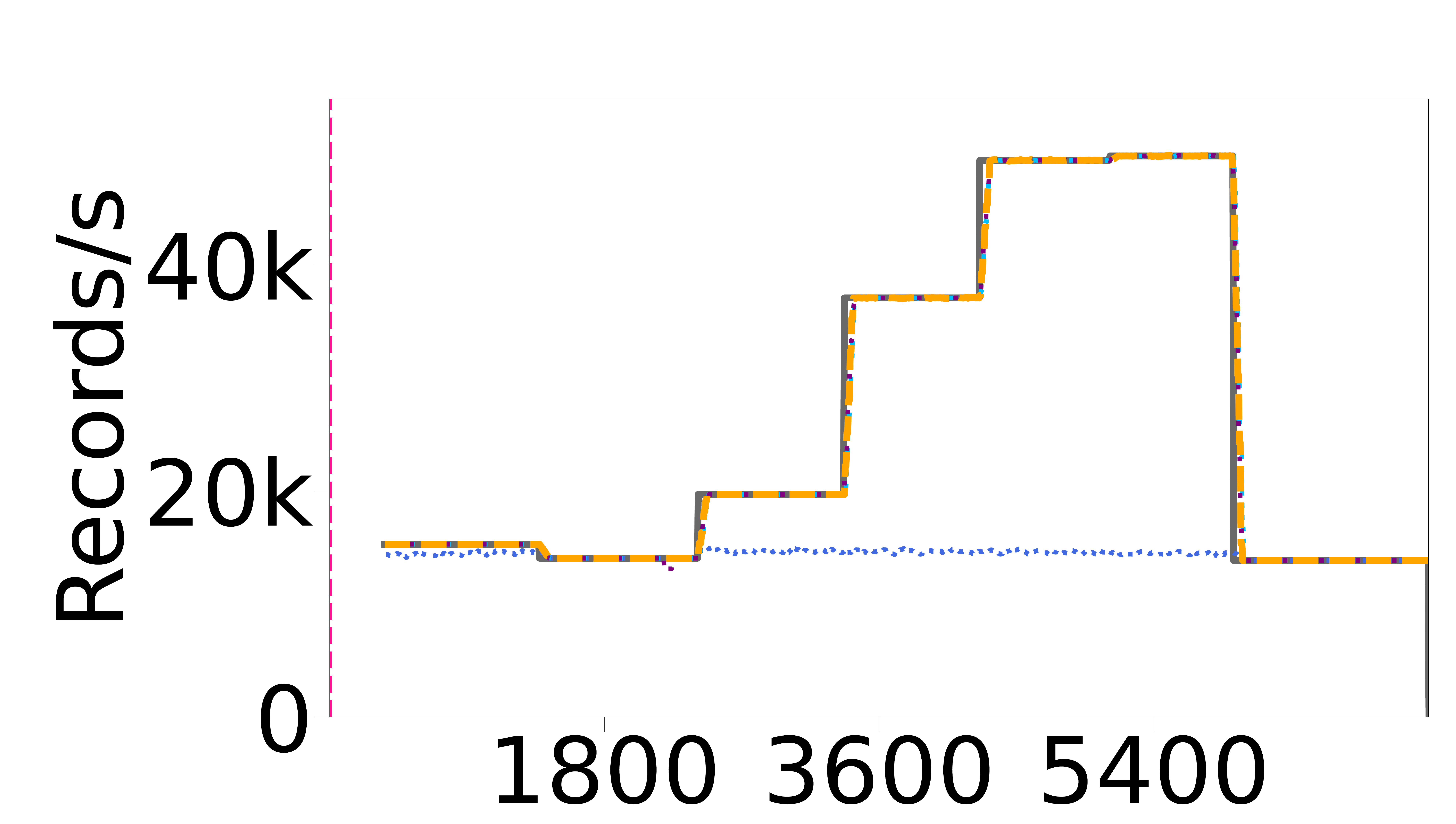}} \\
    \adjustbox{valign=m}{\rotatebox[origin=c]{90}{\textbf{Q5}}} &
    \adjustbox{valign=m}{\includegraphics[width=4.0cm]{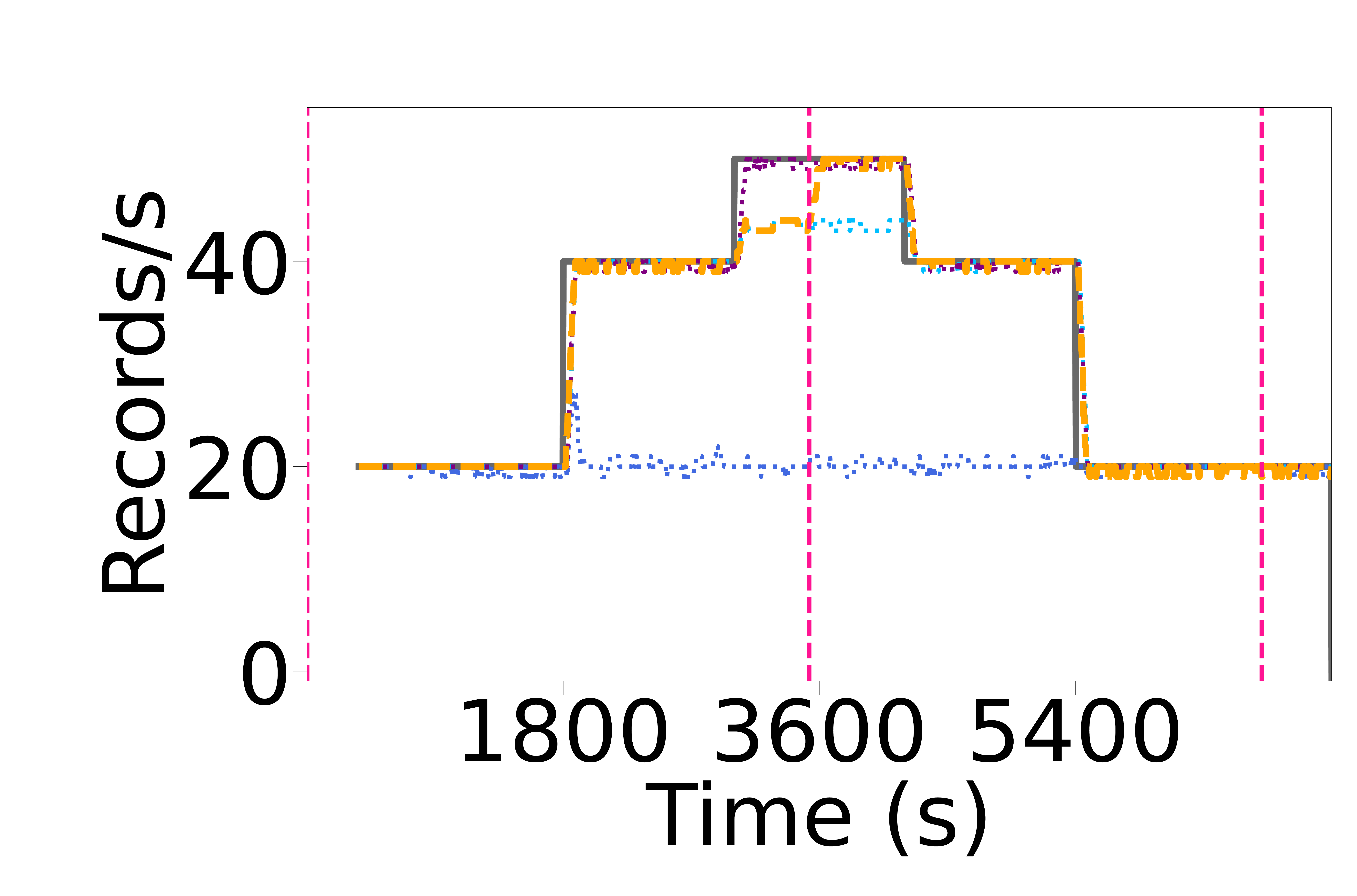}} &
    \adjustbox{valign=m}{\includegraphics[width=4.0cm]{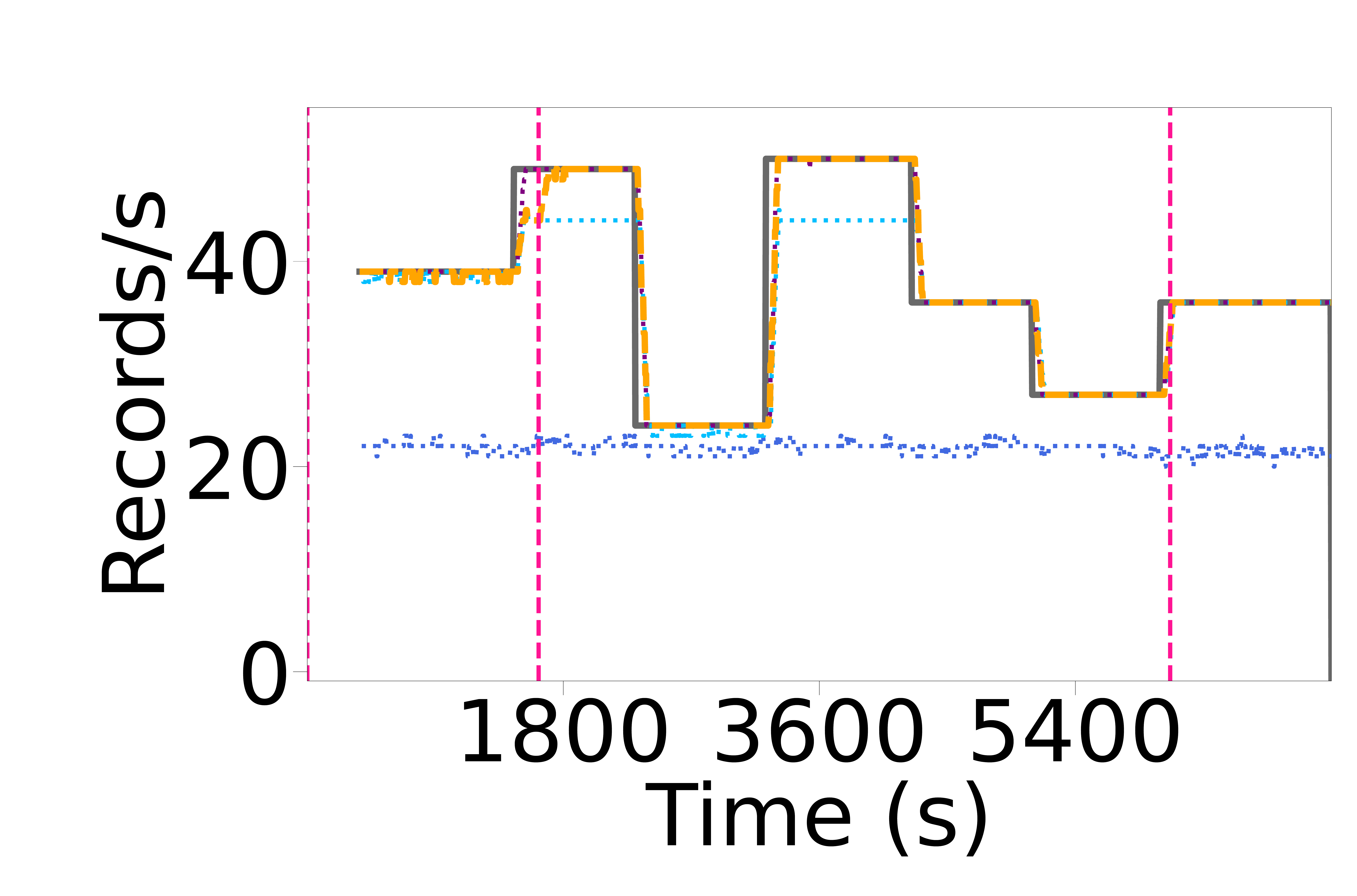}} \\
    & \adjustbox{valign=m}{\refstepcounter{subfigure}  (a) gradual rate pattern} & \adjustbox{valign=m}{\refstepcounter{subfigure}  (b) abrupt rate pattern} \\
\end{tabular}
\captionof{figure}{Impact of input rate - RPi1} 
\label{fig:sup_vary_rate_rpi1}
\end{minipage}\par\vspace{\textfloatsep}

\noindent\begin{minipage}{\columnwidth}
\centering
\begin{minipage}[b]{0.8\linewidth}
\includegraphics[width=\linewidth]{figures/eval/shift/res_video6nw_0.pdf_legend.pdf}
    \vspace{-0.4cm}
\end{minipage}
\setlength{\tabcolsep}{1pt}
\renewcommand{\arraystretch}{1.5}
\begin{tabular}{ccc}
    \adjustbox{valign=m}{\rotatebox[origin=c]{90}{\textbf{Q1}}} &
    \adjustbox{valign=m}{\includegraphics[width=4.0cm]{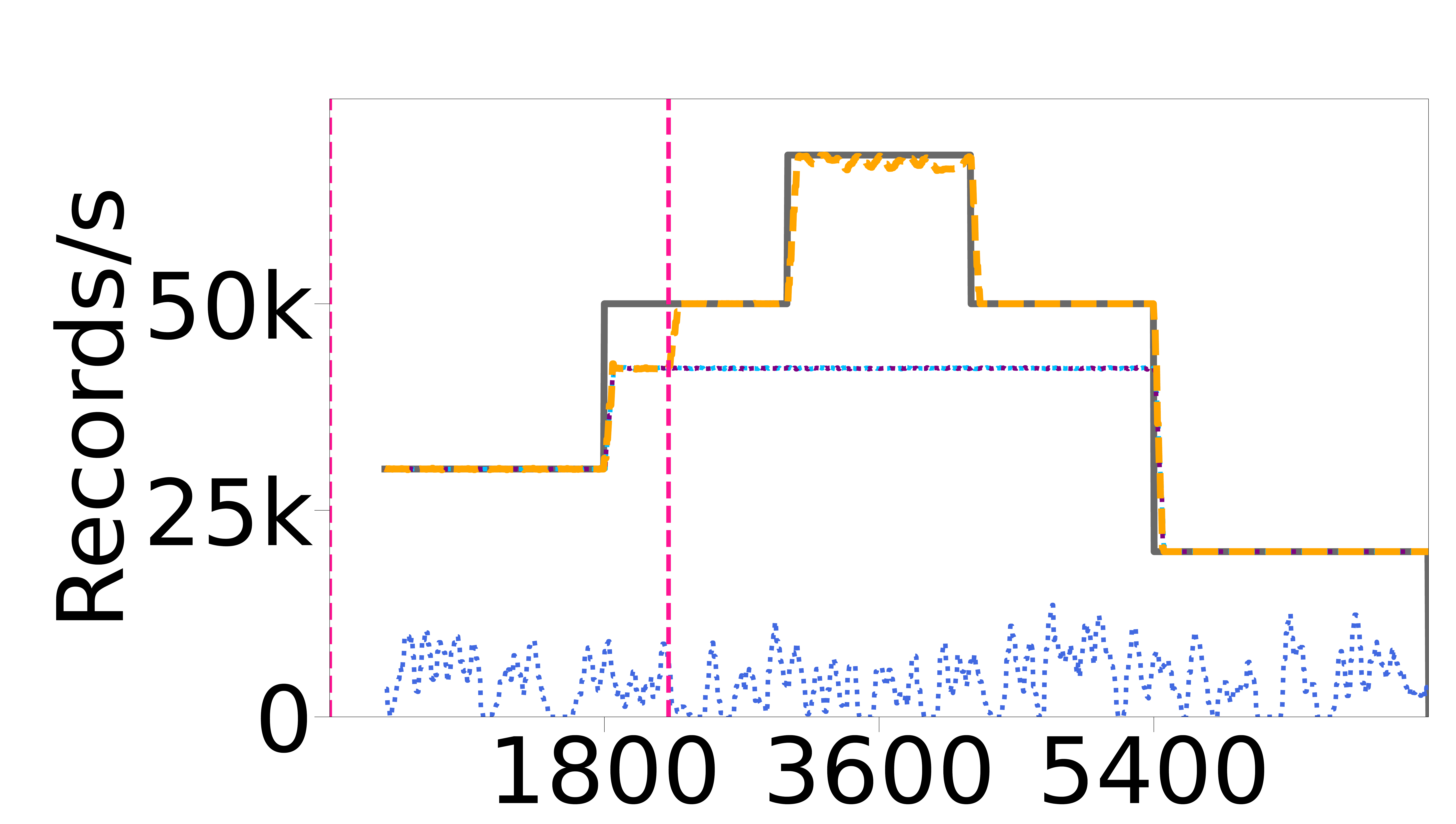}} &
    \adjustbox{valign=m}{\includegraphics[width=4.0cm]{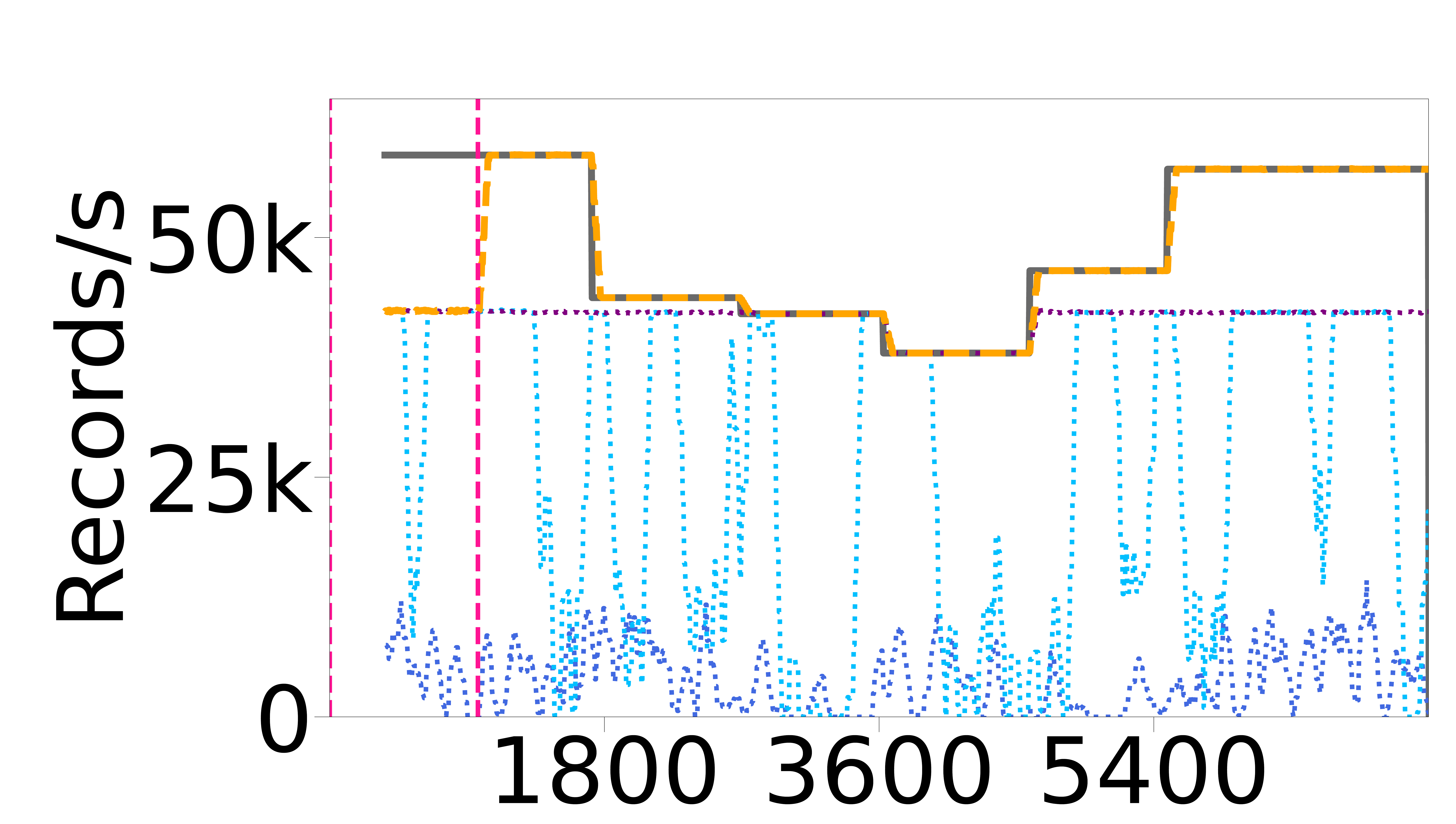}} \\
    \adjustbox{valign=m}{\rotatebox[origin=c]{90}{\textbf{Q2}}} &
    \adjustbox{valign=m}{\includegraphics[width=4.0cm]{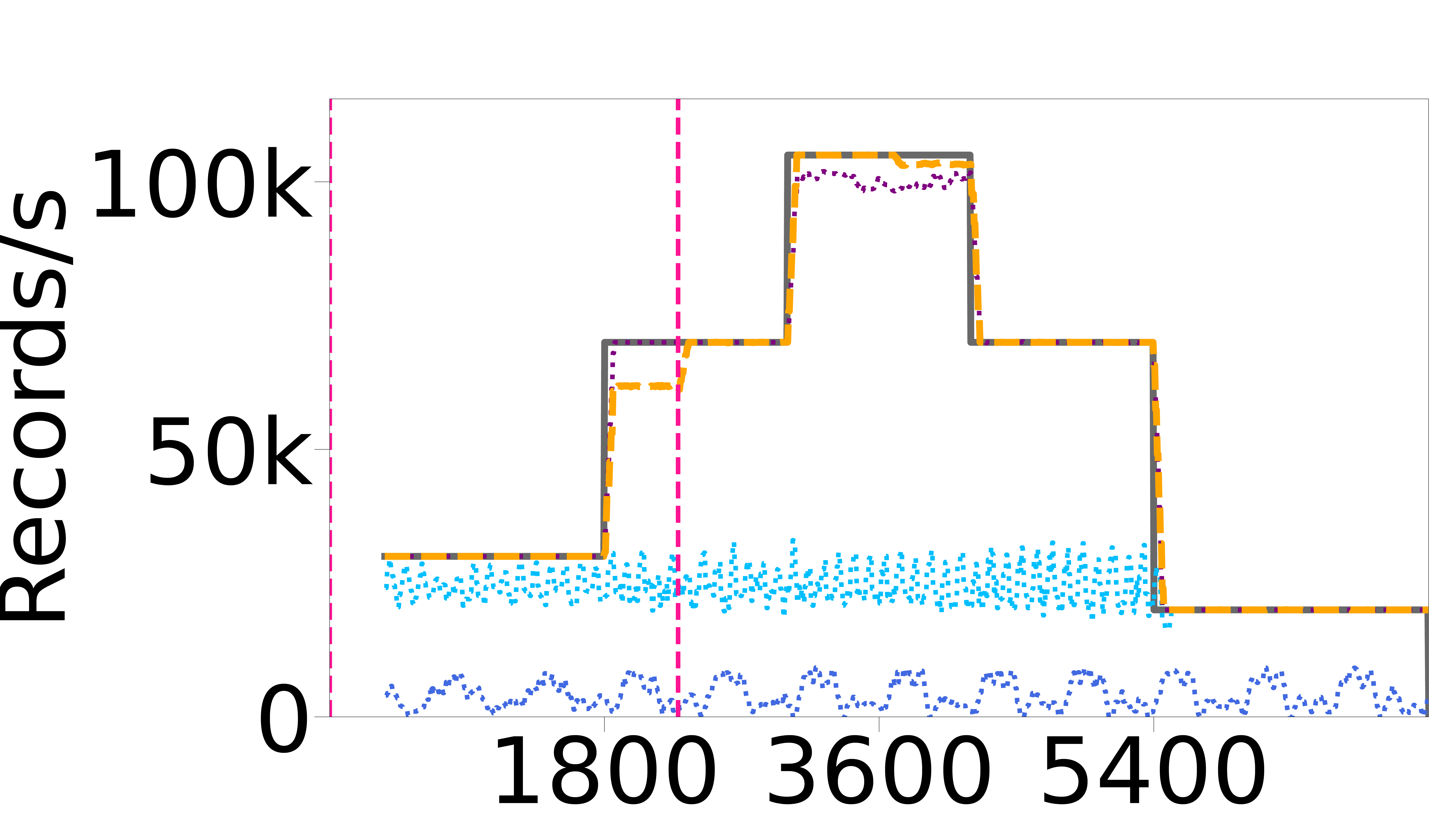}} &
    \adjustbox{valign=m}{\includegraphics[width=4.0cm]{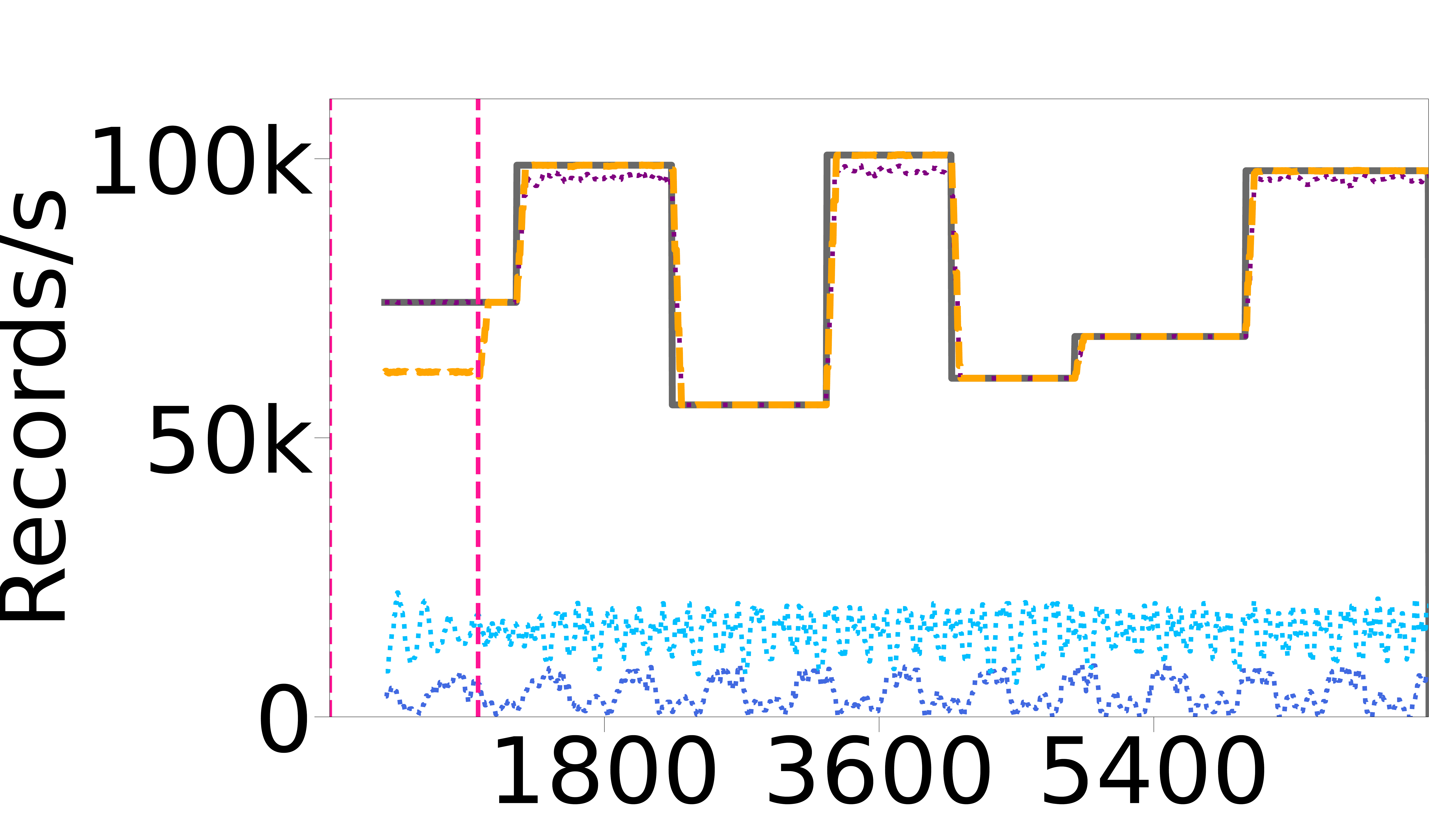}} \\
    \adjustbox{valign=m}{\rotatebox[origin=c]{90}{\textbf{Q3}}} &
    \adjustbox{valign=m}{\includegraphics[width=4.0cm]{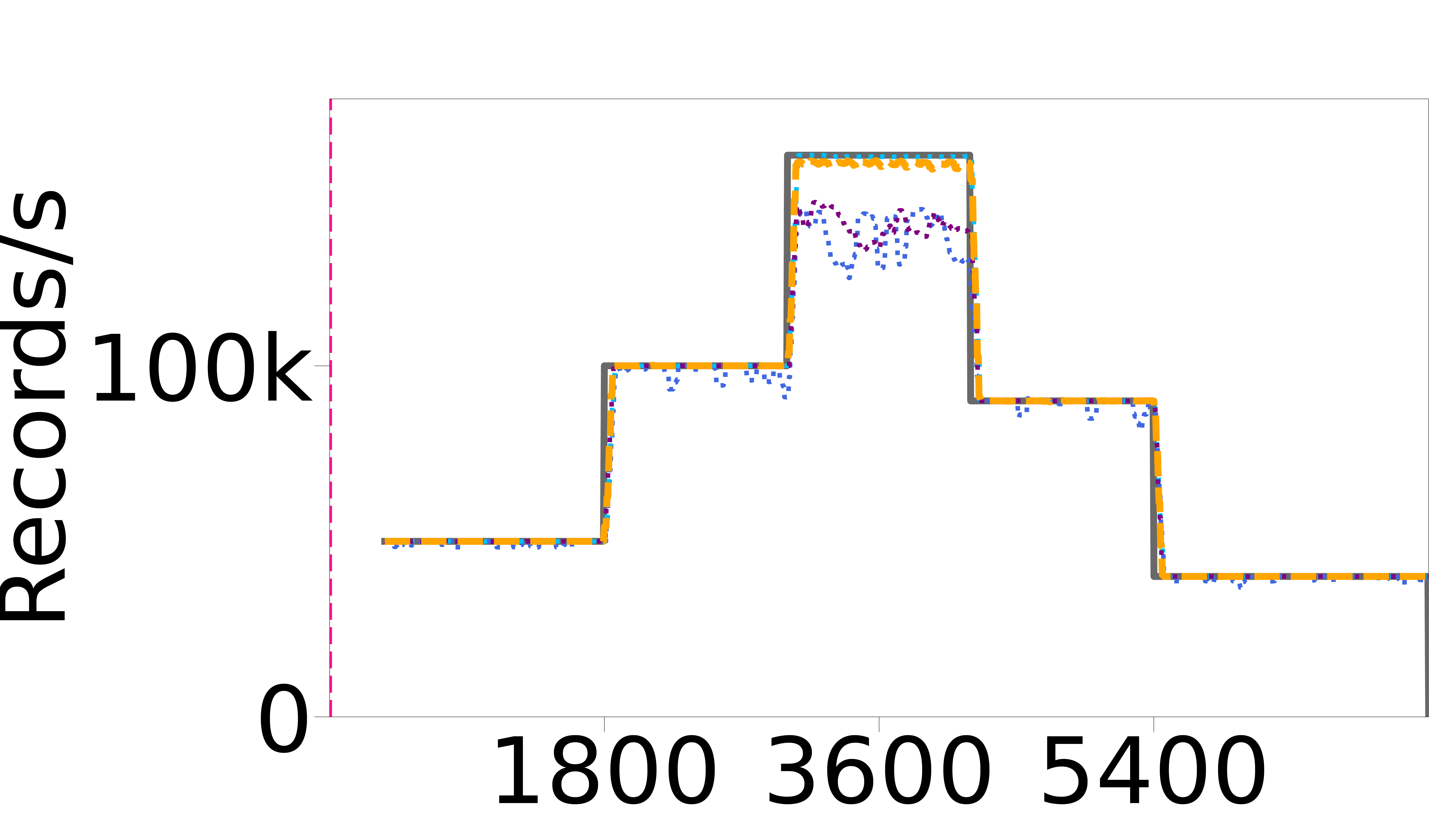}} &
    \adjustbox{valign=m}{\includegraphics[width=4.0cm]{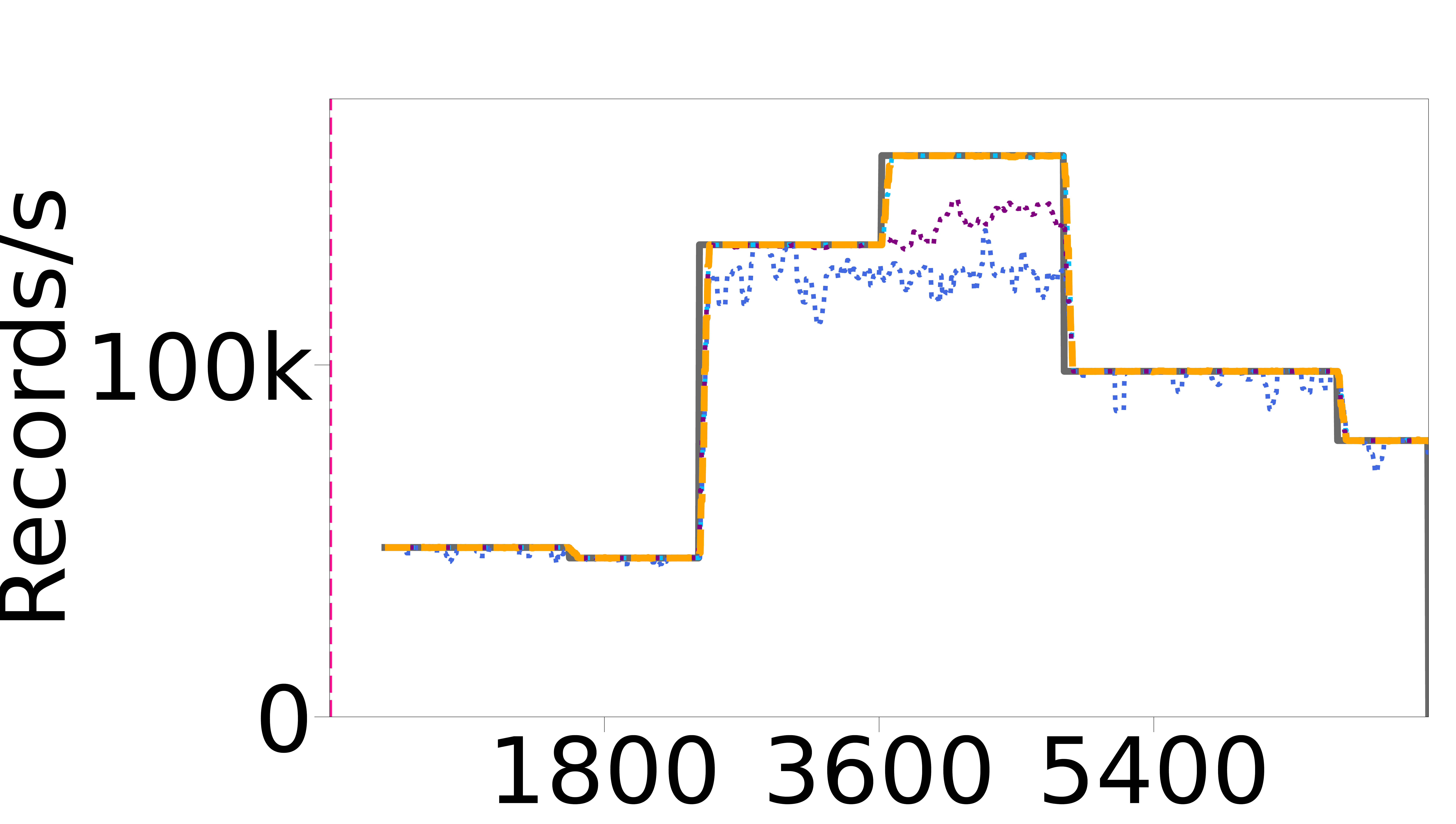}} \\
    \adjustbox{valign=m}{\rotatebox[origin=c]{90}{\textbf{Q4}}} &
    \adjustbox{valign=m}{\includegraphics[width=4.0cm]{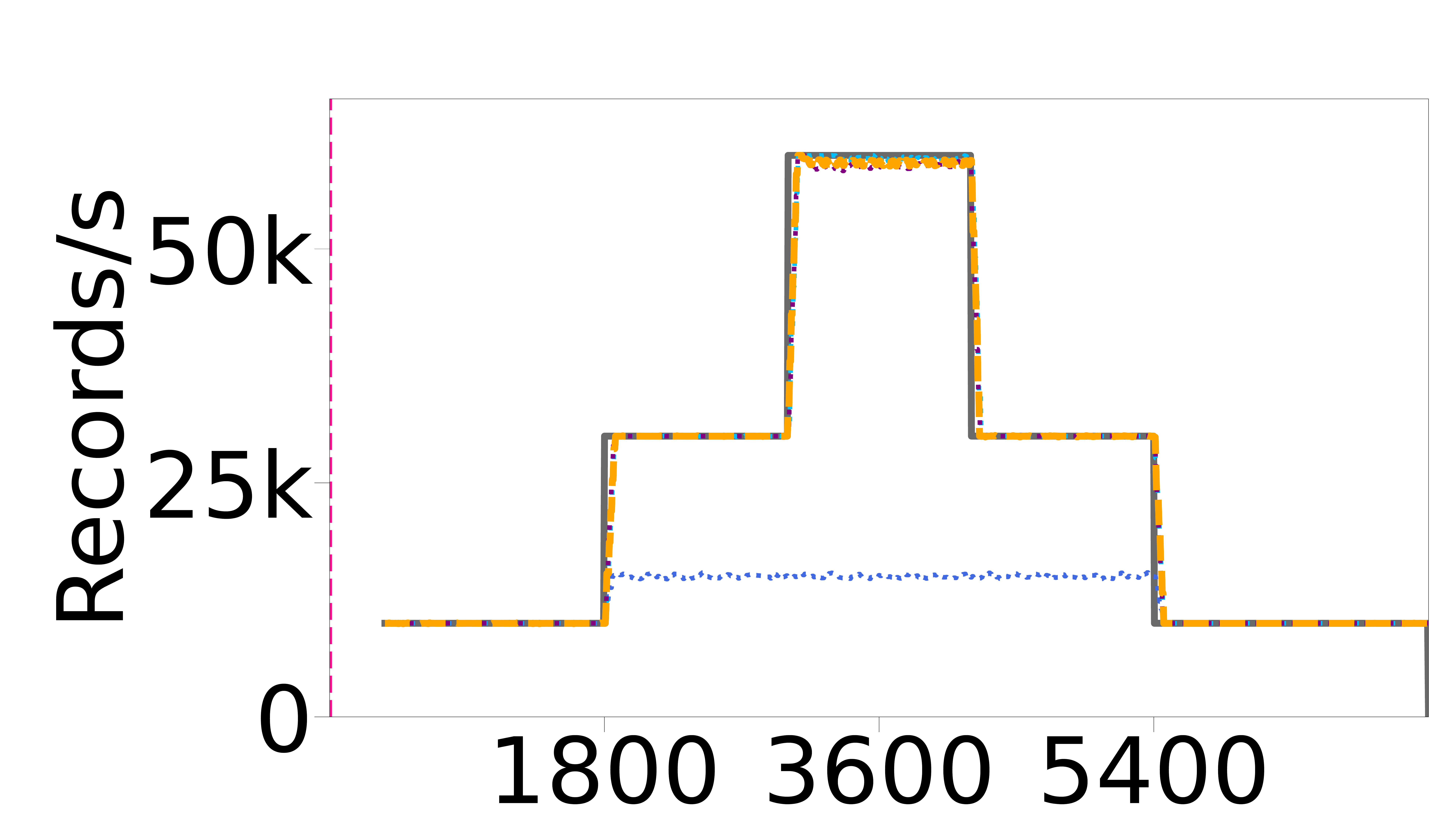}} &
    \adjustbox{valign=m}{\includegraphics[width=4.0cm]{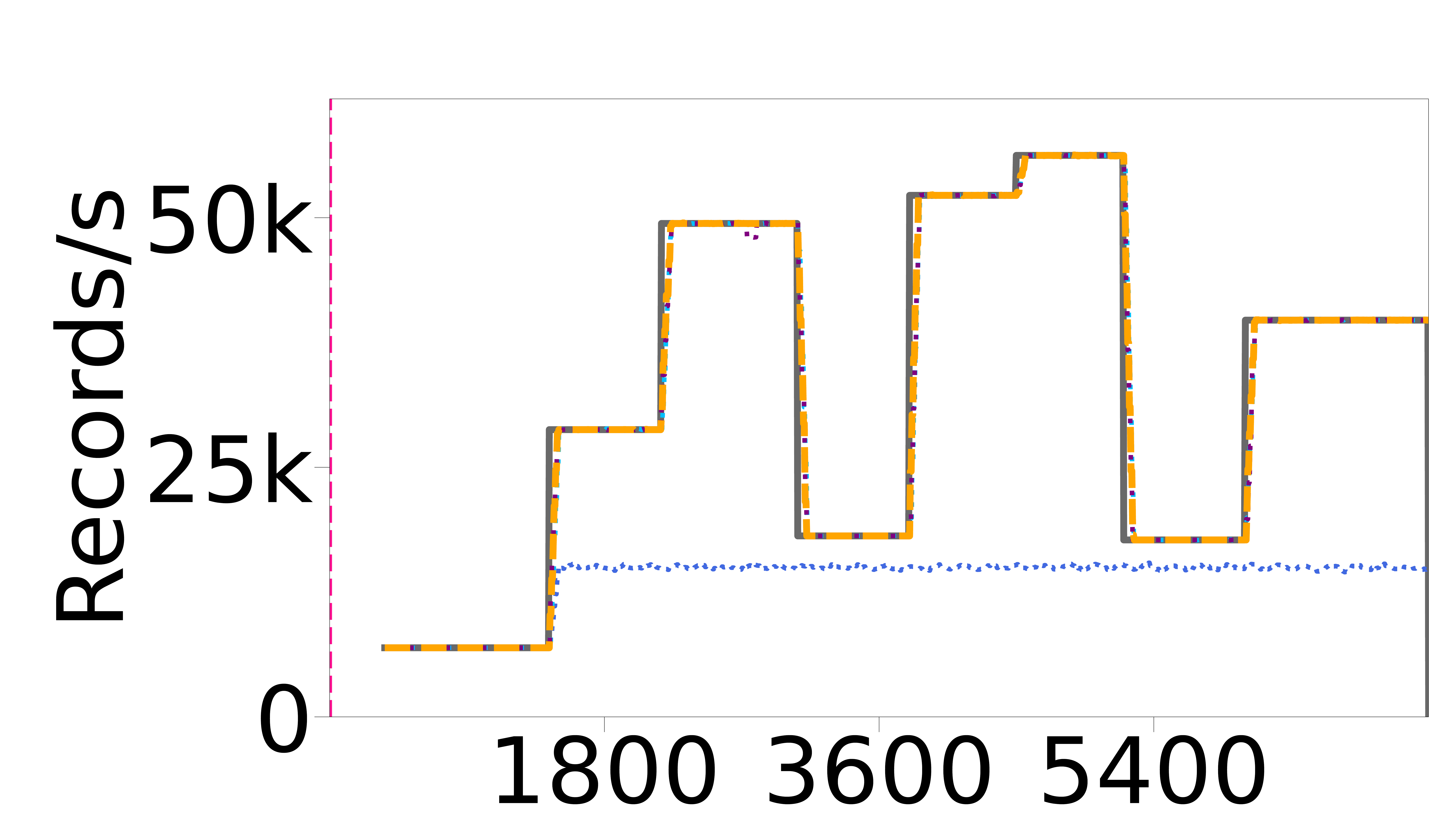}} \\
    \adjustbox{valign=m}{\rotatebox[origin=c]{90}{\textbf{Q5}}} &
    \adjustbox{valign=m}{\includegraphics[width=4.0cm]{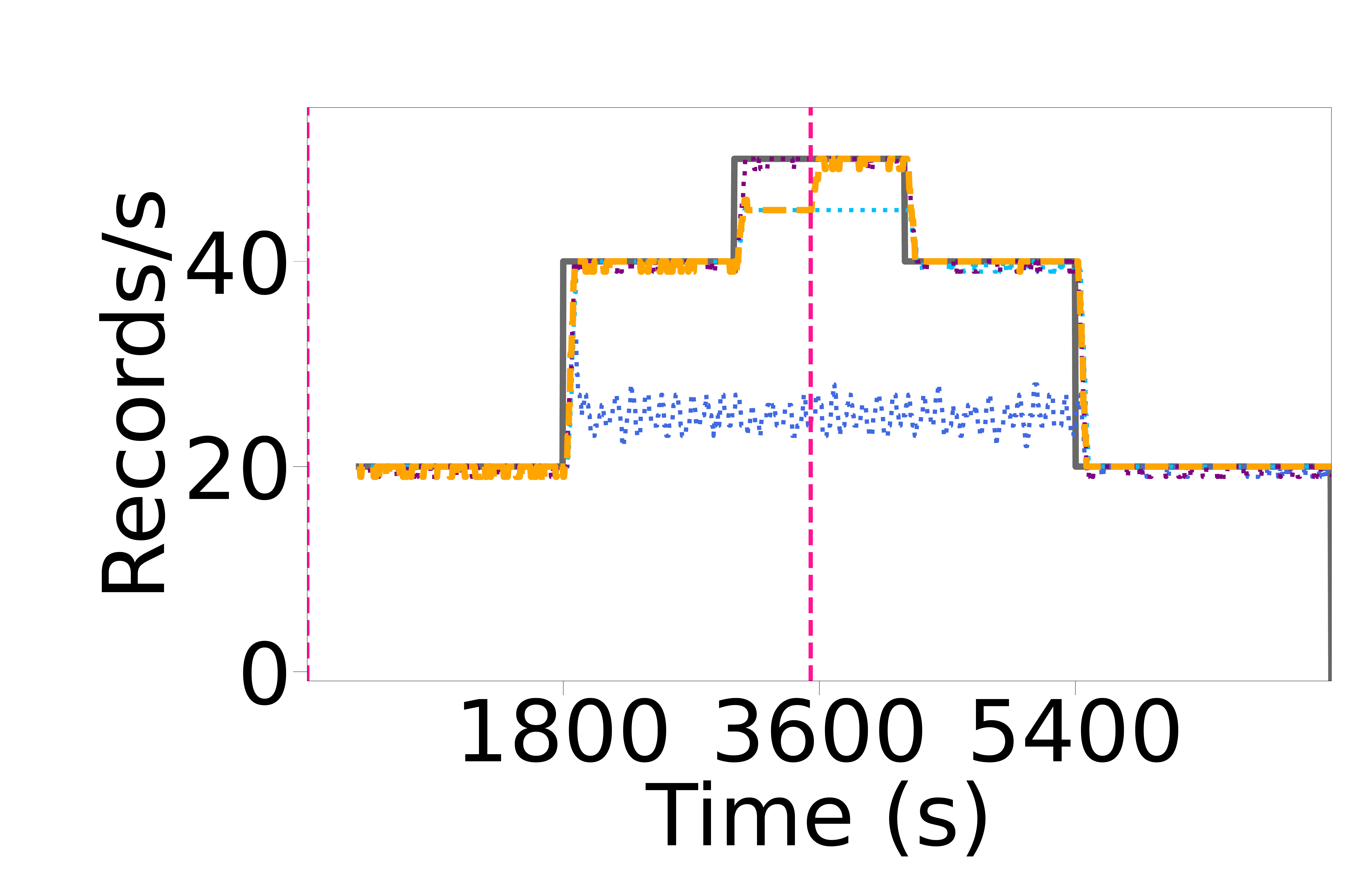}} &
    \adjustbox{valign=m}{\includegraphics[width=4.0cm]{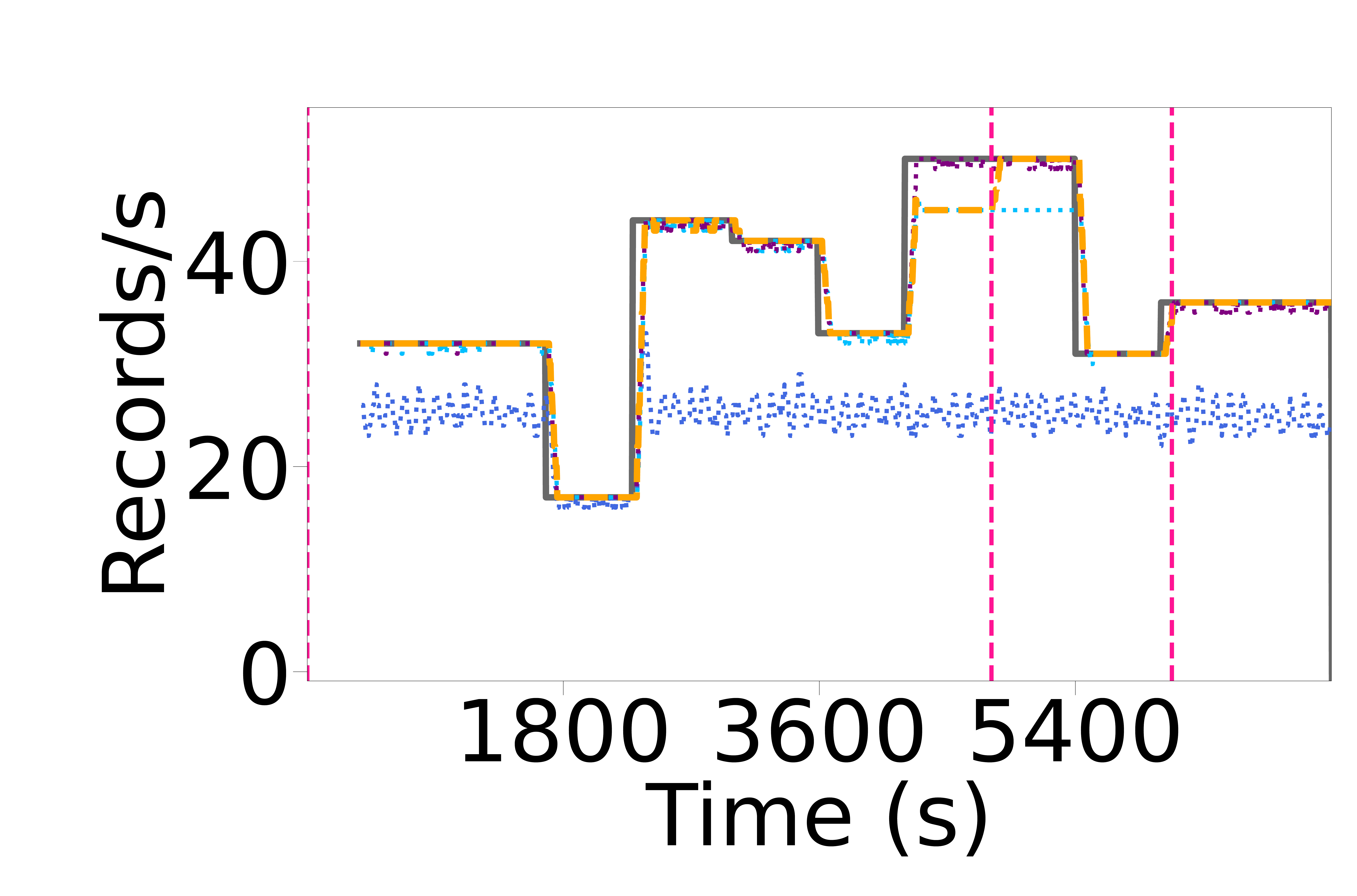}} \\
    & \adjustbox{valign=m}{\refstepcounter{subfigure}  (a) gradual rate pattern} & \adjustbox{valign=m}{\refstepcounter{subfigure}  (b) abrupt rate pattern} \\
\end{tabular}
\captionof{figure}{Impact of input rate - RPi2} 
\label{fig:sup_vary_rate_rpi2}
\end{minipage}\par\vspace{\textfloatsep}

\noindent\begin{minipage}{\columnwidth}
\centering
\begin{minipage}[b]{0.8\linewidth}
\includegraphics[width=\linewidth]{figures/eval/shift/res_video6nw_0.pdf_legend.pdf}
    \vspace{-0.4cm}
\end{minipage}
\setlength{\tabcolsep}{1pt}
\renewcommand{\arraystretch}{1.5}
\begin{tabular}{ccc}
    \adjustbox{valign=m}{\rotatebox[origin=c]{90}{\textbf{Q1}}} &
    \adjustbox{valign=m}{\includegraphics[width=4.0cm]{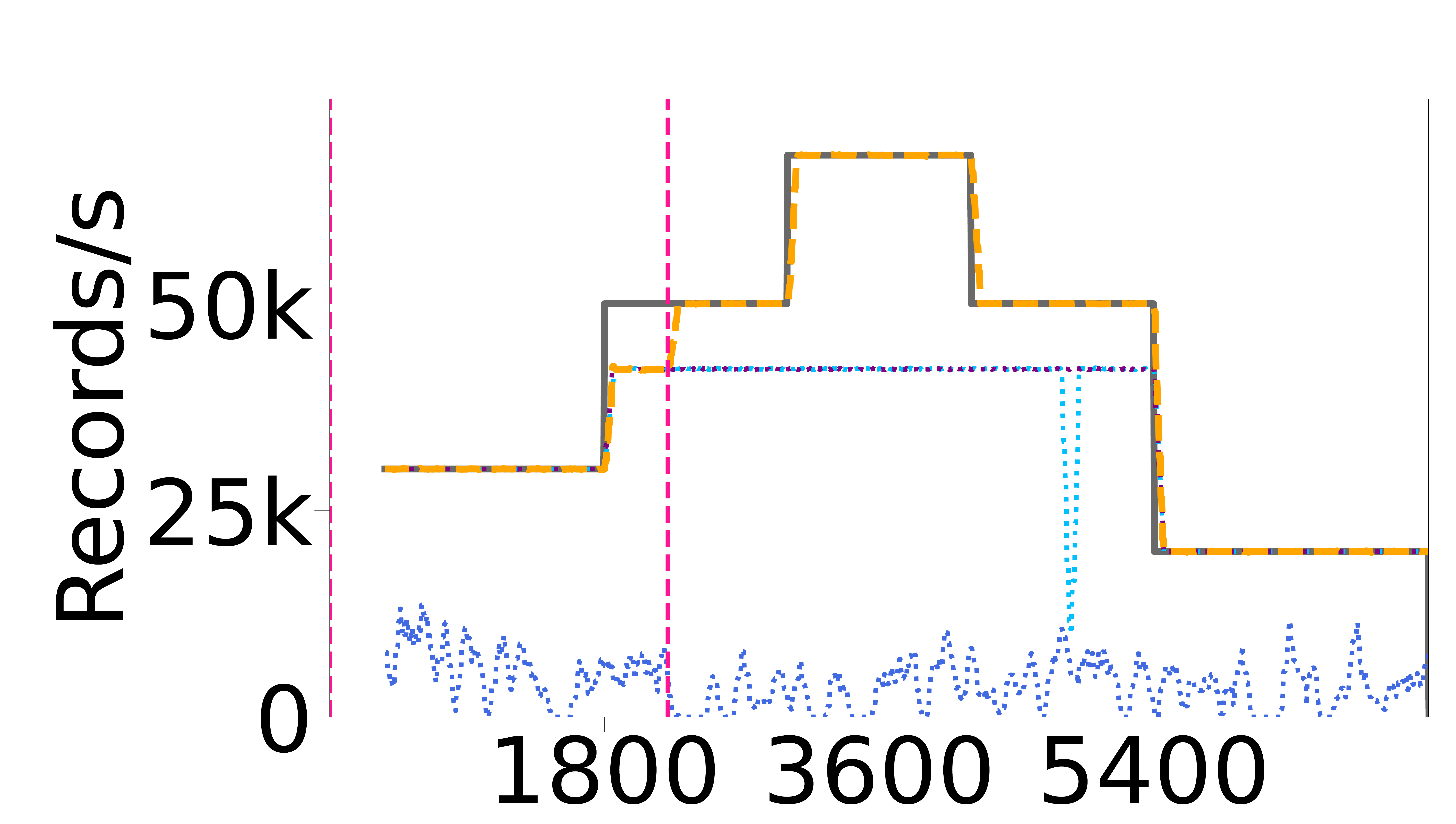}} &
    \adjustbox{valign=m}{\includegraphics[width=4.0cm]{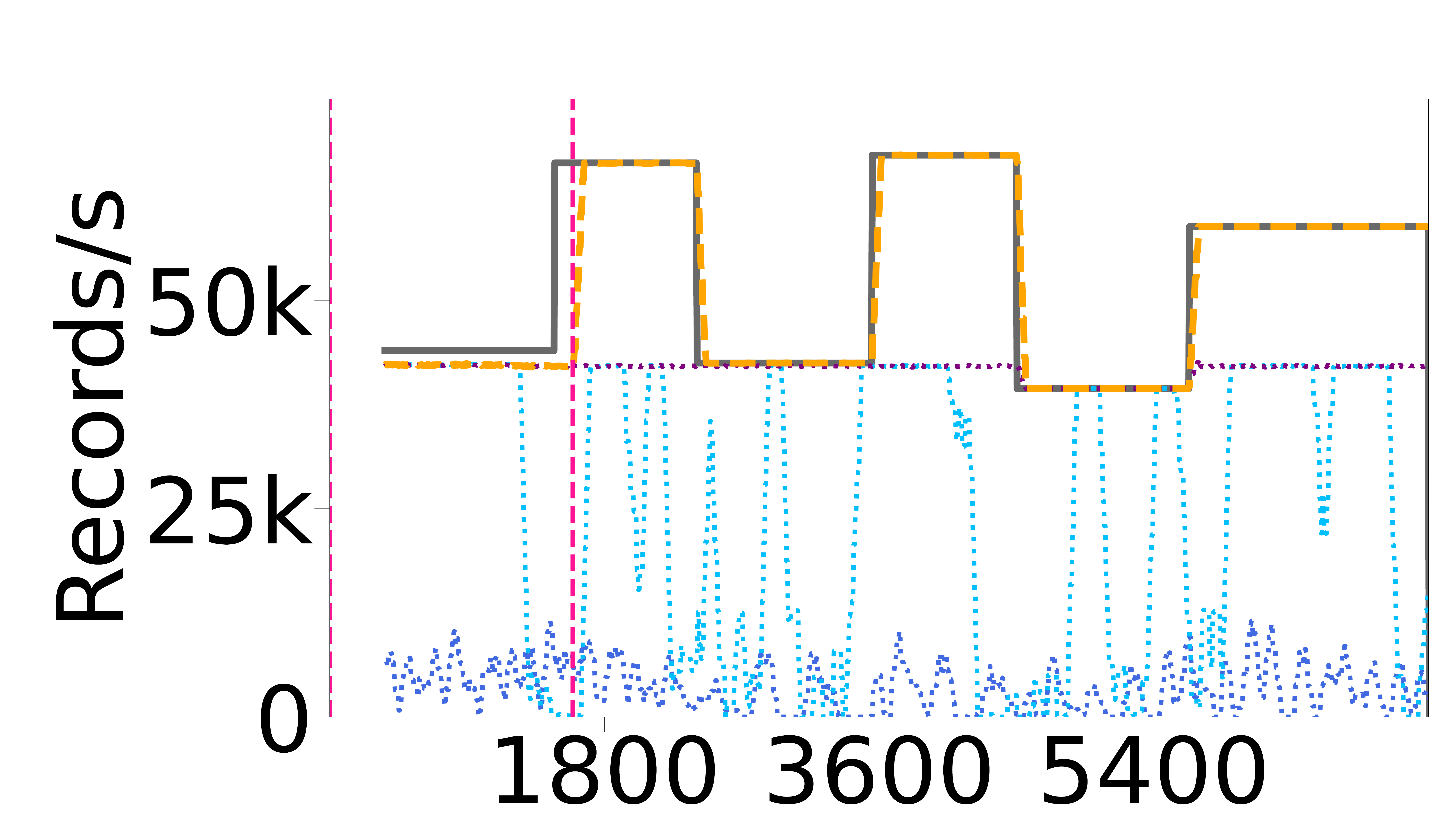}} \\
    \adjustbox{valign=m}{\rotatebox[origin=c]{90}{\textbf{Q2}}} &
    \adjustbox{valign=m}{\includegraphics[width=4.0cm]{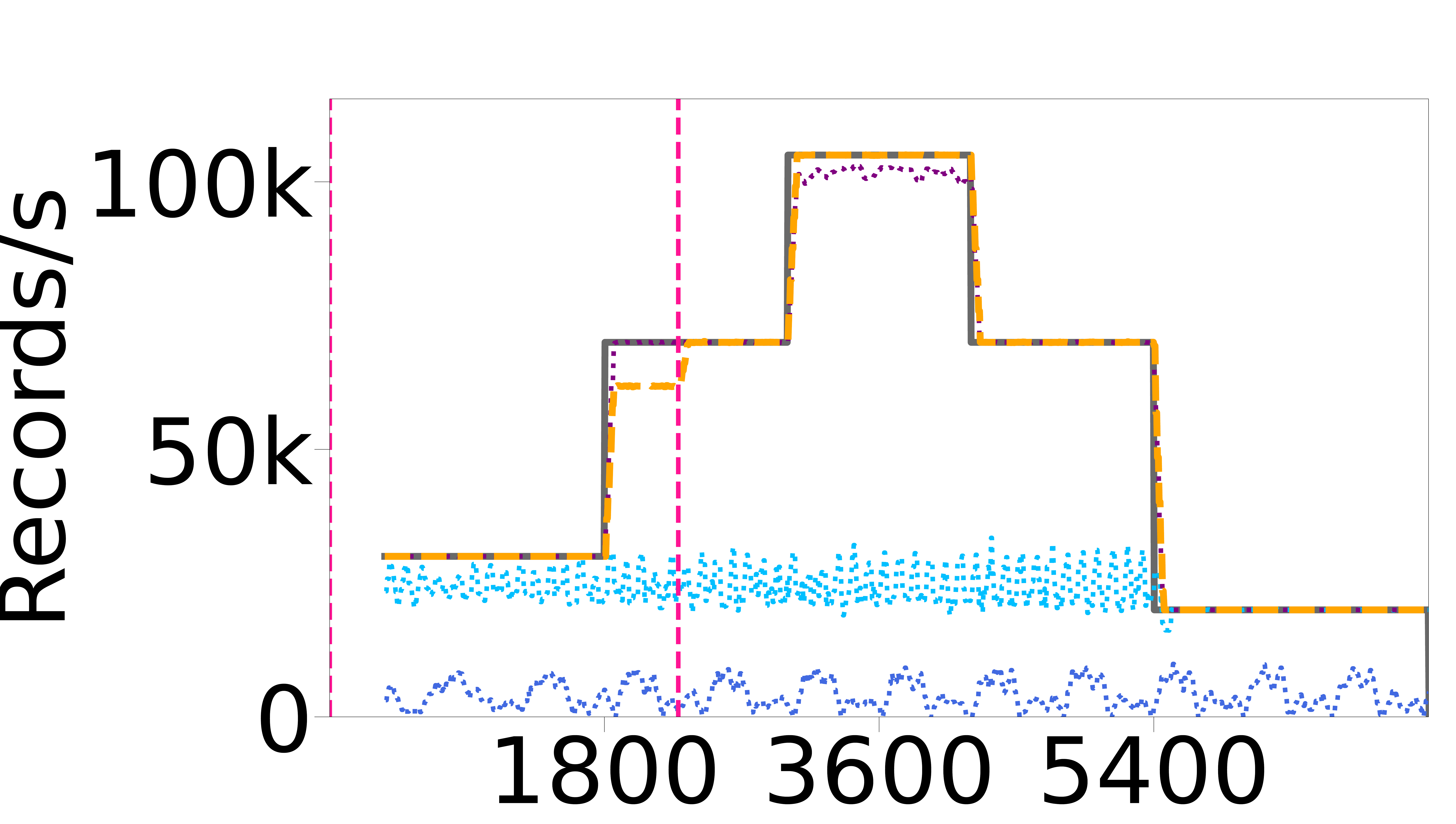}} &
    \adjustbox{valign=m}{\includegraphics[width=4.0cm]{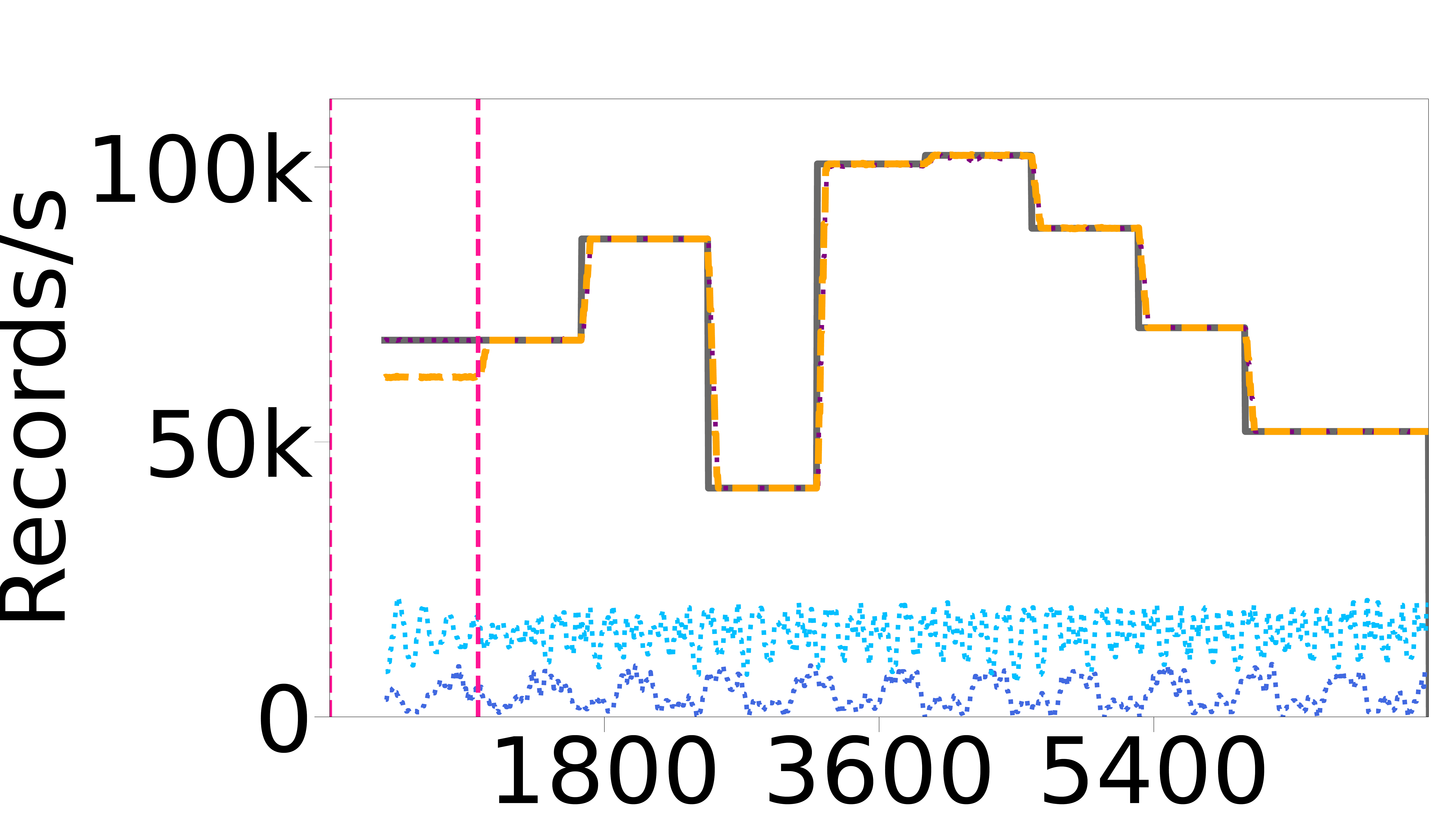}} \\
    \adjustbox{valign=m}{\rotatebox[origin=c]{90}{\textbf{Q3}}} &
    \adjustbox{valign=m}{\includegraphics[width=4.0cm]{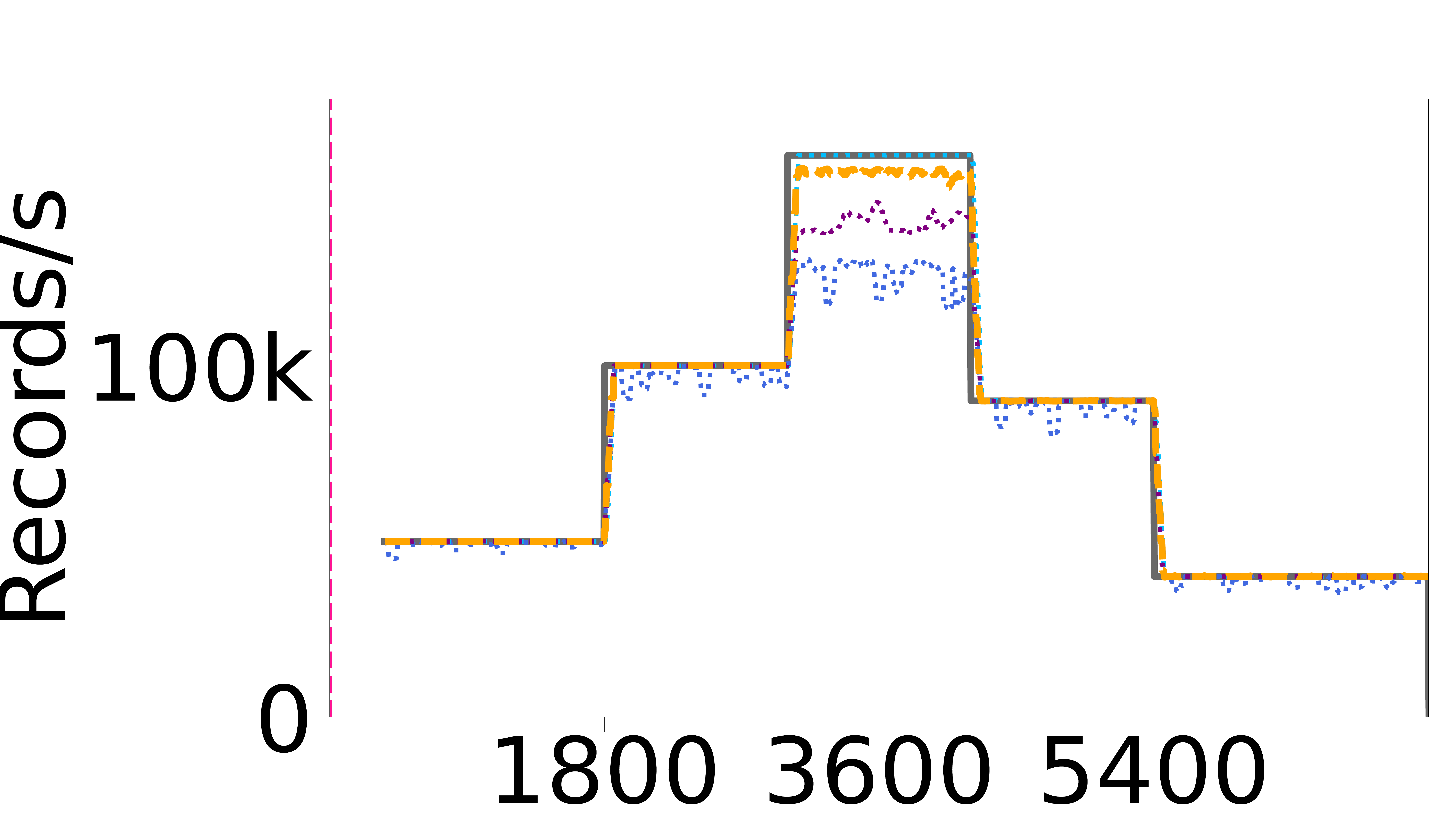}} &
    \adjustbox{valign=m}{\includegraphics[width=4.0cm]{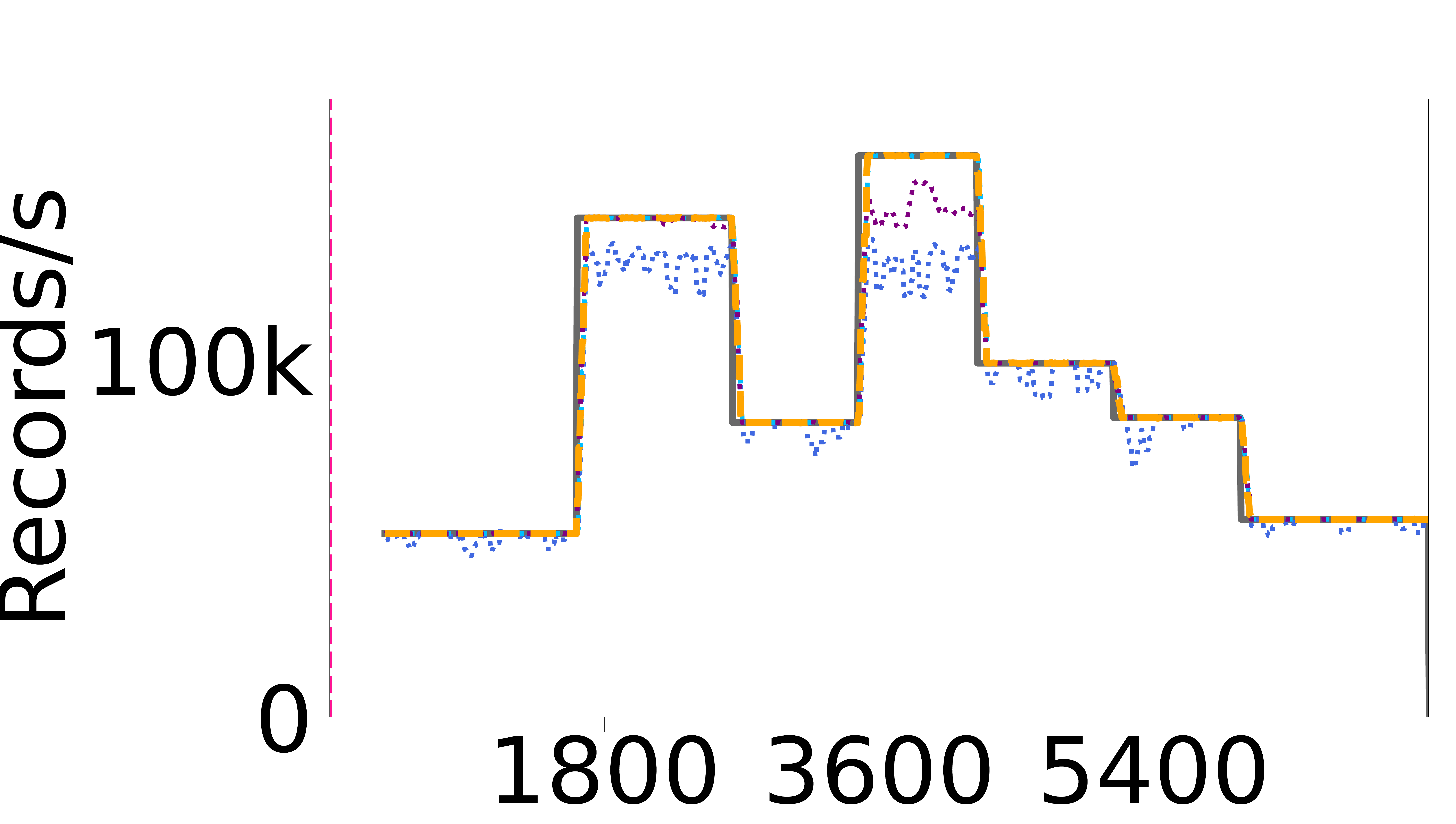}} \\
    \adjustbox{valign=m}{\rotatebox[origin=c]{90}{\textbf{Q4}}} &
    \adjustbox{valign=m}{\includegraphics[width=4.0cm]{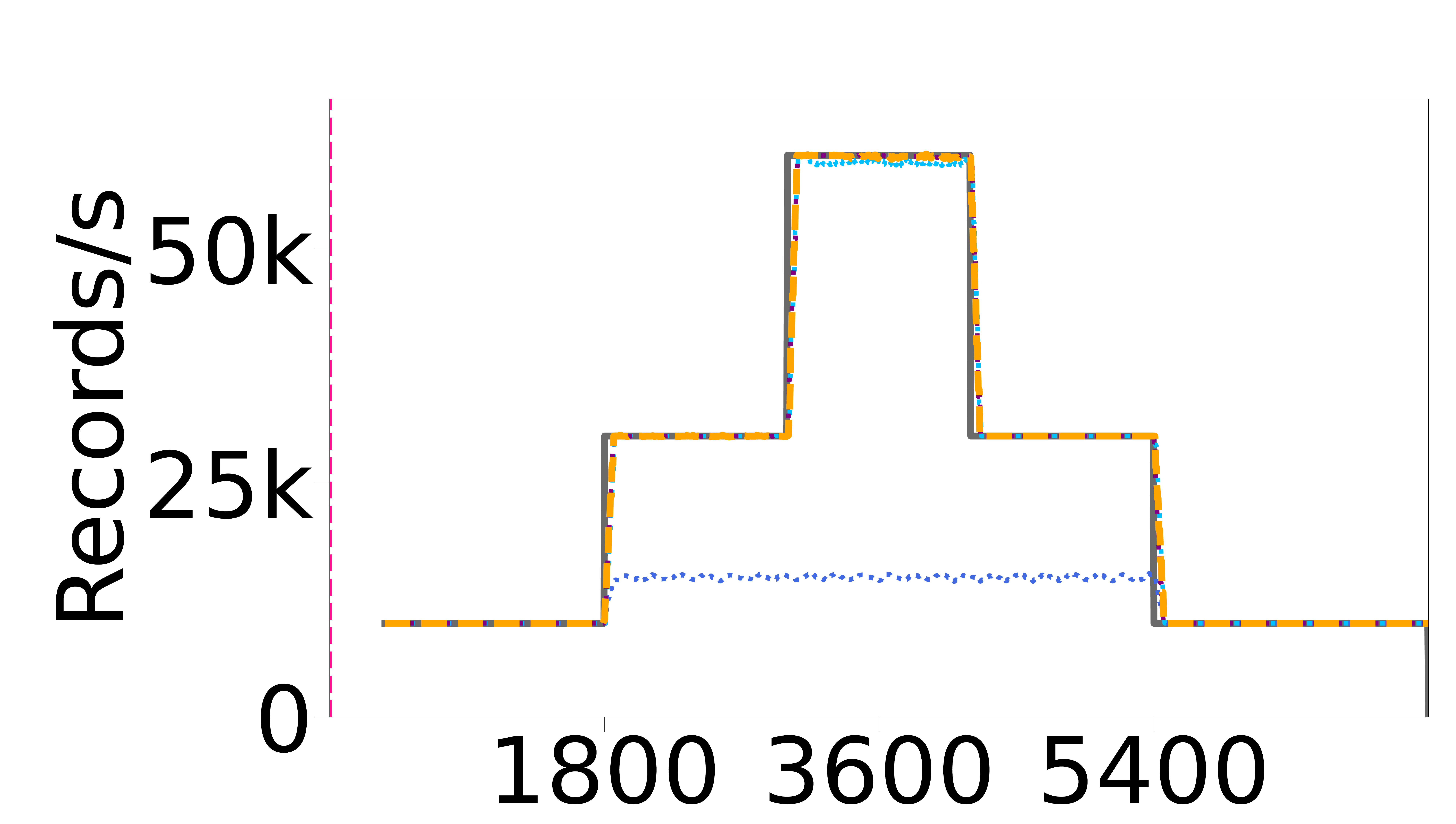}} &
    \adjustbox{valign=m}{\includegraphics[width=4.0cm]{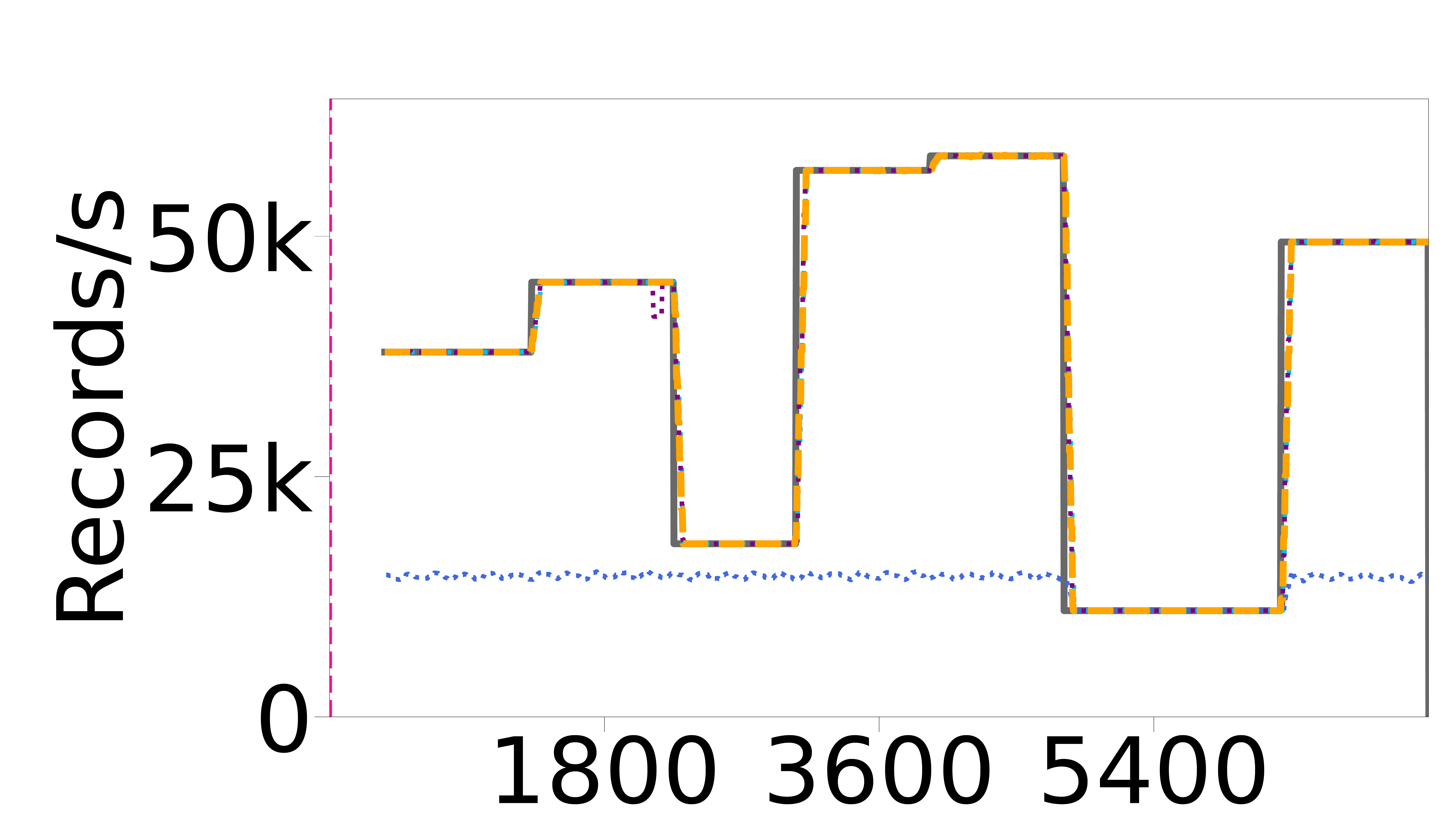}} \\
    \adjustbox{valign=m}{\rotatebox[origin=c]{90}{\textbf{Q5}}} &
    \adjustbox{valign=m}{\includegraphics[width=4.0cm]{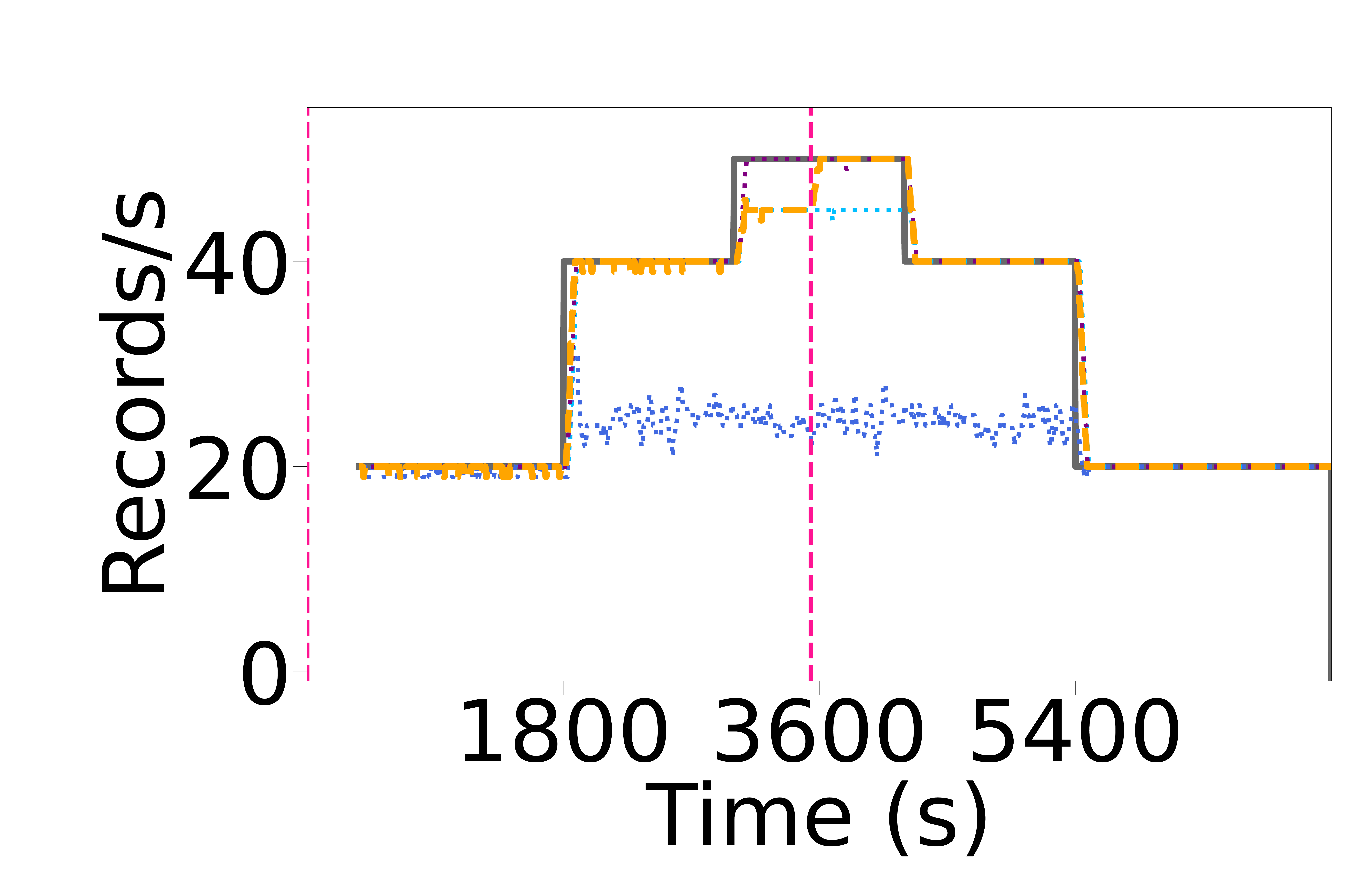}} &
    \adjustbox{valign=m}{\includegraphics[width=4.0cm]{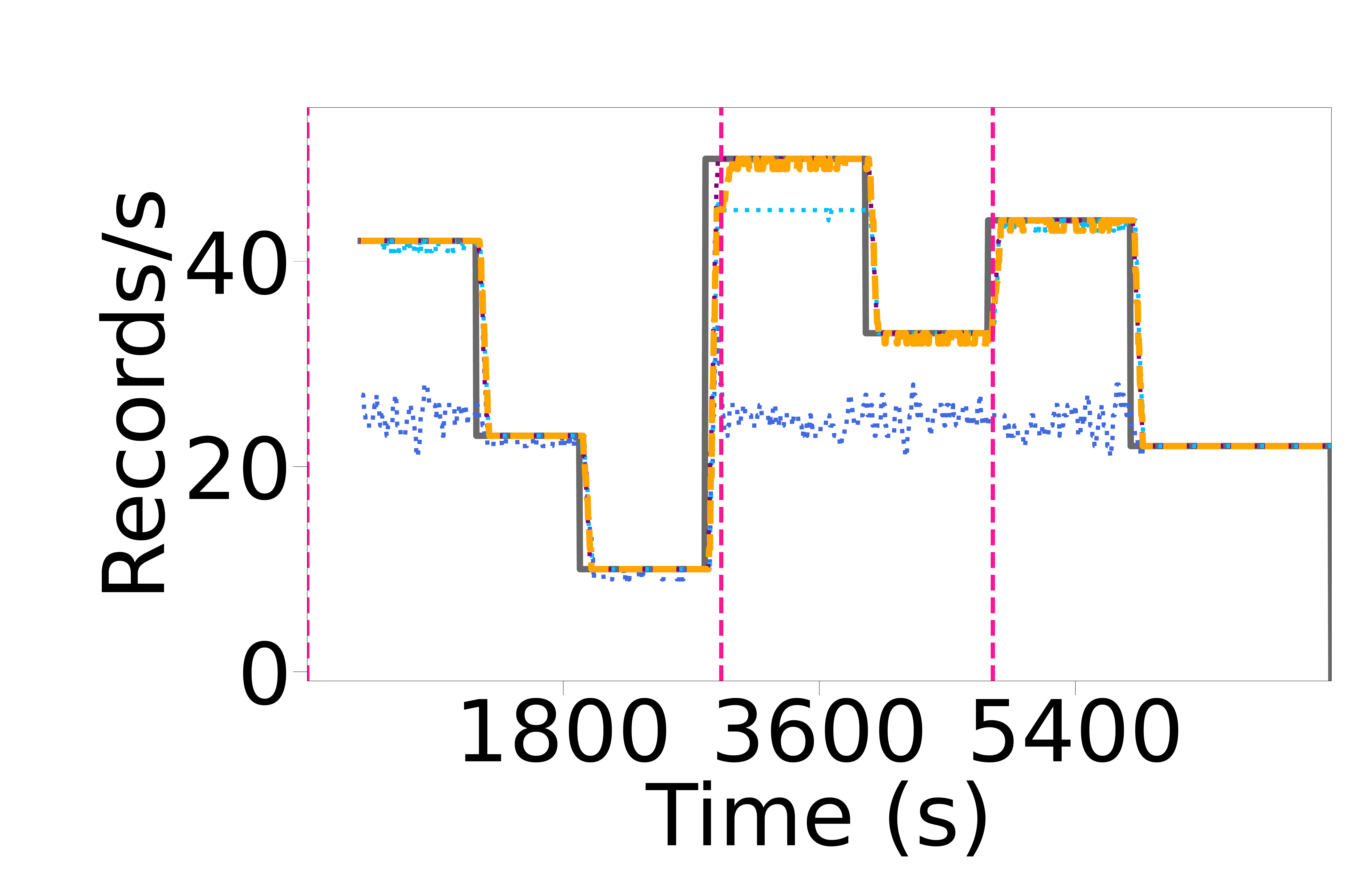}} \\
    & \adjustbox{valign=m}{\refstepcounter{subfigure}  (a) gradual rate pattern} & \adjustbox{valign=m}{\refstepcounter{subfigure}  (b) abrupt rate pattern} \\
\end{tabular}
\captionof{figure}{Impact of input rate - RPi3} 
\label{fig:sup_vary_rate_rpi3}
\end{minipage}\par\vspace{\textfloatsep}

\noindent\begin{minipage}{\columnwidth}
\centering
\begin{minipage}[b]{0.8\linewidth}
\includegraphics[width=\linewidth]{figures/eval/shift/res_video6nw_0.pdf_legend.pdf}
    \vspace{-0.4cm}
\end{minipage}
\setlength{\tabcolsep}{1pt}
\renewcommand{\arraystretch}{1.5}
\begin{tabular}{ccc}
    \adjustbox{valign=m}{\rotatebox[origin=c]{90}{\textbf{Q1}}} &
    \adjustbox{valign=m}{\includegraphics[width=4.0cm]{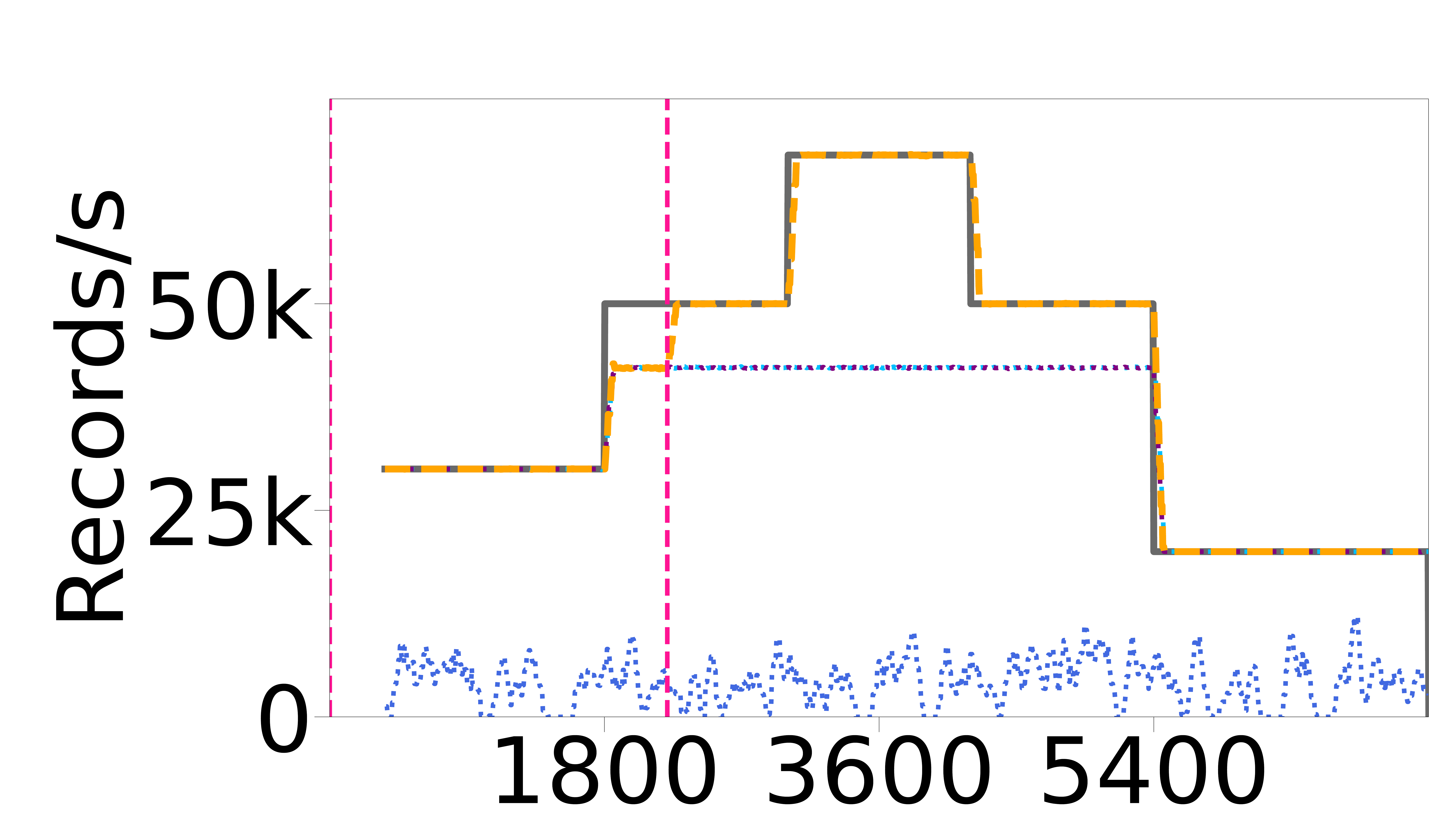}} &
    \adjustbox{valign=m}{\includegraphics[width=4.0cm]{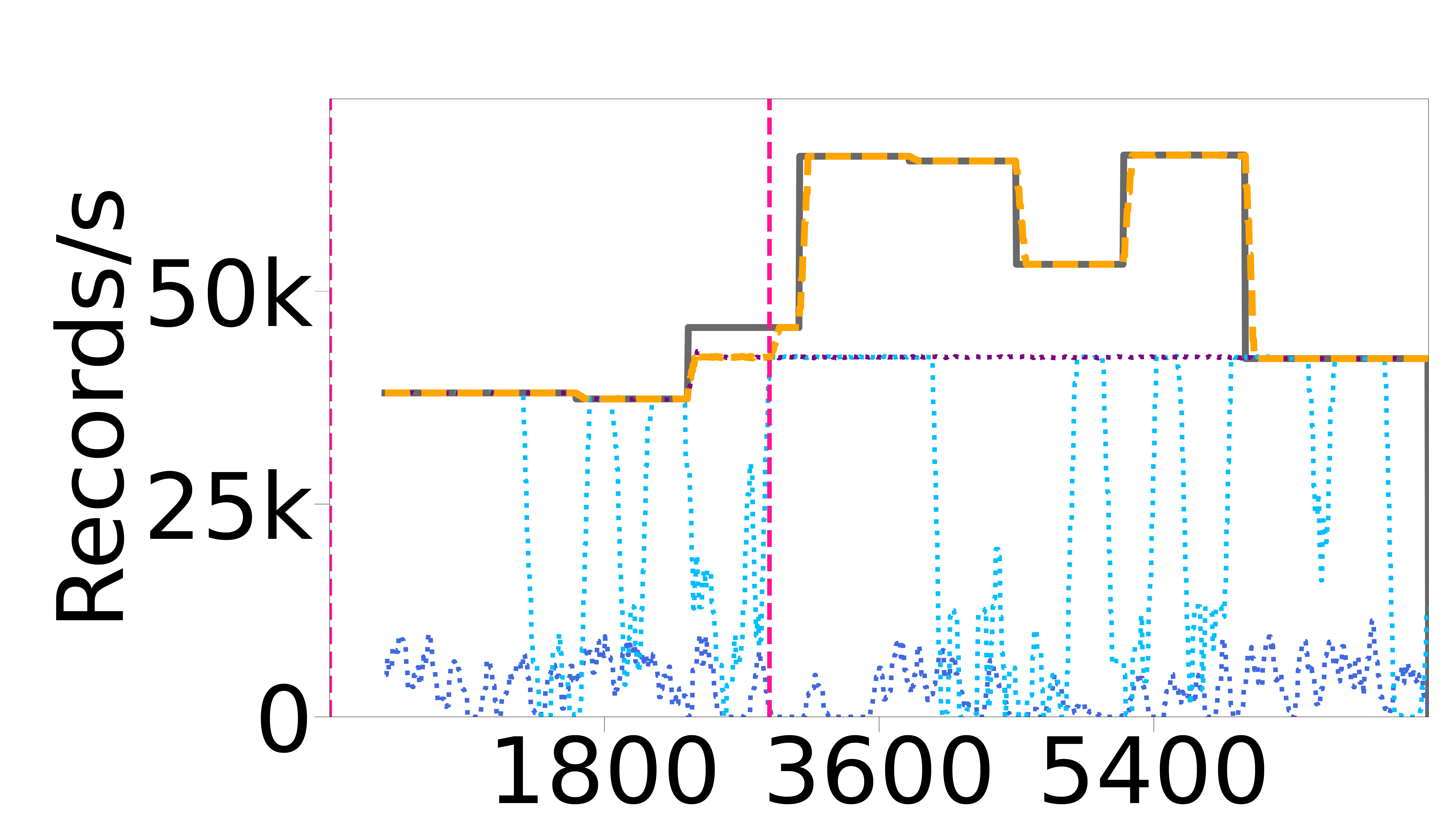}} \\
    \adjustbox{valign=m}{\rotatebox[origin=c]{90}{\textbf{Q2}}} &
    \adjustbox{valign=m}{\includegraphics[width=4.0cm]{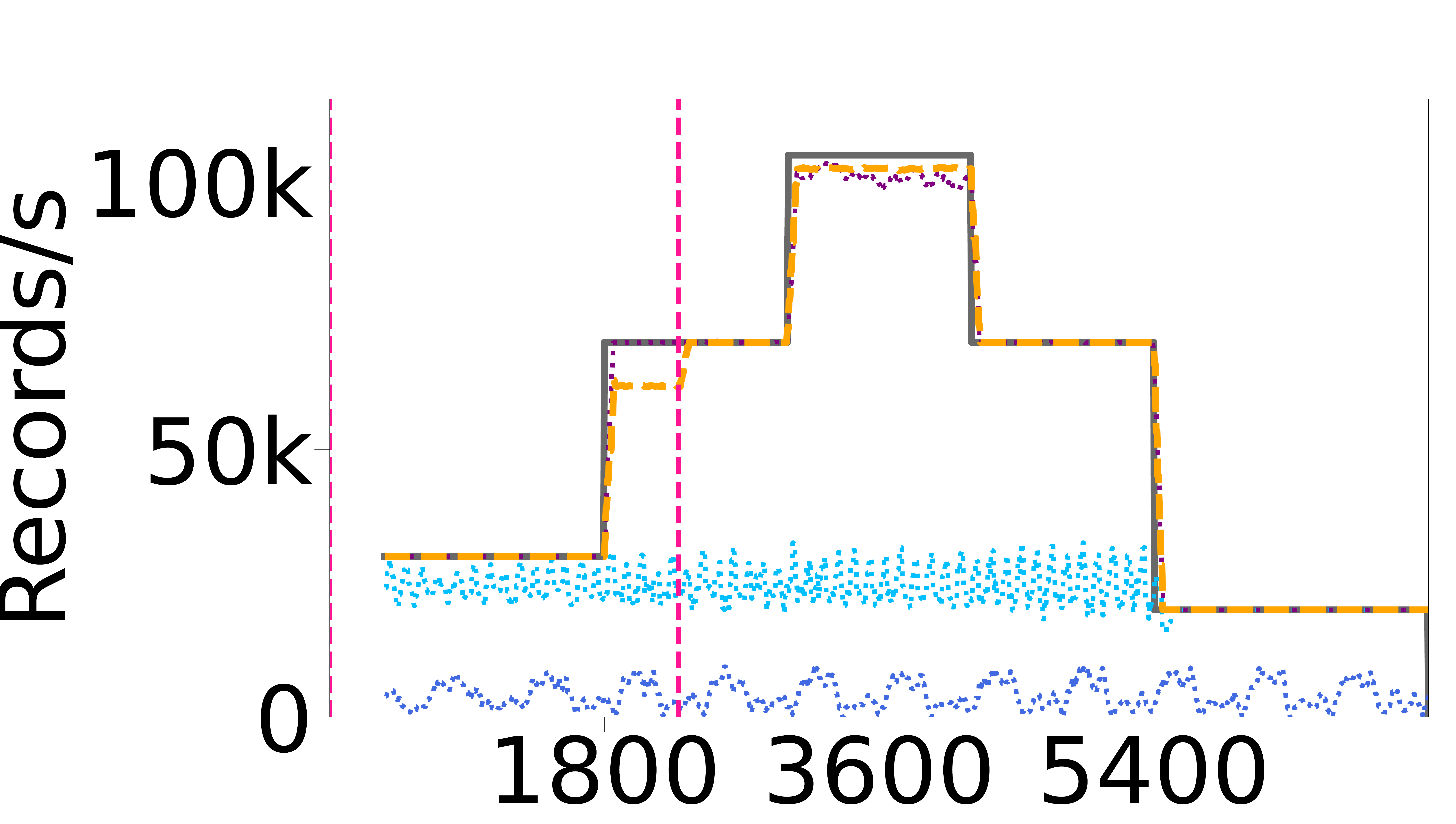}} &
    \adjustbox{valign=m}{\includegraphics[width=4.0cm]{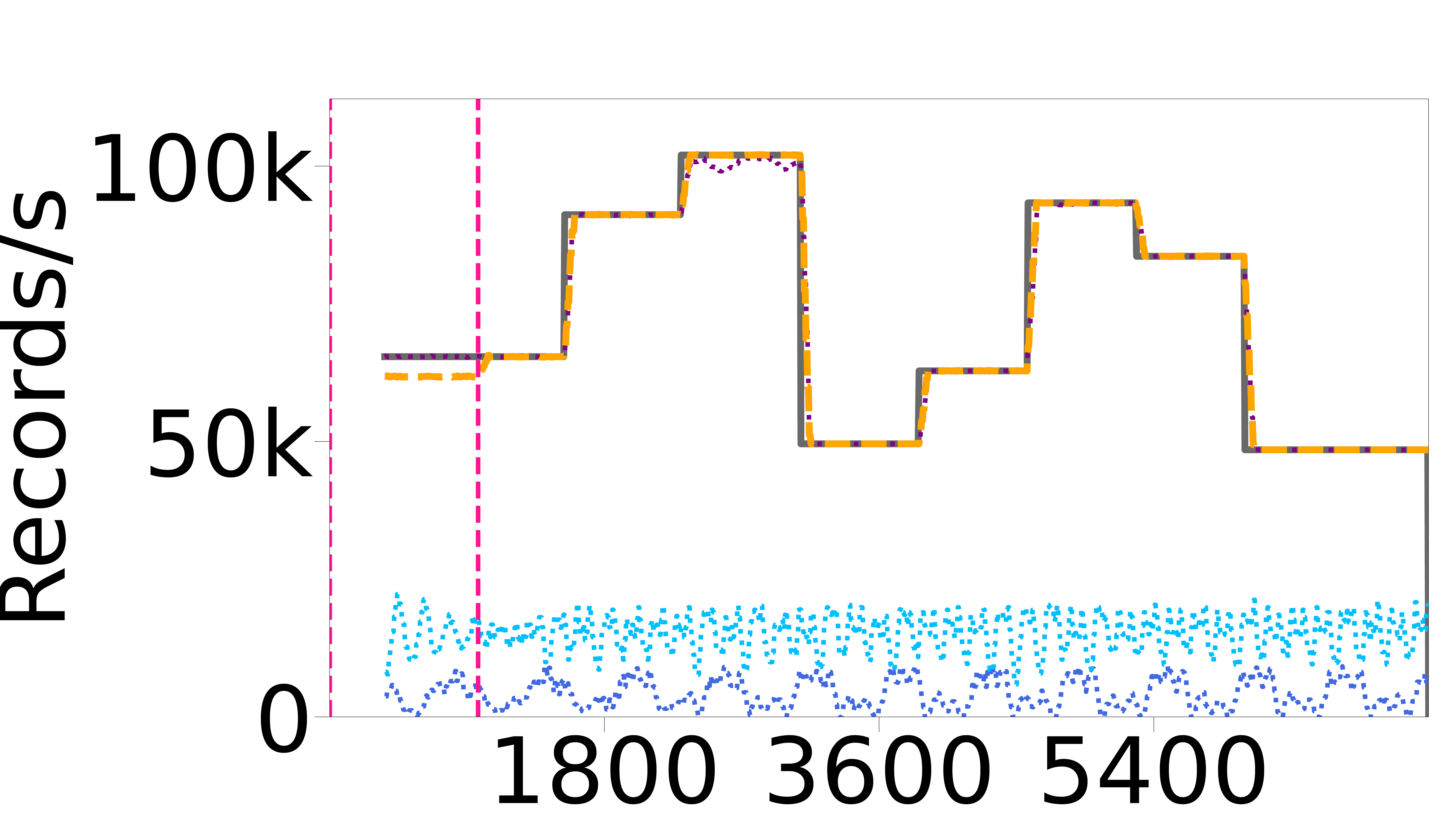}} \\
    \adjustbox{valign=m}{\rotatebox[origin=c]{90}{\textbf{Q3}}} &
    \adjustbox{valign=m}{\includegraphics[width=4.0cm]{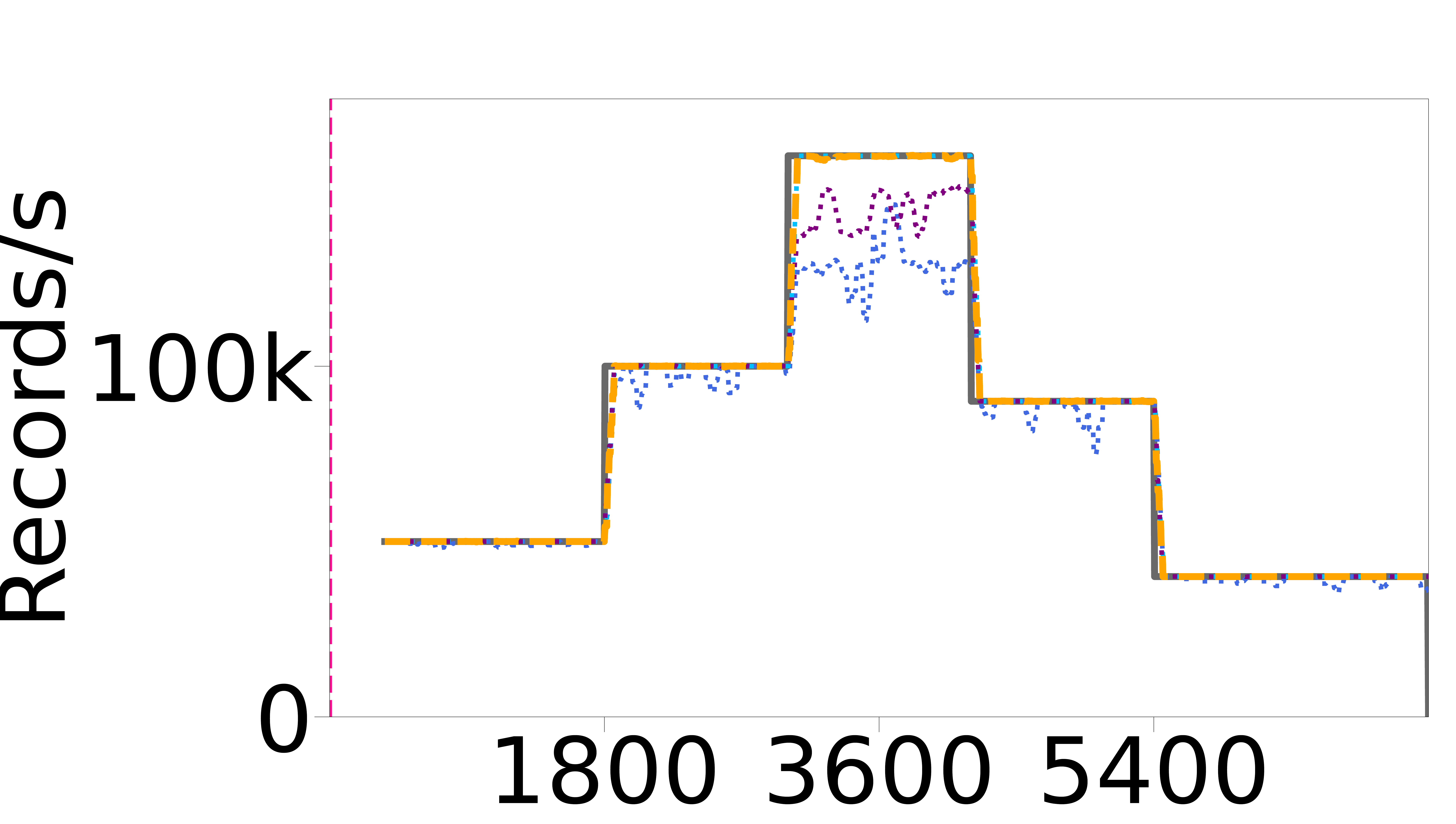}} &
    \adjustbox{valign=m}{\includegraphics[width=4.0cm]{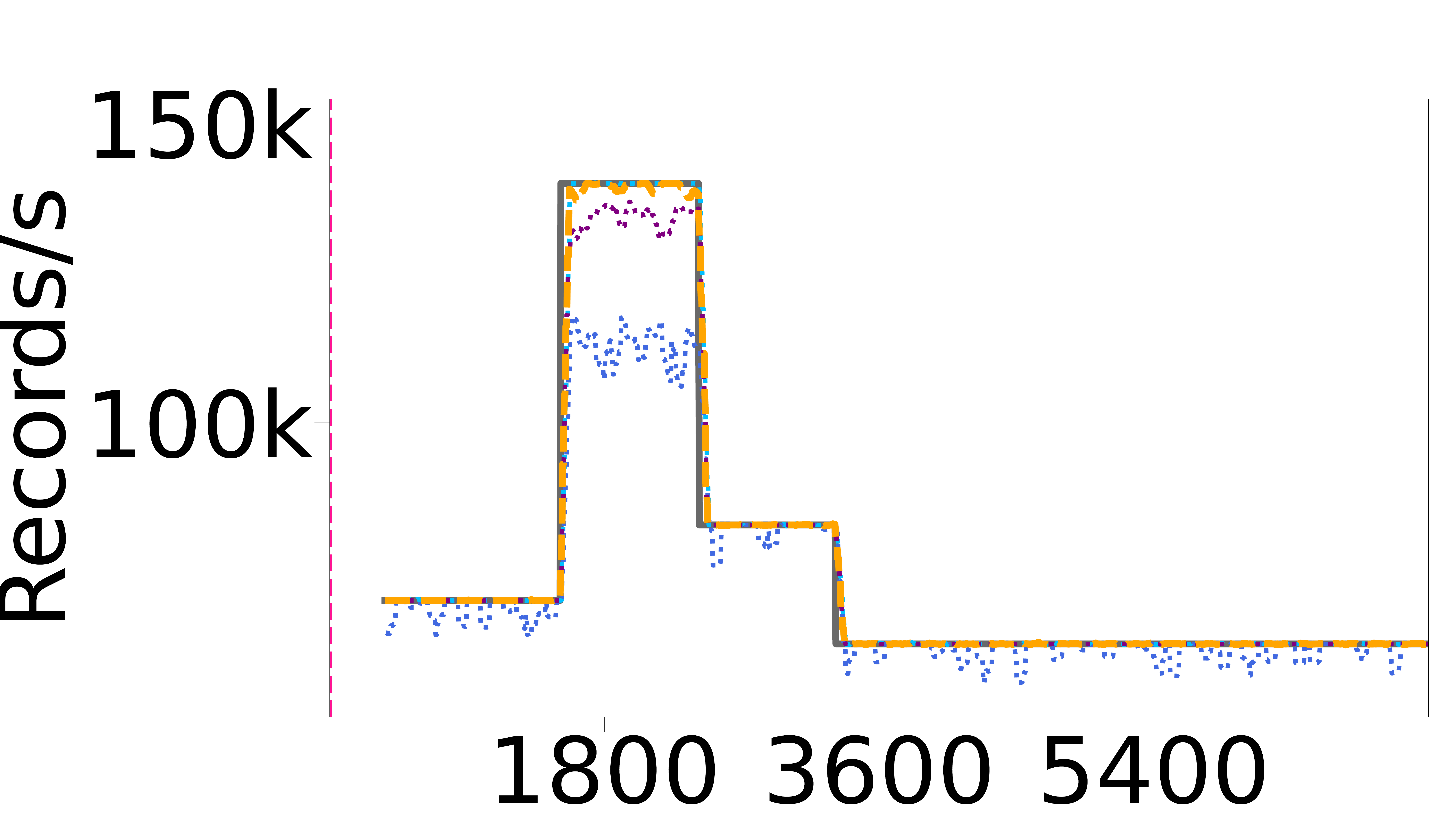}} \\
    \adjustbox{valign=m}{\rotatebox[origin=c]{90}{\textbf{Q4}}} &
    \adjustbox{valign=m}{\includegraphics[width=4.0cm]{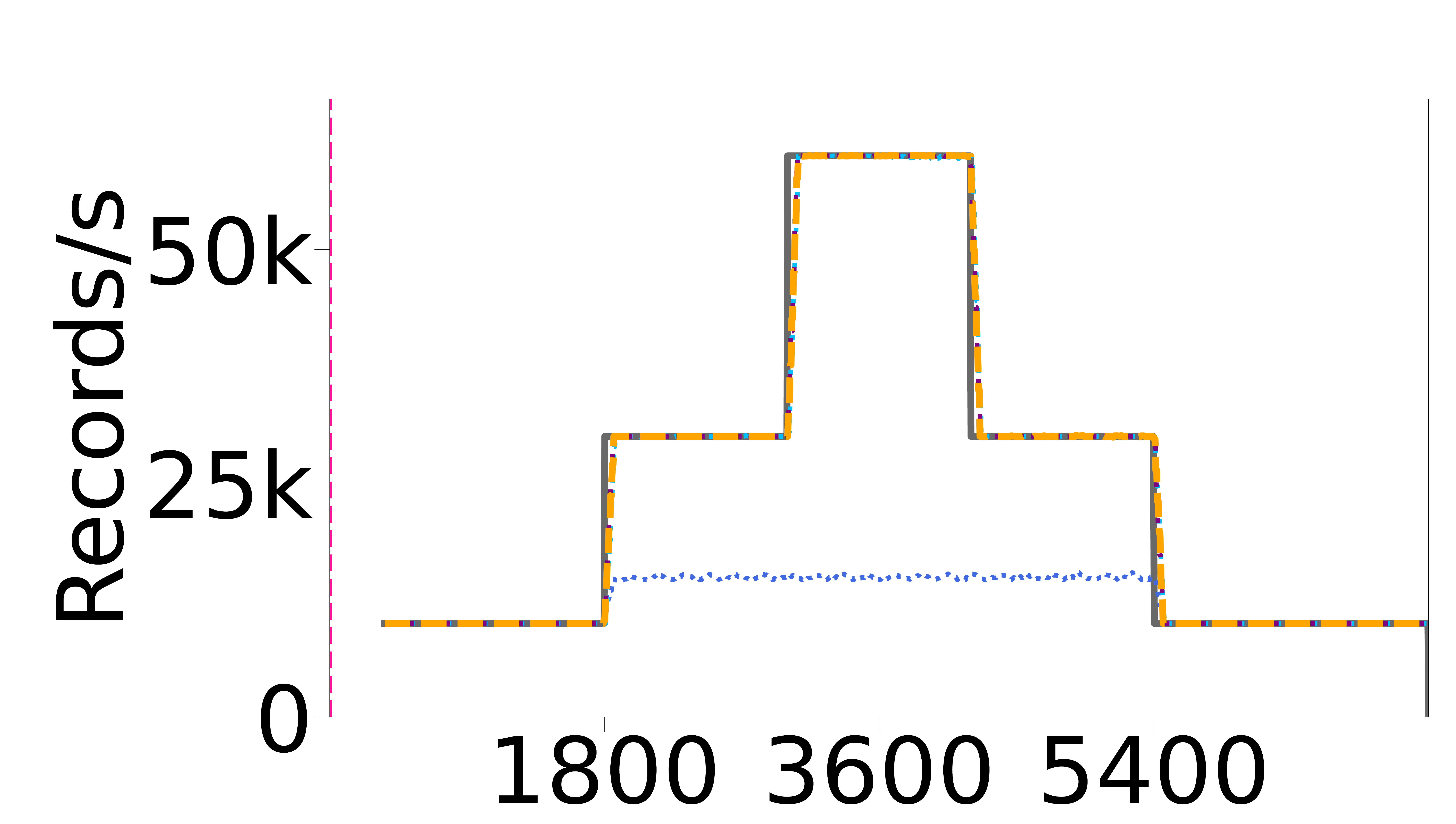}} &
    \adjustbox{valign=m}{\includegraphics[width=4.0cm]{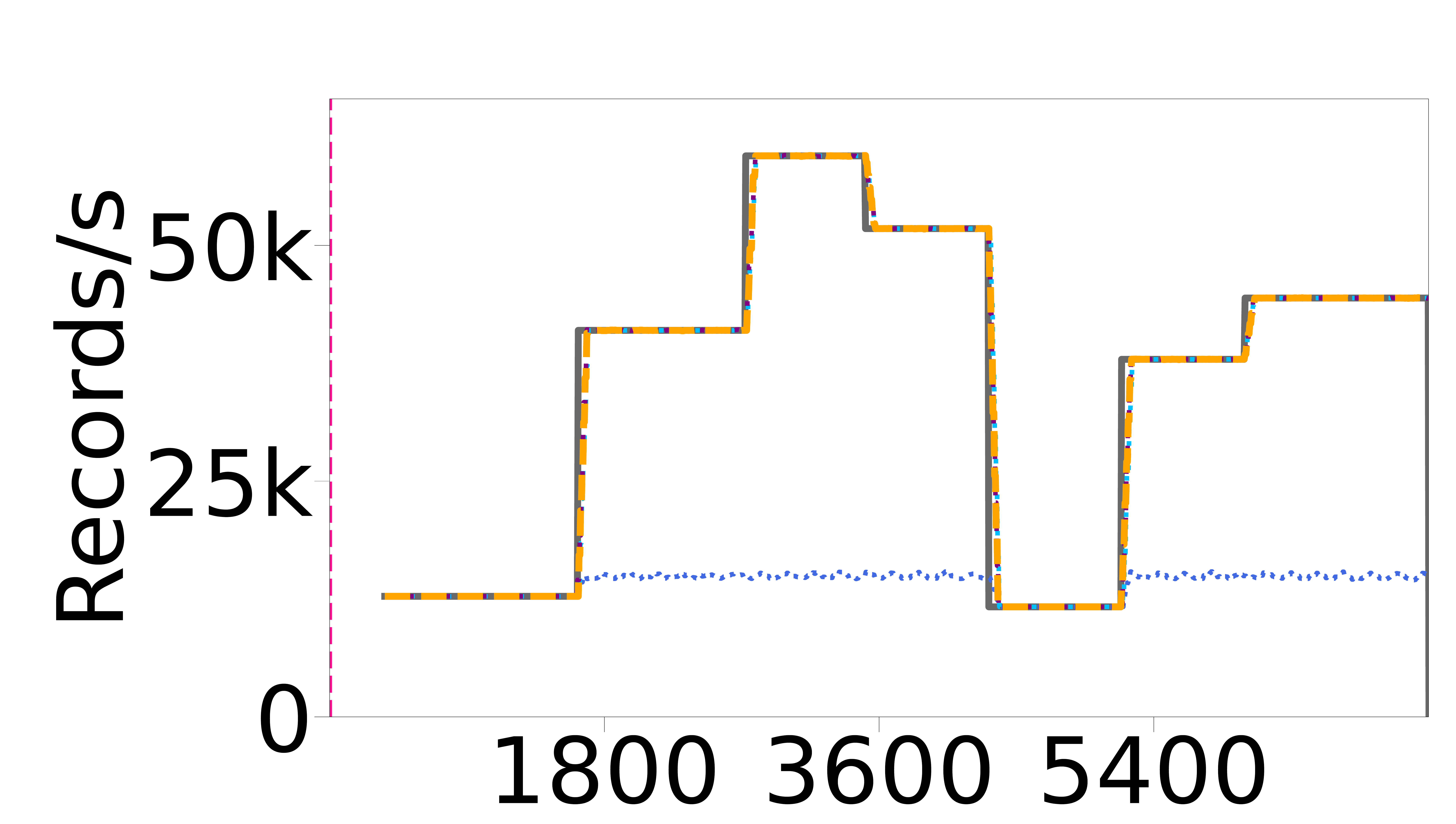}} \\
    \adjustbox{valign=m}{\rotatebox[origin=c]{90}{\textbf{Q5}}} &
    \adjustbox{valign=m}{\includegraphics[width=4.0cm]{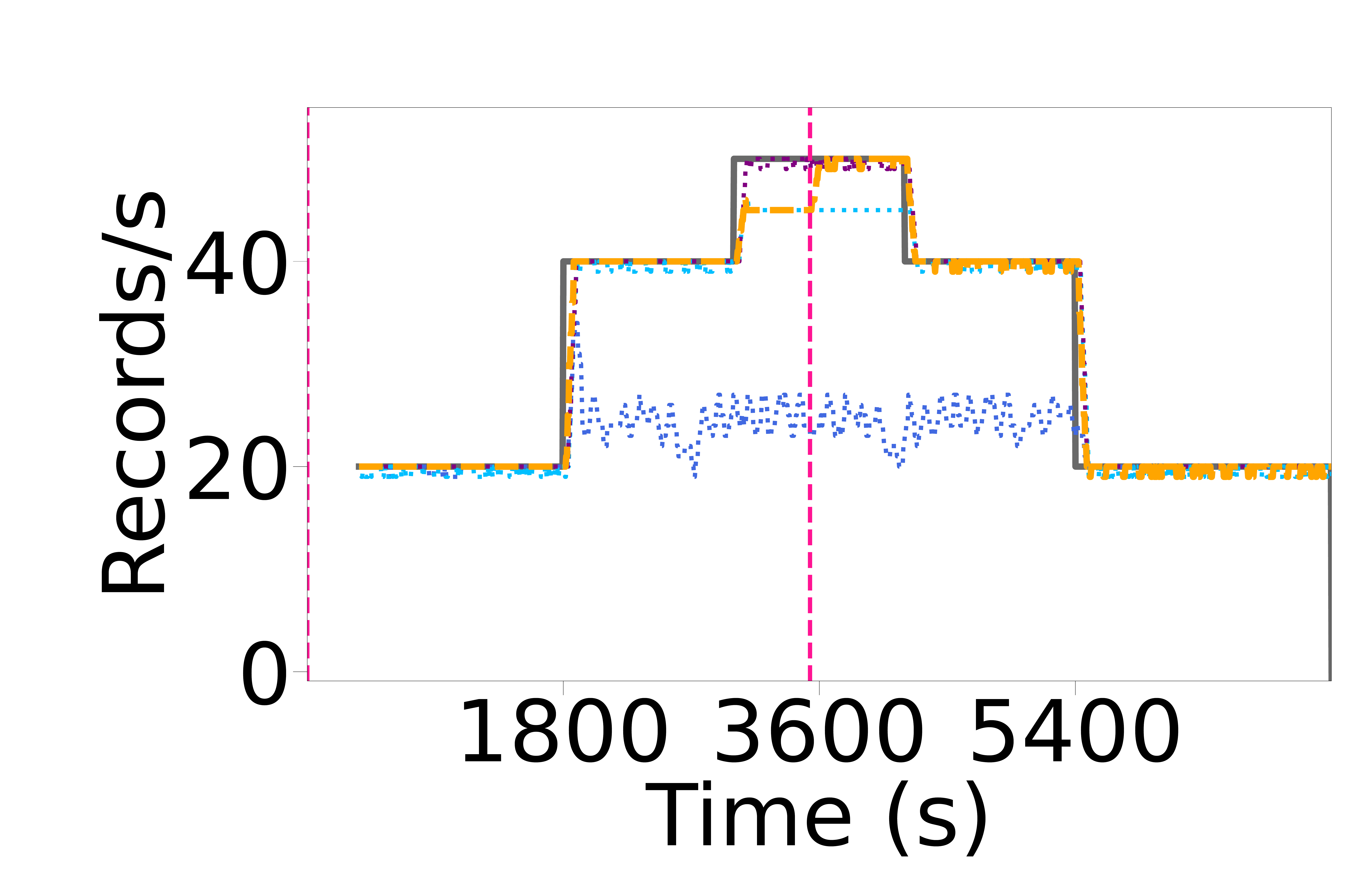}} &
    \adjustbox{valign=m}{\includegraphics[width=4.0cm]{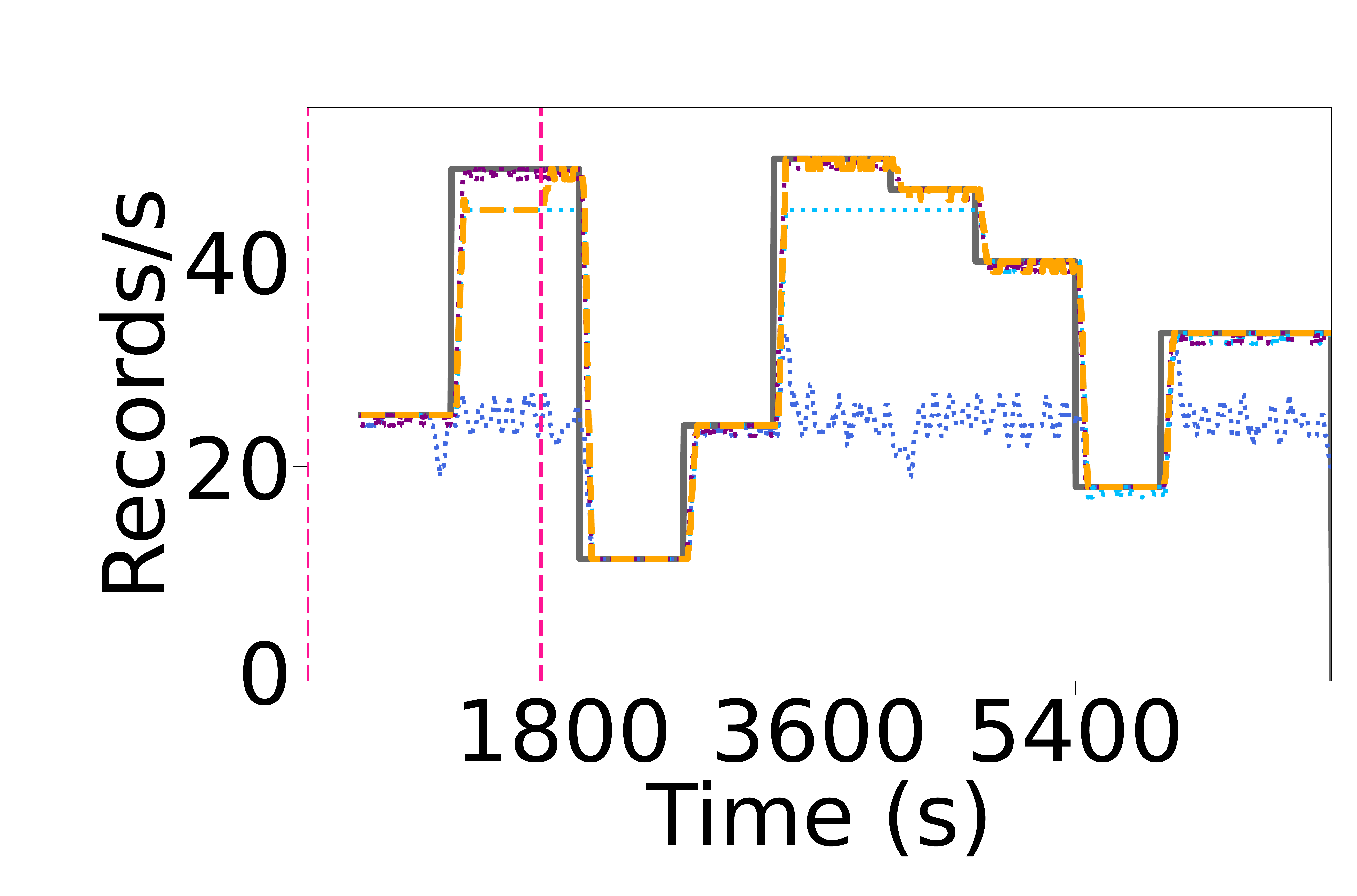}} \\
    & \adjustbox{valign=m}{\refstepcounter{subfigure}  (a) gradual rate pattern} & \adjustbox{valign=m}{\refstepcounter{subfigure}  (b) abrupt rate pattern} \\
\end{tabular}
\captionof{figure}{Impact of input rate - RPi4} 
\label{fig:sup_vary_rate_rpi4}
\end{minipage}\par\vspace{\textfloatsep}

\noindent\begin{minipage}{\columnwidth}
\centering
\begin{minipage}[b]{0.8\linewidth}
\includegraphics[width=\linewidth]{figures/eval/shift/res_video6nw_0.pdf_legend.pdf}
    \vspace{-0.4cm}
\end{minipage}
\setlength{\tabcolsep}{1pt}
\renewcommand{\arraystretch}{1.5}
\begin{tabular}{ccc}
    \adjustbox{valign=m}{\rotatebox[origin=c]{90}{\textbf{Q1}}} &
    \adjustbox{valign=m}{\includegraphics[width=3.8cm]{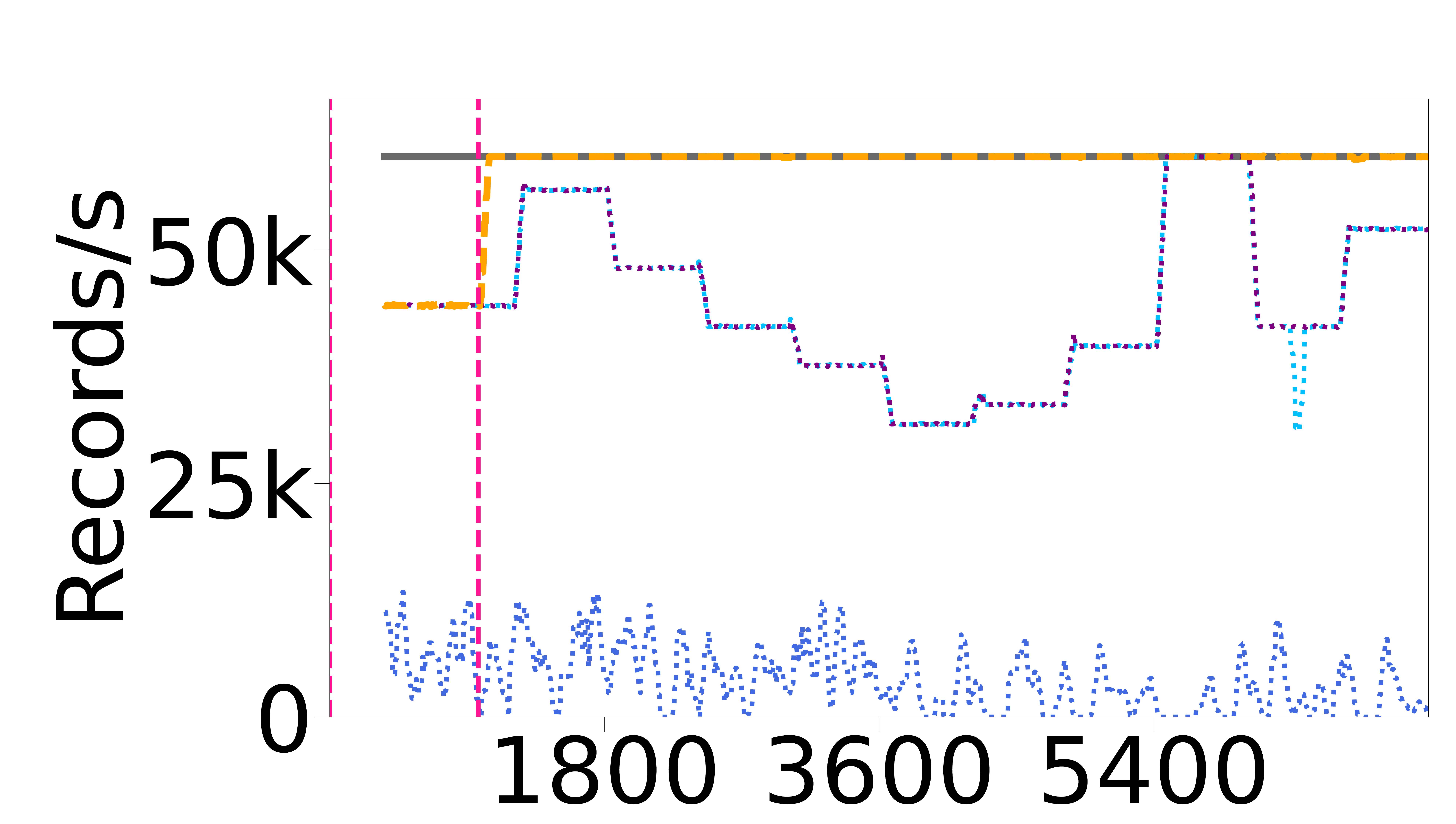}} &
    \adjustbox{valign=m}{\includegraphics[width=3.8cm]{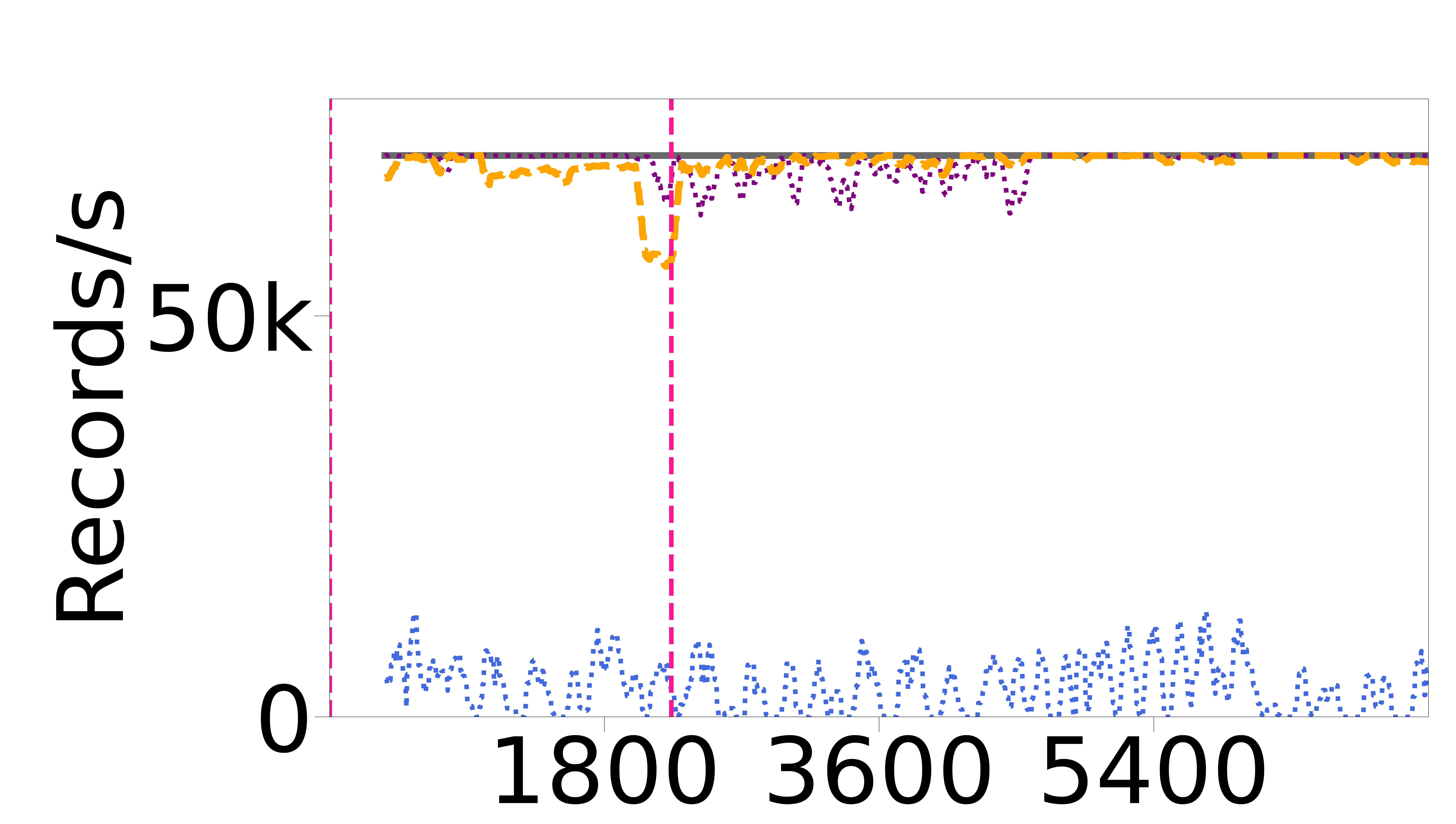}} \\
    \adjustbox{valign=m}{\rotatebox[origin=c]{90}{\textbf{Q2}}} &
    \adjustbox{valign=m}{\includegraphics[width=3.8cm]{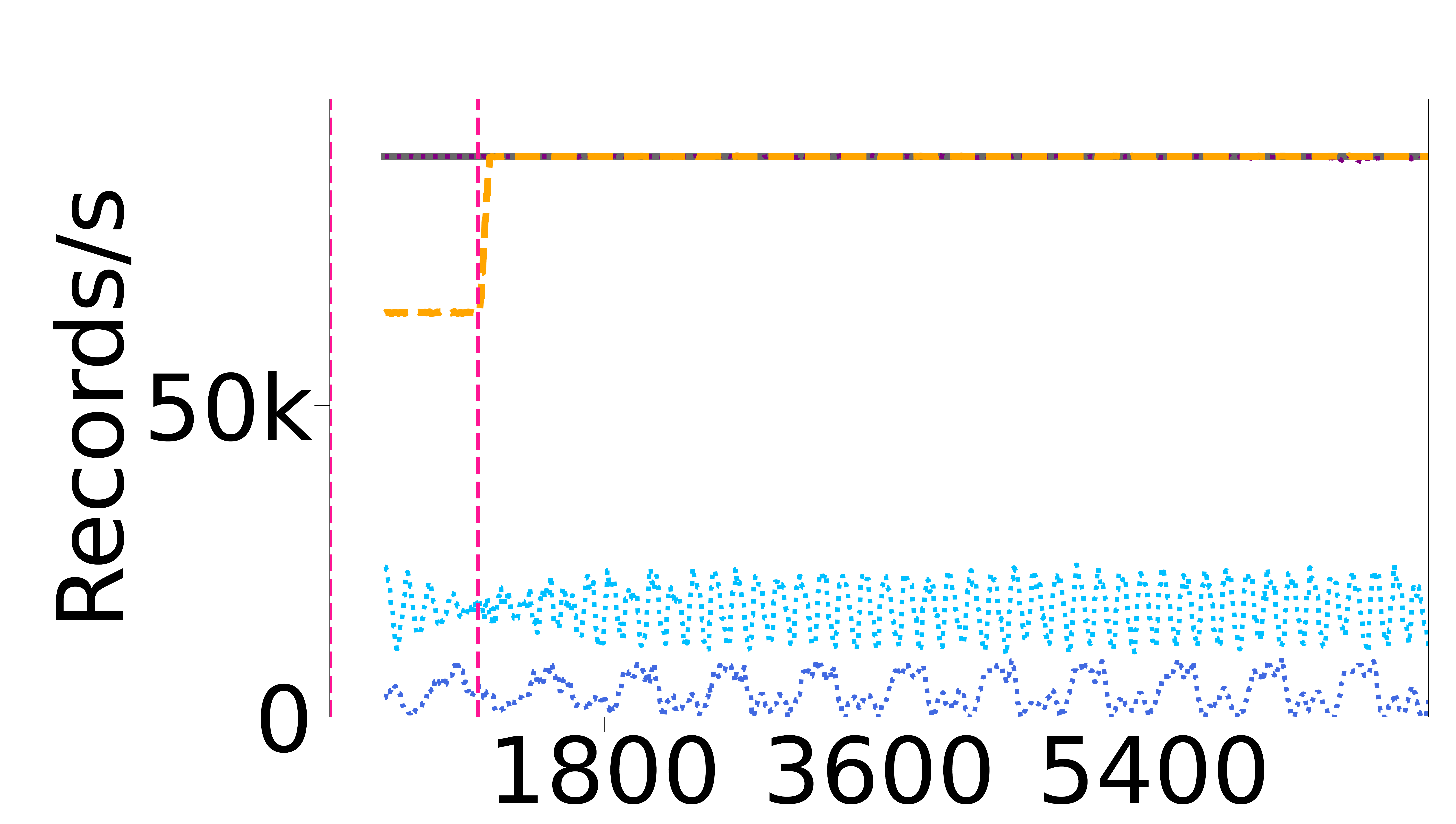}} &
    \adjustbox{valign=m}{\includegraphics[width=3.8cm]{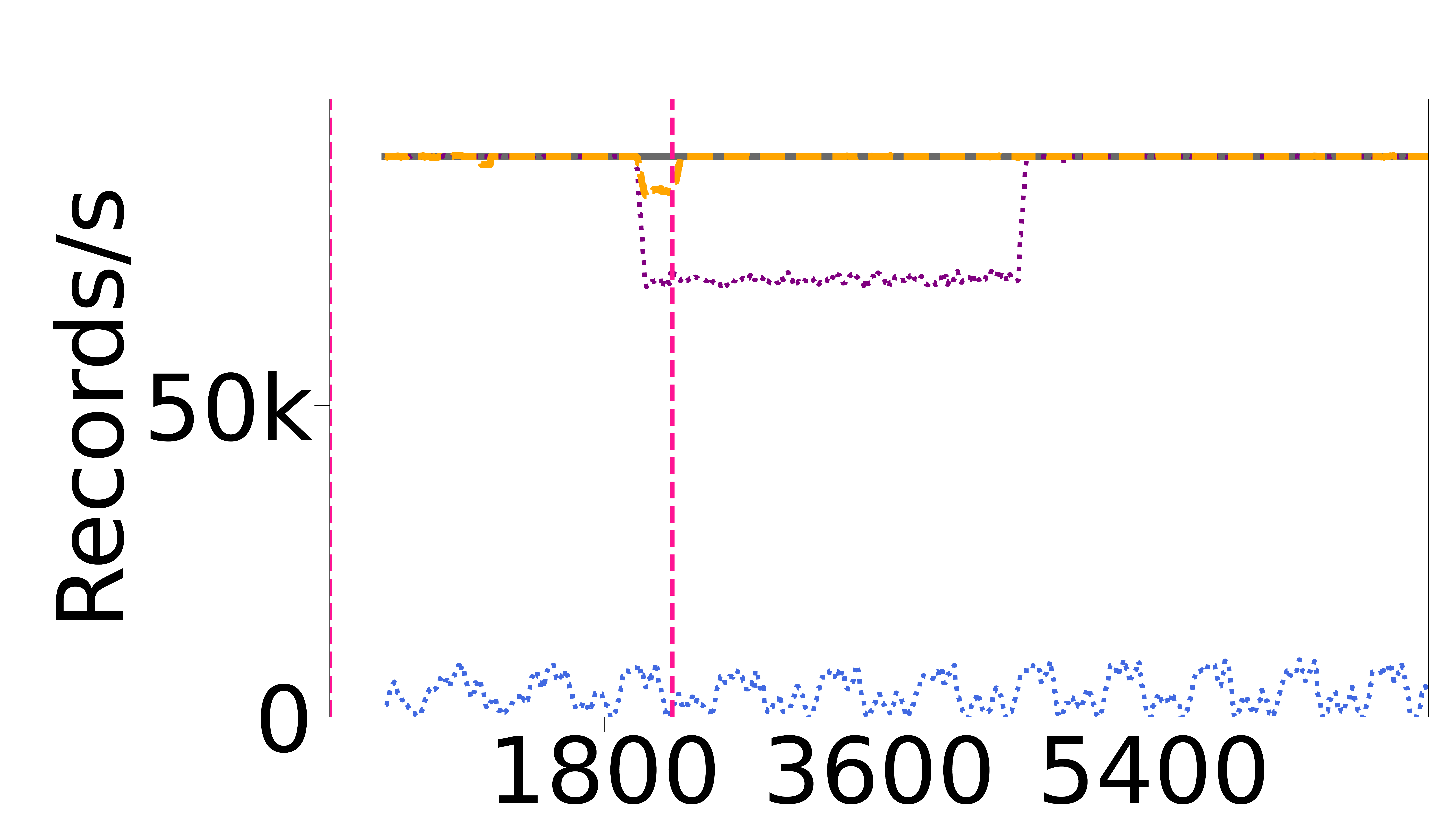}} \\
    \adjustbox{valign=m}{\rotatebox[origin=c]{90}{\textbf{Q3}}} &
    \adjustbox{valign=m}{\includegraphics[width=3.8cm]{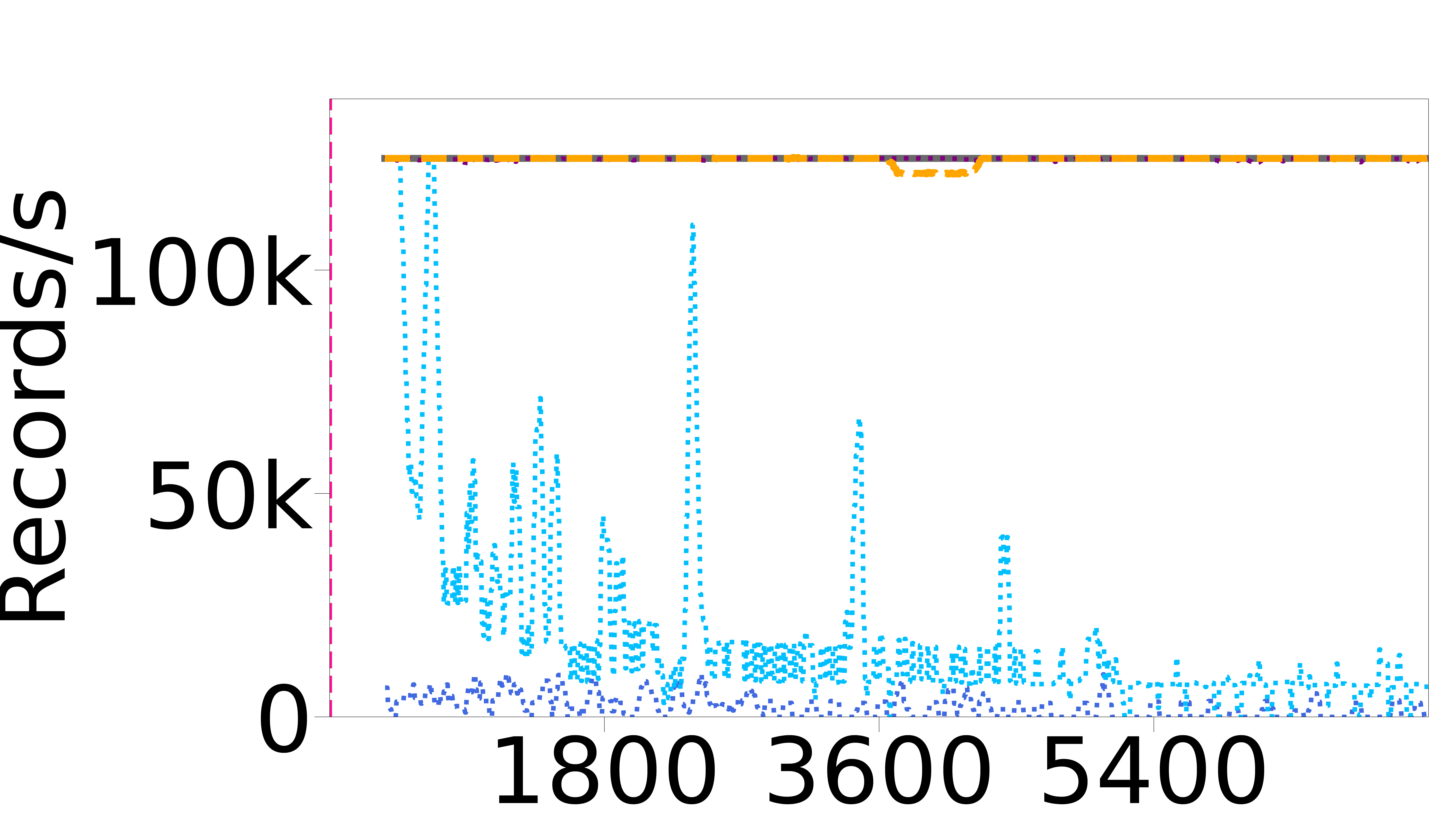}} &
    \adjustbox{valign=m}{\includegraphics[width=3.8cm]{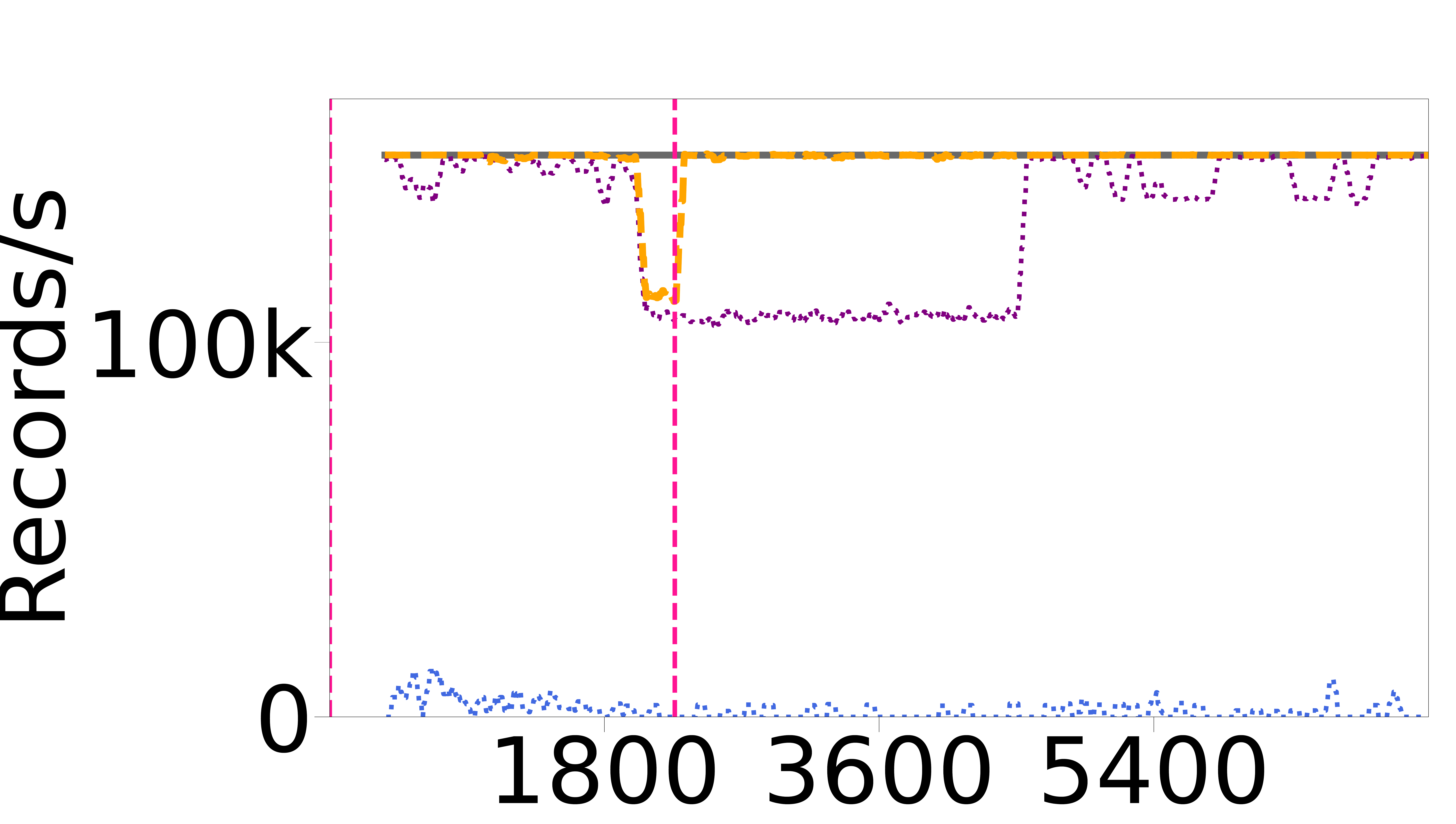}} \\
    \adjustbox{valign=m}{\rotatebox[origin=c]{90}{\textbf{Q4}}} &
    \adjustbox{valign=m}{\includegraphics[width=3.8cm]{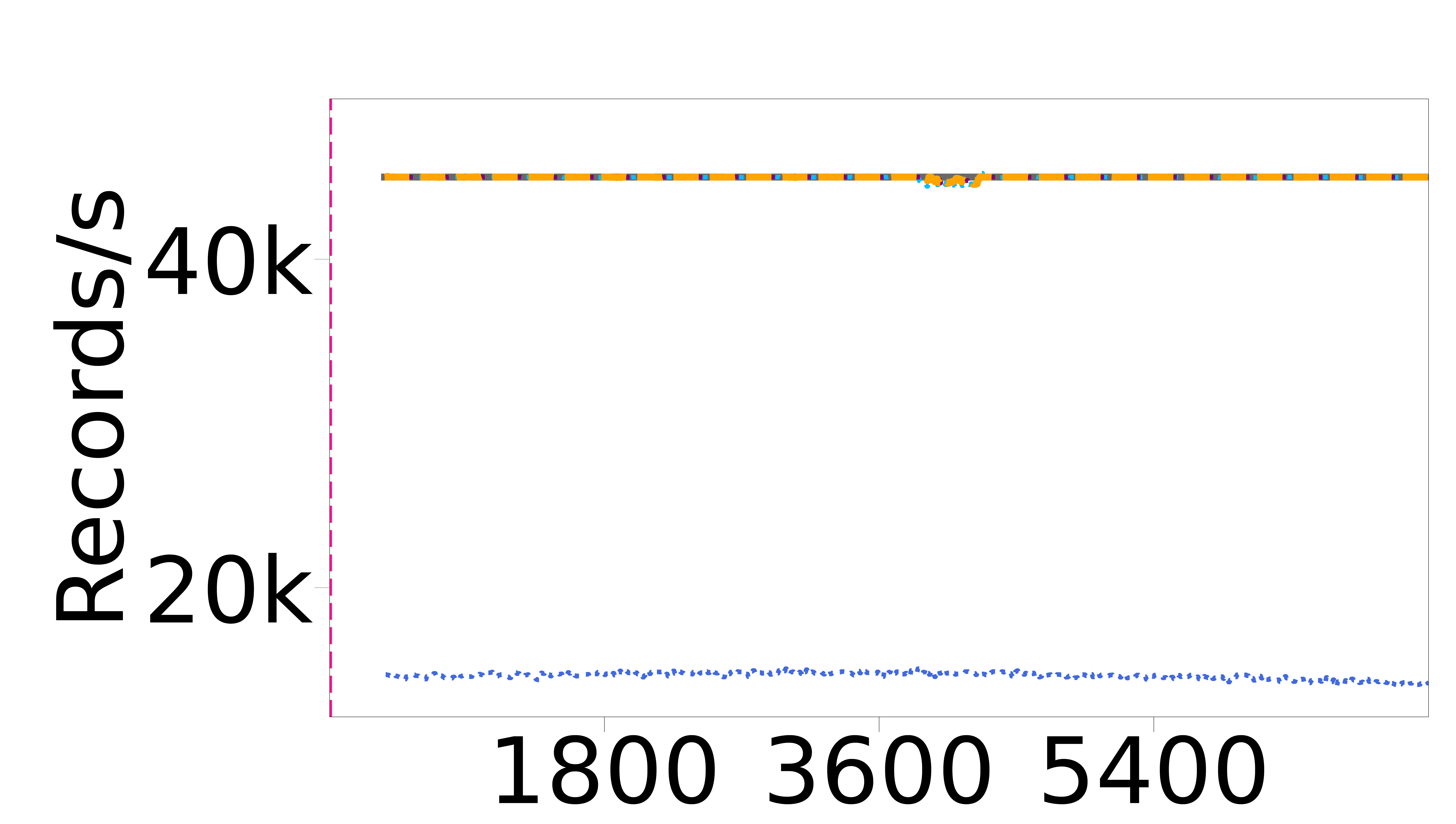}} &
    \adjustbox{valign=m}{\includegraphics[width=3.8cm]{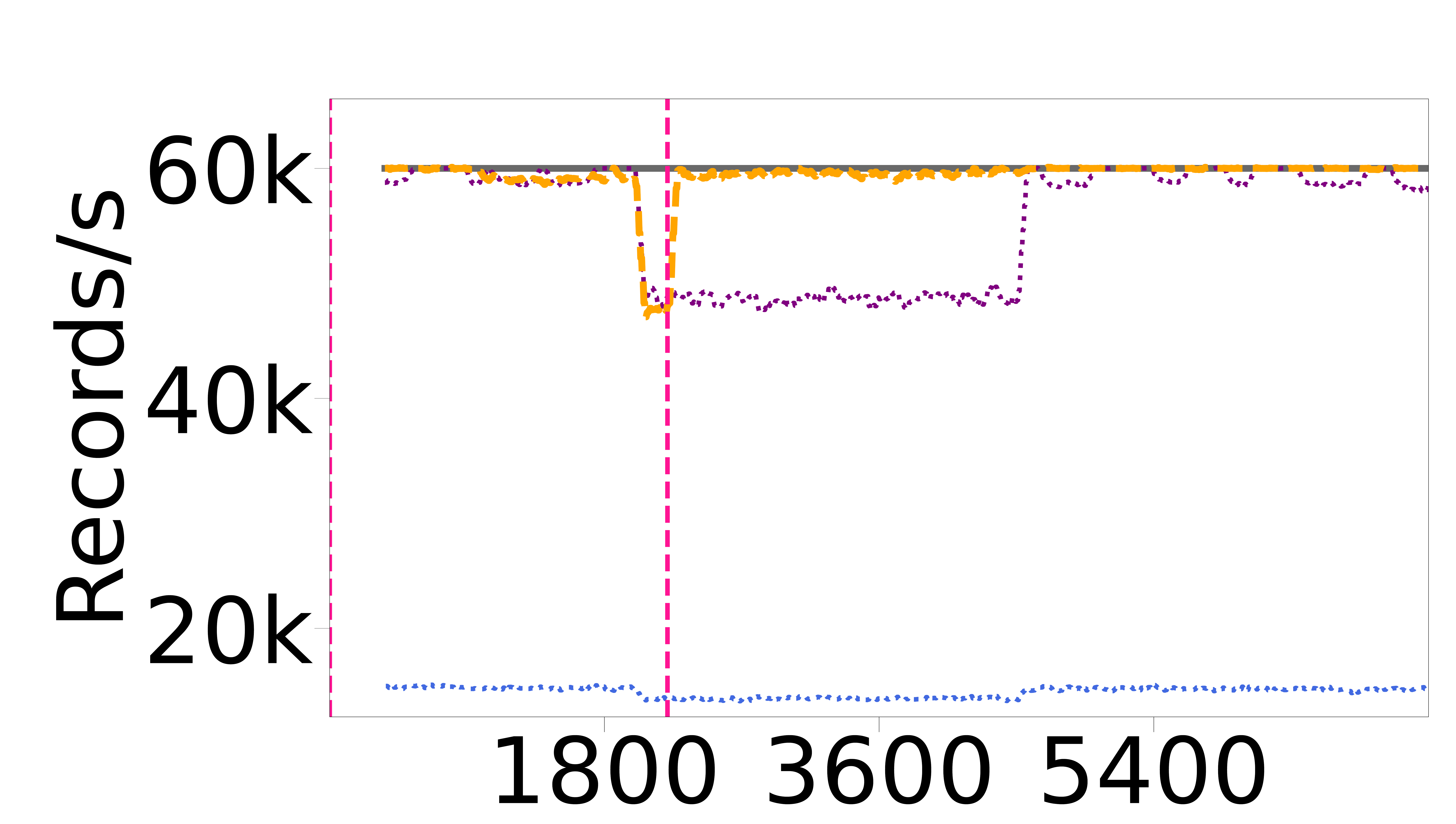}} \\
    \adjustbox{valign=m}{\rotatebox[origin=c]{90}{\textbf{Q5}}} &
    \adjustbox{valign=m}{\includegraphics[width=3.8cm]{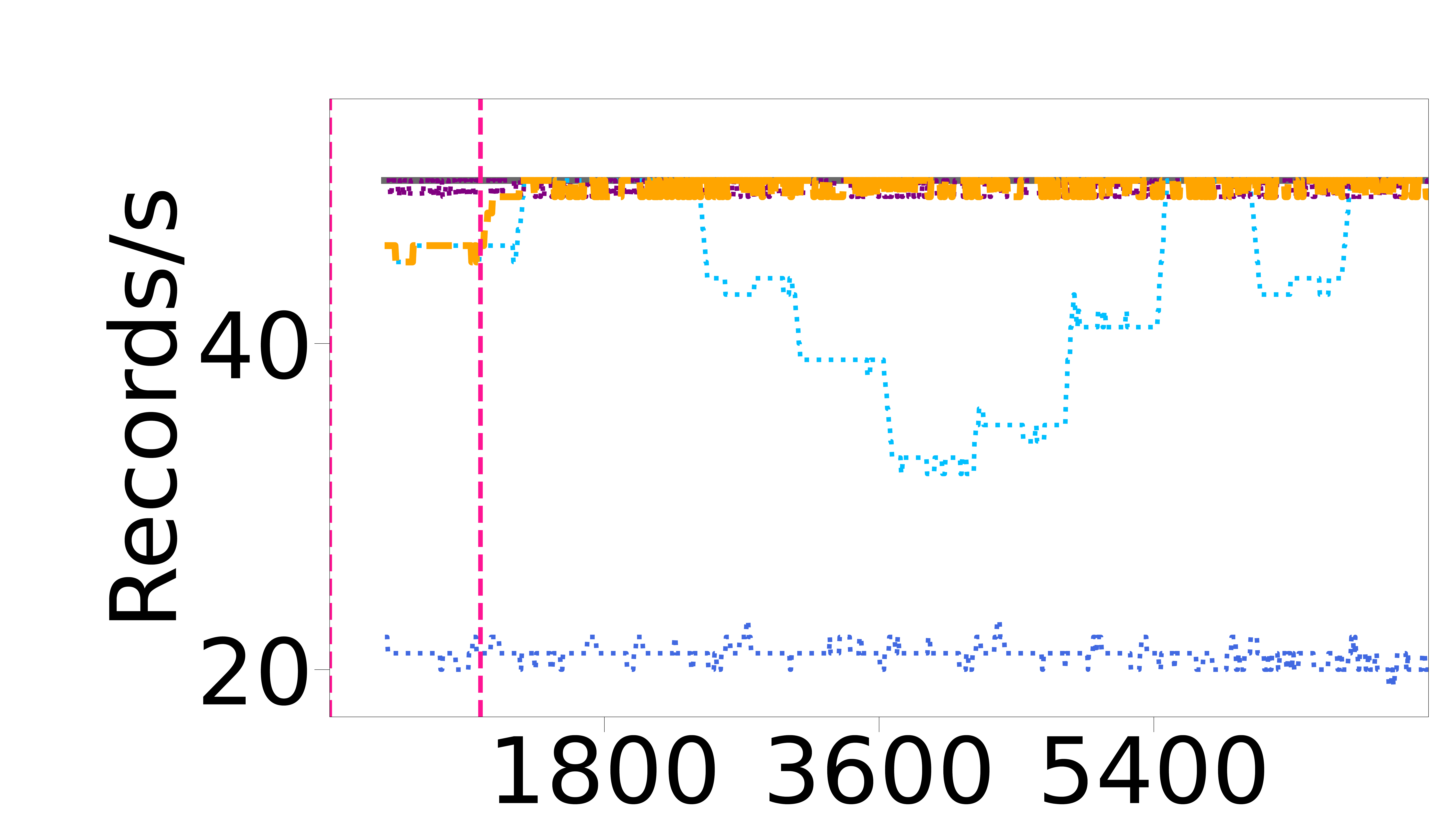}} &
    \adjustbox{valign=m}{\includegraphics[width=3.8cm]{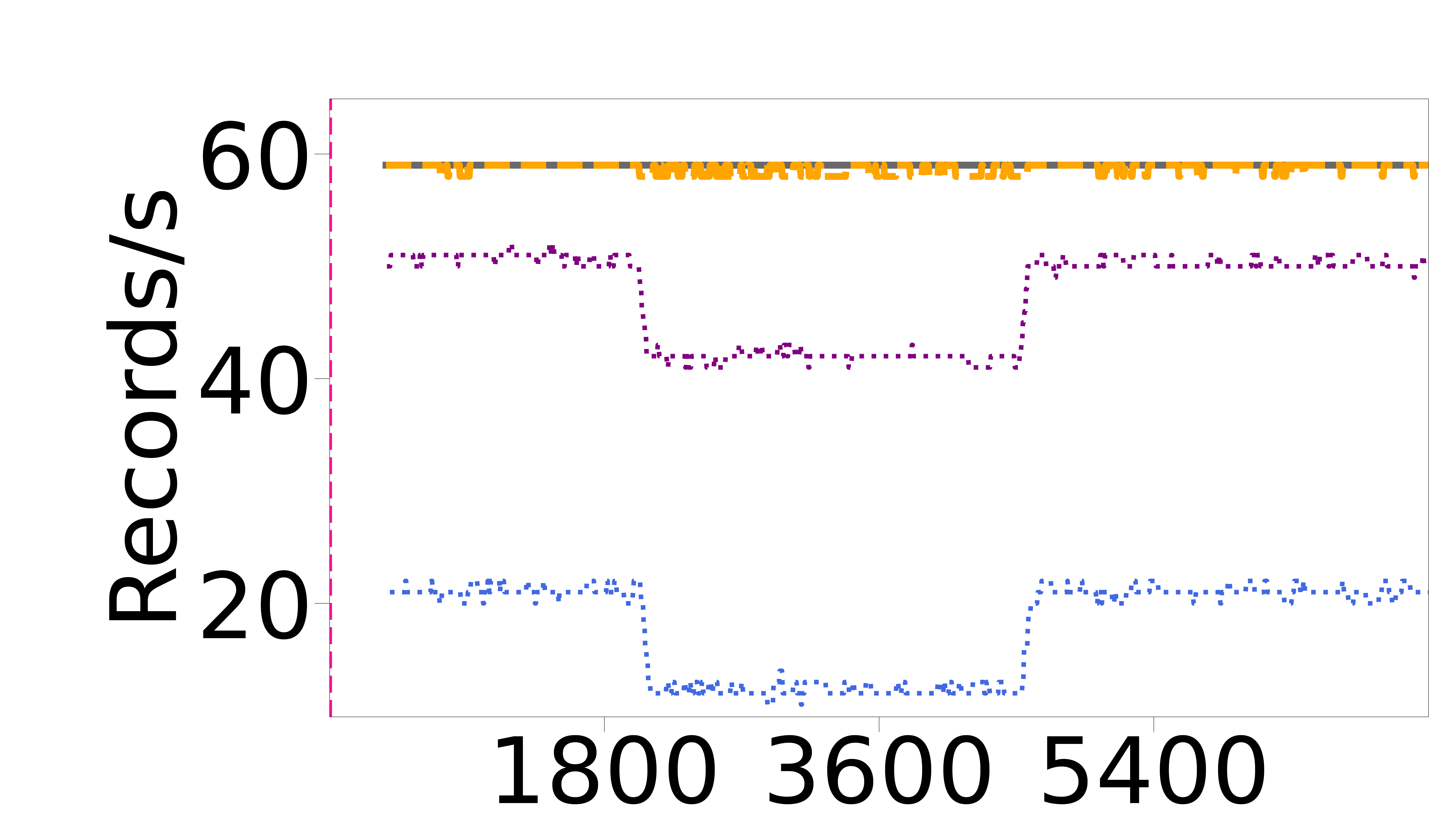}} \\
    \adjustbox{valign=m}{\rotatebox[origin=c]{90}{\parbox{2.5cm}{\centering Operating Conditions}}} &
    \adjustbox{valign=m}{\includegraphics[width=3.8cm]{figures/eval/shift/res_video6nw_0_lower.pdf.pdf}} &
    \adjustbox{valign=m}{\includegraphics[width=3.8cm]{figures/eval/shift/res_video6slow_0_lower.pdf.pdf}} \\
    & \adjustbox{valign=m}{\parbox{3.5cm}{\centering \refstepcounter{subfigure} (a) Network variations }} &
    \adjustbox{valign=m}{\parbox{3.5cm}{\centering \refstepcounter{subfigure} (b) Compute contention }}  \\
\end{tabular}\vspace{-2mm}
\captionof{figure}{Impact of network bandwidth and compute contention - RPi1}
\label{fig:sup_vary_env_rpi1}
\end{minipage}\par\vspace{\textfloatsep}

\noindent\begin{minipage}{\columnwidth}
\centering
\begin{minipage}[b]{0.8\linewidth}
\includegraphics[width=\linewidth]{figures/eval/shift/res_video6nw_0.pdf_legend.pdf}
    \vspace{-0.4cm}
\end{minipage}
\setlength{\tabcolsep}{1pt}
\renewcommand{\arraystretch}{1.5}
\begin{tabular}{ccc}
    \adjustbox{valign=m}{\rotatebox[origin=c]{90}{\textbf{Q1}}} &
    \adjustbox{valign=m}{\includegraphics[width=3.8cm]{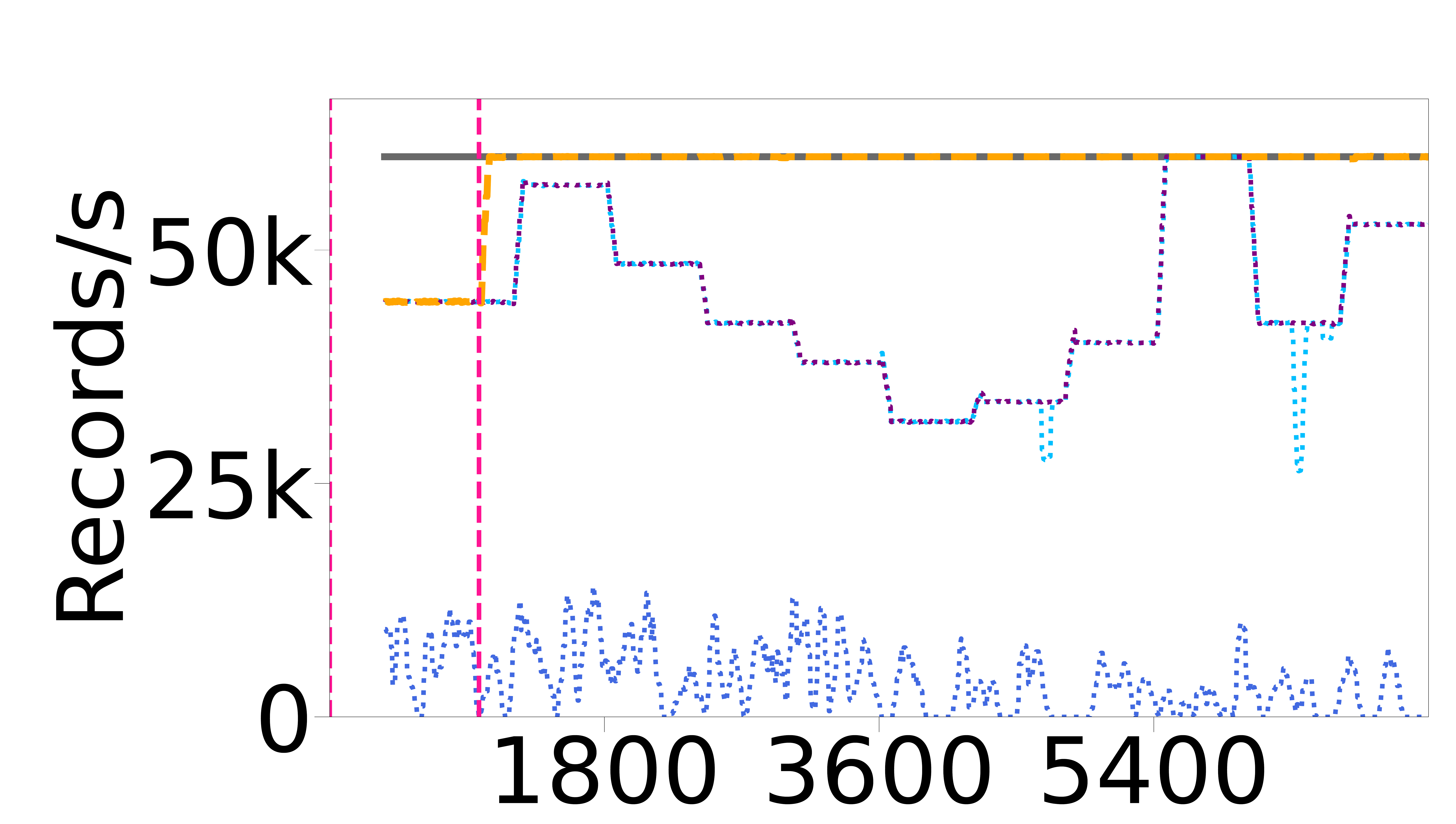}} &
    \adjustbox{valign=m}{\includegraphics[width=3.8cm]{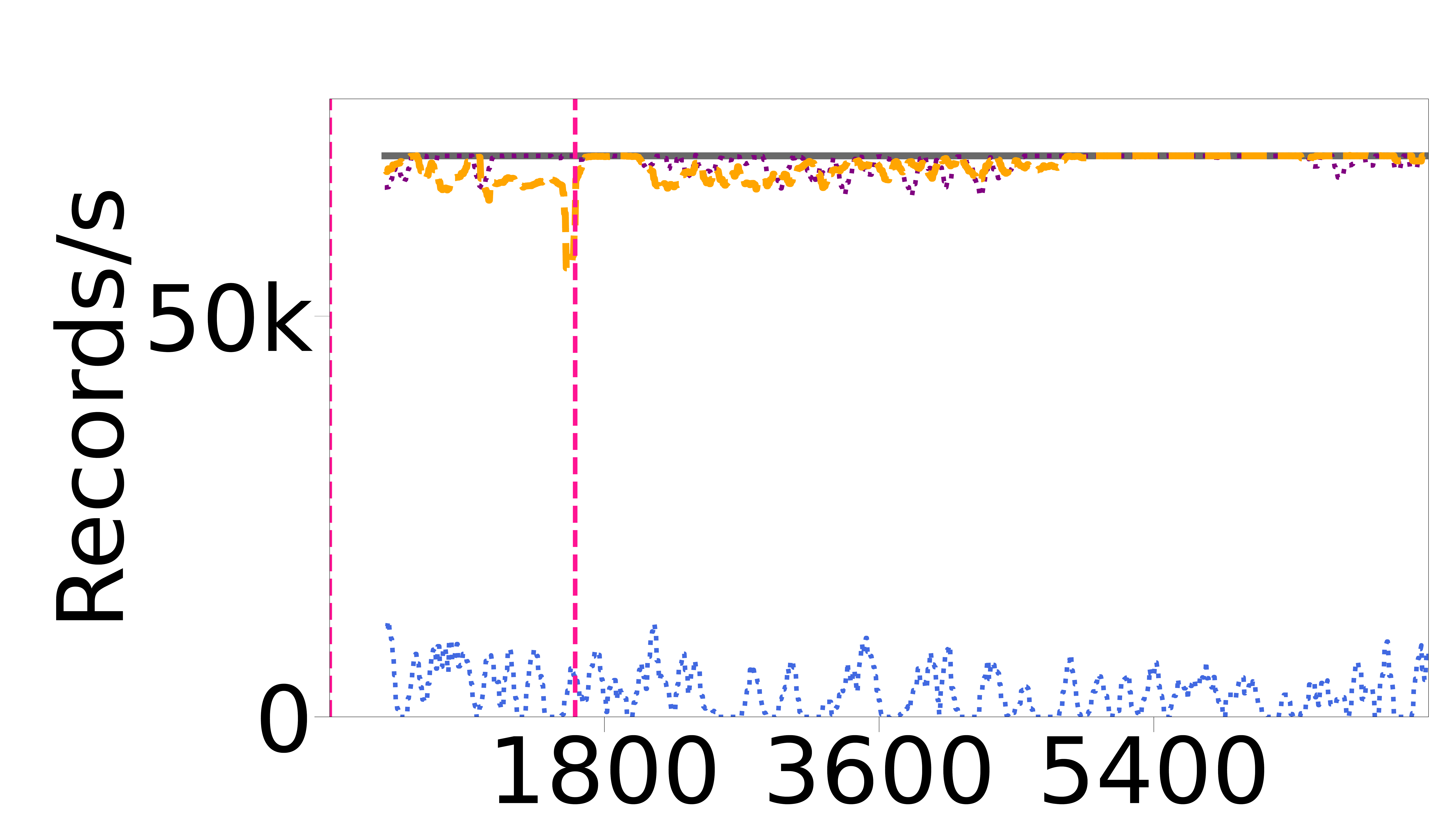}} \\
    \adjustbox{valign=m}{\rotatebox[origin=c]{90}{\textbf{Q2}}} &
    \adjustbox{valign=m}{\includegraphics[width=3.8cm]{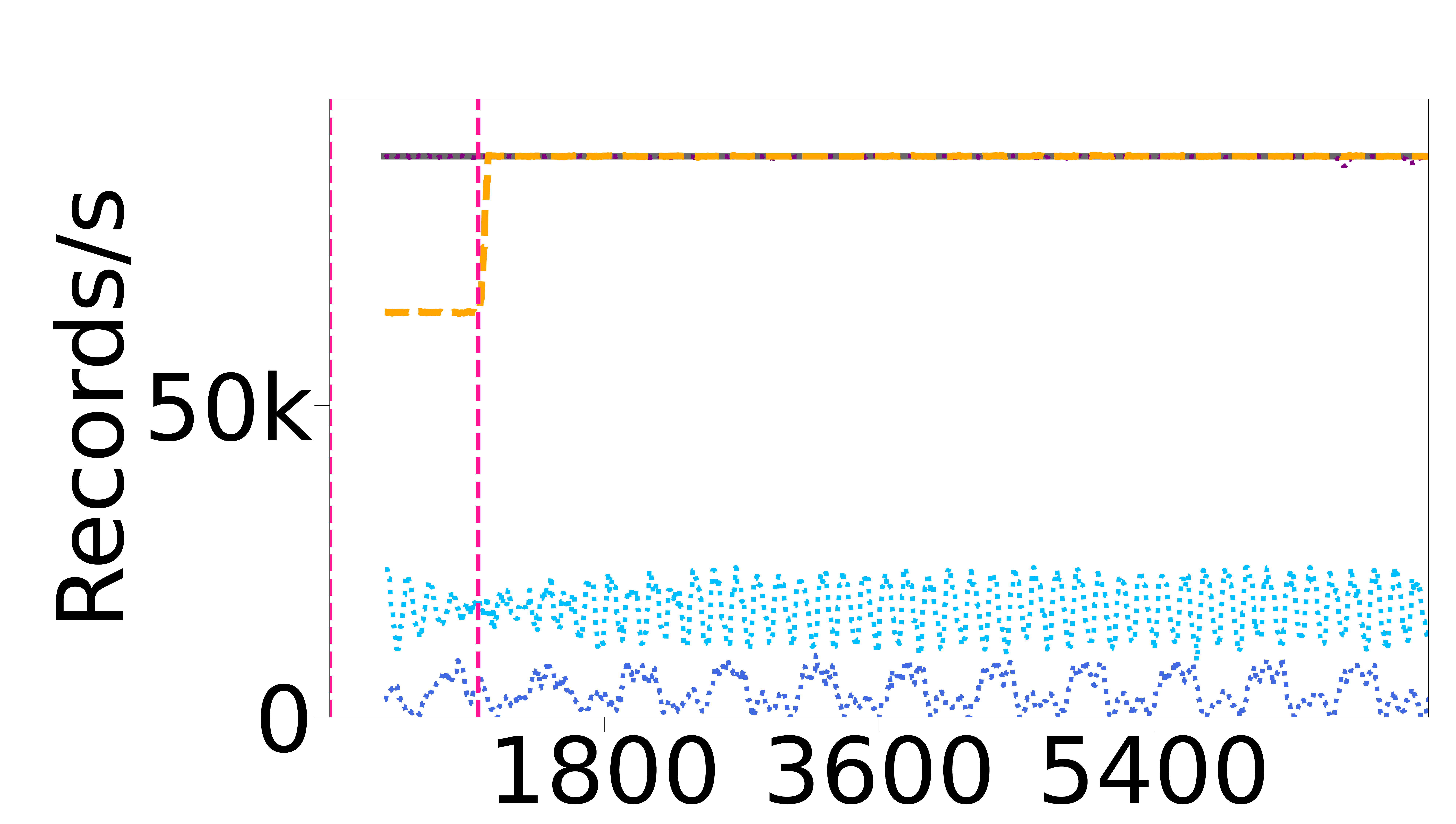}} &
    \adjustbox{valign=m}{\includegraphics[width=3.8cm]{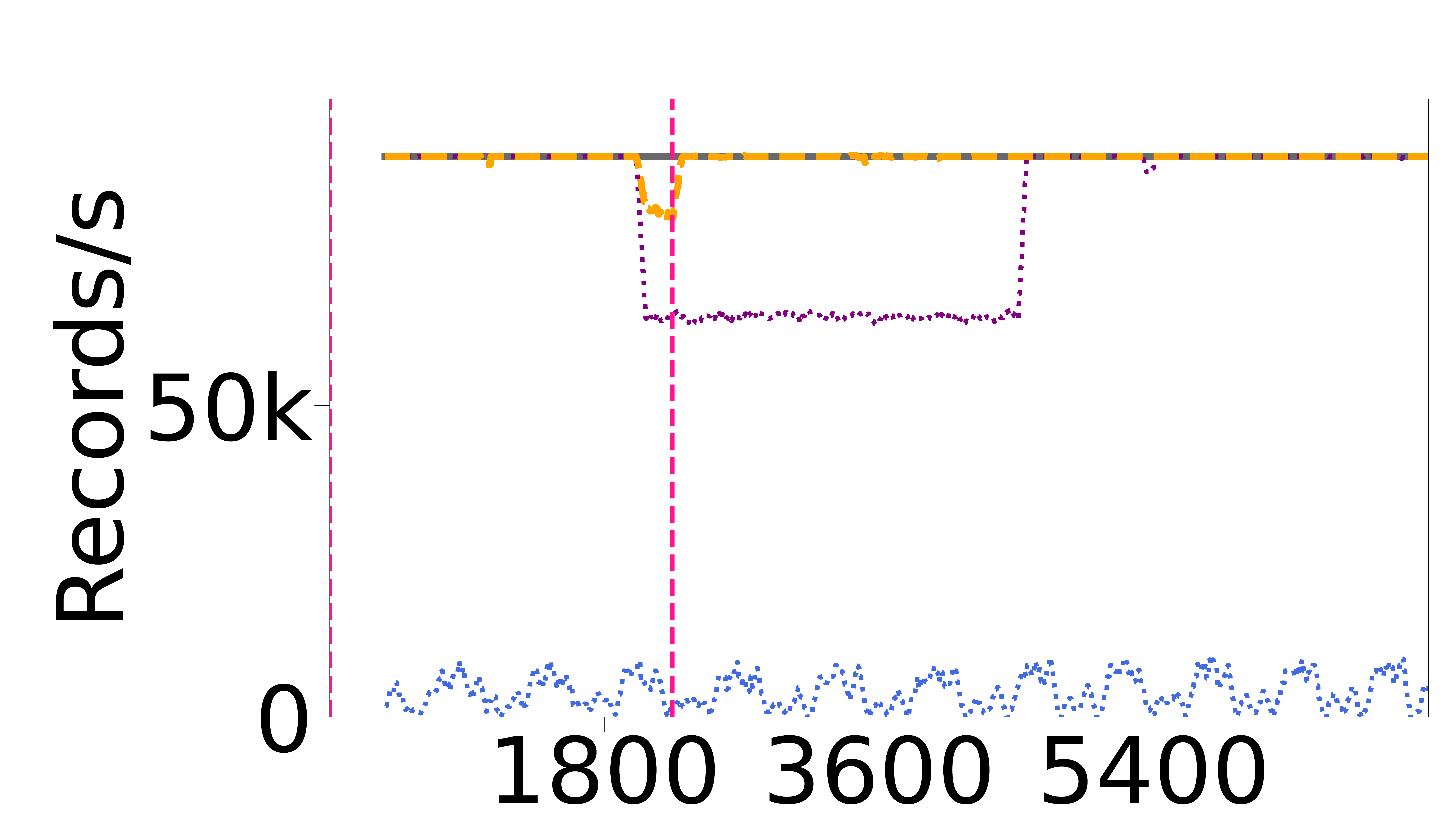}} \\
    \adjustbox{valign=m}{\rotatebox[origin=c]{90}{\textbf{Q3}}} &
    \adjustbox{valign=m}{\includegraphics[width=3.8cm]{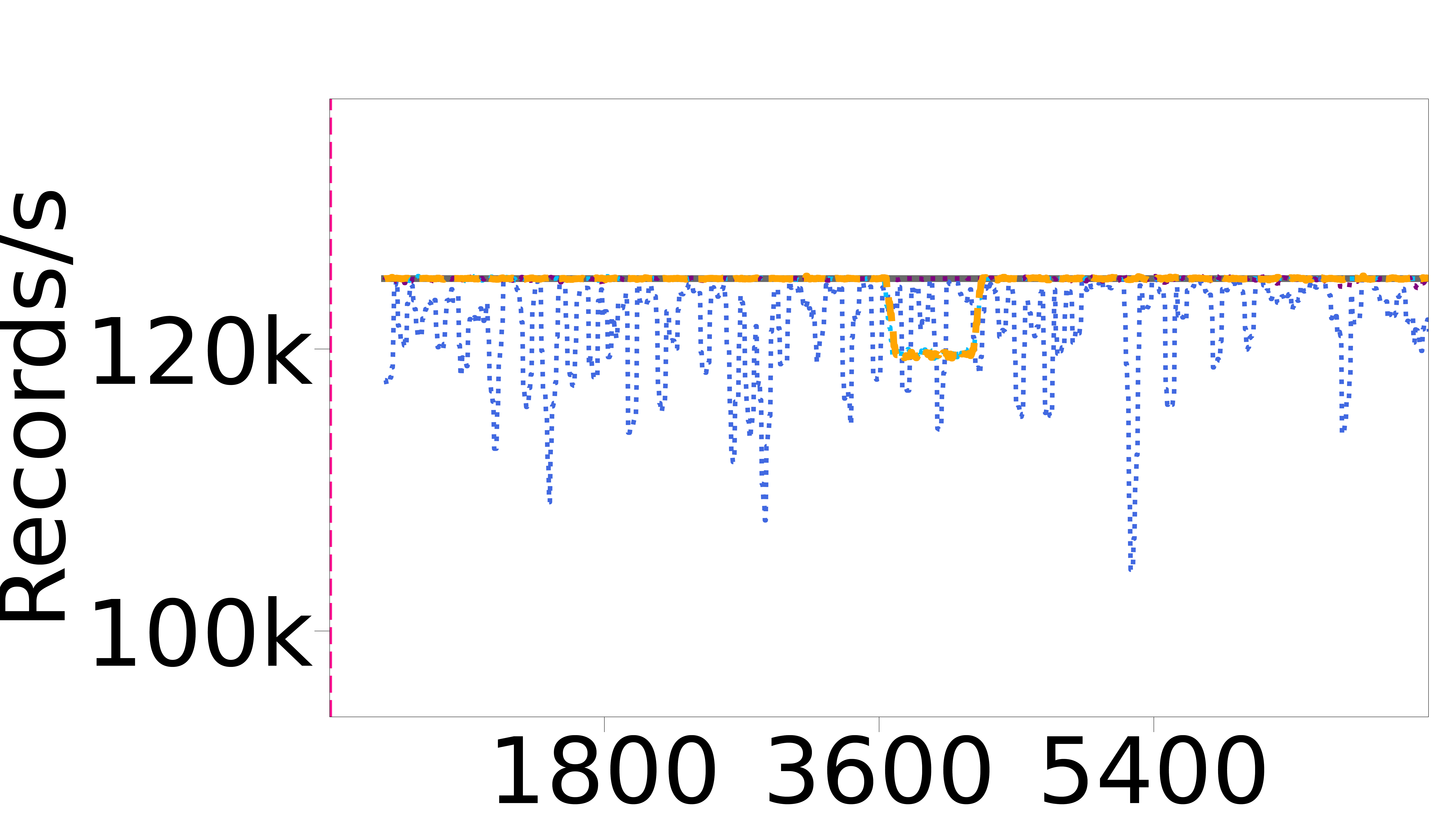}} &
    \adjustbox{valign=m}{\includegraphics[width=3.8cm]{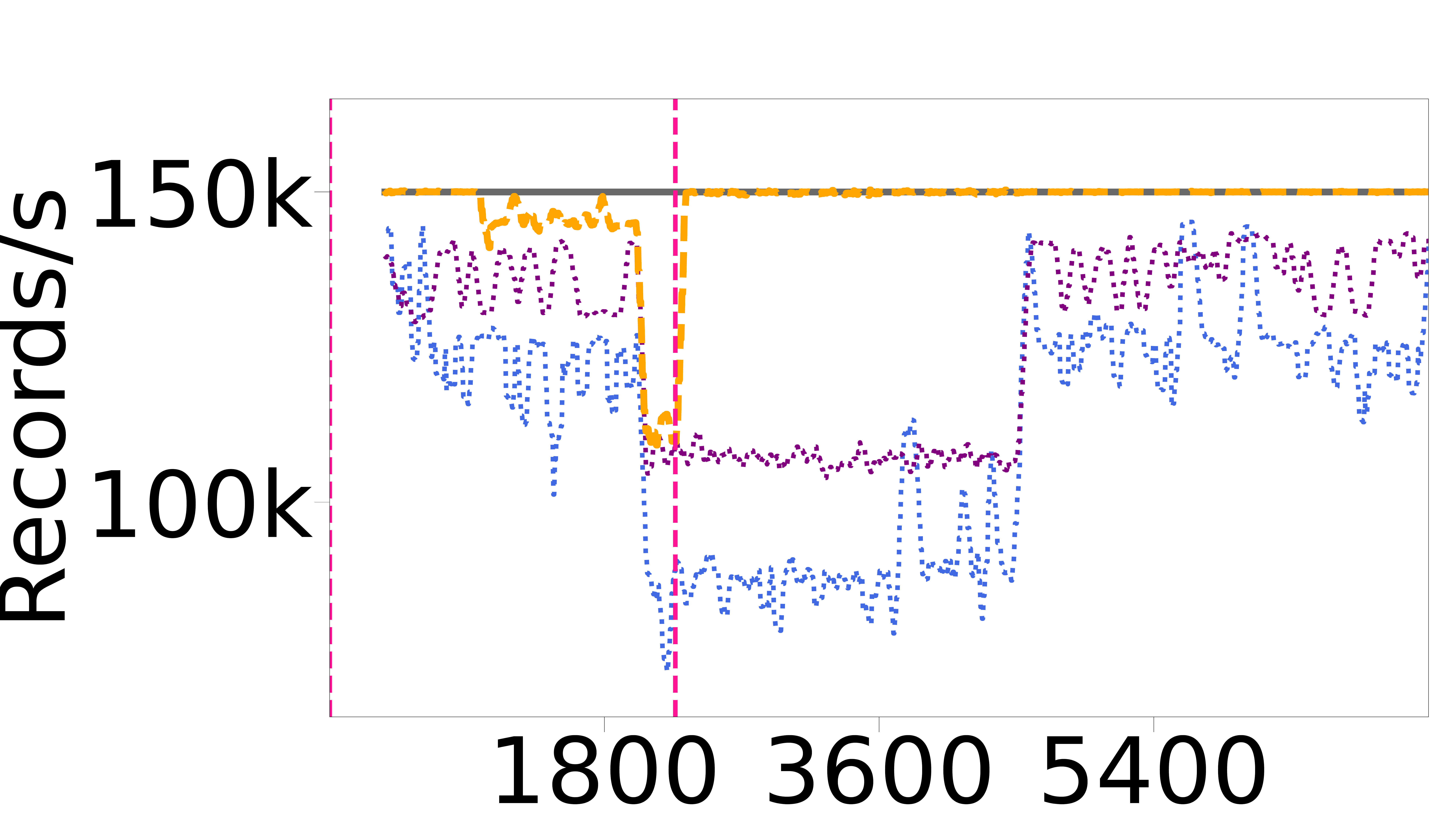}} \\
    \adjustbox{valign=m}{\rotatebox[origin=c]{90}{\textbf{Q4}}} &
    \adjustbox{valign=m}{\includegraphics[width=3.8cm]{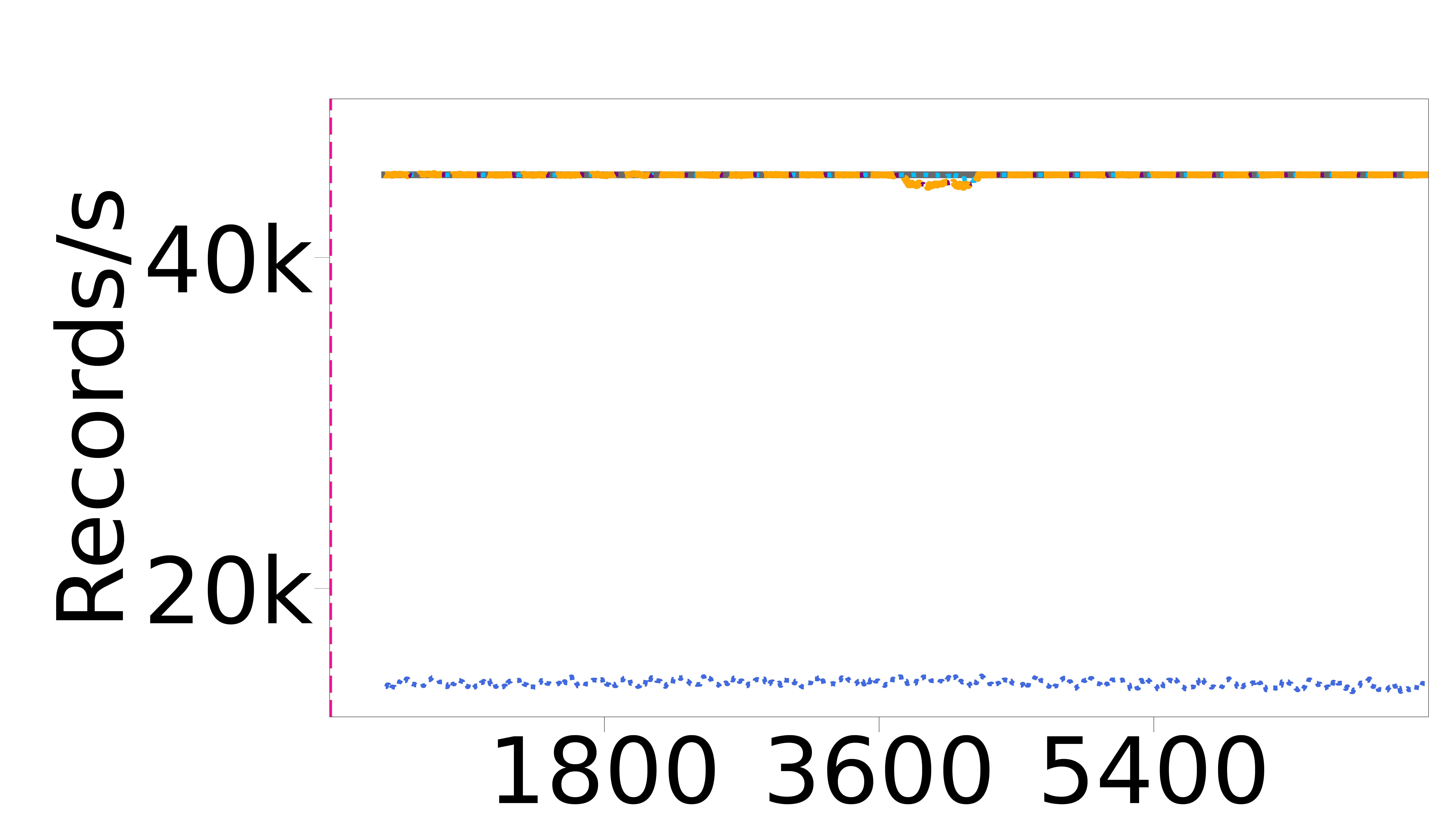}} &
    \adjustbox{valign=m}{\includegraphics[width=3.8cm]{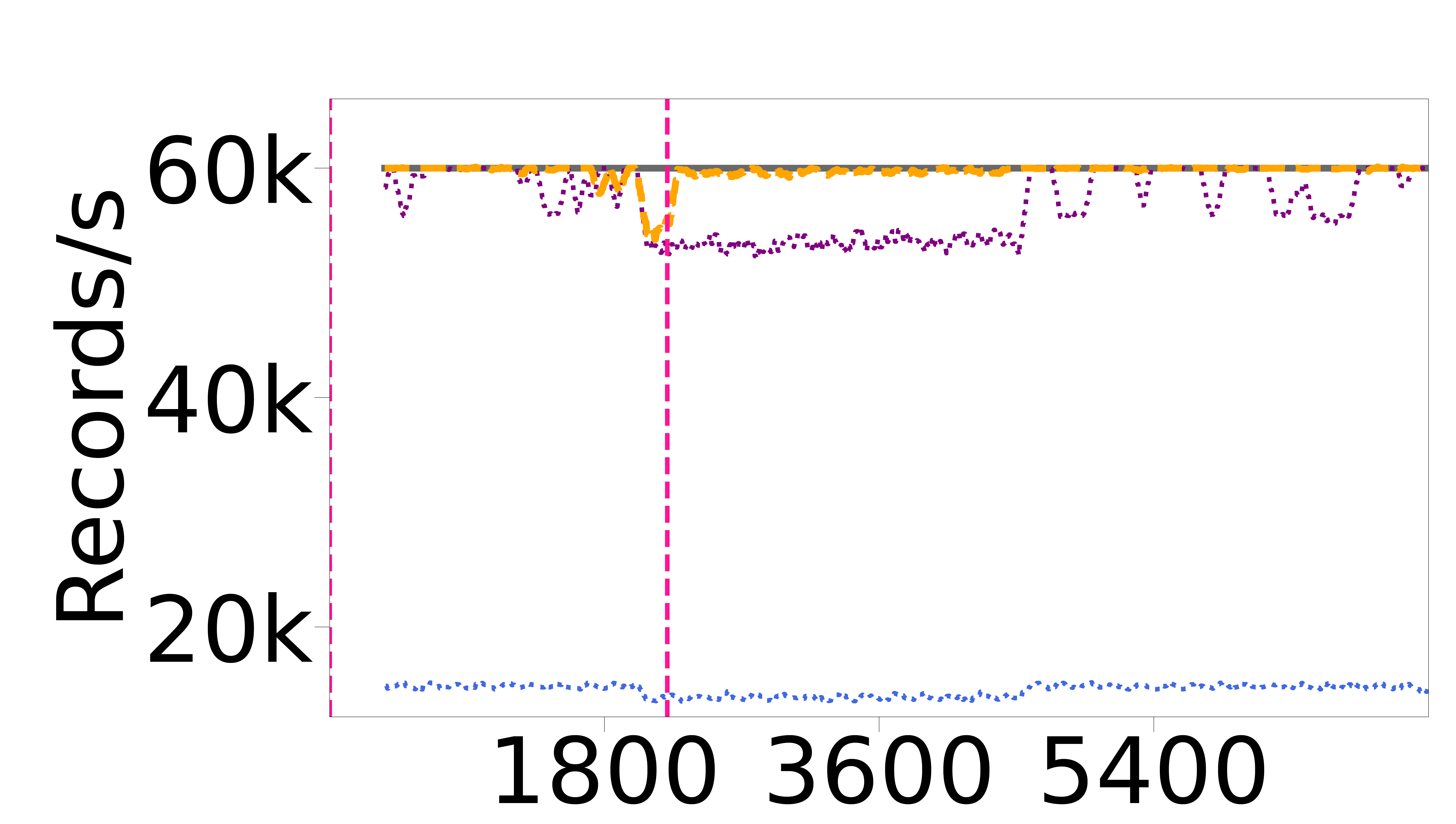}} \\
    \adjustbox{valign=m}{\rotatebox[origin=c]{90}{\textbf{Q5}}} &
    \adjustbox{valign=m}{\includegraphics[width=3.8cm]{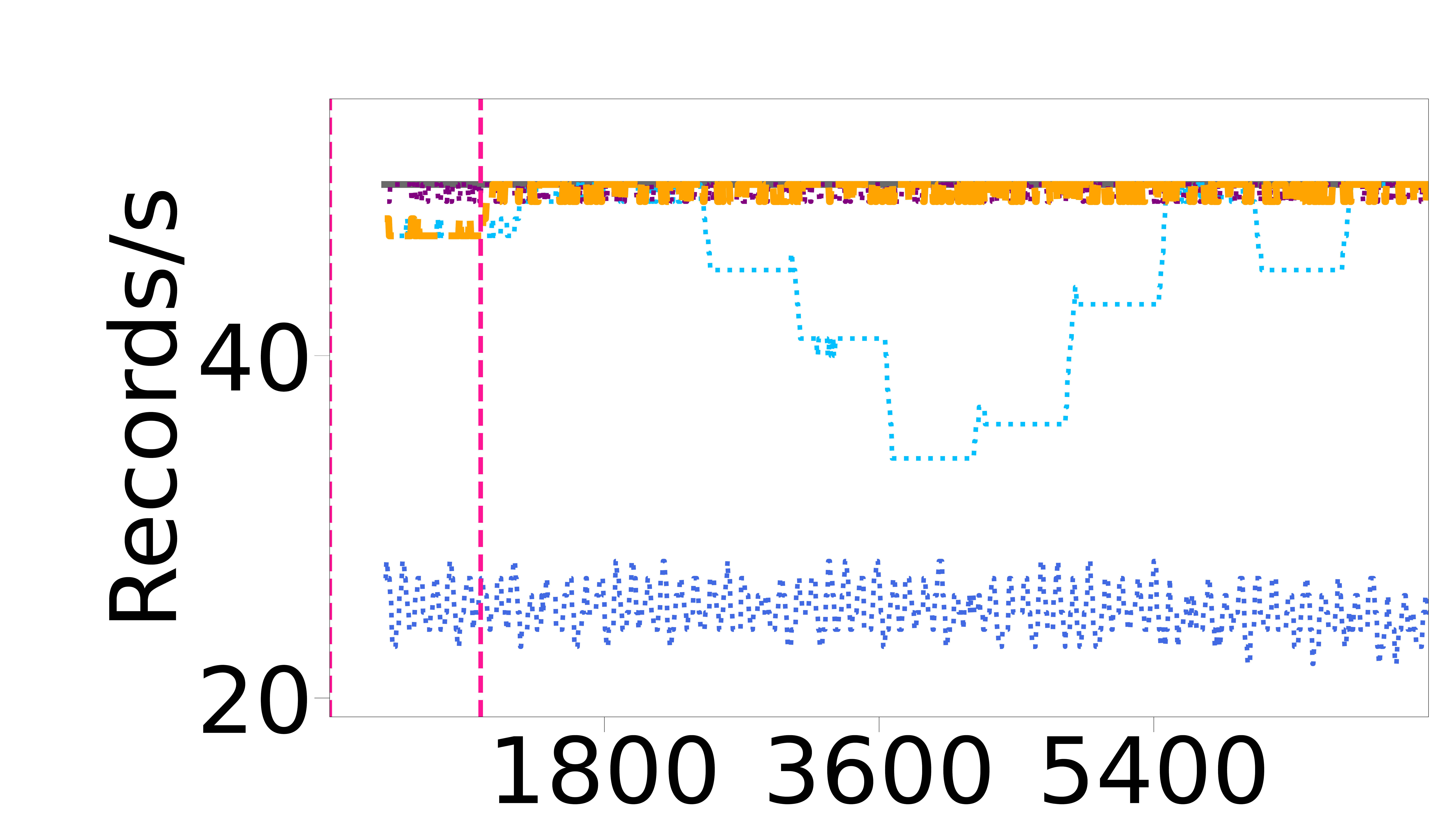}} &
    \adjustbox{valign=m}{\includegraphics[width=3.8cm]{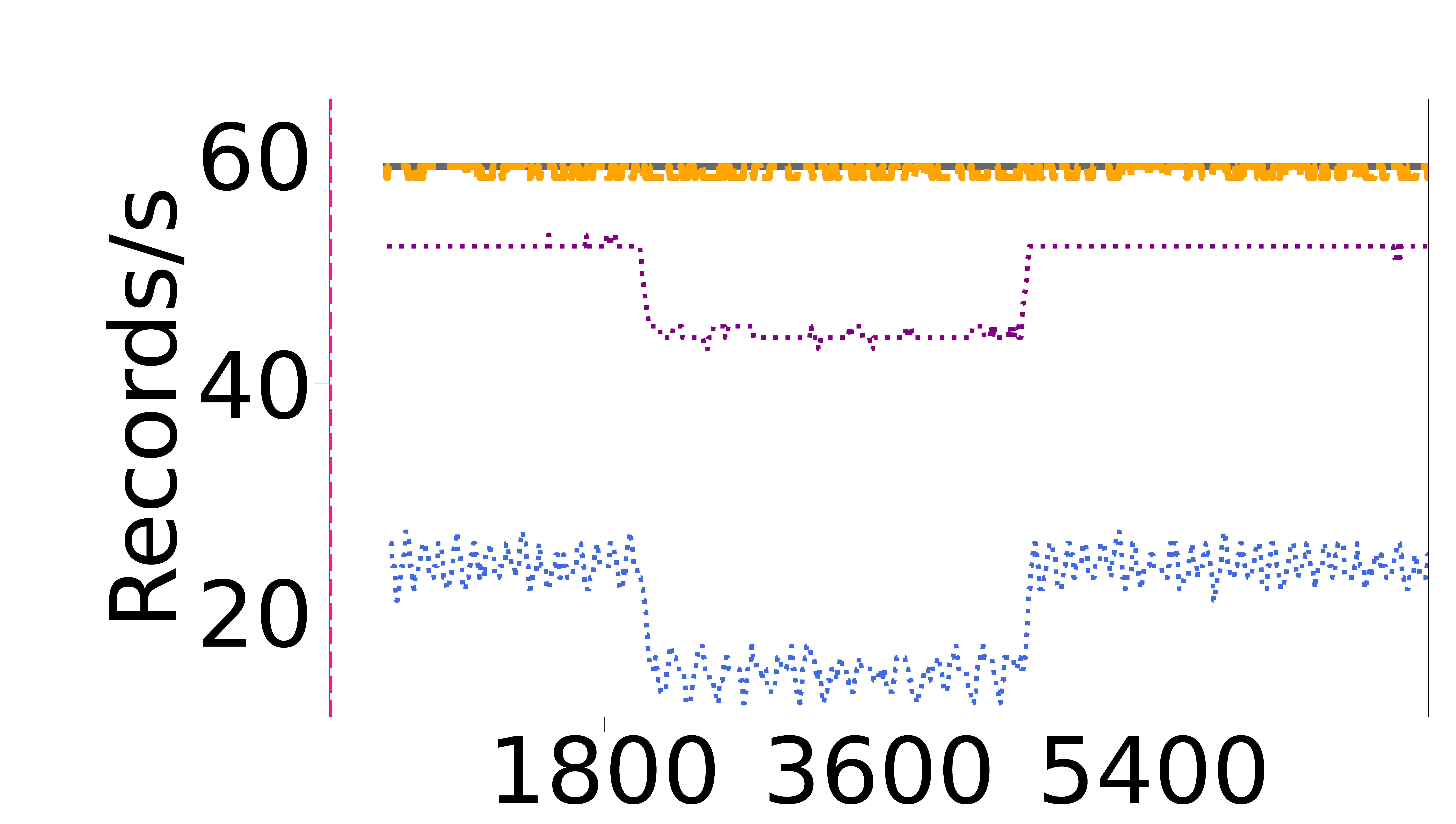}} \\
    \adjustbox{valign=m}{\rotatebox[origin=c]{90}{\parbox{2.5cm}{\centering Operating Conditions}}} &
    \adjustbox{valign=m}{\includegraphics[width=3.8cm]{figures/eval/shift/res_video6nw_0_lower.pdf.pdf}} &
    \adjustbox{valign=m}{\includegraphics[width=3.8cm]{figures/eval/shift/res_video6slow_0_lower.pdf.pdf}} \\
    & \adjustbox{valign=m}{\parbox{3.5cm}{\centering \refstepcounter{subfigure} (a) Network variations }} &
    \adjustbox{valign=m}{\parbox{3.5cm}{\centering \refstepcounter{subfigure} (b) Compute contention }}  \\
\end{tabular}\vspace{-2mm}
\captionof{figure}{Impact of network bandwidth and compute contention - RPi2}
\label{fig:sup_vary_env_rpi2}
\end{minipage}\par\vspace{\textfloatsep}

\noindent\begin{minipage}{\columnwidth}
\centering
\begin{minipage}[b]{0.8\linewidth}
\includegraphics[width=\linewidth]{figures/eval/shift/res_video6nw_0.pdf_legend.pdf}
    \vspace{-0.4cm}
\end{minipage}
\setlength{\tabcolsep}{1pt}
\renewcommand{\arraystretch}{1.5}
\begin{tabular}{ccc}
    \adjustbox{valign=m}{\rotatebox[origin=c]{90}{\textbf{Q1}}} &
    \adjustbox{valign=m}{\includegraphics[width=3.8cm]{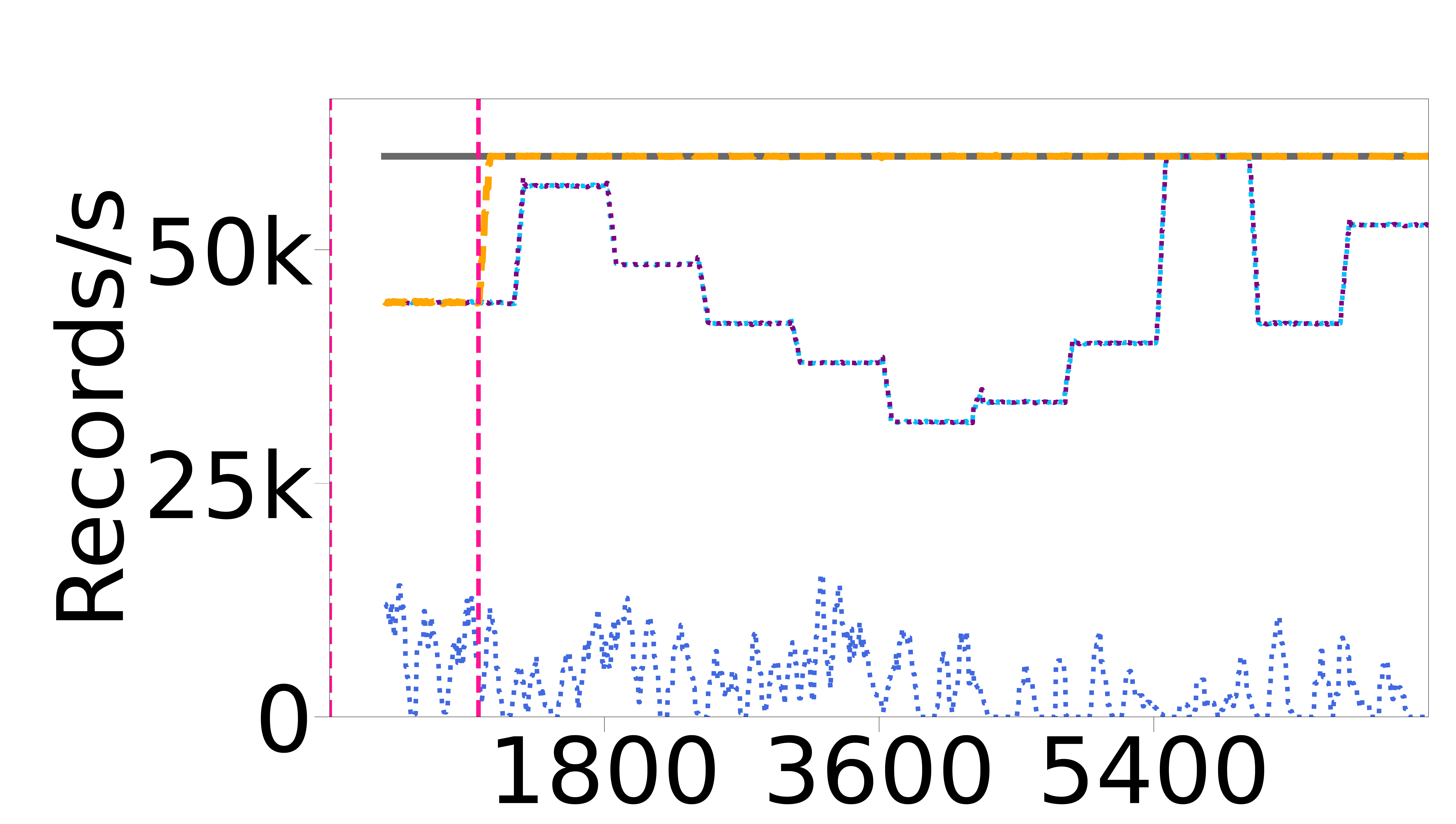}} &
    \adjustbox{valign=m}{\includegraphics[width=3.8cm]{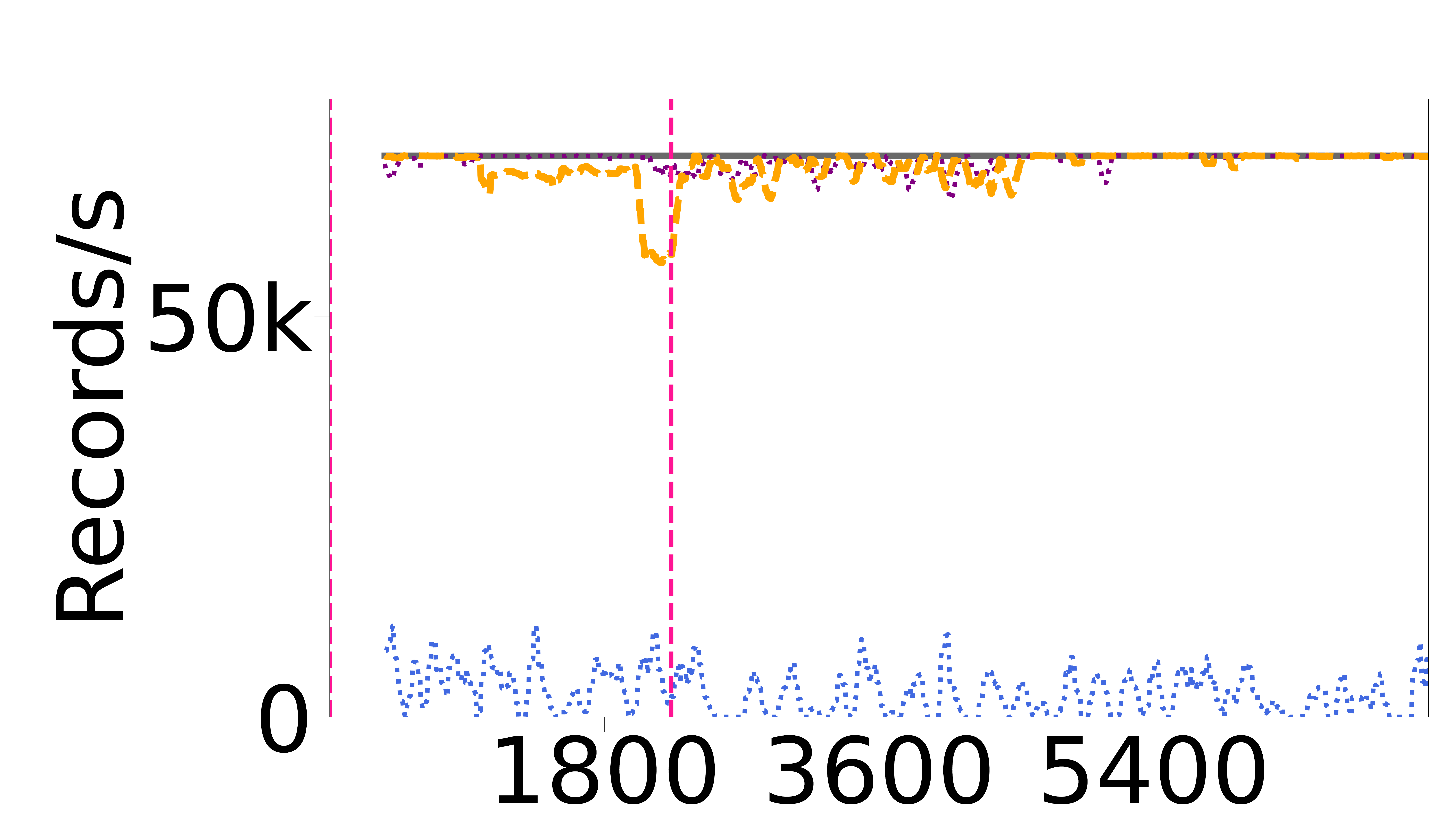}} \\
    \adjustbox{valign=m}{\rotatebox[origin=c]{90}{\textbf{Q2}}} &
    \adjustbox{valign=m}{\includegraphics[width=3.8cm]{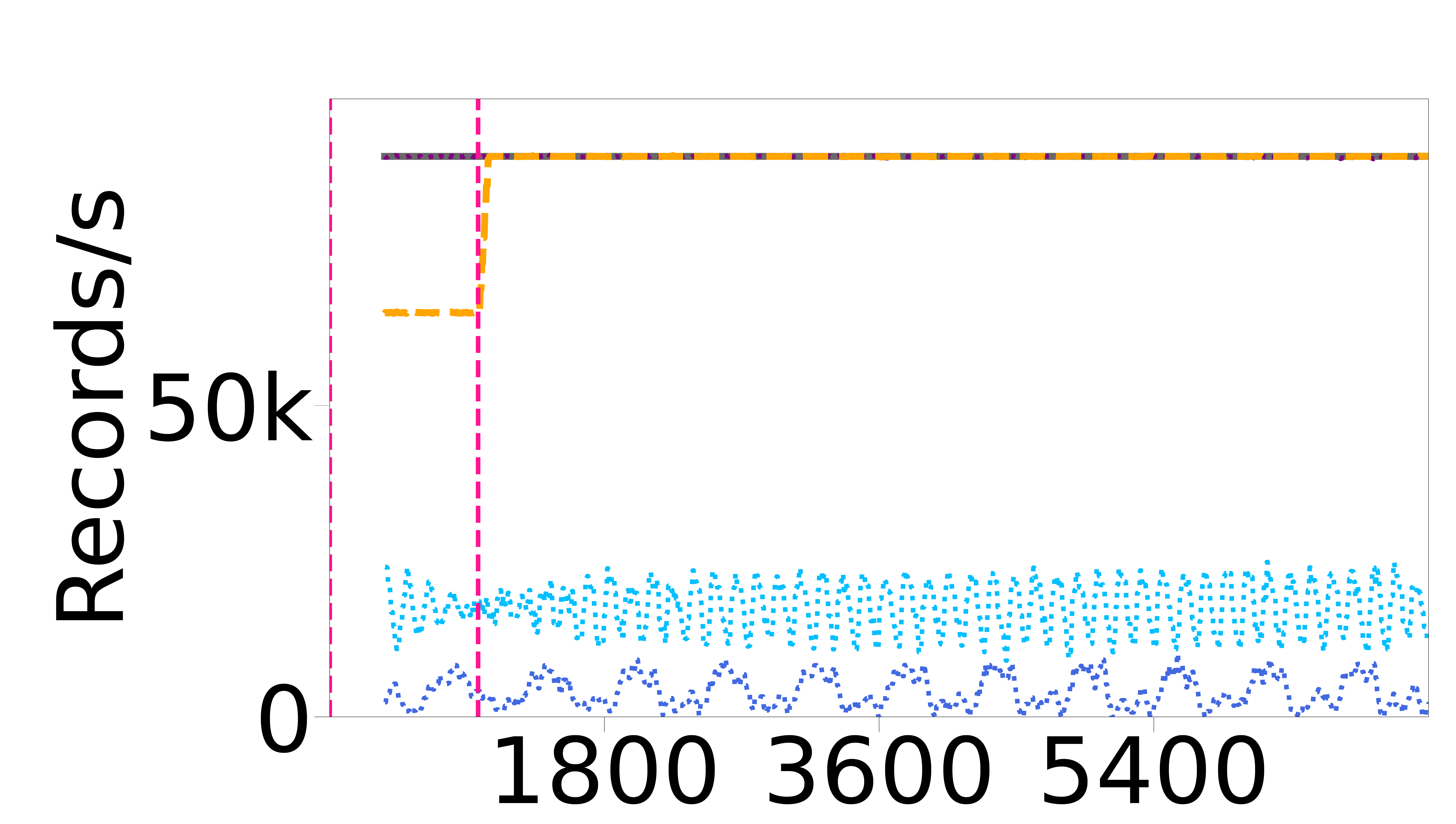}} &
    \adjustbox{valign=m}{\includegraphics[width=3.8cm]{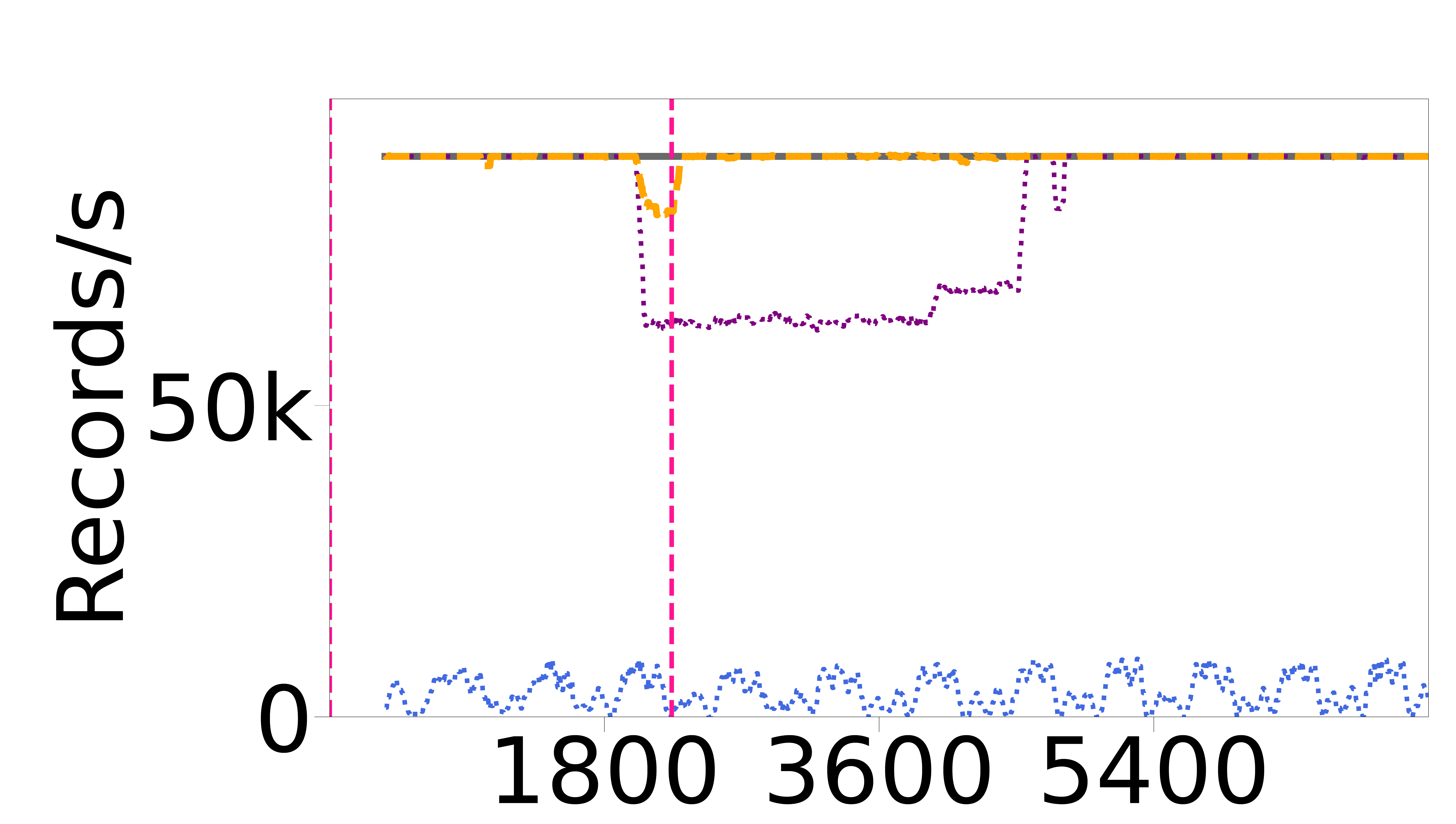}} \\
    \adjustbox{valign=m}{\rotatebox[origin=c]{90}{\textbf{Q3}}} &
    \adjustbox{valign=m}{\includegraphics[width=3.8cm]{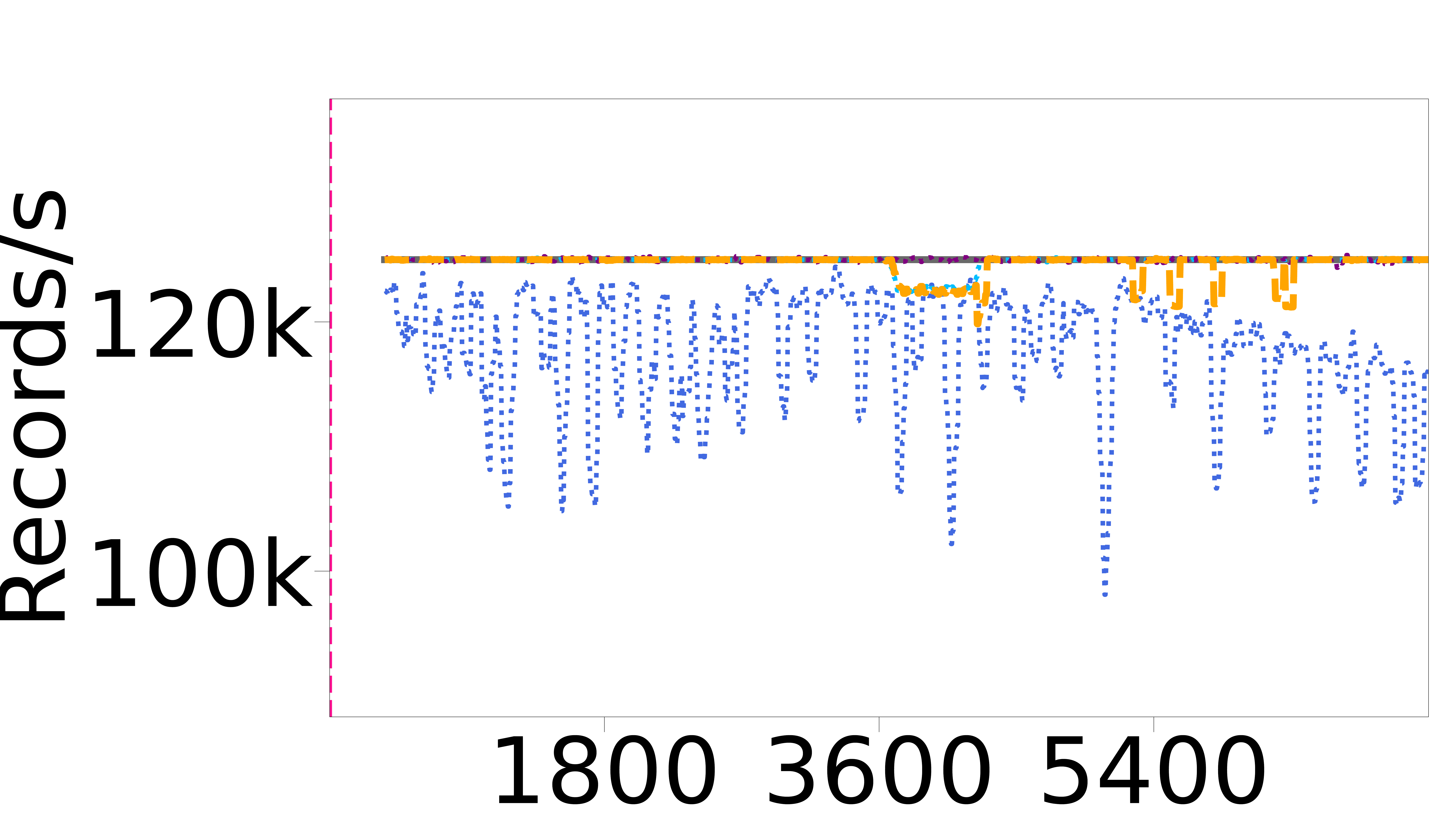}} &
    \adjustbox{valign=m}{\includegraphics[width=3.8cm]{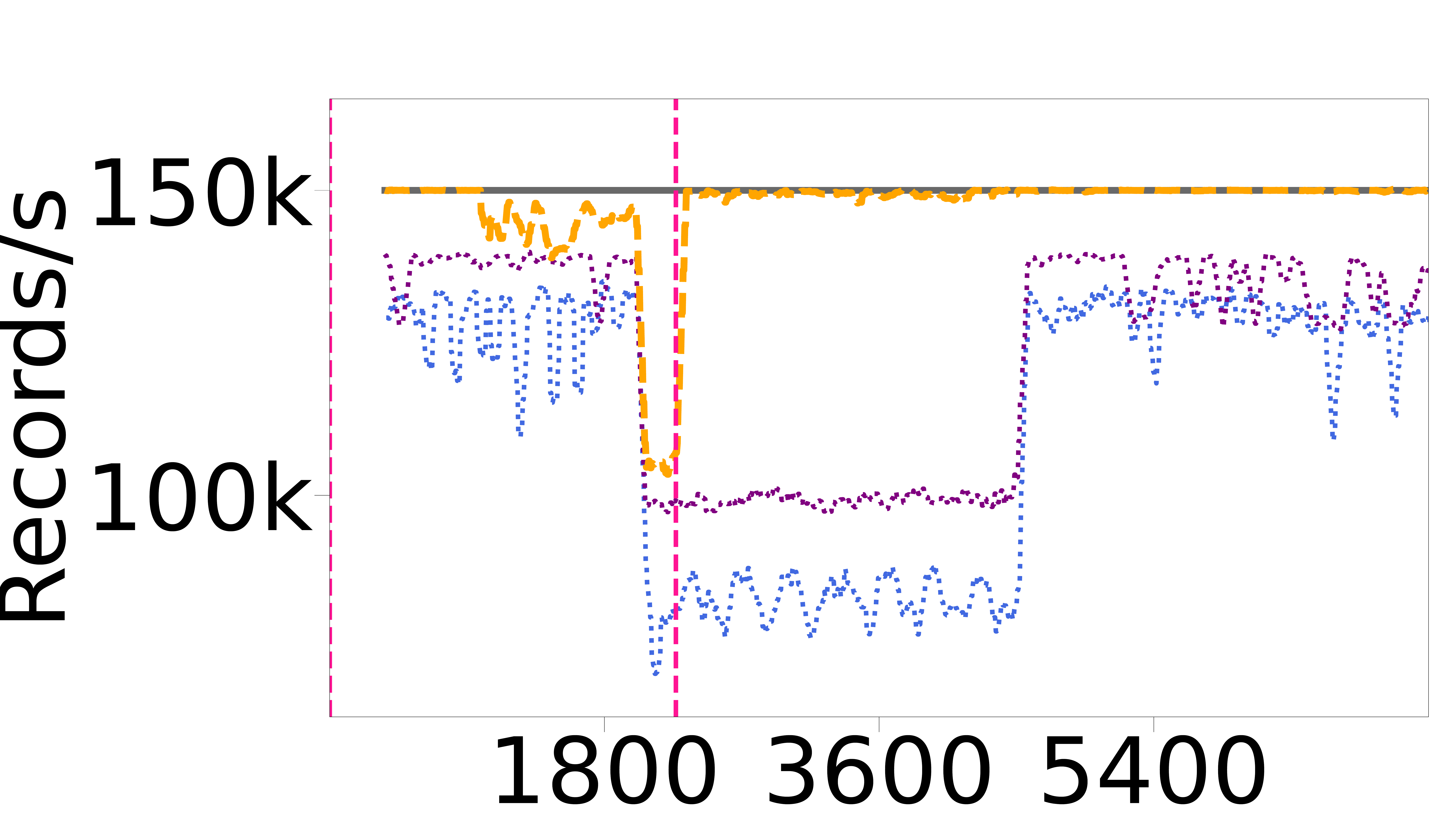}} \\
    \adjustbox{valign=m}{\rotatebox[origin=c]{90}{\textbf{Q4}}} &
    \adjustbox{valign=m}{\includegraphics[width=3.8cm]{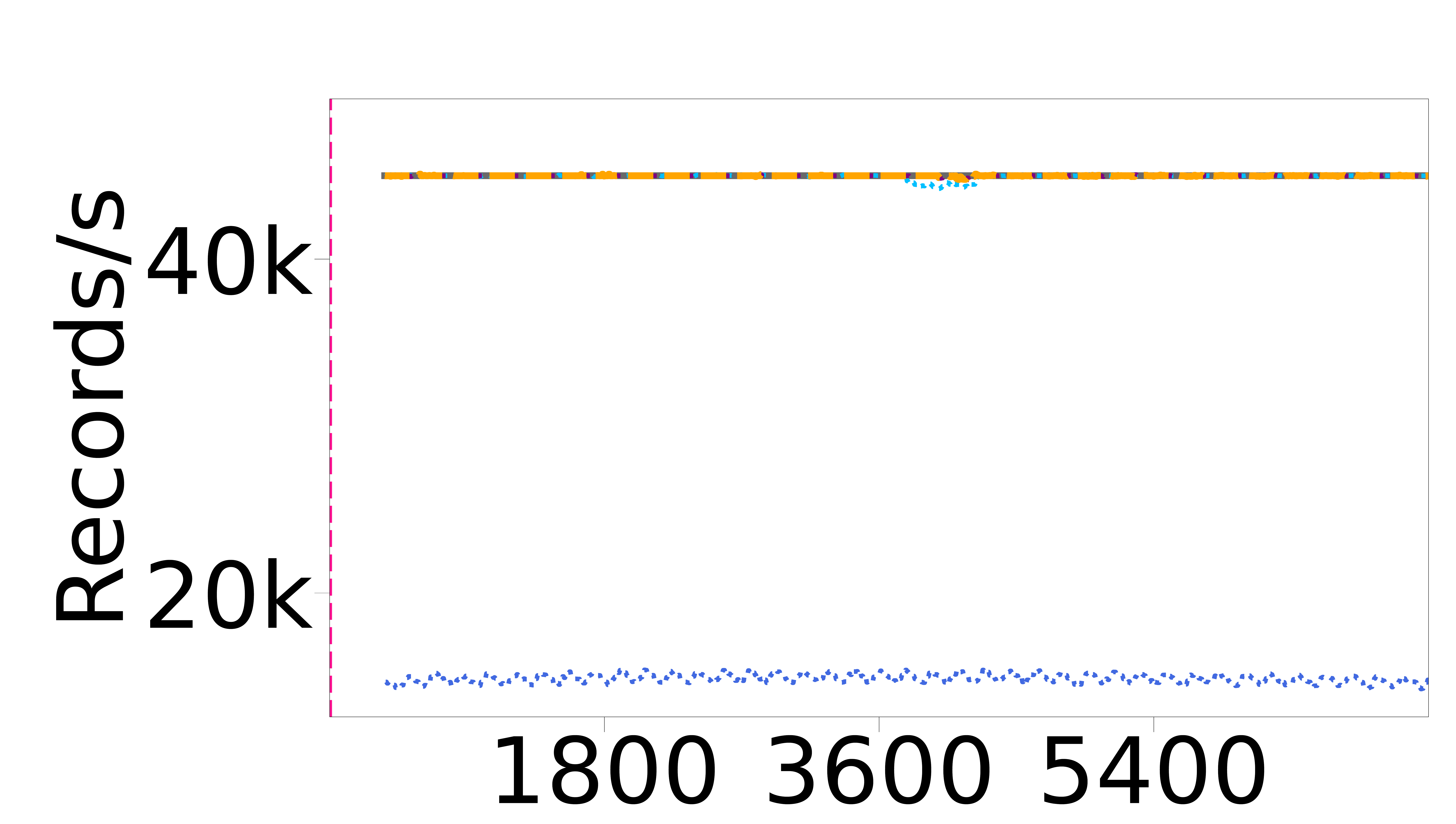}} &
    \adjustbox{valign=m}{\includegraphics[width=3.8cm]{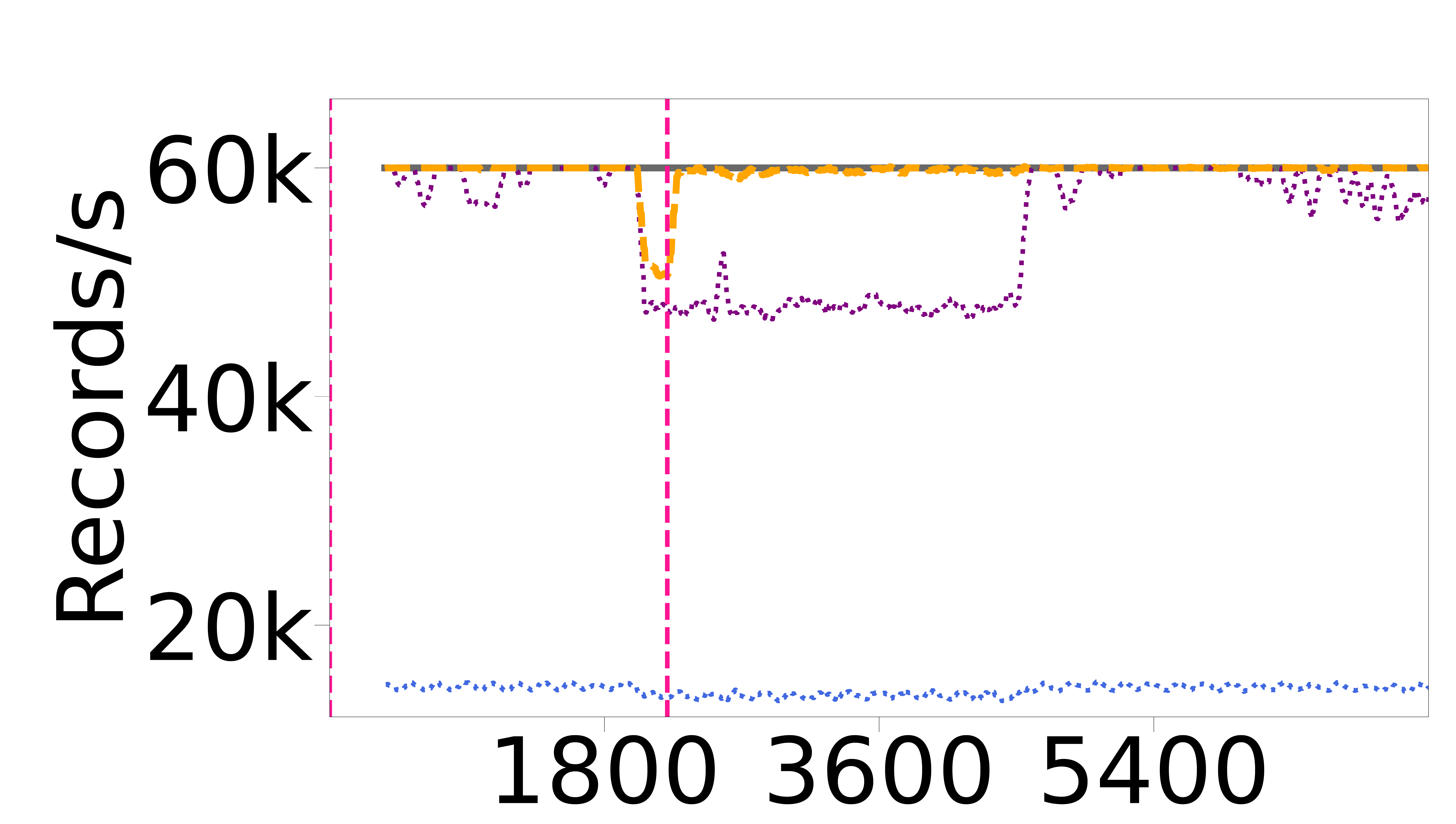}} \\
    \adjustbox{valign=m}{\rotatebox[origin=c]{90}{\textbf{Q5}}} &
    \adjustbox{valign=m}{\includegraphics[width=3.8cm]{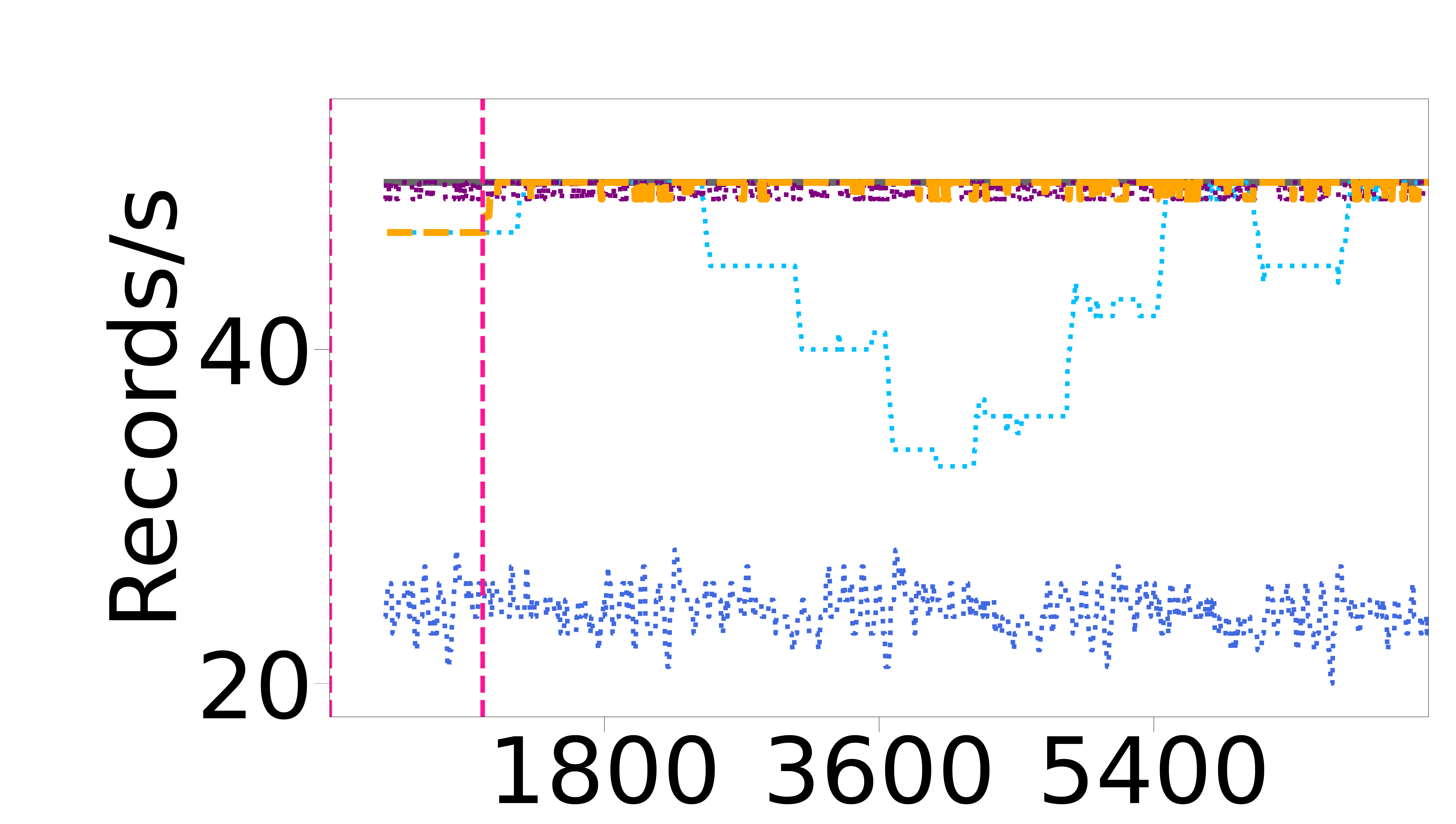}} &
    \adjustbox{valign=m}{\includegraphics[width=3.8cm]{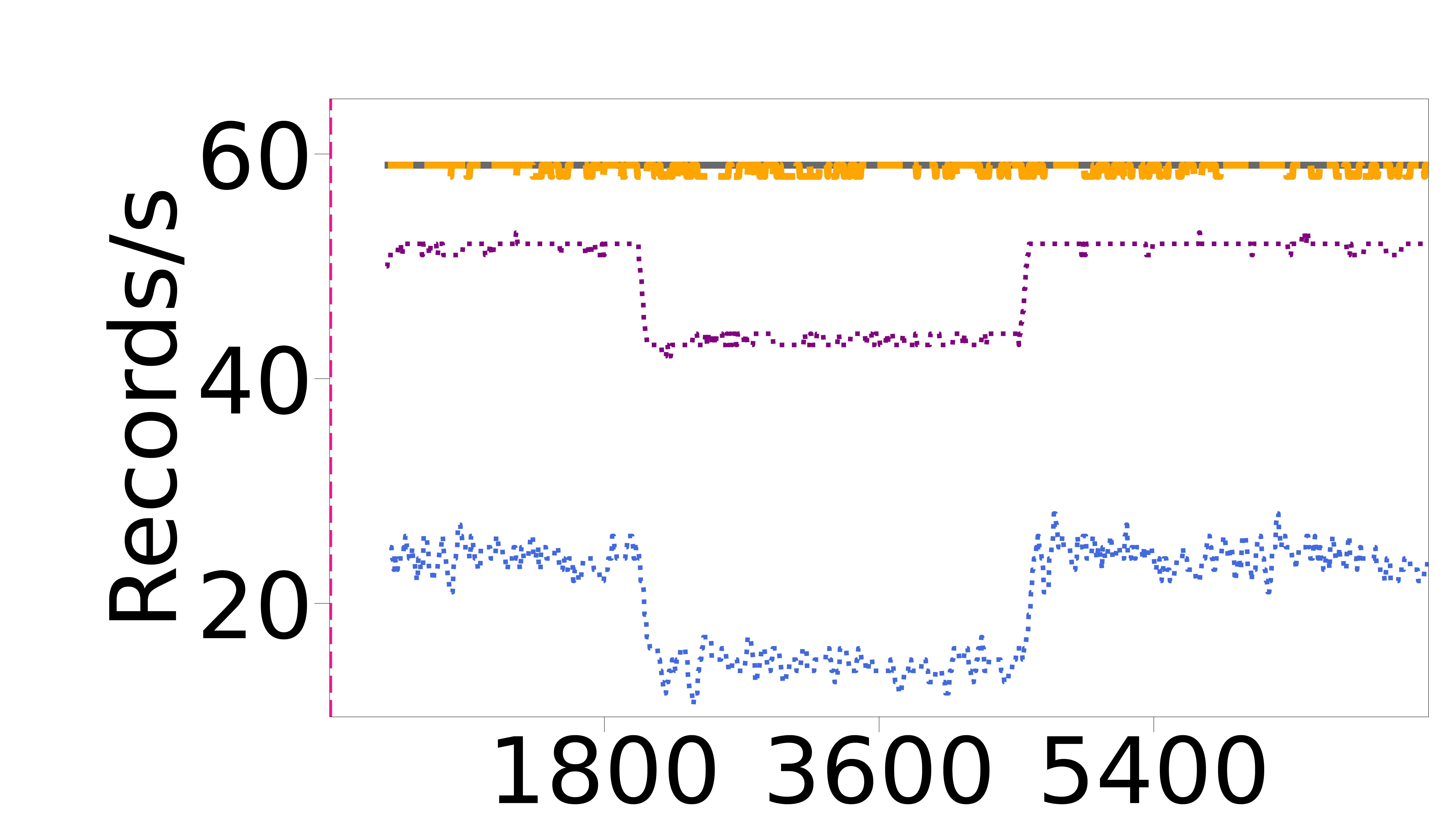}} \\
    \adjustbox{valign=m}{\rotatebox[origin=c]{90}{\parbox{2.5cm}{\centering Operating Conditions}}} &
    \adjustbox{valign=m}{\includegraphics[width=3.8cm]{figures/eval/shift/res_video6nw_0_lower.pdf.pdf}} &
    \adjustbox{valign=m}{\includegraphics[width=3.8cm]{figures/eval/shift/res_video6slow_0_lower.pdf.pdf}} \\
    & \adjustbox{valign=m}{\parbox{3.5cm}{\centering \refstepcounter{subfigure} (a) Network variations }} &
    \adjustbox{valign=m}{\parbox{3.5cm}{\centering \refstepcounter{subfigure} (b) Compute contention }}  \\
\end{tabular}\vspace{-2mm}
\captionof{figure}{Impact of network bandwidth and compute contention - RPi3}
\label{fig:sup_vary_env_rpi3}
\end{minipage}\par\vspace{\textfloatsep}

\noindent\begin{minipage}{\columnwidth}
\centering
\begin{minipage}[b]{0.8\linewidth}
\includegraphics[width=\linewidth]{figures/eval/shift/res_video6nw_0.pdf_legend.pdf}
    \vspace{-0.4cm}
\end{minipage}
\setlength{\tabcolsep}{1pt}
\renewcommand{\arraystretch}{1.5}
\begin{tabular}{ccc}
    \adjustbox{valign=m}{\rotatebox[origin=c]{90}{\textbf{Q1}}} &
    \adjustbox{valign=m}{\includegraphics[width=3.8cm]{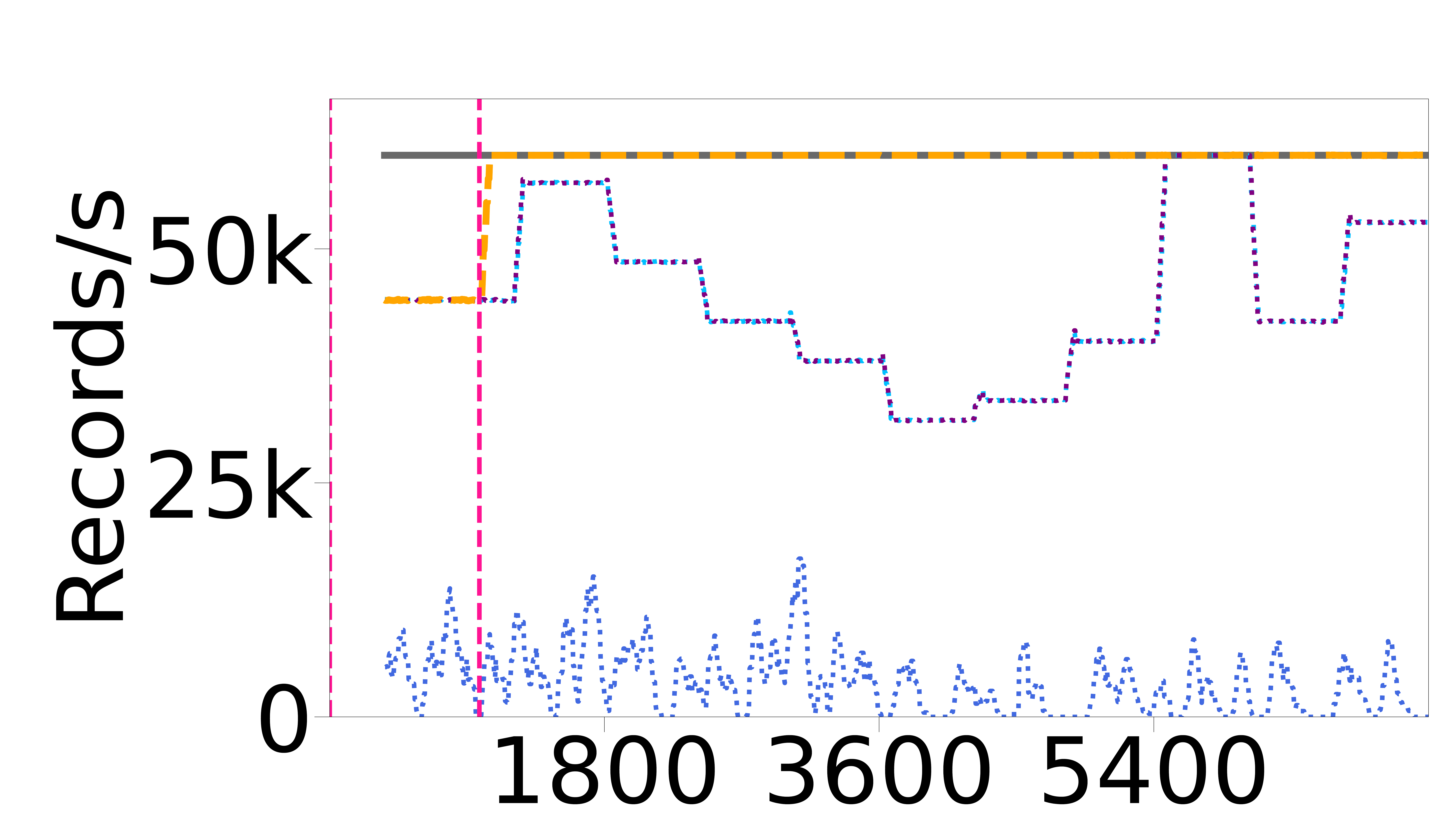}} &
    \adjustbox{valign=m}{\includegraphics[width=3.8cm]{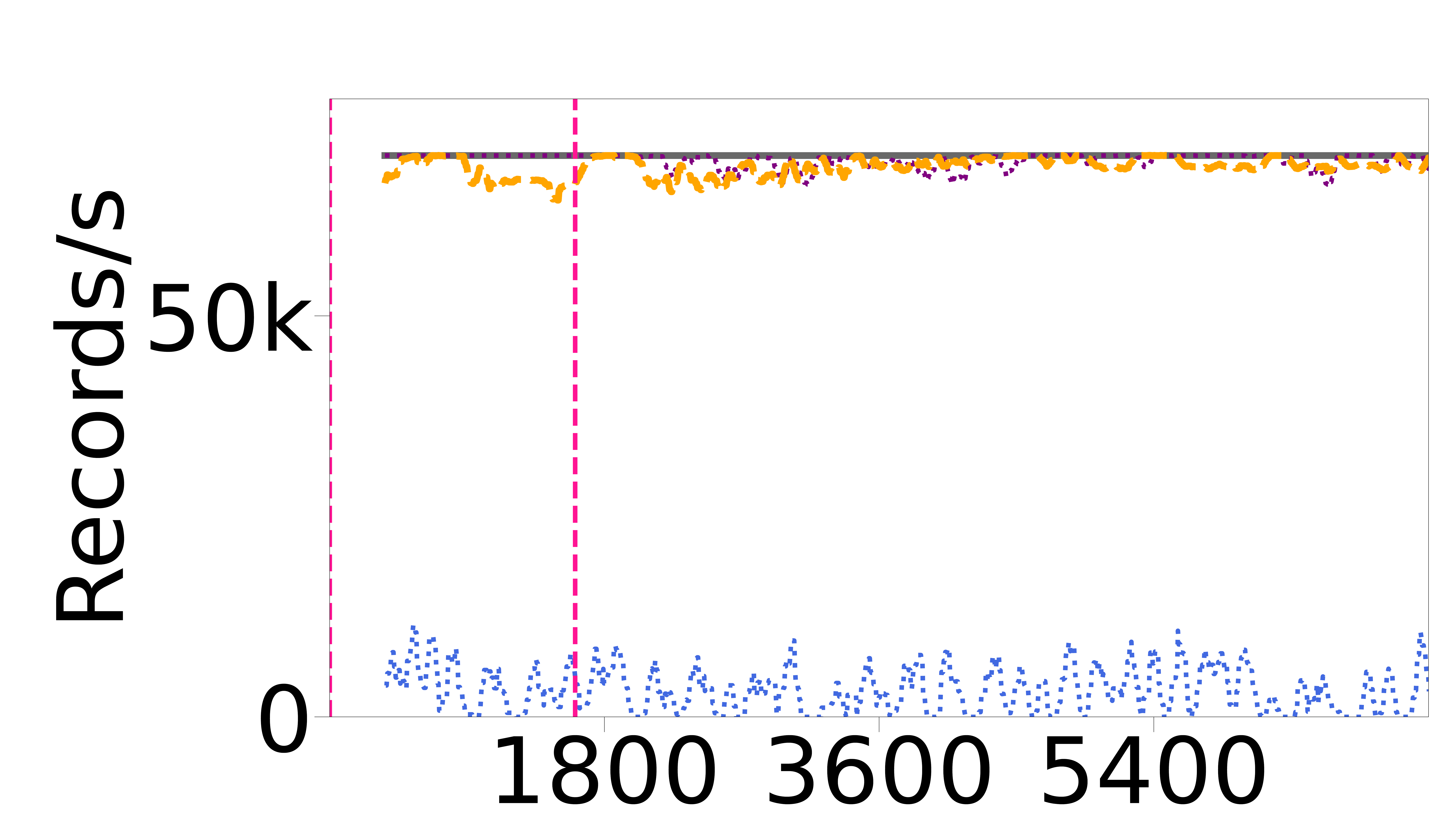}} \\
    \adjustbox{valign=m}{\rotatebox[origin=c]{90}{\textbf{Q2}}} &
    \adjustbox{valign=m}{\includegraphics[width=3.8cm]{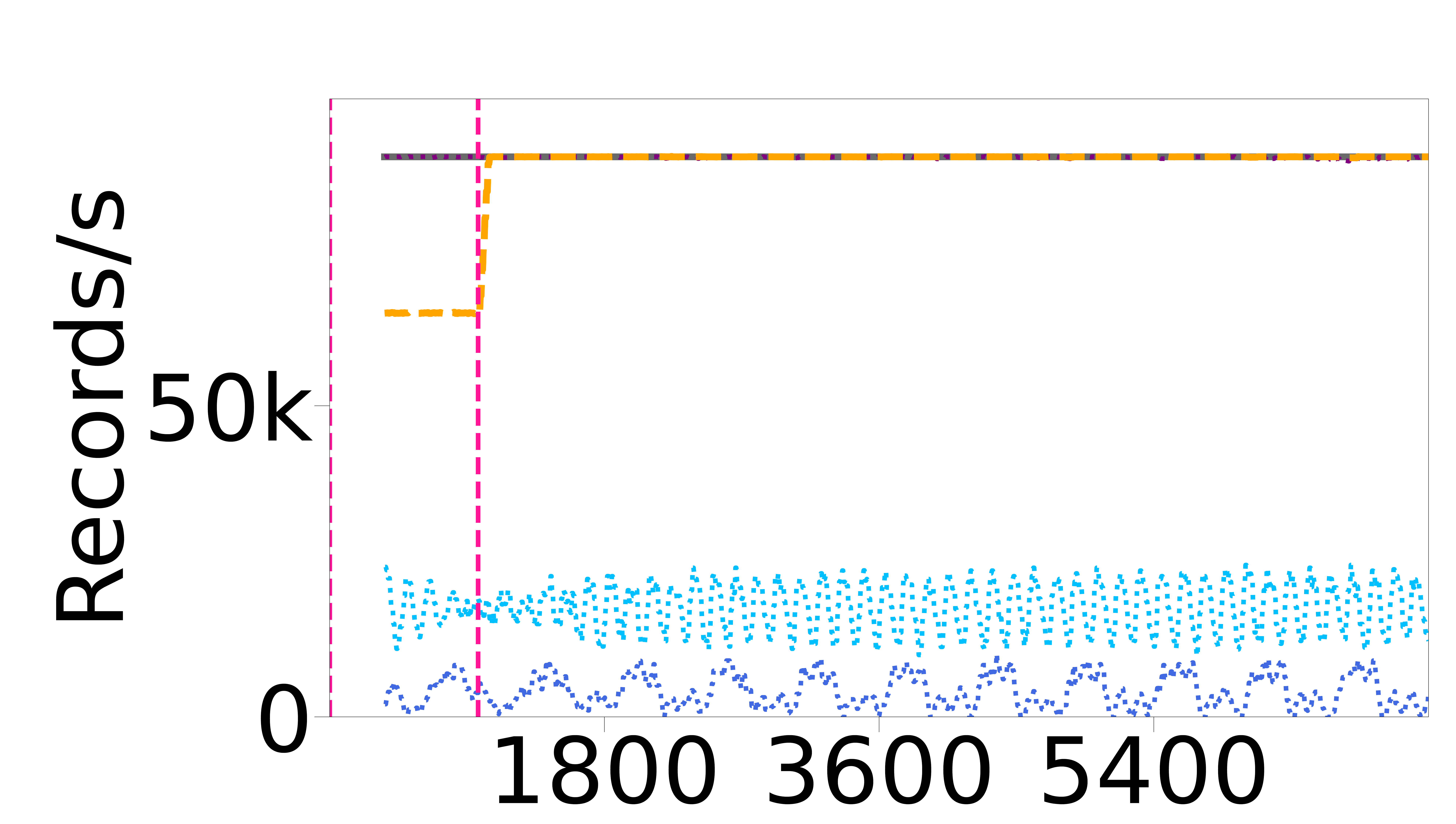}} &
    \adjustbox{valign=m}{\includegraphics[width=3.8cm]{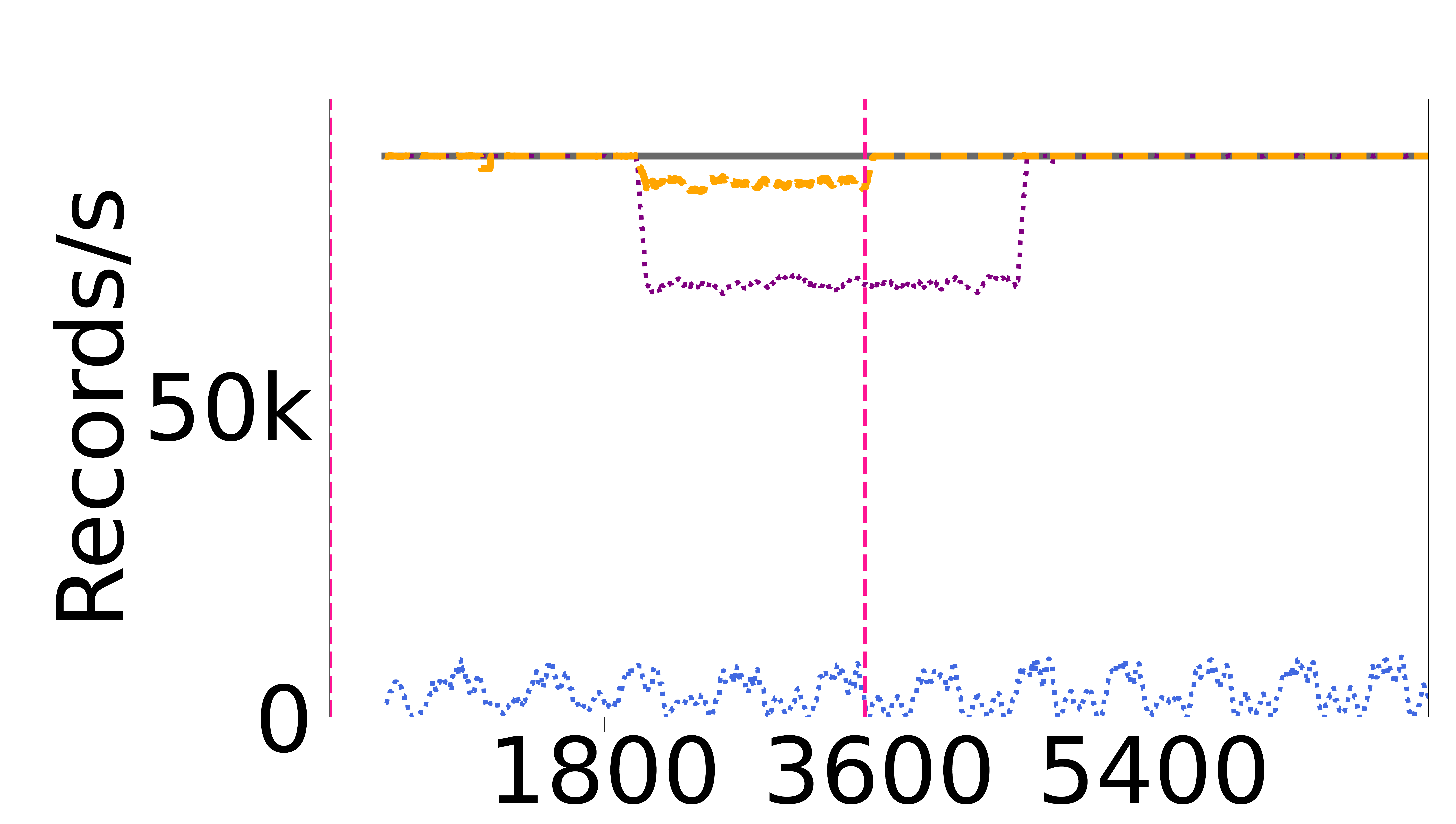}} \\
    \adjustbox{valign=m}{\rotatebox[origin=c]{90}{\textbf{Q3}}} &
    \adjustbox{valign=m}{\includegraphics[width=3.8cm]{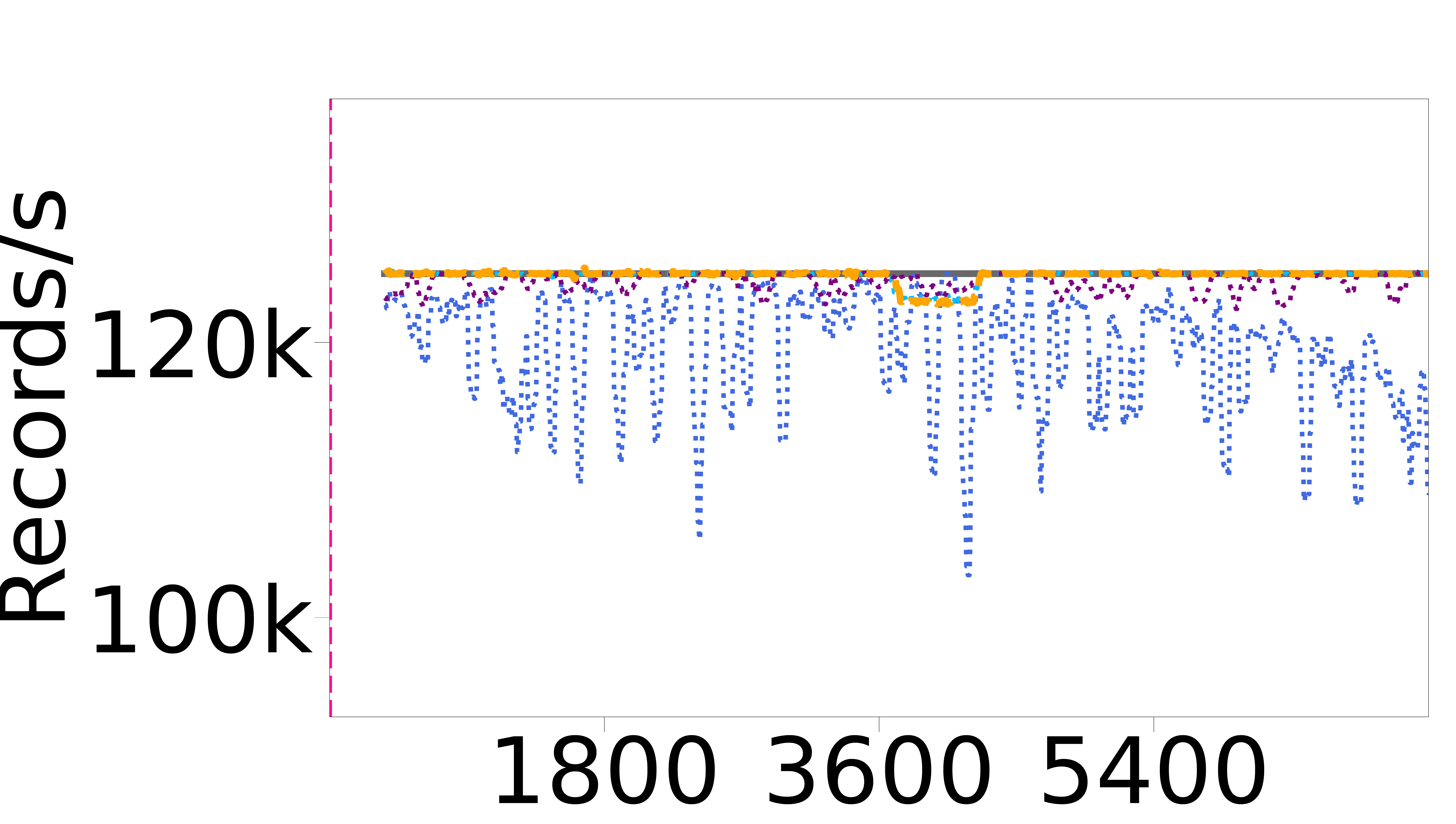}} &
    \adjustbox{valign=m}{\includegraphics[width=3.8cm]{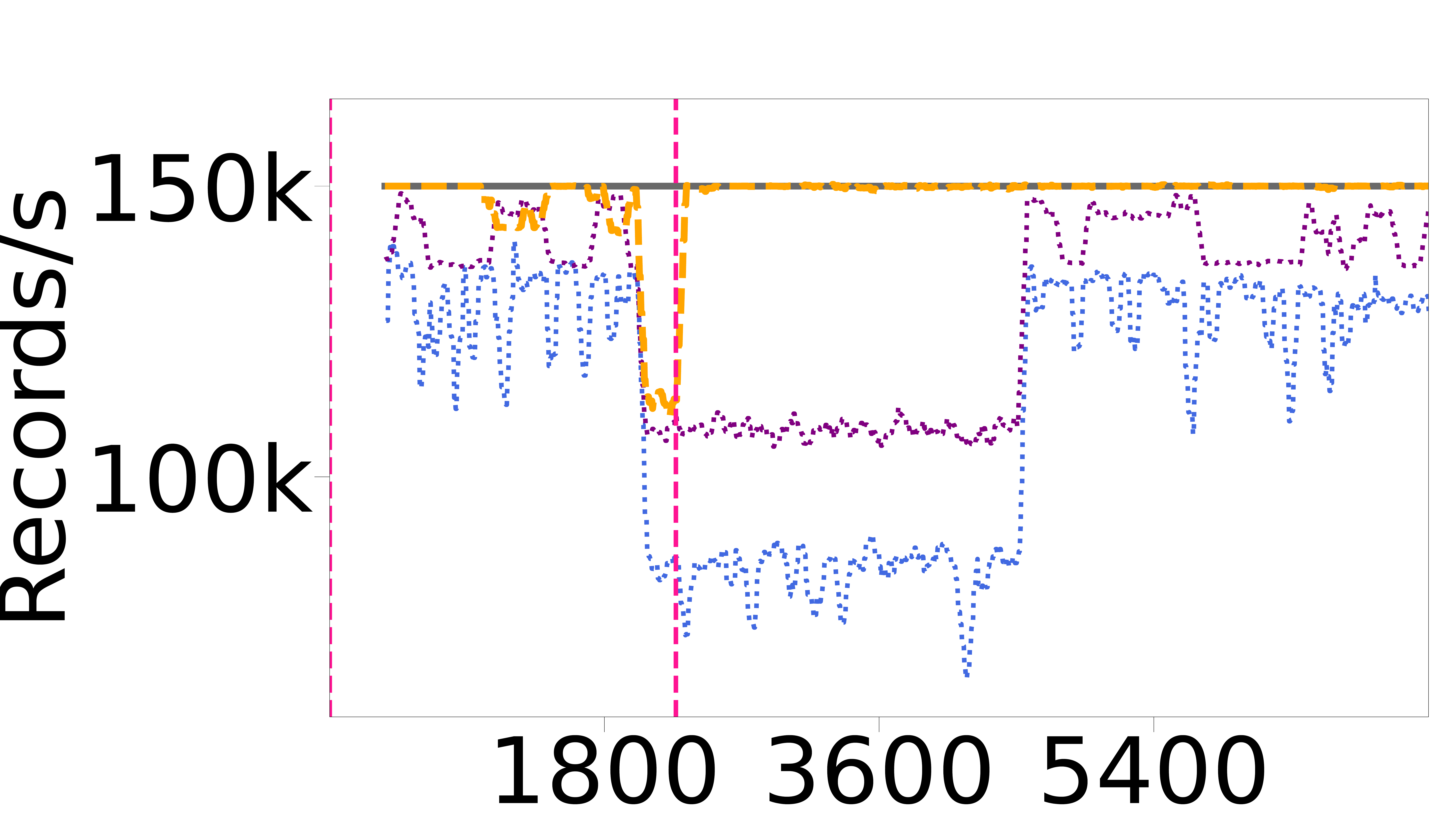}} \\
    \adjustbox{valign=m}{\rotatebox[origin=c]{90}{\textbf{Q4}}} &
    \adjustbox{valign=m}{\includegraphics[width=3.8cm]{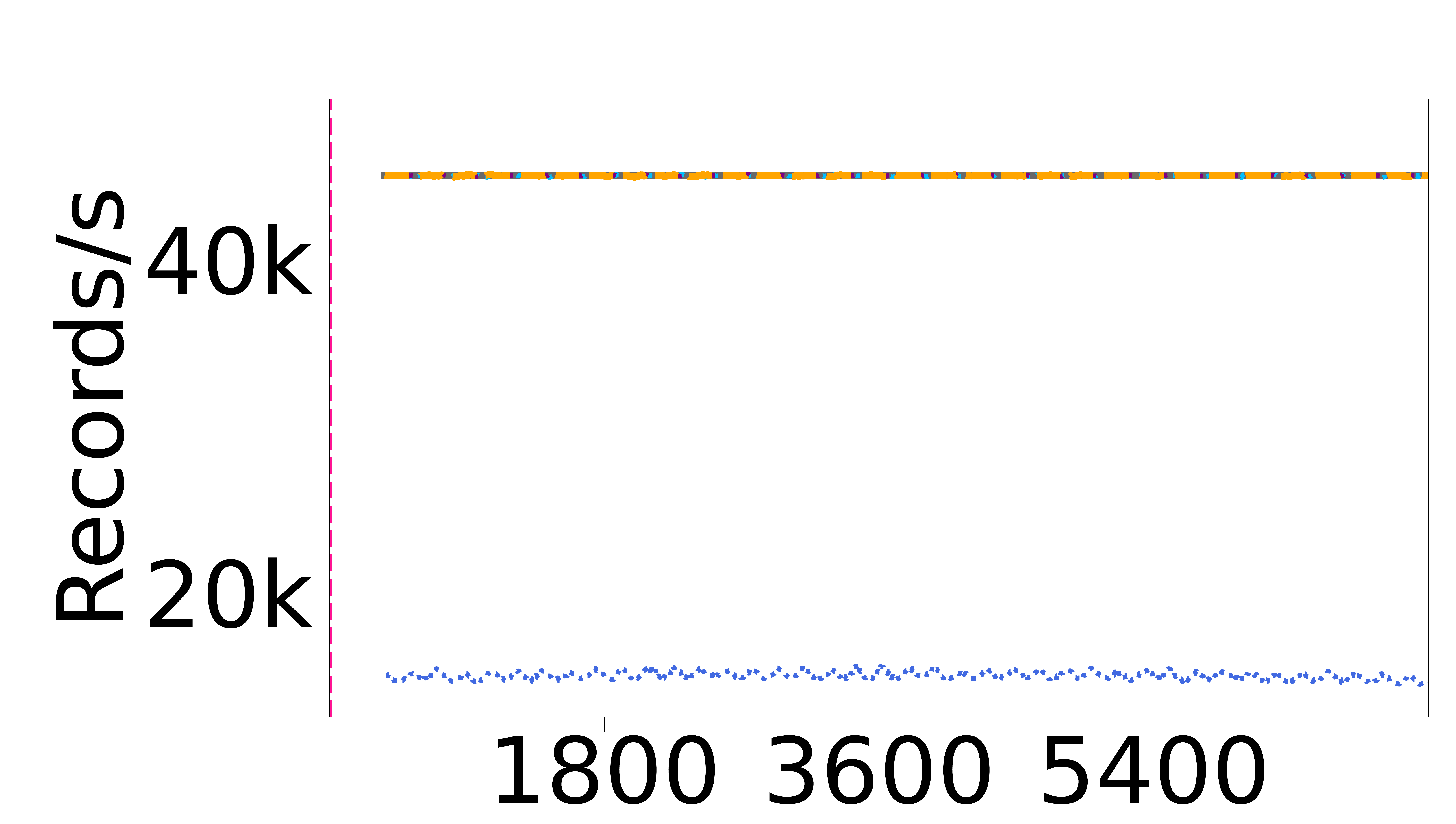}} &
    \adjustbox{valign=m}{\includegraphics[width=3.8cm]{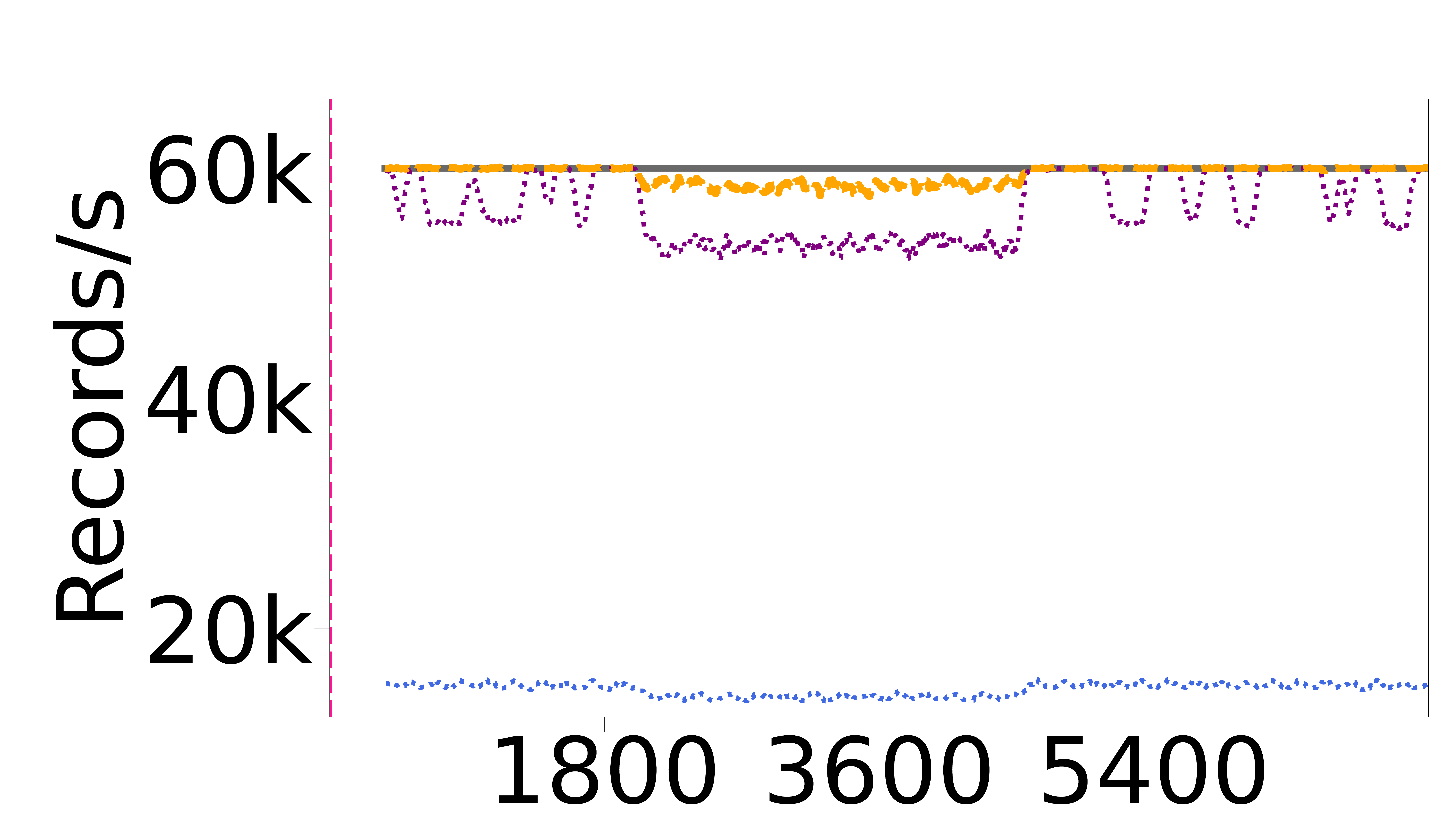}} \\
    \adjustbox{valign=m}{\rotatebox[origin=c]{90}{\textbf{Q5}}} &
    \adjustbox{valign=m}{\includegraphics[width=3.8cm]{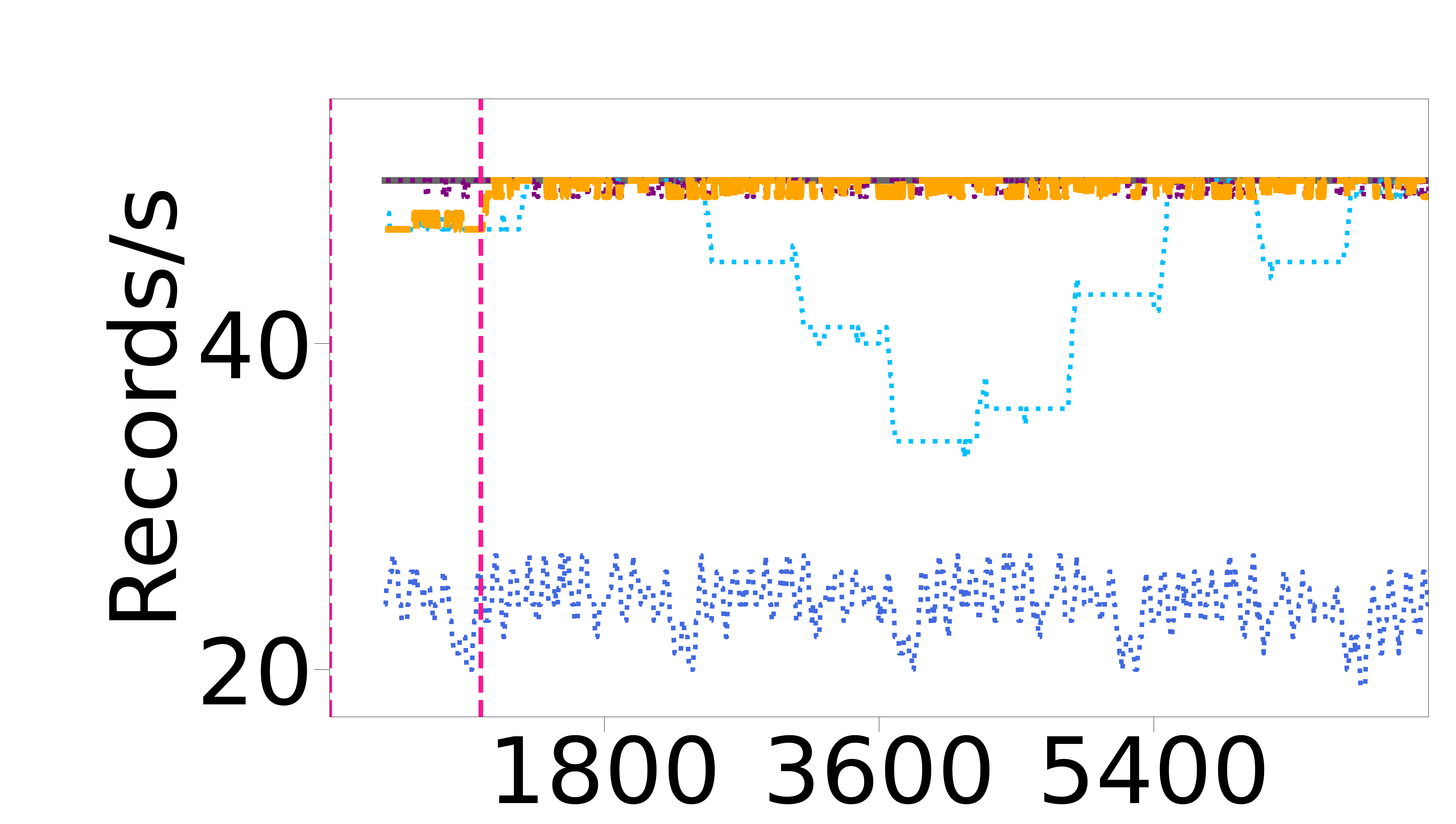}} &
    \adjustbox{valign=m}{\includegraphics[width=3.8cm]{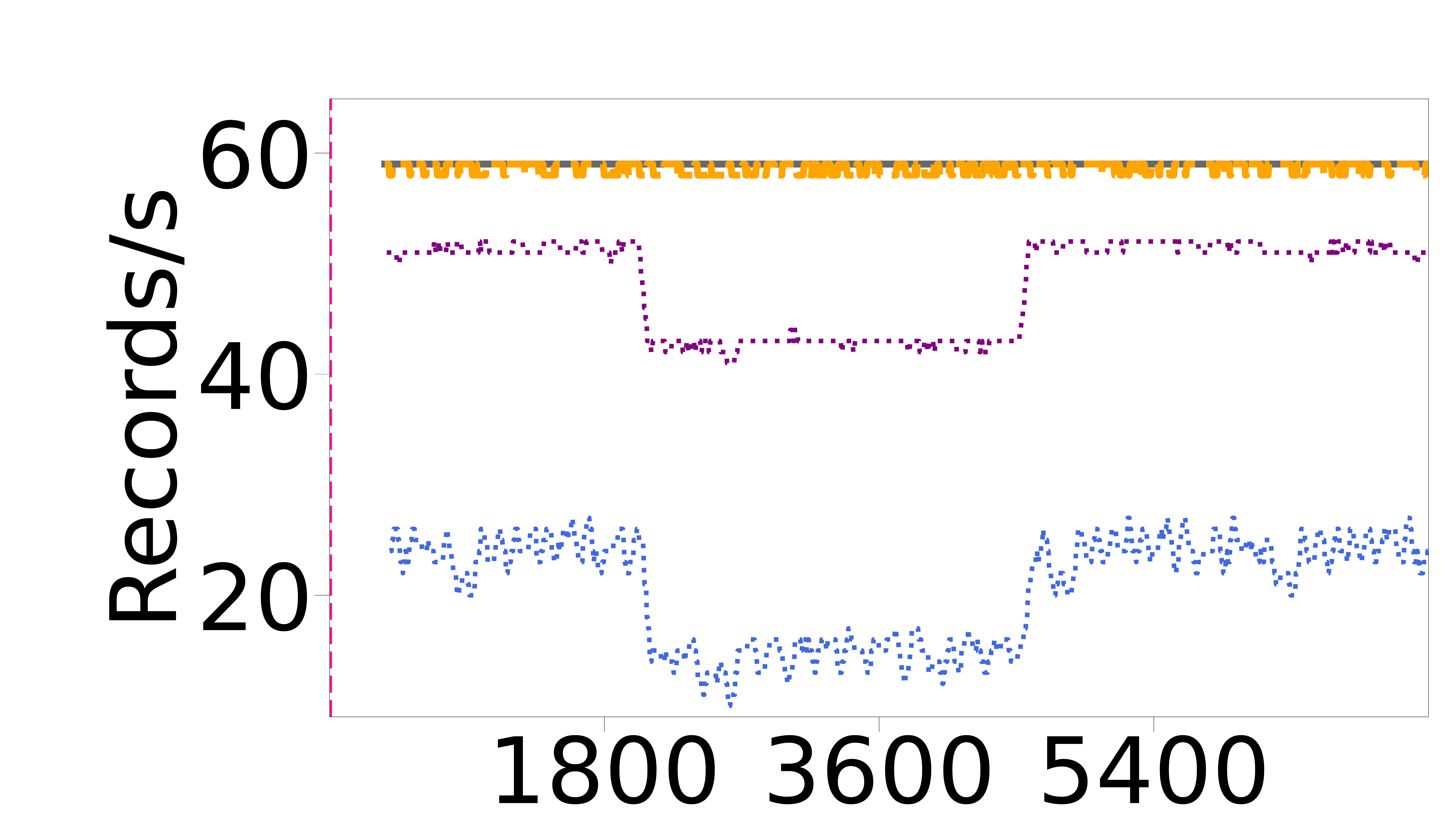}} \\
    \adjustbox{valign=m}{\rotatebox[origin=c]{90}{\parbox{2.5cm}{\centering Operating Conditions}}} &
    \adjustbox{valign=m}{\includegraphics[width=3.8cm]{figures/eval/shift/res_video6nw_0_lower.pdf.pdf}} &
    \adjustbox{valign=m}{\includegraphics[width=3.8cm]{figures/eval/shift/res_video6slow_0_lower.pdf.pdf}} \\
    & \adjustbox{valign=m}{\parbox{3.5cm}{\centering \refstepcounter{subfigure} (a) Network variations }} &
    \adjustbox{valign=m}{\parbox{3.5cm}{\centering \refstepcounter{subfigure} (b) Compute contention }}  \\
\end{tabular}\vspace{-2mm}
\captionof{figure}{Impact of network bandwidth and compute contention - RPi4}
\label{fig:sup_vary_env_rpi4}
\end{minipage}\par\vspace{\textfloatsep}

\end{document}